\documentclass[fleqn,11pt,twoside,openright]{scrbook}

\usepackage{feynmp-pdf}               
\usepackage{amsfonts}                 
\usepackage{amsthm, amssymb, amsmath} 
\usepackage{latexsym}                 
\usepackage{graphicx}                 
\usepackage{array}                    
\usepackage{booktabs}                 
\usepackage{subfigure}                
\usepackage{multind}                  
\usepackage{calc}                     
\usepackage[colorlinks=true,linkcolor=black,urlcolor=blue,citecolor=black]
{hyperref}                            

\newcommand{\ointint}[2]{\mathop{\bigcirc\mkern -30mu} 
  {\int\!\!\!\!\!\int_{#1}^{#2}}}
\newcommand{\ihat}{\hat{\imath}}                       
\newcommand{\jhat}{\hat{\jmath}}
\newcommand{\khat}{\hat{k}}
\newcommand{\rhat}{\hat{r}}

\makeindex{terms}
\newcommand{\term}[1]{{\bf #1}\index{terms}{#1}}

\newcommand{\executeiffilenewer}[3]{\ifnum\pdfstrcmp{\pdffilemoddate{#1}}
  {\pdffilemoddate{#2}}>0{\immediate\write18{#3}}\fi}

\begin{document}

\graphicspath{{Figures/Title/}}

\thispagestyle{empty}

\begin{center}
  \Huge The Hitchhiker's Guide to First Year Physics Labs at UCD\\
  \vspace{2cm}
  \small Philip Ilten\\
  \vspace{4cm}\includegraphics[width=14cm]{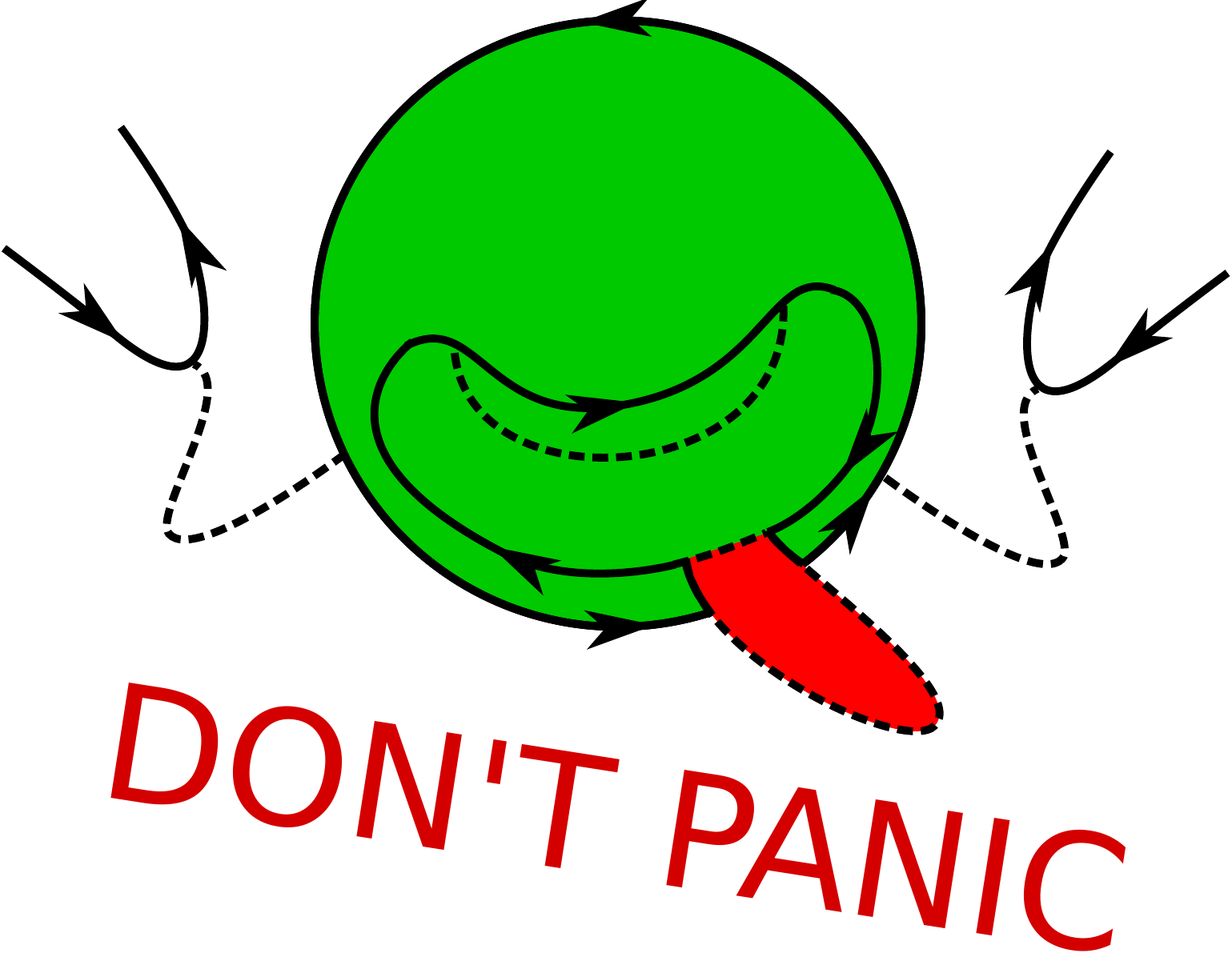}\\
\end{center}

\newpage
\graphicspath{{Figures/Title/}}

\thispagestyle{empty}

\noindent v1.0

\vspace{1cm}

\noindent This book is licensed under the
\href{http://creativecommons.org}{Creative Commons}
\href{http://creativecommons.org/licenses/by/3.0/legalcode}{Attribution
  3.0} license. The contents of this book, in full or in part, can be
copied, distributed, transmitted, or remixed if the material is
attributed to the author.

\vspace{1cm}

\noindent\includegraphics[width=2cm]{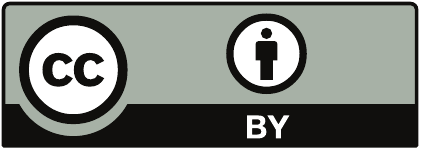} (c) 2010 by Philip Ilten

\newpage
\addcontentsline{toc}{chapter}{Introduction} This book (if this can
actually be called a book) began as a collection of handouts written
by me for the first year undergraduate laboratories at University
College Dublin (UCD), while I was demonstrating during the school year
of $2009-2010$. I realized that perhaps these handouts could be useful
in the future, so in my spare time (primarily during free periods at a
QCD phenomenology conference) I pulled all the source for the handouts
together into this book. That being said, this book requires three
disclaimers.

The first disclaimer is that much of the material should be
understandable by a first year physics student, but some of it can be
very advanced, and perhaps not quite so appropriate. Hopefully I have
managed to point these areas out in the text, so that first years
reading this book don't panic. Of course, it is also possible that I
have written incredibly difficult to understand explanations, in which
case readers of this book should feel free to express their opinions
to me. Of course I might not listen to those opinions, but I would
like to try to make this book better, and the only way to do that is
through revision. I do believe that all the material presented in this
book should be accessible to intrepid first year physics students.

The second disclaimer is that this book might be just as helpful for
demonstrators as it is to undergraduates. As I was demonstrating I
oftentimes wished that concise theoretical refreshers for some of the
topics were available, instead of having to dig through a variety of
text books buried under the dust of neglect. Just because
demonstrators are typically postgraduates does not mean that we
remember every last detail about the wavefunctions of a hydrogen atom,
or BCS theory. Also, just because we are postgraduates doesn't mean
that we necessarily have a good way to explain certain
physics. Hopefully this book will serve both of these needs by
providing a nice overview of the topic, and also present a possible
method of teaching the material.

The third, and hopefully final disclaimer, is that this book contains
mistakes. Despite my best effort, I am certain this book still
contains spelling mistakes, grammar mistakes, and worst of all,
physics mistakes. That means that while reading this book, always
double check my work. If something looks wrong, it could very well be
wrong. If something looks right, it still might be wrong. If you find
a mistake, please let me know, and I will do my best to fix it. The
idea of this book is that it is a growing effort of the community to
provide a useful resource to UCD students. To that end, the source to
this book (written in \LaTeX ~with figures made using Inkscape and
Octave), is available either through the web, or by contacting me.

Enjoy and don't panic.

\vspace{1cm}

\noindent- Philip Ilten

\noindent{\tt philten@lhcb.ucd.ie}
\tableofcontents
\graphicspath{{Figures/Uncertainty/}}

\chapter{Uncertainty}\label{chp:uncertainty}

Uncertainty estimation and propagation is sometimes more of an art
than an exact science, but nevertheless is critical to the scientific
process. Without uncertainty estimation we can never know how well our
theory matches experimental reality. Quite a few books have been
written over the years on uncertainty analysis, but one of the best is
\href{http://www.amazon.com/Introduction-Error-Analysis-Uncertainties-Measurements/dp/093570275X}{\it
  An Introduction to Error Analysis} by John R. Taylor. This book
should be on the book shelf of every physicist, whether an
experimentalist or a theorist.

This chapter focuses on two main areas, the different types of
uncertainty and their estimation, and the propagation of
uncertainty. Before we can delve into either of these areas, we first
need to define uncertainty. Whenever we make a measurement, the
circumstances surrounding that measurement influence the value. For
example, if we measure the value for gravity on earth, $g$, we will
obtain a different value in Dublin than we would in Chicago or
Paris. To indicate this, we must write our central value for $g$
followed by a range in which $g$ might fall. We call this our {\bf
  uncertainty}.
\begin{equation}
  g ~~= \underbrace{9.81}_{
    \begin{array}{c}\mathrm{central~value}\\
      \mathrm{(accuracy)} \end{array}} \pm
  \underbrace{0.21}_{\begin{array}{c}\mathrm{uncertainty}\\
      \mathrm{(precision)} \end{array}}~
  \underbrace{\mathrm{m/s}^2}_{\begin{array}{c}\mathrm{units}\end{array}}
\end{equation}

How close our central value is to the actual value is the {\bf
  accuracy} of the measurement, while the amount of uncertainty
describes the \term{precision} of the measurement. If a measurement has
a very good accuracy, but very low precision, is not very
useful. Conversely, if a measurement has a very poor accuracy, but
very high precision, the measurement is still not useful. The art of
uncertainty is balancing accuracy and precision to provide meaningful
measurements that help confirm or deny theories.

There is one final issue that needs to be discussed regarding the
format of writing uncertainty, and that is \term{significant
  figures}. The number of significant figures on a measurement is the
number of meaningful digits. An uncertainty on a final measurement
should never have more significant figures than the central value, and
should in general have only one or two digits. The number of
significant figures on the central value must always reach the same
precision as the uncertainty. This allows the reader of the
experimental data to quickly see both the accuracy and the precision
of the results.

\section{Types of Uncertainty}

It can be difficult to classify uncertainty; there are many sources,
and oftentimes the cause of uncertainty is unknown. However, we can
broadly classify two types: \term{systematic uncertainty} and {\bf
  random uncertainty.} Systematic uncertainties are types of random
uncertainty, but are caused by calibration within the
experiment. First we will discuss the types of random uncertainty, and
then use these to understand systematic uncertainty.

Random uncertainty can be caused by a variety of sources, but these
can be classified in three general areas. The first is \term{apparatus
  uncertainty}. This uncertainty arises when the design of the
experiment limits the precision of the measurements being made. For
example, consider trying to measure the pressure of a basketball,
using a pressure gauge. On cold days the pressure gauge might leak
because the gaskets have contracted, while on warm days the pressure
gauge might not leak at all. Taking measurements from different days
will yield a range of results caused by this uncertainty. In this
example the measurement from the warm days are more reliable, as there
is no leaking, but in some experimental situations this is not so
readily apparent.

The next type of random uncertainty is \term{inherent
  uncertainty}. Some measured quantities just are different each time
they are measured. This oftentimes is found in the biological
sciences, especially in population analysis. When measuring the
average weight of all deer in Ireland, we don't expect to find each
deer has the same weight, but rather that the weights are spread over
a range of values. This means if we wish to quote the average weight
of a deer in Ireland, we will need to associate an uncertainty with
it. Another place in physics where inherent uncertainty is found is in
particle physics and quantum mechanics. Here, the value of the
measurement isn't decided until the observer actually makes the
measurement.

The final and most common type of random uncertainty is {\bf
  instrumental uncertainty}. Here the instruments being used to take
the measurement have a limited precision, and we can only quote the
value to the precision of the instrument. In many of the labs done in
this course, the primary source of uncertainty will be from
instrumentation. An example of instrumental uncertainty is reading the
voltage from a circuit using a multimeter. The multimeter can only
read to $0.1$ V, and so our uncertainty in the voltage must be $\pm
0.1$ V. Similarly, if we are measuring a length with a ruler, the
instrumentation of the ruler limits are precision to usually a
millimeter, and so we have an uncertainty of $\pm 1$ mm.

Sometimes all three types of random uncertainty can be combined. If we
return to the voltage example, consider what would happen if the
reading on the multimeter fluctuated between $2$ and $3$ V. Now our
apparatus uncertainty (caused by fluctuations in the power supply, or
for some other reason) is larger than our instrumental uncertainty,
and so now our uncertainty is $\pm 0.5$ V instead of $\pm 0.1$ V. The
important point to remember about random uncertainty is that we never
know which direction the uncertainty is. In the voltage example we
could measure a value of $2.5$ V, but not be certain if that voltage
was actually $2$ V, $3$ V, or any values in between.

The second broad category of uncertainty, systematic uncertainty, is
caused by a \term{calibration error} in the experiment. Let us return
to the basketball example, where we are trying to measure the pressure
of a basketball. When we read the pressure gauge while it is not
attached to the basketball we obtain a value of $0$ bar. This is
because the gauge is calibrated incorrectly; we need to account for
atmospheric pressure which is around $1$ bar. If we want to correct
all of our pressure measurements from different days, we must add
something around $1$ bar to all the measurements. The only problem is
that the atmospheric pressure changes from day to day, and so we have
an associated random inherent uncertainty on $1$ bar of around
$\pm0.1$ bar. We call this random uncertainty, from a calibration
adjustment of the data, the systematic uncertainty. When we quote
systematic uncertainty we add another $\pm$ symbol after the random
uncertainty. Let us say that we have measured the basketball pressure
to $1.3\pm0.2$ bar without our calibration adjustment. Now, when we
add on the atmospheric pressure we quote the measurement as
$2.3\pm0.2\pm0.1$ bar.

\section{Propagating Uncertainty}

While we now know how to estimate uncertainty on individual
experimental measurements, we still do not know how to \term{propagate}
the uncertainty. If we have measured the length, $L$, and width, $W$,
of a rectangle of paper, and we have an associated uncertainty on each
measurement, what is the uncertainty on the area, $A$, of the paper?
Finding this from our uncertainties on $L$ and $W$ is called
propagation of uncertainty.

\begin{figure}
  \begin{center}
    \subfigure[Central Value]{
      \executeiffilenewer{Figures/Uncertainty/rectangle.svg}
  {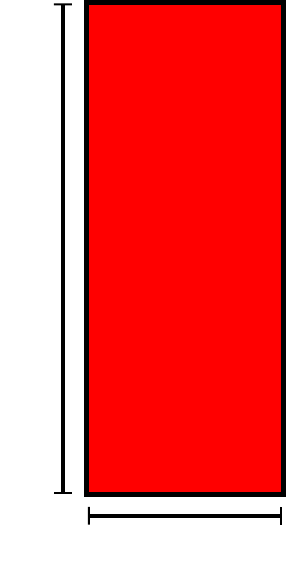}{inkscape-0.48pre1 -z -D --file=Figures/Uncertainty/rectangle.svg 
    --export-pdf=Figures/Uncertainty/rectangle.pdf --export-latex} \def\svgwidth{2.9cm}
  \input{Figures/Uncertainty/rectangleLabel.tex}
      \label{fig:uncertainty:rectangle}}\nolinebreak
    \subfigure[Extremum Method]{
      \executeiffilenewer{Figures/Uncertainty/extremum.svg}
  {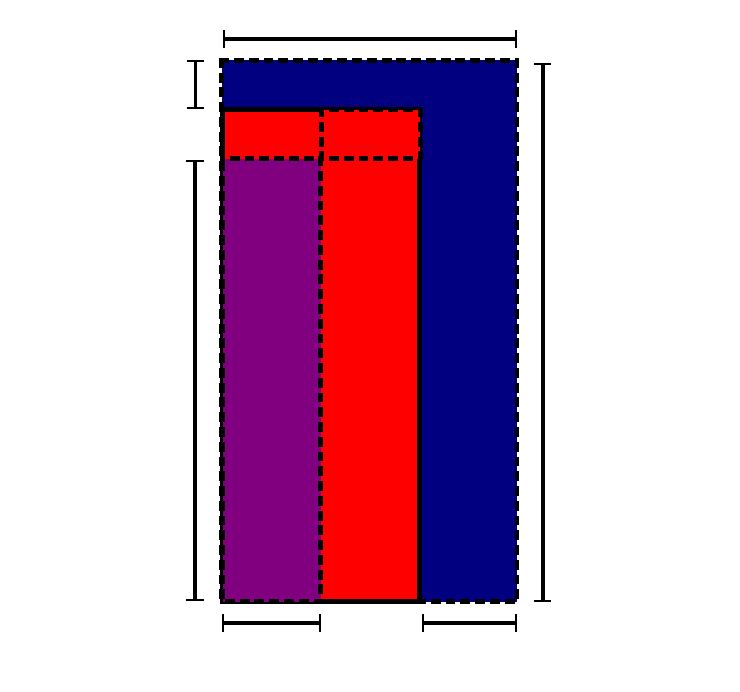}{inkscape-0.48pre1 -z -D --file=Figures/Uncertainty/extremum.svg 
    --export-pdf=Figures/Uncertainty/extremum.pdf --export-latex} \def\svgwidth{7.51cm}
  \input{Figures/Uncertainty/extremumLabel.tex}
      \label{fig:uncertainty:extremum}} 
    \subfigure[Relative Method]{
      \executeiffilenewer{Figures/Uncertainty/relative.svg}
  {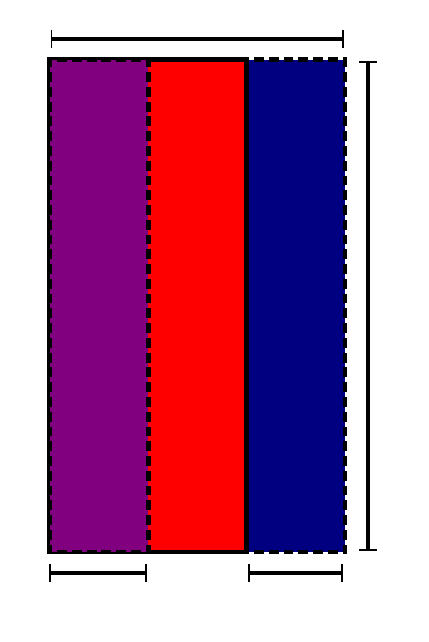}{inkscape-0.48pre1 -z -D --file=Figures/Uncertainty/relative.svg 
    --export-pdf=Figures/Uncertainty/relative.pdf --export-latex} \def\svgwidth{4.36cm}
  \input{Figures/Uncertainty/relativeLabel.tex}
      \label{fig:uncertainty:relative}}
    \caption{Geometric representations for the different methods of
      propagating uncertainty for the area of a
      rectangle.\label{fig:uncertainty:uncertainty}}
  \end{center}
\end{figure}

In Figure \ref{fig:uncertainty:uncertainty} we graphically show how we
calculate the uncertainty. In this example we have measured
$L=5.0\pm0.5$ cm, and $W=2.0\pm1.0$ cm (perhaps we used a very poor
ruler for measuring the width). As a matter of notation we often
represent uncertainty with the the Greek letter $\delta$ (a lower case
delta), or sometimes $\sigma$ (the Greek letter sigma).\footnote{We
  use $\sigma$ when referring to uncertainty taken from a large number
  of measurements as it refers to the standard deviation, which is
  explained more later on in this chapter.} We calculate our central
value for $A$ by,
\begin{equation}
  A = L\times W
  \label{equ:uncertainty:area}
\end{equation}
which is the formula for the area of a rectangle, shown in Figure
\ref{fig:uncertainty:rectangle}. At first glance, propagating our
uncertainty to $A$ might seem simple; we can just multiply the
uncertainties, just as we multiplied the central values to obtain $A$.

Figure \ref{fig:uncertainty:extremum} shows why this does not work. The red
rectangle is still the central value for $A$, $10.0$
$\mathrm{cm^2}$. The dashed purple rectangle is the smallest possible
value for $A$ we can obtain within the range of uncertainty for $L$
and $W$, while the dashed blue rectangle is the largest possible value
for $A$ we can obtain. If we just multiply $\delta_L$ and $\delta_W$
we obtain the small dashed rectangle in red. This uncertainty is
clearly much too small.

The possible values for the area of the rectangle range from the
purple rectangle to the blue rectangle. These are the extremum
(maximum and minimum) of the area, and by taking their difference and
dividing by two, we can find the average range about the central value
for $A$.
\begin{equation}
  \begin{aligned}
    \delta_A &= \frac{A_\mathrm{max}-A_\mathrm{min}}{2}\\
    &= \frac{(L+\delta_L)(W+\delta_W)-(L-\delta_L)(W-\delta_W)}{2}\\
    &= \frac{\delta_L\delta_W+2W\delta_L+2L\delta_L}{2}\\
  \end{aligned}
\end{equation}

Plugging in our values for $L$, $W$, $\delta_L$, and $\delta_W$, we
arrive at $\delta_A = 6.2$ $\mathrm{cm}^2$, a much larger uncertainty
than the incorrect $\delta_L\times\delta_W = 0.50$
$\mathrm{cm}^2$. This gives us a value for $A$ of $10.0\pm6.2$
$\mathrm{cm}^2$; the actual area for the rectangle could range
anywhere from around $4$ to $16$ $\mathrm{cm}^2$. Notice how the upper
and lower bounds match nicely with the blue and purple rectangles
respectively.

The uncertainty propagation method outlined above can be applied to
any formula and is called the \term{extremum uncertainty} method. This
is because to find the uncertainty on the calculated quantities, we
find the largest value for the quantity possible, and the smallest
value possible, take the difference, and divide by two. The tricky
part of this method is finding the maximum and the minimum values for
the calculated quantity. In the example above it is easy to visualize,
as we see the rectangle is the largest when we add uncertainty onto
$L$ and $W$, and smallest when we subtract the uncertainty. But what
happens when we look at a more complicated function? Let us consider
the following function $f(x,y,z)$ which is dependent upon the measured
quantities $x$, $y$, and $z$ (analogous to $L$ and $W$ in the example
above).
\begin{equation}
  f = x^2-2y+\frac{1}{z}
\end{equation}

Now we need to see what happens to $f$ when we change either $x$, $y$,
or $z$. For $x$ we see that $f$ is maximized whenever $x$ is as large
as possible (both positive and negative) and minimized for $x$ near
zero. For $y$, we see $f$ is maximized for large negative $y$ and
minimized for large positive $y$. For $z$, the behavior of $f$ is even
trickier. Large positive and negative $z$ make $1/z$ very
close to zero. Very small positive values of $z$ make $1/z$ a
very large positive number, and very small negative values of $z$ make
$1/z$ a very large negative number. From this behavior we can
see that to maximize $f$, we want $z$ to be as close to zero as
possible while still being positive. To minimize $f$, we want $z$ to
be as close to zero as possible while being negative.

From the example above it is readily apparent that the extremum method
for propagating uncertainty can quickly become very complicated, and
also a little tedious. Luckily, in some cases we can bypass the
extremum method and propagate the uncertainty using \term{relative
  uncertainty}. Relative uncertainty is the uncertainty on a quantity
$\delta_x$ divided by that measurement $x$, i.e. $\delta_x/x$. If we
return to the rectangle example of Figure \ref{fig:uncertainty:uncertainty}, we
see that the relative uncertainty on W is $0.50$ (or $50\%$), and the
relative uncertainty on $L$ is $0.10$ (or $10\%$).

If one relative uncertainty for a measurement is much larger than the
relative uncertainties for the other measurements, we can just focus
on the largest relative uncertainty, and assume that our calculated
quantity will have approximately the same relative uncertainty. In
this example the relative uncertainty on $W$ is much larger than the
relative uncertainty on $L$, and so we assume that $A$ will have a
relative uncertainty of approximately $0.50$.
\begin{equation}
  \frac{\delta_A}{A} = \frac{\delta_W}{W} ~~~\Rightarrow~~~ \delta_A =
  A\left(\frac{\delta_W}{W}\right)
\end{equation}
Looking at Figure \ref{fig:uncertainty:relative} we can see this method in
action. Now we have $A=10.0\pm5.0$ $\mathrm{cm}^2$ which is very close
to the extremum method which supplied an uncertainty of $10.0\pm6.2$
$\mathrm{cm}^2$. The uncertainty from the extremum method is larger of
course, as with the relative method we are ignoring our uncertainty on
$L$. We can also apply this to any general function $f$ dependent upon
multiple measurements, but with the main source of relative
uncertainty from the variable $x$.
\begin{equation}
  \delta_f = f\left(\frac{\delta_x}{x}\right)
\end{equation}
Notice that this method is much faster (and simpler) than the extremum
method, but is only valid when the relative uncertainty on $x$ is much
larger than the relative uncertainty on the other variables.

There is one final method for propagating uncertainty, the \term{normal
  uncertainty} method, which is the most common method used in
physics. However, understanding this method can be a bit challenging,
and understanding when and when not to use this method is not
trivial. The remainder of this chapter is devoted to attempting to
explain the motivation behind this method, but here, a brief example
will be given just to demonstrate the method.

If we have a function $f(x_1,x_2,\ldots,x_n)$ dependent upon $x_1$ up
to $x_n$ independent measurements (or more generally, measurements
$x_i$), we can propagate the uncertainty as follows.
\begin{equation}
  \sigma_f^2 = \left(\frac{\partial f}{\partial
      x_1}\right)^2\sigma_1^2 + \left(\frac{\partial f}{\partial
      x_2}\right)^2\sigma_2^2 + \cdots + \left(\frac{\partial f}{\partial
      x_n}\right)^2\sigma_n^2
  \label{equ:uncertainty:normalMethod}
\end{equation}
Here we have replaced our traditional $\delta$ letter for uncertainty
with the letter $\sigma$ where $\sigma_i$ corresponds to the
uncertainty associated with measurement $x_i$. The symbol $\partial$
denotes a partial derivative; this is where we keep all other
variables in $f$ constant, and just perform the derivative with
respect to $x_i$.

We can now use this method of propagation on the rectangle example of
Figure \ref{fig:uncertainty:uncertainty}. First we must apply Equation
\ref{equ:uncertainty:normalMethod} to Equation \ref{equ:uncertainty:area}.
\begin{equation}
  \begin{aligned}
    \sigma_A^2 &= \left(\frac{\partial A(L,W)}{\partial
        L}\right)^2\sigma_L^2 +
    \left(\frac{A(L,W)}{\partial W}\right)^2\sigma_L^2\\
    &= L^2\sigma_W^2+W^2\sigma_L^2\\
  \end{aligned}
\end{equation}
Plugging in the uncertainties we arrive at a value of $A=10.0\pm5.0$
$\mathrm{cm}^2$, the same uncertainty we arrived at using the relative
uncertainty method!

\section{Probability Density Functions}

In the previous sections we have seen what uncertainty is, and how it
is possible to propagate it using two different methods, the extremum
method, and the relative uncertainty method. However, the two methods
described in detail above are very qualitative and do not work well
when a single value is repeatedly measured. As an example, let us
consider a race car driving around a track, and trying to measure the
velocity each time it passes. The race car driver is trying to keep
the velocity of the car as constant as possible, but of course this is
very difficult. Therefore, we expect that the velocities we record for
each lap will be similar, but not exactly the same. If we make a
histogram\footnote{A histogram is just a bar graph that plots the
  number of events per value range of that variable.}  of these values
and divide the histogram by the number of measurements we have made,
we will have created a \term{probability density function} for the
velocity of the race car.

This histogram describes the probability of measuring the velocity of
the car to be within a certain range of velocities. For example we can
find the area under the entire histogram, which should return a value
of one. This tells us that we always expect to measure a velocity
within the ranges we have previously measured. All probability density
functions when integrated from negative infinity to infinity should
yield an area of one; when the area under any curve is one we call the
curve \term{normalized}. Similarly, if we want to find the probability
of measuring a certain velocity instead of a range of velocities, we
see the probability is zero. This is because we do not expect to be
able to measure an arbitrarily precise value for the velocity of the
car.

Probability density functions can be described by a variety of
properties, but two of the most important are the \term{mean} and the
\term{variance}. The mean for a probability density function is exactly
the same as the mean average taught in grade school, and is often
denoted by the Greek letter $\mu$ (spelled mu).

Before we can understand exactly what the variance of a probability
density function is, we need to introduce the \term{expectation value}
of a probability density function. If we have a probability density
function $PDF(x)$ dependent upon the variable $x$ we define the
expectation value for a function of $x$, $f(x)$, as
\begin{equation}
  E\left[f(x)\right] = \int_{-\infty}^\infty f(x)PDF(x)\,dx
  \label{equ:uncertainty:expectationValue}
\end{equation}
where we just integrate the instantaneous probability of $x$,
$PDF(x)$, times the function of $x$ for which we are trying to find
the expectation value, $f(x)$. The mean of a probability density
function is just the expectation value of the function $f(x) = x$,
$E\left[x\right]$. In other words, we expect that if we measured the
variable $x$ many times, we would find an average value of
$E\left[x\right]$.

The variance of a probability density function dependent on the
variable $x$ is,
\begin{equation}
  \sigma^2(x) = E\left[\left(x-\mu\right)^2\right] = E(x^2)-\mu^2
  \label{equ:uncertainty:variance}
\end{equation}
or the expectation value of $(x-\mu)^2$ and is denoted by the symbol
$\sigma^2$ or sometimes the letters Var. The \term{standard deviation}
of a probability density function is just the square root of the
variance.
\begin{equation}
  \sigma(x) = \sqrt{\sigma^2(x)}
  \label{equ:uncertainty:standardDeviation}
\end{equation}
The standard deviation, given by the Greek letter $\sigma$ (spelled
sigma), measures how far most measured values for a variable $x$
deviate from $\mu$, the mean of $x$. When the uncertainty for a
measurement with a known probability density function is quoted, the
uncertainty is usually just one standard deviation as calculated
above.

\section{Normal Uncertainty}

\begin{figure}
  \begin{center}
    \executeiffilenewer{Code/Uncertainty/normal.m}{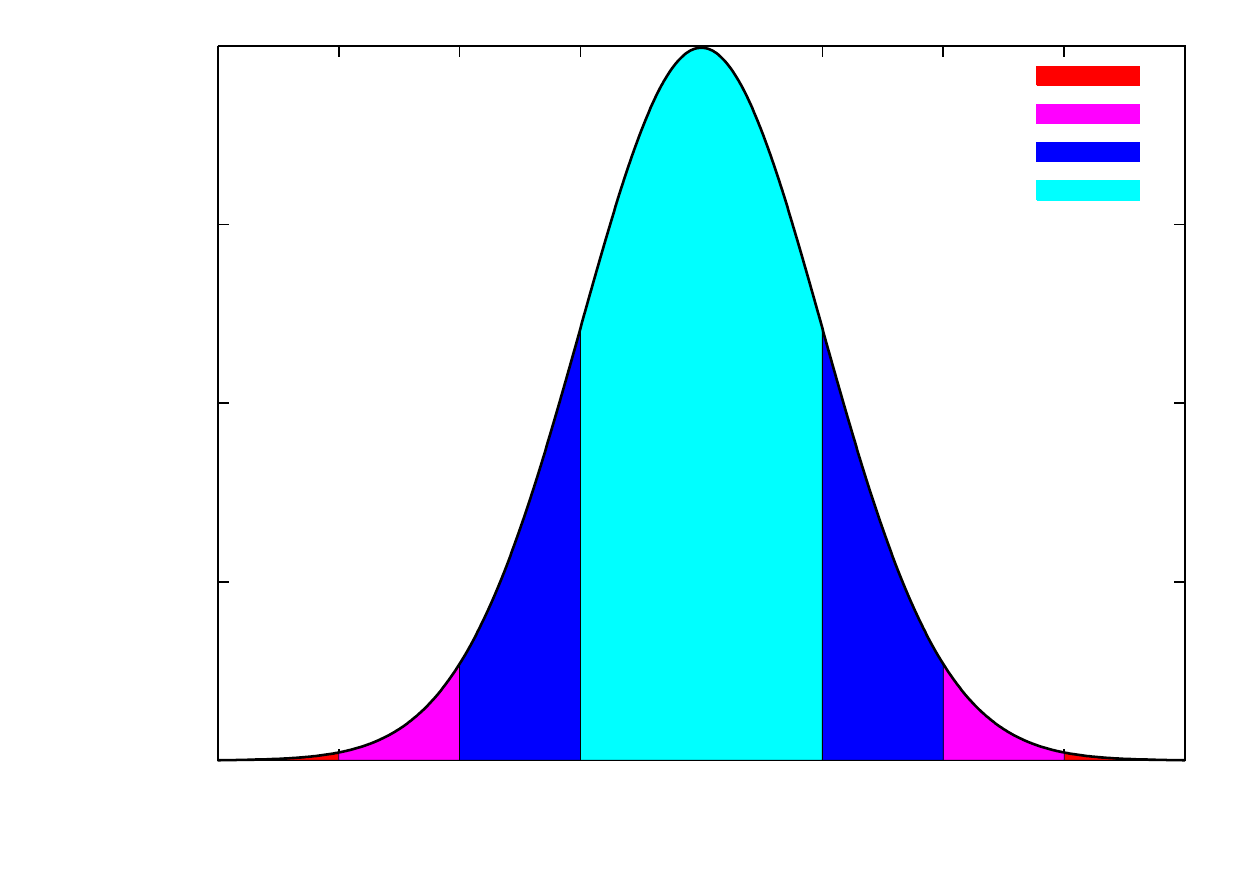}
  {octave --silent --eval "addpath([pwd(),'/Code']); 
    addpath([pwd(),'/Code/Uncertainty']); normal(1);"} 
  \setlength{\unitlength}{\columnwidth*\real{0.00013}}
  \input{Figures/Uncertainty/normalLabel.tex}
    \caption{The normal probability density function of Equation
      \ref{equ:uncertainty:normal} is given with one standard deviation in light
      blue, two standard deviations in dark blue, three standard
      deviations in purple, and four standard deviations in
      red.\label{fig:uncertainty:normal}}
  \end{center}
\end{figure}

In physics the probability density functions of most measurements are
described by the \term{normal distribution}.\footnote{This distribution
  also goes under the names of Gaussian distribution, normal curve, or
  bell curve.} The formula for the normal distribution is,
\begin{equation}
  PDF(x) =
  \frac{1}{\sqrt{2\pi\sigma^2}}e^{\left(\frac{x-\mu}{\sqrt{2}\sigma}\right)^2}
  \label{equ:uncertainty:normal}
\end{equation}
and is plotted in Figure \ref{fig:uncertainty:normal}. Here $\sigma$ is the
standard deviation of the curve, and $\mu$ the mean as defined in the
previous section. For a normal distribution, $68\%$ of all measured
values of $x$ are expected to fall within $\sigma$ of $\mu$, while
$95\%$ are expected to fall within $2\sigma$.

When a physics measurement is stated with an uncertainty, the
uncertainty is assumed to represent one standard deviation of the
data, unless indicated otherwise. Because most measurements in physics
are described by the normal distribution, $68\%$ of the values
measured by the experimenter fell within this uncertainty range.

But this leads to the question, why are most physics measurements
described by a normal distribution? The reason for this is what is
known as the \term{central limit theorem}. The central limit theorem
can be interpreted many ways, but dictates that under the correct
initial conditions, most probability density functions when sampled
many times converge to the normal distribution.

The above statement of the central limit theorem is very general and
not the most intuitive to understand, so it may be more helpful to
illustrate a consequence of the central limit theorem. Let us consider
a box which we fill with different colored marbles. First we
place $N_r$ red marbles into the box, where $N_r$ is a random number
chosen from a \term{uniform distribution} between $0$ and $10$. In a
uniform distribution each number is equally likely to be picked, and
so the distribution is just a rectangle from $0$ to $10$ with area
$1$. We have chosen this distribution as it is clearly not a normal
distribution.

After placing the red marbles into the box, we count the number of
marbles in the box, and record this number for this first trial as
$N_r$.  Next we add $N_o$ orange marbles, where again we determine
the random number $N_o$ from a uniform distribution between $0$ and
$10$. The number of marbles in the box will now just be $N_r+N_o$. We
record this value for our first trial as well. Next we add a random
number of yellow marbles, $N_y$, between $0$ and $10$ and again record
the total number of marbles in the box. We continue this experiment by
adding green, blue, and purple marbles in the exact same fashion to
the box, recording the total number of marbles in the box after each
new color is added.

\begin{figure}
  \begin{center}
    \tiny 
    \executeiffilenewer{Code/Uncertainty/clt.m}{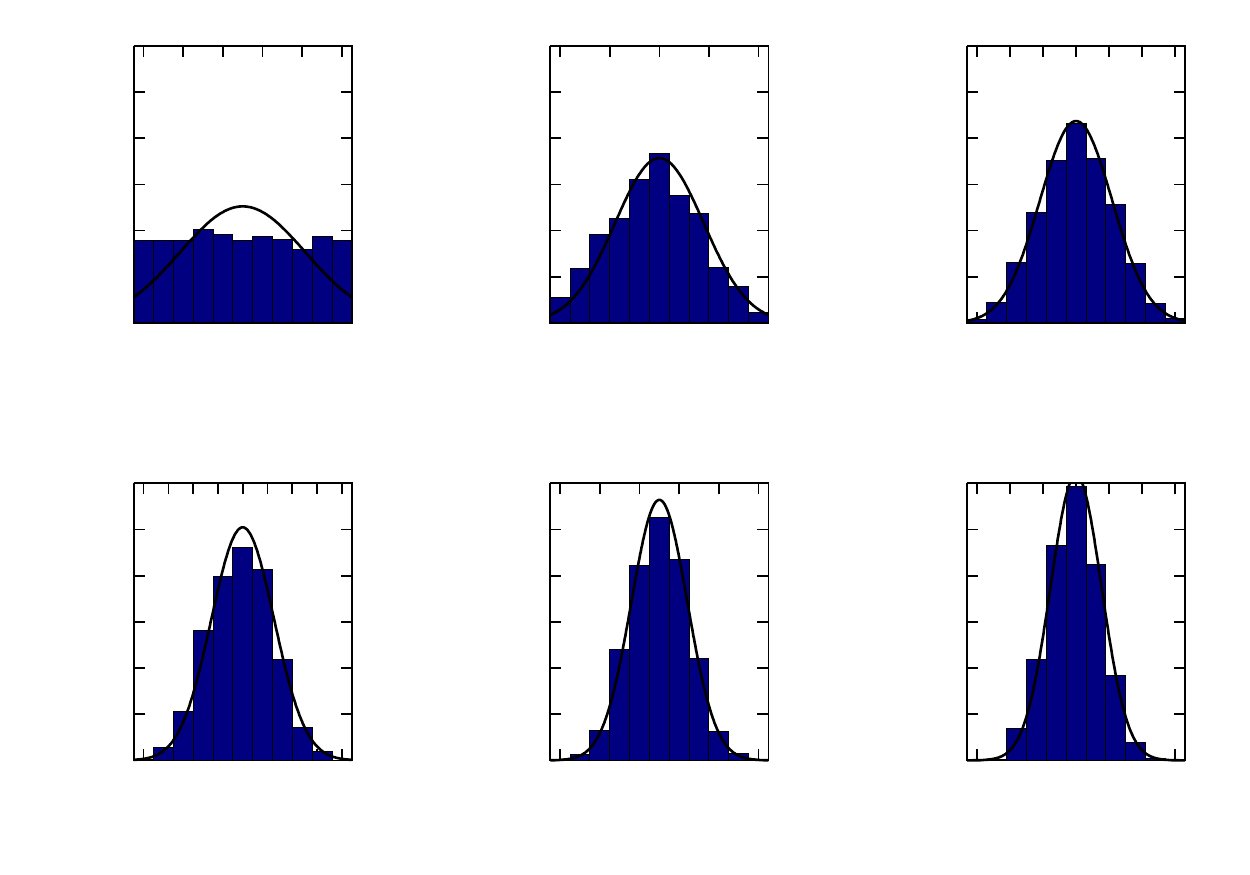}
  {octave --silent --eval "addpath([pwd(),'/Code']); 
    addpath([pwd(),'/Code/Uncertainty']); clt(1);"} 
  \setlength{\unitlength}{\columnwidth*\real{0.00013}}
  \input{Figures/Uncertainty/cltLabel.tex}
    \caption{An illustration of the central limit theorem. The black
      curve is the normal curve the distribution is
      approaching.\label{fig:uncertainty:clt}}
  \end{center}
\end{figure}

We now perform this experiment thousands of times and tabulate the
number of marbles after each step of adding a new color. We take all
these numbers and make a histogram for each step. In Figure
\ref{fig:uncertainty:clt} we have simulated the experiment $2000$
times. The histogram for the first step is a uniform distribution, as
we expect. This distribution just tells us the probability of finding
$N_r$ red marbles in the box after each step which is $1/11$.

In the histogram for the second step, something unexpected happens.
Here we have added the uniform distribution for the red marbles with
the uniform distribution for the orange marbles, and have obtained a
non-uniform distribution for the total number of marbles in the box!
This is a direct consequence of the central limit theorem. As we add
on more uniform distributions from the yellow, green, blue, and purple
marbles, the distribution for the total number of marbles in the box
after each step looks more and more like a normal curve. The result is
striking. What began as a flat distribution now resembles a normal
distribution; the total number of marbles in the box, after adding all
the colors, is normally distributed!

There are a few important features to notice about the steps performed
in Figure \ref{fig:uncertainty:clt}. The first is that the mean of the
distribution, $\mu$, is just the addition of the means of the
component distributions. For example, we know that on average we will
pick $5$ red marbles, $5$ orange marbles, etc. Subsequently in the first
histogram the mean for the histogram is $\mu_r$, while in the second histogram
the mean is $\mu_r+\mu_o$. The second important point to notice is
that for each histogram the variances add just like the means. The
variance for a discrete uniform distribution is just,
\begin{equation}
  \sigma^2 = \frac{N^2-1}{12}
\end{equation}
where $N$ is the number of discrete values available. The distribution
in the first histogram of Figure \ref{fig:uncertainty:clt} is sampled between $0$ and
$10$, and so with $N=11$ the variance is $10$. Subsequently the
following distributions have variances of $20$, $30$, $40$, $50$, and
$60$. In Figure \ref{fig:uncertainty:clt} the values given for $\mu$ and
$\sigma^2$ do not match exactly what is written above. This is because
the histograms were made by simulating the experiment outlined above
$2000$ times.\footnote{This is what we call a \term{Monte Carlo}
  experiment. We use a random number generator with a computer to
  simulate the experiment.} This is similar to when we flip a coin
$10$ times; we don't expect exactly $5$ heads and $5$ tails, but
instead numbers near $5$.

From the example above, we can see the power of the central limit
theorem. If we think of taking measurements in physics as adding
together many different probability density functions (like in the
marble example), we see that the end result is a normal
distribution. Whether this approximation is valid or not depends upon
the situation, but in general, normal distributions model physics
measurements well.

\section{Normal Uncertainty Propagation}

Let us assume that we have experimentally measured $n$ different
variables, $x_i$, in an experiment. Also, let us assume that for each
measured variable we have taken a large number of data points, $N$,
and verified that the data points for each variable are normally
distributed with a variance of $\sigma_i^2$. Now we wish to calculate
the standard deviation, $\sigma_f$, for a function dependent upon the
measured variables, $f(x_1,x_2,\ldots,x_n)$.

We can't just add the standard deviation of the variables together, as
we have seen in the previous sections. Instead we must use normal
uncertainty propagation, which was briefly demonstrated earlier. We
will begin by looking at the full propagation method and then trying
to understand it, first through an intuitive argument and then through
a more rigorous proof. The full formula for propagating normal
uncertainty is given below in all of its glory.

\begin{equation}
  \begin{aligned}
    \sigma_f^2 = &\begin{array}{cccc} \left[ \frac{\partial
          f}{\partial x_1} \right.& \frac{\partial f}{\partial x_2} &
      \ldots &
      \left. \frac{\partial f}{\partial x_n} \right]\\
      \\
      \\
      \\
      \\
      \\
      \\
    \end{array}
    \left[\begin{array}{cccc} \sigma_1^2 & \rho_{1,2} & \ldots &
        \rho_{1,n} \\
        \\
        \rho_{1,2} & \sigma_2^2 &\ldots &
        \rho_{2,n} \\
        \\
        \vdots & \vdots & \ddots & \vdots \\
        \\
        \rho_{1,n} & \rho_{2,n} &\ldots &
        \sigma_n^2 \\
      \end{array}\right]
    \left[\begin{array}{c}
        \frac{\partial f}{\partial x_1} \\
        \\
        \frac{\partial f}{\partial x_2} \\
        \\\vdots \\
        \\
        \frac{\partial f}{\partial x_n} \\
      \end{array}\right]\\
    = &\left(\frac{\partial f}{\partial x_1}\right)^2\sigma_1^2 +
    \left(\frac{\partial f}{\partial x_2}\right)^2\sigma_2^2 + \ldots
    + \left(\frac{\partial f}{\partial x_n}\right)^2\sigma_n^2\\
    &+ 2\rho_{1,2}^2\left(\frac{\partial f}{\partial x_1}\right)
    \left(\frac{\partial f}{\partial x_2}\right) + \ldots +
    2\rho_{n-1,n}^2\left(\frac{\partial f}{\partial x_{n-1}}\right)
    \left(\frac{\partial f}{\partial
        x_n}\right)\\
    \label{equ:uncertainty:covarianceMatrix}
  \end{aligned}
\end{equation}

Needless to say the above method for propagation looks very nasty, and
it is. The matrix in the middle of the equation, consisting of
$\sigma$'s and $\rho$'s is called the \term{variance-covariance matrix}
and essentially relates the correlations of all the variables
$x_i$. The coefficient $\rho_{i,j}$ is defined as the \term{correlation
  coefficient} and represents how strongly the variables $x_i$ and
$x_j$ are correlated. For variables that are completely correlated
(i.e. they are the same variable) $\rho_{i,j}$ is just one. For the
case that the variables are completely uncorrelated $\rho_{i,j}$ is
just zero. The general form of $\rho_{i,j}$ is,
\begin{equation}
  \rho_{i,j} =
  \frac{E\left[\left(x_i-\mu_i\right)\left(x_j-\mu_j\right)\right]}
  {\sigma_i\sigma_j}
  \label{equ:uncertainty:correlation}
\end{equation}
where $E$ is the expectation value explained above, and $\mu_i$ is the
mean value of the variable $x_i$.

Luckily for us, the majority of experiments in physics consist of
measuring independent variables, variables where $\rho_{i,j} = 0$, and
so Equation \ref{equ:uncertainty:covarianceMatrix} is greatly simplified.
\begin{equation}
  \sigma_f^2 = \left(\frac{\partial f}{\partial x_1}\right)^2\sigma_1^2 +
  \left(\frac{\partial f}{\partial x_2}\right)^2\sigma_2^2 + \ldots
  + \left(\frac{\partial f}{\partial x_n}\right)^2\sigma_n^2
  \label{equ:uncertainty:independent}
\end{equation}
This unfortunately is not a very intuitive equation, and so a bit more
explanation is necessary.

\begin{figure}
  \begin{center}
    \subfigure[]{
      \executeiffilenewer{Figures/Uncertainty/slope.svg}
  {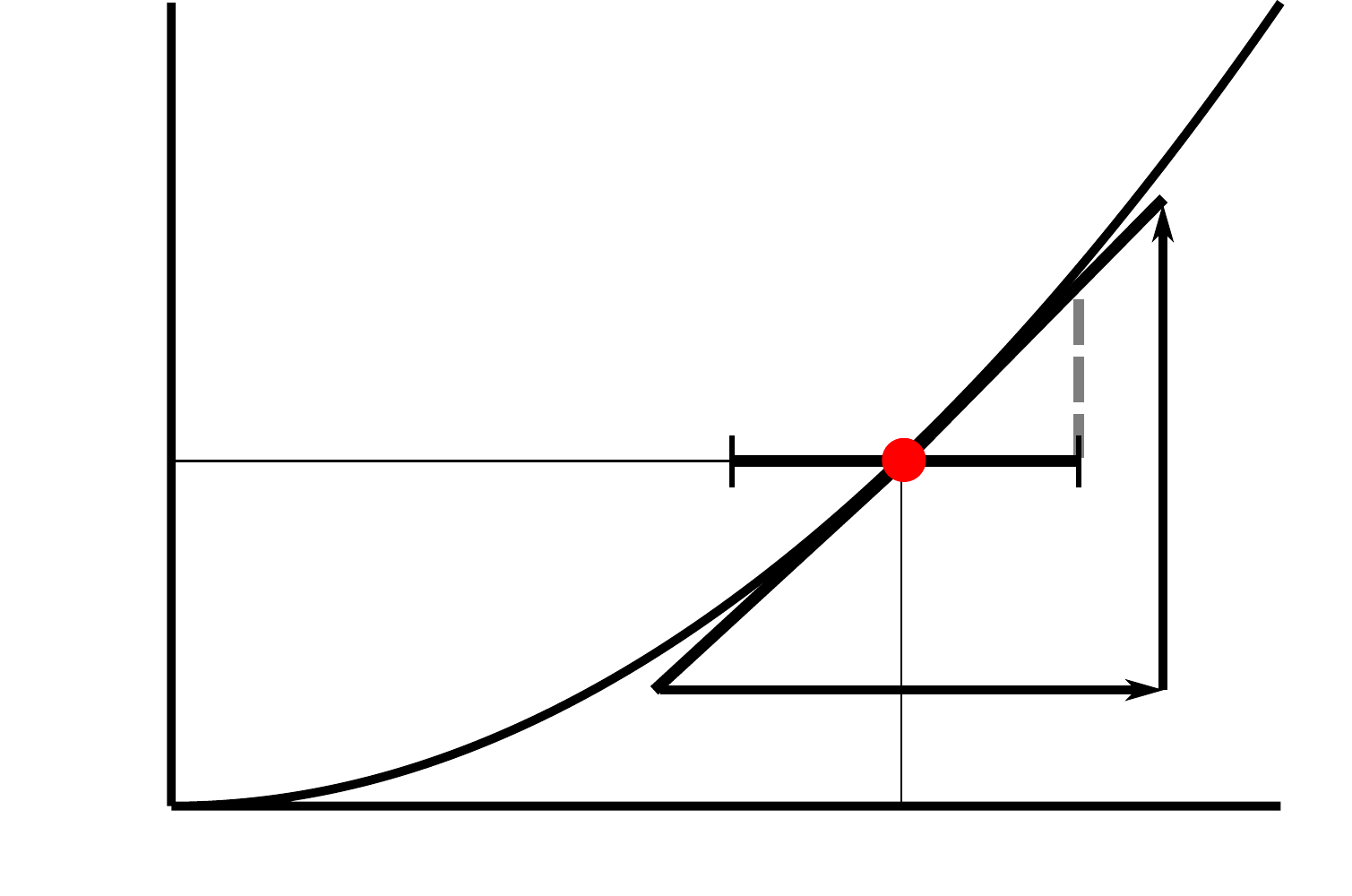}{inkscape-0.48pre1 -z -D --file=Figures/Uncertainty/slope.svg 
    --export-pdf=Figures/Uncertainty/slope.pdf --export-latex} \def\svgwidth{8cm}
  \input{Figures/Uncertainty/slopeLabel.tex}
      \label{fig:uncertainty:slope}}
    \subfigure[]{
      \executeiffilenewer{Figures/Uncertainty/triangle.svg}
  {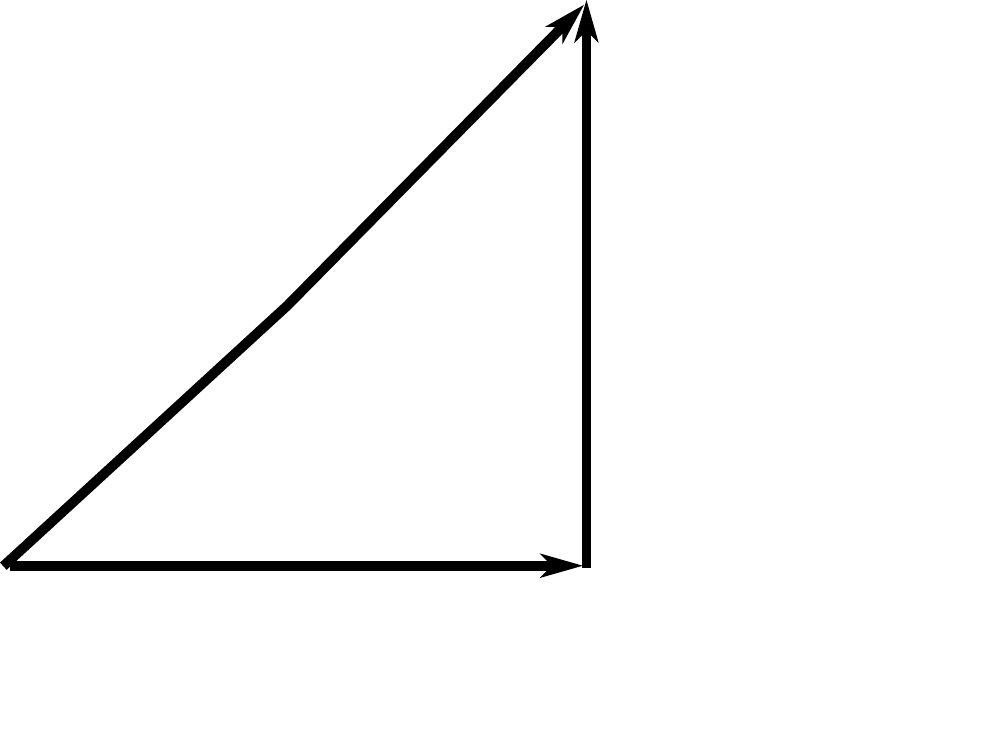}{inkscape-0.48pre1 -z -D --file=Figures/Uncertainty/triangle.svg 
    --export-pdf=Figures/Uncertainty/triangle.pdf --export-latex} \def\svgwidth{6cm}
  \input{Figures/Uncertainty/triangleLabel.tex}
      \label{fig:uncertainty:triangle}}
    \caption{Figure \ref{fig:uncertainty:slope} geometrically
      illustrates how the uncertainty $\sigma_f$ is propagated from an
      uncertainty $\sigma_1$. Figure \ref{fig:uncertainty:triangle}
      demonstrates how the components of uncertainty must be added in
      quadrature.}
  \end{center}
\end{figure}

Let us first consider a simple example where we have one variable
$x_1$ and a function, $f(x_1)$, dependent upon only $x_1$. We can make
a plot of $x_1$ on the $x$-axis and $f(x_1)$ on the $y$-axis. Let us
choose a specific $x_1$ and label this point $p$. All points of $x_1$
have an associated uncertainty of $\sigma_1$ and so we know the
uncertainty for $p$ is $\sigma_1$. Looking at Figure \ref{fig:uncertainty:slope}
we see that our uncertainty $\sigma_1$ is in the $x$-direction, and
that we want $\sigma_f$ which should be in the $y$-direction.

To find $\sigma_f$ we can just use simple geometry. It seems
reasonable that at point $p$ the change in $f(x_1)$ over the change in
$x_i$ should equal the change in the uncertainty on $f(x_i)$ over the
change in uncertainty on $x_i$.
\begin{equation}
  \frac{\Delta f(x_1)}{\Delta x_1} = \frac{d f(x_1)}{dx_1} =
  \frac{\sigma_f}{\sigma_1}
  ~~~\Rightarrow~~~
  \sigma_f = \frac{d f(x_1)}{dx_1}\sigma_1
  \label{equ:uncertainty:slope}
\end{equation}
In the first step we have taken the limit of $\Delta f(x_i)/\Delta
x_i$ as $\Delta x_i$ grows very small which just gives us the
derivative, or the slope exactly at point $p$ rather than in the
general vicinity of $p$. In the next step we set this equal to the
change in uncertainty and in the final step we just solve for
$\sigma_f$.

Now we have found how to propagate the uncertainty for just one
variable $x_1$, but we need to be able to do this for $n$ variables
$x_i$. The first change we need to make is substitute the derivative
of Equation \ref{equ:uncertainty:slope} with a \term{partial
  derivative}.\footnote{For those not familiar with a partial
  derivative, we denote it with the symbol $\partial$. To take a
  partial derivative such as $\partial f(x,y,z)/\partial x$ just
  differentiate $f(x,y,z)$ with respect to $x$ and think of $y$ and
  $z$ as constants.} The second change we need to make is how we think
of our uncertainty. The uncertainties don't just add linearly like
numbers, but rather are components of an uncertainty vector. We don't
care about the direction of the uncertainty vector but we do care
about the magnitude of the vector, as this gives us our uncertainty on
$f(x_i)$. To find the magnitude of a vector, we just take the square
root of the sum of all the components squared. This process is called
\term{adding in quadrature}.

\begin{equation}
  \sigma_f = \sqrt{\left(\frac{d f}{dx_1}\sigma_1\right)^2 +
    \left(\frac{d f}{dx_2}\sigma_2\right)^2 + \ldots +
    \left(\frac{d f}{dx_n}\sigma_n\right)^2}
  \label{equ:uncertainty:independentPropagation}
\end{equation}

For those readers not familiar with taking the magnitude of a vector,
think of the Pythagorean theorem where we find the length of the
hypotenuse of a triangle, $c$, from the sides of the triangle, $a$ and
$b$, by the formula $c=\sqrt{a^2+b^2}$. In Figure
\ref{fig:uncertainty:triangle} we are now considering an example where
$n=2$ and $f$ is dependent on two variables, $x_1$ and $x_2$, with
uncertainties of $\sigma_1$ and $\sigma_2$. Using Equation
\ref{equ:uncertainty:slope} we replace sides $a$ and $b$ with the
$x_1$ and $x_2$ components of $\sigma_f$, and the hypotenuse, $c$,
with $\sigma_f$. Using Pythagoras' theorem we arrive back at Equation
\ref{equ:uncertainty:independentPropagation}.

Unfortunately, Equation \ref{equ:uncertainty:independentPropagation}
only matches Equation \ref{equ:uncertainty:covarianceMatrix} when
$\rho_{i,j} =0$, or when the variables $x_i$ are independent of each
other. To understand how we can introduce the $\rho_{i,j}$ terms we
must leave the geometric derivation for uncertainty propagation
outlined above and turn to a more mathematically rigorous
derivation. The math here can get a little complicated, but is given
for the curious.

A \term{Taylor series} is an expansion of a function $f$ about a
certain point $a$ and is given by an infinite sum. For the $n$
dimensional case the function $f(x_1,x_2,\ldots,x_n)$ is expanded in
infinite sums about the points $a_1,a_2,\ldots,a_n$.
\begin{equation}
  \begin{aligned}
    T\left(f(x_1,\ldots,x_n)\right) = \sum_{m_1 =
      0}^\infty\cdots\sum_{m_n = 0}^\infty
    &\left[\left(\frac{\partial^{m_1+\cdots+m_n}f(x_1,\ldots,x_n)}
        {\partial^{m_1}x_1\cdots\partial^{m_n}{x_n}}\right)\right.\\
    &\times\left.\left(\frac{\left(x_1-a_1\right)^{m_1}
          \cdots\left(x_n-a_n\right)^{m_n}}
        {m_1!\cdots m_2!}\right)\right]\\
    \label{equ:uncertainty:taylor}
  \end{aligned}
\end{equation}

Let us assume now that we have made $N$ measurements and that for the
$N^\mathrm{th}$ measurement we have the measured values $x_{1,N}$
through $x_{n,N}$. We can now expand our function $f(x_1,\ldots,x_n)$
about its average $\mu_f$ for each measurement $N$, and truncate the
expansion at first order (i.e. we only look at terms with first
derivatives). The Taylor series for measurement $N$ to first order is
as follows.
\begin{equation}
  \begin{aligned}
    f_N-\mu_f = &\left(\frac{\partial f}{\partial
        x_1}\right)\left(x_{1,N}-\mu_1\right) + \left(\frac{\partial f}{\partial
        x_2}\right)\left(x_{2,N}-\mu_2\right) + \cdots\\ 
    &+ \left(\frac{\partial f}{\partial
        x_n}\right)\left(x_{n,N}-\mu_n\right)\\
  \end{aligned}
  \label{equ:uncertainty:expansion}
\end{equation}
Using our definition for variance given in Equation
\ref{equ:uncertainty:variance} in combination with the limit of a
discrete version of Equation \ref{equ:uncertainty:expectationValue},
we can then write the variance of $f$.
\begin{equation}
  \sigma_f^2 = \lim_{N\rightarrow\infty}\frac{1}{N}\sum_{m=0}^N
  \left(f_N - \mu_f\right)
\end{equation}
Plugging Equation \ref{equ:uncertainty:expansion} into the above we arrive at,
\begin{equation}
  \begin{aligned}
    \sigma_f^2 = \lim_{N\rightarrow\infty}\frac{1}{N}\sum_{m=0}^N
    &\left[ \left(\frac{\partial f}{\partial
          x_1}\right)\left(x_{1,N}-\mu_1\right) + \cdots +
      \left(\frac{\partial f}{\partial
          x_n}\right)\left(x_{n,N}-\mu_n\right)\right.\\
    &+2\left(x_{1,N}-\mu_1\right)\left(x_{2,N}-\mu_2\right)
    \left(\frac{\partial f}{\partial x_1}\right)
    \left(\frac{\partial f}{\partial x_2}\right)\\
    &\left.+2\left(x_{n-1,N}-\mu_{n-1}\right)\left(x_{n,N}-\mu_n\right)
      \left(\frac{\partial f}{\partial x_{n-1}}\right)
      \left(\frac{\partial f}{\partial
          x_n}\right)\right]\\
  \end{aligned}
\end{equation}
where we have expanded the square. We can break this into the
individual sums, and using the definition of variance along with a
discrete definition of the correlation coefficient given in Equation
\ref{equ:uncertainty:correlation}, recover Equation
\ref{equ:uncertainty:covarianceMatrix}!
\begin{equation}
  \begin{aligned}
    \sigma_f^2 = &\left(\frac{\partial f}{\partial
        x_1}\right)^2\sigma_1^2 + \left(\frac{\partial f}{\partial
        x_2}\right)^2\sigma_2^2 + \ldots
    + \left(\frac{\partial f}{\partial x_n}\right)^2\sigma_n^2\\
    &+ 2\rho_{1,2}^2\left(\frac{\partial f}{\partial x_1}\right)
    \left(\frac{\partial f}{\partial x_2}\right) + \ldots +
    2\rho_{n-1,n}^2\left(\frac{\partial f}{\partial x_{n-1}}\right)
    \left(\frac{\partial f}{\partial
        x_n}\right)\\
  \end{aligned}
\end{equation}

\graphicspath{{Figures/Plots/}}

\chapter{Plots}\label{chp:plots}

The old adage ``a picture is worth a thousand words'' is true for many
things but could not be more relevant to the field of physics,
although perhaps the saying should be changed to ``a plot is worth a
thousand words''. Plots quickly and easily allow readers to assimilate
data from experiment, compare data to theory, observe trends, the list
goes on and on. Because plots are so important in physics, it is
critical that plots are made correctly. This chapter attempts to cover
the basics of plotting, fitting data, and the methods behind fitting
of data.

\section{Basics}

\begin{figure}[h]
  \begin{center}
    \executeiffilenewer{Code/Plots/example.m}{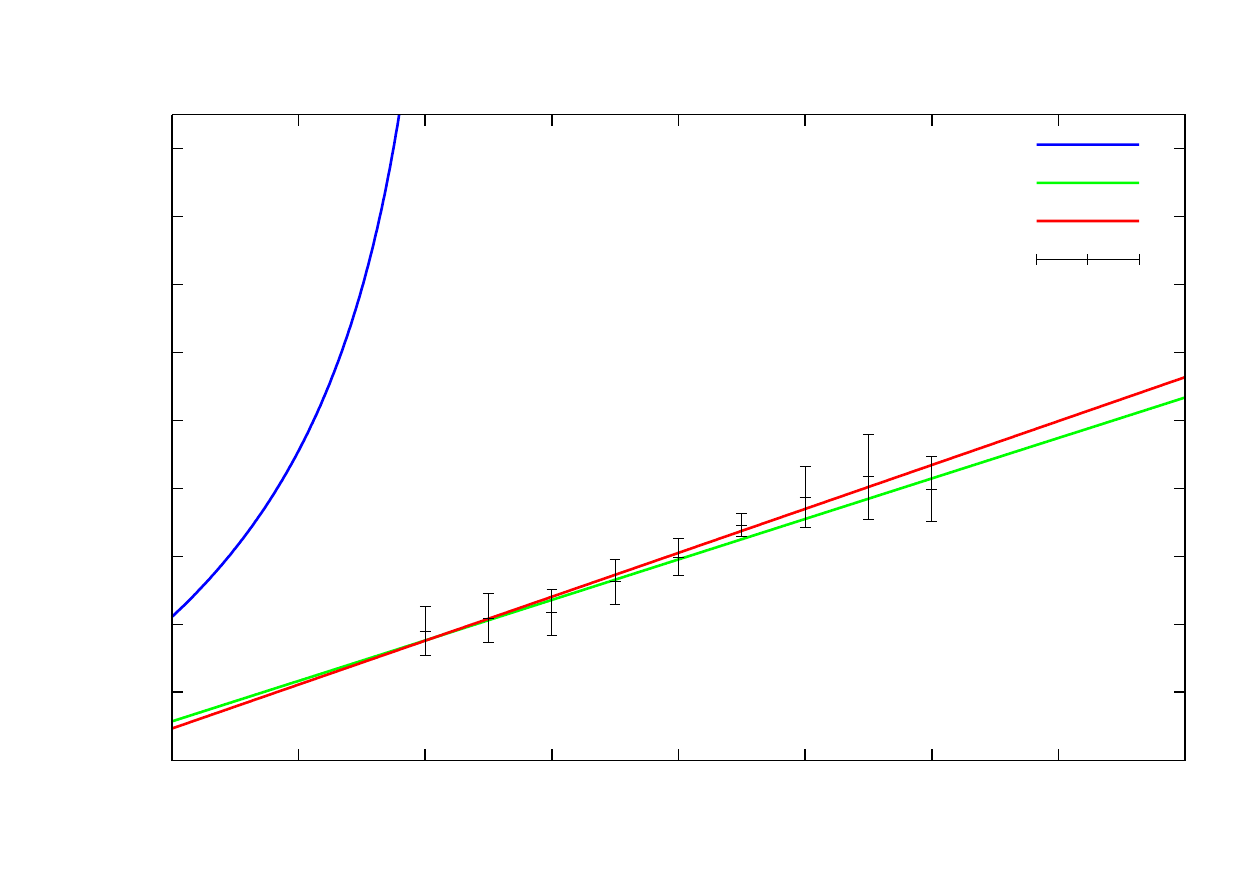}
  {octave --silent --eval "addpath([pwd(),'/Code']); 
    addpath([pwd(),'/Code/Plots']); example(1);"} 
  \setlength{\unitlength}{\columnwidth*\real{0.00013}}
  \input{Figures/Plots/exampleLabel.tex}
    \caption{An example of a well scaled plot with a title, axis
      labels, units, a legend, theory curves, a curve of best fit
      (with associated reduced $\chi^2$), and data points with error
      bars.\label{fig:plots:example}}
  \end{center}
\end{figure}

Figure \ref{fig:plots:example} gives an example of a well-made plot. The first
point to notice about Figure \ref{fig:plots:example} is the \term{title}. Every
plot should have a title; this allows the reader to quickly understand
what the plot is attempting to show, without having to read pages of
accompanying text. Sometimes if the figure is accompanied by a caption
the title is neglected, but in general, plots should always have
titles. Additionally, the title should be relatively short, but also
convey meaning. In this example the title is ``Velocity of an Electron
in a Magnetic Field''. The meaning here is clear, we expect to find
information within the plot pertaining to the velocity of an electron
as it moves through a magnetic field. Notice that titles such as
``Graph 1'' are not helpful. Such a title tells us absolutely nothing
about the content of the plot.

The next items to notice about the plot in Figure \ref{fig:plots:example} are
the \term{labels} for the $x$-axis and $y$-axis. It is important that
the labels are clear about what is being plotted along each axis. In
this plot the product of two quantities $\beta$ and $\gamma$, both
calculated from the velocity of an electron is plotted on the $y$-axis
while the intensity of a surrounding magnetic field is plotted on the
$x$-axis. The labels clearly and concisely summarize this information
without being ambiguous. Each label is followed by square brackets
filled with the units of the axis. Notice that the quantity
$\beta\gamma$ is unitless and so the label is followed with the word
``unitless'' in square brackets. While this is not necessary, it
informs the reader the plotted quantity has no units, and a unit label
was not just forgotten. Without units it is nearly impossible to guess
what information the plot is trying to depict.

In the example of Figure \ref{fig:plots:example} an electron is
passing through a magnetic field and its velocity is being measured
using a velocity selector.\footnote{This is an actual experiment using
  real data. The experiment was performed at MIT's Junior Lab, and the
  full write up is available at
  \url{http://severian.mit.edu/philten/physics/dynamics.pdf}.} From
the plot it is clear that the magnetic field is being adjusted, and
for each adjustment in the field the velocity of the electron is
measured and the value $\beta\gamma$ calculated. We know this because
traditionally the variable quantity is plotted along the $x$-axis
while the measured quantity is plotted along the $y$-axis. Similarly,
for three dimensional plots the measured quantity is plotted on the
$z$-axis, while the variable quantities are plotted along the $x$ and
$y$-axis.

Now that we have discussed all the important aspects of labeling we
can focus on the actual contents of the plot, but it is important to
remember without proper labels the content of a plot is
meaningless. Looking at the contents of Figure \ref{fig:plots:example}
we see three lines, a blue line, a green line, and a red line along
with black data points. When multiple items are plotted, a
\term{legend}, given in the left hand corner of the plot, is a
necessity. A quick glance at the legend tells us that the blue line is
a theoretical prediction made using classical mechanics, the green
line is a theoretical prediction made using relativistic mechanics,
the red line is a line of best fit, and the black data points are the
data from the experiment.

Above and below each data point a line is vertically extended; this is
called an \term{error bar} which represents the uncertainty on the
data point. For example, the first data point at $B = 80$ Gauss has a
measured value of $0.981$ with an uncertainty of $0.073$ and so the
lower error bar extends to a value of $0.908$ while the upper error
bar extends to a value of $1.054$. In this experiment the uncertainty
was arrived at using normal uncertainty, and so we know that $68\%$ of
the experimenter's measurements fell within the range given by the
error bars, or in this case the first measurement can be written as
$0.981\pm0.073$.\footnote{See Chapter \ref{chp:uncertainty} for more
  details on how uncertainty is estimated and propagated.} Uncertainty
can be displayed along the $x$-axis, and can also be simultaneously
displayed along both $x$ and $y$, although traditionally uncertainty
is propagated so that it is only displayed along the
$y$-axis. Additionally, sometimes an \term{error band} is used
instead, which is just a continuous version of error bars.

\section{Fitting Data}

One important part of Figure \ref{fig:plots:example} that was not discussed in
the previous section were the blue, green, and red lines. Lines such
as these, especially the red line, can be found in most scientific
plots as they help the reader understand how well different theories
match with the experimental data. Before we can fully explain the plot
we need a small amount of theory. In classical (or Newtonian)
mechanics, the mechanics taught in this course, the velocity of an
object has no upper limit. Einstein, at the beginning of the
$20^\mathrm{th}$ century postulated that this actually is not true,
and that very fast moving objects (nothing that we will observe in the
lab) cannot exceed the speed of light.

In the creation of this theory Einstein introduced two new quantities
that can be calculated from the velocity of an object. The first,
\begin{equation}
  \beta = \frac{v}{c}
\end{equation}
is represented by the Greek letter $\beta$ (spelled beta), and is just
the velocity of an object divided by the speed of light in a vacuum,
$c$. Notice that according to Einstein because $v < c$, $\beta$ must
always be less than or equal to $1$. The second quantity,
\begin{equation}
  \gamma = \frac{1}{\sqrt{1-\beta^2}}
  \label{equ:plots:gamma}
\end{equation}
is called the Lorentz $\gamma$-factor is and is represented by the
Greek letter $\gamma$ (spelled gamma).

Without getting bogged down in details, the data for Figure
\ref{fig:plots:example} was gathered by firing electrons at a very high
velocity through a magnetic field and recording their velocity. Using
both classical and relativistic theory it is possible to predict
$\beta$ for the electron (the subscript c designates classical theory,
and the subscript r relativistic theory).
\begin{equation}
  \begin{aligned}
    \beta_\mathrm{c} &= \frac{\rho B e}{m_e c^2}\\
    \beta_\mathrm{r} &= \frac{1}{\sqrt{1+\left(
          \frac{m_ec^2} {\rho B e}\right)^2}}\\
  \end{aligned}
  \label{equ:plots:beta}
\end{equation}
Here the quantity $\rho$ is a physical constant of the experiment. The
quantity $e$ is the fundamental charge of an electron, $m_e$ the mass
of an electron, and $c$ the speed of light. The quantity $B$ is the
strength of the magnetic field through which the electron is
traveling, and is the variable in this experiment which we know
already, as $B$ is plotted on the $x$-axis of Figure \ref{fig:plots:example}.

We can find $\gamma$ for both classical and relativistic theory by
plugging the respective values for $\beta$ given in Equation
\ref{equ:plots:beta} into Equation \ref{equ:plots:gamma}. From Equation
\ref{equ:plots:beta} we notice that classical theory predicts that $\beta$
is given by a linear relationship in $B$. We can write a general
linear relationship between $y$ and $x$ in the \term{slope-intercept
  form} of
\begin{equation}
  y = mx+b
  \label{equ:plots:linear}
\end{equation}
where $m$ is the slope of the line and $b$ is the $y$-intercept of the
line. By definition the $y$-intercept is where the line crosses the
$y$-axis, which occurs when $x$ is zero.

In general, physicists like working with linear relationships. They
are easier to fit than complicated curves and oftentimes are able to
provide just as much information. The only problem we have with
Equation \ref{equ:plots:beta} is that we expect relativistic theory to
be correct, not classical theory, and it is clear that $\beta$ is not
given by a linear relationship in $B$ for relativistic theory. To
circumvent this problem and allow us to still make a linear plot, we
\term{recast} Equation \ref{equ:plots:beta} into a format where we
have a linear relationship between some quantity and the variable
$B$. Here we multiply $\beta$ by $\gamma$.
\begin{subequations}
  \begin{equation}
    \beta_\mathrm{c}\gamma_\mathrm{c} = \left(\frac{\rho B e}{m_e
        c^2}\right)\left(1-\left(\frac{\rho B
          e}{m_e c^2}\right)^2\right)^{-\frac{1}{2}}
    \label{equ:plots:betaGammaClassical}
  \end{equation}
  \begin{equation}
    \beta_\mathrm{r}\gamma_\mathrm{r} = \left(\frac{\rho e}{m_e c^2}\right)B
    \label{equ:plots:betaGammaRelativistic}
  \end{equation}
  \label{equ:plots:betaGamma}
\end{subequations}
Now we have a linear relationship for our relativistic theory, from
which we can predict the charge to mass ratio of the electron from the
slope of the line!

With Equation \ref{equ:plots:betaGamma} we can better understand the curves
presented in Figure \ref{fig:plots:example}. Looking at our classical
prediction from Equation \ref{equ:plots:betaGammaClassical} we no longer
expect a linear relationship, and expect that for large $B$ values,
the value for $\beta\gamma$ will explode. This behavior can be seen in
the blue line of Figure \ref{fig:plots:example}. As $B$ grows large the value
$\beta\gamma$ grows rapidly, and most certainly in a non-linear
fashion.

The green curve, corresponding to relativistic theory, is linear as we
expect from Equation \ref{equ:plots:betaGammaRelativistic}. Notice that our
data points match very well with relativistic theory; all except one
data point falls within its uncertainty on the green line!

The final red curve in Figure \ref{fig:plots:example} gives the best linear fit
of the data points. What we mean by ``best'' will be discussed in the
following section. The general idea, however, is that a linear fit
matches the data points well, and allows us to calculate the mass to
charge ratio of the electron from the slope of our fit. For this fit
we have obtained values of,
\begin{equation}
  m = 0.0129\pm0.0021,~~~b = -0.08\pm0.21
\end{equation}
where $m$ is slope and $b$ is $y$-intercept as defined previously in
Equation \ref{equ:plots:linear}. If we plug in our value for $\rho$ and
$c^2$ we can then calculate out the charge to mass ratio of the
electron.
\begin{equation}
  \frac{e}{m_e} = 0.0129\left(\frac{c^2}{\rho}\right)
  \label{equ:plots:parameters}
\end{equation}

The core idea to come away with from the discussion above is how easy
it is to represent theory and experiment, and their subsequent
agreement with just a single plot. In Figure \ref{fig:plots:example}
we have shown how drastically different the theoretical predictions of
classical and relativistic mechanics are, and we have shown that the
data matches the relativistic prediction, not the classical
prediction. Furthermore, by fitting the data we have managed to
calculate the ratio of two fundamental constants of nature, the mass
of the electron and the charge of the electron. Remember that to
obtain a linear best fit we had to recast the equations in such a way
as to provide a linear relationship. This is a technique that will be
used throughout this book.

\section{Fitting Methods}

In the previous section we discussed the curve of ``best'' fit for the
data points in Figure \ref{fig:plots:example}. But what exactly do we mean by
``best'', and how do we determine the fit? To fully explain this we
must first introduce some new definitions.

To begin, we define a \term{residual}.\footnote{Residuals are not just
  used for experimental data but are also very important in numerical
  analysis. In reality, fitting methods and their theory is more in
  the realm of numerical analysis than physics. Because of this it is
  important that physicists have a strong grasp of numerical
  analysis.} If we have some observed value $y_\mathrm{o}$ but expect
the value of $y_\mathrm{e}$ then the residual is just
$y_\mathrm{o}-y_\mathrm{e}$. We can apply this definition to the plot
of Figure \ref{fig:plots:example}. We can think of the best fit line
as the expected values $y_e$, and the actual data points as the
observed values $y_o$. Using common sense we can define what a best
fit line is; it is the line that minimizes the sum of the residuals
between the line and the data points. In other words, we want every
data point to fall as closely as possible to the best fit line. We can
adjust the parameters $m$ and $b$ for the line accordingly until we
have the smallest possible sum of residuals for the line.

Before we apply this idea to the plot we must take into consideration
two problems. The first problem is that the residuals can be both
positive and negative, and so we could minimize the total residuals to
zero, while still having very large positive and negative residuals
that cancel each other out. We can negate this effect by adding the
residuals together in \term{quadrature}, denoted by the $\oplus$
symbol. By this, we mean that instead of adding the residuals
together, we add the squares of the residuals.\footnote{There is a
  more detailed mathematical explanation as to why we add the
  residuals in quadrature, and this stems from the theory behind {\bf
    chi-squared distributions}. A more intuitive way to think of the
  residuals is to think of them as components of a vector, and we are
  trying to find the magnitude of the vector. The same method is used
  for explaining why we add uncertainties in quadrature, as is
  described in Chapter \ref{chp:uncertainty} of this book.}

The second problem is that we don't want data points with very large
uncertainty to effect the placement of the line just as much as data
points with small uncertainty. For example, consider an example where
we have made three measurements with associated uncertainty of
$1.0\pm0.1$, $2.0\pm0.1$, and $20\pm10$. For the last data point the
experimenter was distracted by a ninja and so the uncertainty is huge
(nearly $50\%$). We might not want to throw out the last data point,
but we certainly don't want our line of best fit to consider the last
data point equally with the first two. To ensure the first two data
points are considered more than the final data point, we must {\bf
  weight} the residuals by the associated uncertainty for that
specific measurement. We do this by dividing the residual by the
uncertainty. Subsequently, large uncertainty makes the residual
smaller (and it matters less), and small uncertainty makes the
residual larger (it matters more). Summing these weighted residuals in
quadrature yields a value called the \term{chi-squared} value, denoted
by the Greek letter $\chi^2$.
\begin{equation}
  \chi^2 = \sum_{i=1}^N \left(\frac{y_i-f(x_i)}{\sigma_i}\right)^2
  \label{equ:plots:chiSquared}
\end{equation}
Here we have $N$ data points $y_i$ measured at the variable $x_i$ with
an associated uncertainty of $\sigma_i$. Additionally, the value
$f(x_i)$ is the value calculated by the curve we are fitting to the
data points at the variable $x_i$.

We now have a method by which we can find the line of best fit for a
given set of data points; we minimize $\chi^2$ as is given in Equation
\ref{equ:plots:chiSquared}. But what if we wish to compare the {\bf
  goodness} of our fit for one set of data with another set of data?
By looking at Equation \ref{equ:plots:chiSquared} for $\chi^2$ we see that
if the number of data points increase but the uncertainty on each data
point remains constant, our $\chi^2$ will increase. This means that by
our definition of best fit, more data points means a worse fit. This
of course does not make sense, and so we need to modify Equation
\ref{equ:plots:chiSquared} slightly.

Before we modify Equation \ref{equ:plots:chiSquared}, we need to note this
equation does not just apply to fitting data points with lines, but
with any arbitrary curve, as $f(x_i)$ can be determined by any
arbitrary function! If we are going to modify Equation
\ref{equ:plots:chiSquared} so we can compare the goodness of different
linear fits, we may as well modify it so that we can compare the
goodness of any arbitrary curve. To do this we define the \term{number
  of degrees of freedom} (commonly abbreviated as NODF) for a fit as,
\begin{equation}
  \nu = N-n-1
  \label{equ:plots:df}
\end{equation}
where $N$ is the number of data points being fitted, $n$ is the number
of parameters of the curve being fit, and the Greek letter $\nu$
(spelled nu) is the number of degrees of freedom. The number of
parameters for a curve is the number of variables that need to be
determined. For the line there are two variables, the slope, $m$, and
the $y$-intercept $b$. For a second degree polynomial such as,
\begin{equation}
  y = a_0+a_1x+a_2x^2
  \label{equ:plots:polynomial}
\end{equation}
there are three parameters, $a_0$, $a_1$, and $a_2$. Now, by using the
number of degrees of freedom from Equation \ref{equ:plots:df} and $\chi_2$
from Equation \ref{equ:plots:chiSquared} we can define the \term{reduced
  chi-squared} of a fit which allows for the comparison of the
goodness of a fit using any number of data points and any arbitrary
curve.
\begin{equation}
  \chi^2_\nu = \frac{\chi^2}{\nu} = \frac{1}{N-n-1}\sum_{i=1}^N
  \left(\frac{y_i-f(x_i)}{\sigma_i}\right)^2
  \label{equ:plots:reducedChiSquared}
\end{equation}

The number of degrees of freedom for a curve must always be greater
than $0$ to fit that curve to the data points using the minimization
of $\chi^2_\nu$ method. We can think of this intuitively for the case
of a line. A line is defined by a minimum of two points, so if we are
trying to fit a line to a data set with two data points we see there
is only one line we can draw, and $\nu = 0$. This means we don't need
to bother with minimizing $\chi^2_\nu$ because we already know the
solution. Another way to think of the number of degrees of freedom for
a fit is for larger $\nu$ the fit is more free and for smaller $\nu$
the fit is more confined. If we add ten more data points to our data
set in the example with the line, $\nu$ becomes $9$ and the fit of the
line is more free because it has more data points to consider when
finding $m$ and $b$. If we change the type of curve we are fitting to
the second degree polynomial of Equation \ref{equ:plots:polynomial} $\nu$
now becomes $8$ because we have added an extra parameter. This
decreases the freedom of the fit because the fit must now use the same
number of data points to find more parameters; there are essentially
less options for the parameters of the curve.

Looking at Equation \ref{equ:plots:reducedChiSquared} we can now make some
general statements about what the $\chi_\nu^2$ for a curve of best fit
indicates. We see that $\chi_\nu^2$ becomes very large, $\chi_\nu^2
\gg 1$, when either the values for $\sigma_i$ are very small, or the
differences between the observed and expected results are very
large. In the first case the experimenter has underestimated the
uncertainty but in the second case the fit just does not match the
data and most probably the theory attempting to describe the data is
incorrect. If the $\chi_\nu^2$ is very small, $\chi_\nu^2 \ll 1$,
either the uncertainties are very large, or the residuals are very
small. If the uncertainties are too large this just means the
experimenter overestimated the uncertainty for the
experiment. However, if the residuals are too small this means that
the data may be \term{over-fit}. This can happen when the number of
parameters for the fitting curve is very near the number of data
points, and so the best fit curve is able to pass very near most of
the points without actually representing a trend in the data.

The above discussion is a bit complicated but the bottom line is that
a $\chi_\nu^2$ near one usually indicates a good fit of the data. If
$\chi_\nu^2$ is too small either the fitting curve has too many
parameters or the uncertainty was overestimated. If $\chi_\nu^2$ is
too large the fitting curve is either incorrect, or the uncertainty
was underestimated. In Figure \ref{fig:plots:example} the $\chi_\nu^2$
value, displayed beneath the line of best fit, is given as
$0.29$. This value is less than $1$ but still close (same order of
magnitude) and so the fit is a good fit. Perhaps the uncertainty was
slightly overestimated for the data points, but not by much.

\begin{table}[h]
  \begin{center}
    \begin{tabular}{>$c<$>$c<$>$c<$>$c<$}
      \toprule
      x_i\mbox{ [Gauss]} & y_i\mbox{ [unitless]} & \sigma_i \mbox{
        [unitless]} & f(x_i) \mbox{ [unitless]}\\
      \midrule
      80  & 0.981 & 0.073 & 0.952 \\
      85  & 1.018 & 0.072 & 1.017 \\
      90  & 1.036 & 0.069 & 1.081 \\
      95  & 1.126 & 0.066 & 1.146 \\
      100 & 1.198 & 0.054 & 1.211 \\
      105 & 1.292 & 0.034 & 1.275 \\
      110 & 1.374 & 0.089 & 1.340 \\
      115 & 1.435 & 0.125 & 1.404 \\
      120 & 1.398 & 0.096 & 1.469 \\
      \bottomrule
    \end{tabular}
    \caption{Data points used in Figure \ref{fig:plots:example} with associated
      uncertainty. Column one gives $x_i$, the magnetic field $B$ in
      Gauss. Column two gives $y_i$, the $\beta\gamma$ of the
      electron, and column three its associated uncertainty,
      $\sigma_i$. Column four gives the value calculated for the point
      $x_i$ using the best fit parameters of Equation
      \ref{equ:plots:parameters} in Equation
      \ref{equ:plots:linear}.\label{tab:plots:data}}
  \end{center}
\end{table}

Now let us consider in more detail the fit of Figure
\ref{fig:plots:example}. The curve of best fit has $\chi_{\nu}^2 = 0.29$, which
we can calculate explicitly by using the best fit parameters of Figure
\ref{equ:plots:parameters}, the $x$ and $y$ values for each data point, and
their associated uncertainty. The $x_i$ and $y_i$ values can be read
from Figure \ref{fig:plots:example} and then $f(x_i) = mx_i+b$ calculated using
$m$ and $b$ from Equation \ref{equ:plots:parameters}. However, this can be a
bit tedious and so the numbers $x_i$, $y_i$, $\sigma_i$, and $f(x_i)$
are provided in Table \ref{tab:plots:data}.\footnote{Feel free to check
  them. It is not only a good exercise to just do the calculations,
  but also to check the author's work.}

We have $9$ data points so $N=9$. We are fitting with a line so $n=2$
and consequently $\nu = 6$. Using the numbers from Table
\ref{tab:plots:data} we can calculate $\chi_\nu^2$ from Equation
\ref{equ:plots:reducedChiSquared} for the line of best fit in Figure
\ref{fig:plots:example} in all of its gory detail.
\begin{equation}
  \begin{aligned}
    \chi^2_\nu &=&
    \frac{1}{6}&\left[\left(\frac{0.981-0.952}{0.073}\right)^2 +
      \left(\frac{1.018-1.017}{0.072}\right)^2 +
      \left(\frac{1.036-1.081}{0.069}\right)^2 \right. \\
    &&&+\left(\frac{1.126-1.146}{0.066}\right)^2 +
    \left(\frac{1.198-1.211}{0.054}\right)^2 +
    \left(\frac{1.292-1.275}{0.034}\right)^2 \\
    &&&\left. + \left(\frac{1.374-1.340}{0.089}\right)^2 +
      \left(\frac{1.435-1.404}{0.125}\right)^2 +
      \left(\frac{1.398-1.469}{0.096}\right)^2 \right]\\
    &= &\frac{1}{6}&\Bigl[(0.397)^2+(0.014)^2+(-0.652)^2
    +(-0.303)^2 +(-0.240)^2 \\
    &&&+( 0.500)^2 +( 0.382)^2 +( 0.248)^2
    +(-0.740)^2 \Bigr]\\
    &= &\frac{1}{6}&\Bigl[0.158+0.000+0.425+0.092+0.058+0.25\\
    &&&+0.146+0.062+0.548\Bigr]\\
    &= &&\frac{1.739}{6}\\
    &= &&0.290\\
  \end{aligned}
  \label{equ:plots:calculate}
\end{equation}
We can also check to ensure that the values for $m$ and $b$ from
Equation \ref{equ:plots:parameters} give the minimum $\chi_\nu^2$ by
changing the best fit values of $m$ and $b$ slightly. We do not
explicitly write out the calculations, but the results in Table
\ref{tab:plots:minimize} can be checked using the exact same method as
Equation \ref{equ:plots:calculate} but now using different $f(x_i)$ values
based on the change in $m$ and $b$.

\begin{table}[h]
  \begin{center}
    \begin{tabular}{>$l<$|>$c<$|>$c<$|>$c<$}
      \toprule
      & b-0.01 & b & b+0.01 \\
      \midrule
      m-0.01 & 401.6 & 393.8 & 386.0 \\
      m      & 0.329 & 0.290 & 0.329 \\
      m+0.01 & 386.0 & 393.8 & 401.6 \\
      \bottomrule
    \end{tabular}
    \caption{The reduced chi-squared for small perterbations around
      the best fit parameters given in Equation \ref{equ:plots:parameters}
      for Figure \ref{fig:plots:example}.\label{tab:plots:minimize}}
  \end{center}
\end{table}

\begin{figure}[h]
  \begin{center}
    \executeiffilenewer{Code/Plots/minimize.m}{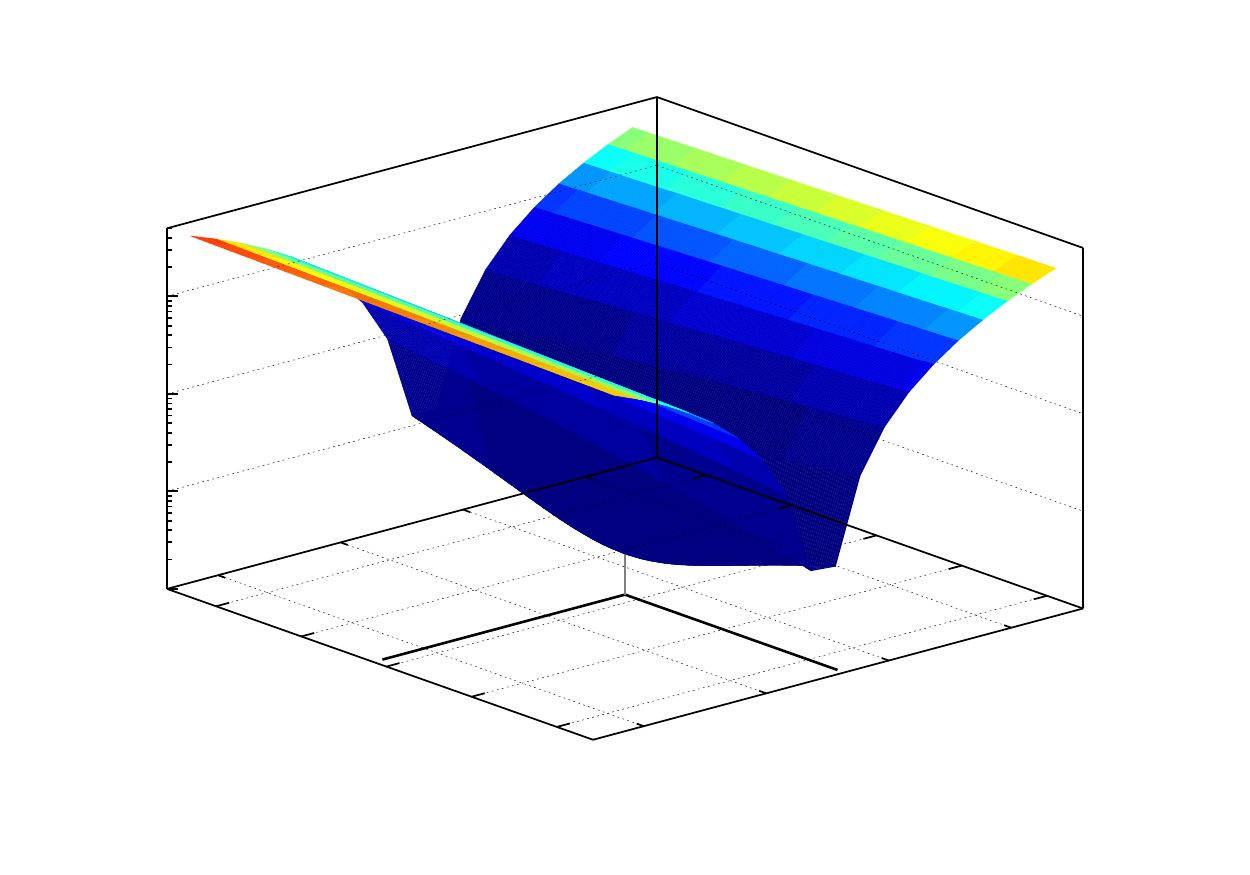}
  {octave --silent --eval "addpath([pwd(),'/Code']); 
    addpath([pwd(),'/Code/Plots']); minimize(1);"} 
  \setlength{\unitlength}{\columnwidth*\real{0.00013}}
  \input{Figures/Plots/minimizeLabel.tex}
    \caption{The surface plot for the minimization of the reduced
      chi-squared of Figure \ref{fig:plots:example}. Notice how the minimum
      occurs near $b=-0.08$ and $m=0.0125$, corresponding to the best
      fit parameters given in Equation
      \ref{equ:plots:parameters}.\label{fig:plots:minimize}}
  \end{center}
\end{figure}

From Table \ref{tab:plots:minimize} we can see that the parameters given in
Equation \ref{equ:plots:parameters} actually do minimize the $\chi_\nu^2$
for data points of Figure \ref{fig:plots:example}. More importantly we see how
much the values change over small variations in the parameters. Figure
\ref{fig:plots:minimize} shows the same behaviour of $\chi_\nu^2$ as Table
\ref{tab:plots:minimize} but now visually represents the change of
$\chi_\nu^2$ with a three dimensional surface. The lowest point on the
plot corresponds to the best fit parameters of Equation
\ref{equ:plots:parameters} and the lowest $\chi_\nu^2$ value of $0.29$.

If a marble was placed on the curve of Figure \ref{fig:plots:minimize} it
would roll to the lowest point. This is exactly how the minimum
reduced chi-squared for a line of best fit is found.\footnote{The
  method for minimizing $\chi_\nu^2$ varies from numerical package to
  numerical package, but most use what is known as the
  Levenberg-Marquardt algorithm which is a combination of the
  Gauss-Newton algorithm and the steepest gradient method.} For a
curve with more than two parameters (take for example the polynomial
of Equation \ref{equ:plots:polynomial}) the marble is just placed on an
$n$-dimensional surface. This is of course more difficult to
visualize, but the principal is exactly the same as for the line.

Looking back at Equation \ref{equ:plots:parameters} we see that there
are uncertainties associated with the best fit parameters. There is a
rigorous deriviation for these uncertainties, but it is also possible
to visualize them using the marble analogy above. We now take the
marble and place it at the minimum that we previously found by
dropping the marble, and push the marble with an equal amount of force
in both the $m$ and $b$ directions. The marble will begin oscillating
back and forth around the minimum point, but the oscillations will not
be the same in both the $m$ and $b$ directions. Figure
\ref{fig:plots:minimize} looks somewhat like a channel, and so we
expect to see rather large osciallations along the direction of $b$,
but very small ones along the direction of $m$. The size of these
oscillations correspond directly with the size of the uncertainties
associated with $m$ and $b$. Looking back at the uncertainties in
Equation \ref{equ:plots:parameters} we see that indeed we do have very
small uncertainty on $m$, and a much larger uncertainty on $b$!

\graphicspath{{Figures/Newton/}}

\chapter{Newton's Laws}\label{chp:newton}

In 1687 Isaac Newton published his {\it Philosophi\ae Naturalis
  Principia Mathematica} and revolutionized the field of physics (some
might even say create). Within the {\it Principia} Newton postulated
his famous three \term{laws of motion}.\footnote{{\it Newton's
    Principia : the mathematical principles of natural philosophy}
  translated by Daniel Adee from the original Latin. 1846. The full
  text can be downloaded
  \href{http://ia310836.us.archive.org/2/items/newtonspmathema00newtrich/newtonspmathema00newtrich.pdf}{here}
  (it is a very large pdf).}
\begin{enumerate}
\item {\it Every body perseveres in its state of rest, or of uniform
    motion in a right line, unless it is compelled to change that
    state by forces impressed thereon.}
\item {\it The alteration of motion is ever proportional to the motive
    force impressed; and is made in the direction of the right line
    in which that force is impressed.}
\item {\it To every action there is always opposed an equal reaction :
    or the mutual actions of two bodies upon each other are always
    equal, and directed to contrary parts.}
\end{enumerate}
Perhaps a more succinct and modern version will help convey the simplicity of
the laws.
\begin{enumerate}
\item An object at rest will remain at rest and an object in motion
  will remain in motion unless acted upon by an external force.
\item Force is equal to mass times acceleration.
\item For every action there is an equal and opposite reaction.
\end{enumerate}

The laws themselves are deceptively simple. It seems obvious that an
object at rest will remain at rest, yet before Newton, no one had ever
considered this as a physical law! Because people saw this behavior
every day they took it for granted, and did not even realize it was a
general rule. Oftentimes the quote ``the exception proves the rule''
is used. In the case of Newton's three laws there were no exceptions,
and so no one realized the rule.

These three laws make up what is known as \term{Newtonian mechanics}
and remained undisputed for over $200$ years until the advent of
\term{relativity} and \term{quantum mechanics}. Relativity describes
objects going very close to the speed of light, much faster than
anything seen in day-to-day life. Quantum mechanics describes objects
that are much smaller than day-to-day life, on the same order
magnitude as the size of the atom, or smaller. The combination of
these two fields of physics is \term{relativistic quantum
  mechanics}. The beauty of both relativity and quantum mechanics is
that the limits of these theories (very slow for relativity, or very
large for quantum mechanics) approach Newtonian mechanics.

One of the primary tools in Newtonian mechanics is the use of
\term{free-body diagrams}. In these diagrams all the relevant forces
for a system (usually involving some ridiculous combination of sliding
blocks and pulleys) are represented as force vectors, and the motion
of the system can then be calculated. These are also sometimes called
\term{force diagrams} and help us visually apply the three laws of
motion. Consider for example a block being pushed across a
frictionless plane. By the first law we know the block must be moving
because it is experiencing an external force. By the second law we
know the force being exerted on the block is equal to its mass times
its acceleration. And by the the third law we know the block is not
falling, as the frictionless plane is providing an equal and opposite
force countering the force of gravity on the block.

Sometimes free-body diagrams can become very complicated, and it is
simpler to use what is called \term{Lagrangian mechanics}. Here, a much
more mathematical approach is used to derive the \term{equations of
  motion} for the system, rather than visualizing the physical
reality. Because of this, free-body diagrams help provide a much more
intuitive approach to physical problems, while Lagrangian mechanics
sometimes provide a simpler method.

\section{Experiment}

Perhaps the second law of motion is the least intuitive of all the
laws. We know the first law is true because objects sitting on desks
or lab benches don't just get up and walk away unless some sort of
force is applied. The third law is clearly true or else when sitting
down, we would fall right through our chair. The second law however is
not quite so obvious, and so in the experiment associated to this
chapter, the second law is verified using two methods. In the first
method we vary the mass of the object, while in the second method we
vary the force acting on the object. The setup for the first method
consists of a cart moving along an almost frictionless track. Attached
to this cart is a string which runs over a pulley and is attached to a
small weight. Gravity pulls the weight downwards which subsequently
causes a tension in the string which acts on the cart.

\begin{figure}
  \begin{center}
    \subfigure[Pulley Method]{
      \executeiffilenewer{Figures/Newton/pulley.svg}
  {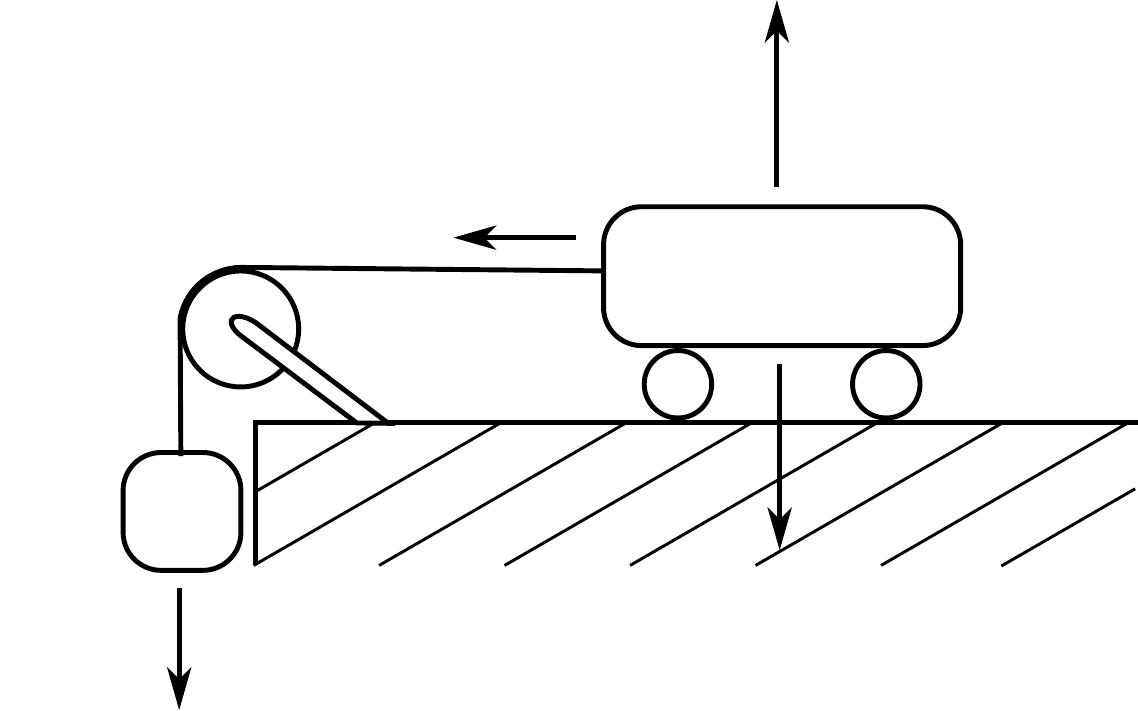}{inkscape-0.48pre1 -z -D --file=Figures/Newton/pulley.svg 
    --export-pdf=Figures/Newton/pulley.pdf --export-latex} \def\svgwidth{0.7\columnwidth}
  \input{Figures/Newton/pulleyLabel.tex}
      \label{fig:newton:pulley}}
    \subfigure[Ramp Method]{
      \executeiffilenewer{Figures/Newton/incline.svg}
  {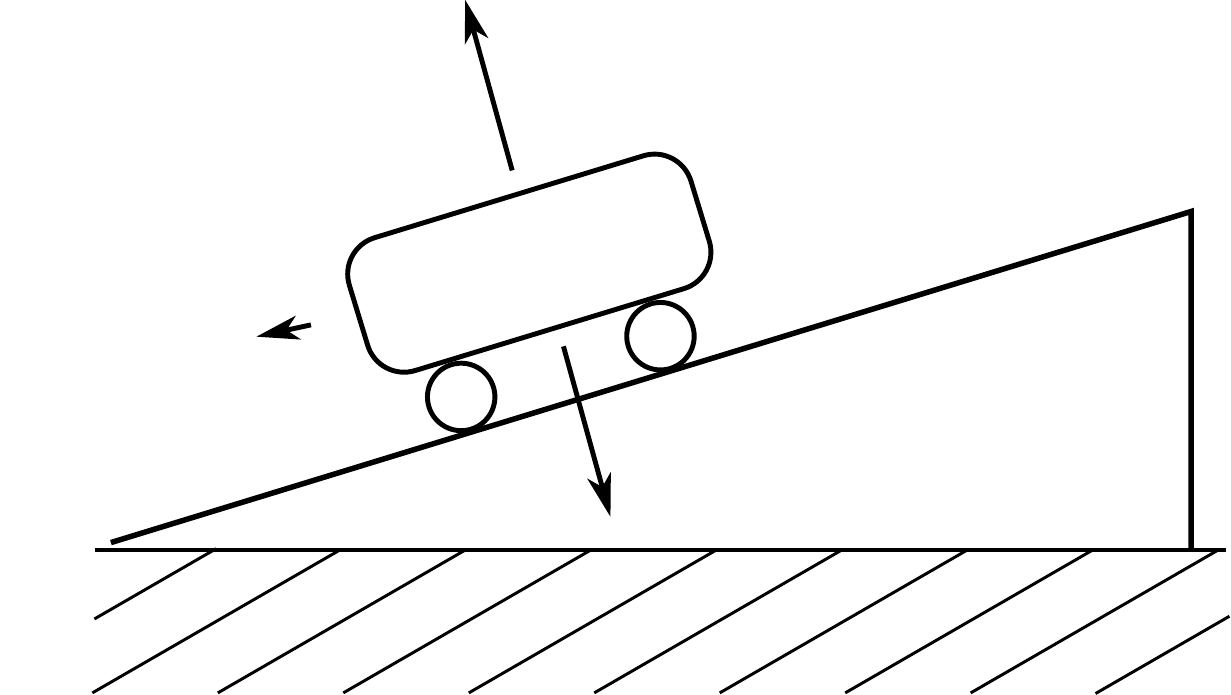}{inkscape-0.48pre1 -z -D --file=Figures/Newton/incline.svg 
    --export-pdf=Figures/Newton/incline.pdf --export-latex} \def\svgwidth{0.8\columnwidth}
  \input{Figures/Newton/inclineLabel.tex}
      \label{fig:newton:incline}}
    \caption{Free-body diagrams for the two methods used to verify
      Newton's second law in the experiment associated with this
      chapter.\label{equ:newton:force}}
  \end{center}
\end{figure}

The first step in approaching this problem is to draw a free-body
diagram, as is done in Figure \ref{fig:newton:pulley}. From this diagram we
can see that the tension on the string, $T$, must equal the force on
the hanging weight $m_w$ which is just $m_wg$. We also can see that
the only force acting on the cart is the tension, so by Newton's
second law we have $m_ca = m_wg$. Here $m_c$ is the mass of the
cart. From this we know the acceleration of the cart.
\begin{equation}
  a = \frac{m_wg}{m_c}
  \label{equ:newton:pulley}
\end{equation}
We see that if we increase the mass of the cart, the acceleration will
decrease. If we verify Equation \ref{equ:newton:pulley}, then we have
verified Newton's second law. We can do this by changing the mass of
the cart $m_c$ while applying the same force, and measuring the
acceleration of the cart.

With the second method, we roll the cart down an inclined plane as
shown in the free-body diagram of Figure \ref{fig:newton:incline}. Now, the
force acting on the cart is just,
\begin{equation}
  F = m_cg\sin\theta
\end{equation}
and so by varying $\theta$ we can change the force on the cart. From
this we can theoretically calculate the acceleration using Newton's
second law and compare this to our experimentally determined values.
\begin{equation}
  a = g\sin\theta
\end{equation}
Looking at this equation we see that if we plot $\sin\theta$ on the
$x$-axis and our measured $a$ on the $y$-axis we should obtain a
straight line. Furthermore, the slope of this straight line should be
$g$!
\graphicspath{{Figures/Momentum/}}

\chapter{Momentum}\label{chp:momentum}

Oftentimes the question ``what's the point, what can I actually do
with physics?''\footnote{Oftentimes with slightly stronger language.}
is asked by frustrated physics students (or non-physics students
forced to study physics). The answer is almost anything, although this
is not always readily apparent. However, one area where the use of
physics is clear is in the modeling of \term{kinematics} or the
interactions between objects. But how does modeling kinematics help
us? Everything from video game physics engines to crime scene
investigators use kinematics and the fundamental laws of physics to
recreate realistic physical realities.

Another extremely useful\footnote{Useful is highly dependent upon the
  eye of the beholder.} application of kinematics is the analysis of
movie scenes to determine if they are physically possible. Consider
the opening scene to the 2010 ``Star Trek'' movie where a young
Captain Kirk has stolen his step-father's vintage Corvette and is
driving through a corn field in Iowa. A police officer tries to pull
Kirk over, but he refuses and accidentally drives the Corvette off a
cliff during the ensuing pursuit. Luckily he manages to jump out of
the Corvette just in time and grabs the edge of the
cliff.\footnote{Sorry about the spoiler, I know everyone was hoping
  Kirk would die.} Is this scene physically possible?

First we need to make a few assumptions and rough estimates. We can
estimate the mass of the car to be approximately $1400$ kg, $m_c =
1400$, and the mass of Kirk to be around $55$ kg, $m_k = 55$. From an
earlier shot in the scene we know that the car (and Kirk) has been
traveling near $75$ mph or $33.5$ m/s, $v_c^i = 33.5$. After the
collision we know that Kirk has a velocity of zero as he hangs onto
the cliff face, $v_k^f = 0$.

By using \term{conservation of momentum} which requires the momentum in
a system to be conserved (the same before as it is after), we can find
the velocity of the car after Kirk jumps out.
\begin{equation}
  p_i = 33.5\times1400+33.5\times55 = v_c^f\times1400+0\times55
  ~~\Rightarrow~~ v_c^f = 34.8\mbox{ m/s} = 125\mbox{ km/h}
\end{equation}
The action of Kirk jumping out of the car increases the speed of the
car by $5$ km$/$h. But we can go even further. By \term{Newton's second
  law} we know that force is equal to mass times acceleration,
\begin{equation}
  F = m\vec{a} = \frac{\Delta\vec{p}}{\Delta t}
\end{equation}
which we can also write in terms of change of momentum. We already
know the change of momentum for Kirk, so to find the force all we need
is the time period over which the momentum changed. Let us assume he
managed to perform his entire jump within a second, so $\Delta t = 1$
s.
\begin{equation}
F_k = \frac{33.5\times55-0~\mbox{ kgm/s}}{1\mbox{ s}} = 1842.5\mbox{ N}
\end{equation}
The average adult can lift over a short period around $135$ kg with
their legs, equivalent to a force of $\approx 1350$ N. Assuming that
Kirk has the strength of an adult, we see that Kirk would miss by
about $500$ N and plunge to his death over the cliff.\footnote{Yes,
  this example was written shortly after having just watched ``Star
  Trek''. In all honesty this scene falls much closer to reality than
  quite a few other scenes from the movie and in terms of popular
  films is about as close to a realistic physical situation as you
  will see.}

\section{Collisions}

In kinematics there are two types of collisions, \term{inelastic} and
\term{elastic}. In an inelastic collision kinetic energy is not
conserved, while in an elastic collision kinetic energy is
conserved. Furthermore, there are two types of inelastic collisions,
\term{total} and \term{partial}. At the end of a totally inelastic
collision all of the individual objects are moving with the same
velocity. In a partially inelastic collision, this does not
occur. Explosions, similar to the example given above, correspond to
the reverse of a totally inelastic collision. We begin with a single
object moving at a single velocity, and end with a multitude of
objects at different velocities (in the example above, Kirk and the
car). Conservation of energy still occurs for inelastic collisions;
the lost kinetic energy is usually converted into potential energy
through the deformation of the colliding objects.

\begin{figure}[h]
  \begin{center}
    \subfigure[Initial]{
      \executeiffilenewer{Figures/Momentum/collisionInitial.svg}
  {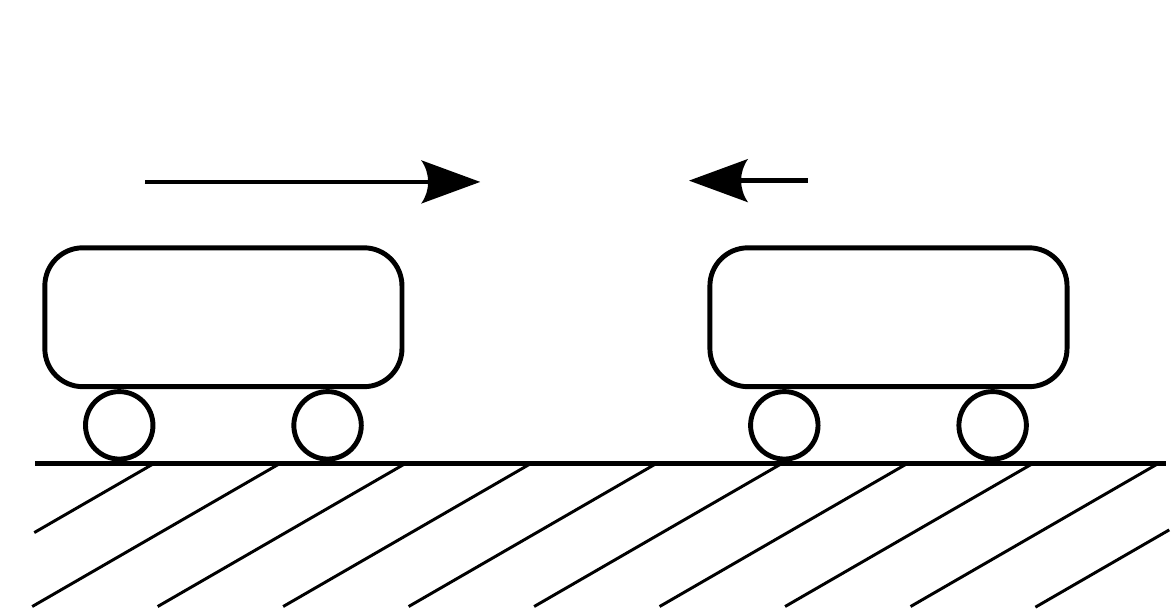}{inkscape-0.48pre1 -z -D --file=Figures/Momentum/collisionInitial.svg 
    --export-pdf=Figures/Momentum/collisionInitial.pdf --export-latex} \def\svgwidth{0.4\columnwidth}
  \input{Figures/Momentum/collisionInitialLabel.tex}
      \label{fig:momentum:collisionInitial}}
    \subfigure[Final]{
      \executeiffilenewer{Figures/Momentum/collisionFinal.svg}
  {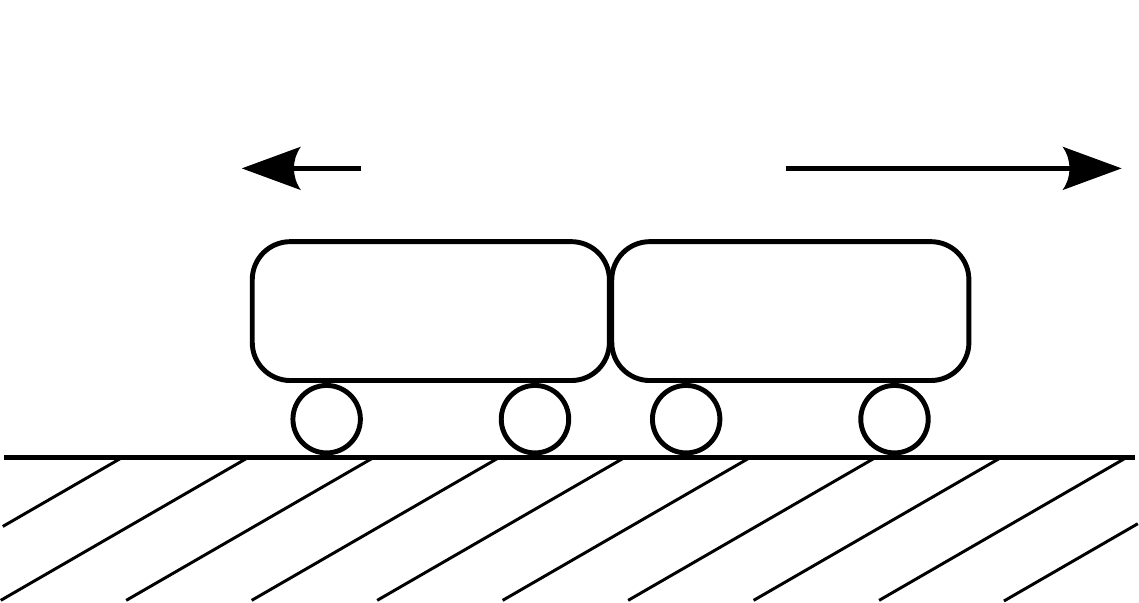}{inkscape-0.48pre1 -z -D --file=Figures/Momentum/collisionFinal.svg 
    --export-pdf=Figures/Momentum/collisionFinal.pdf --export-latex} \def\svgwidth{0.4\columnwidth}
  \input{Figures/Momentum/collisionFinalLabel.tex}
      \label{fig:momentum:collisionFinal}}
    \caption{An example collision of two objects, $A$ and $B$ with
      conservation of momentum.\label{fig:momentum:collision}}
  \end{center}
\end{figure}

As an example let us consider a simple setup, shown in Figure
\ref{fig:momentum:collision}. Object $B$ of mass $m_B$ is moving with an
initial velocity $v_B^i$ on a frictionless track when object $A$ of
mass $m_A$ is fired with an initial velocity of $v_A^i$ at object
$B$. After the collision object $A$ has final velocity $v_A^f$ and
object $B$ has final velocity $v_B^f$. Note that in Figure
\ref{fig:momentum:collision} the velocities are not necessarily in the
direction indicated, but are drawn as an example.

If the collision between object $A$ and object $B$ is elastic, kinetic
energy is conserved.
\begin{equation}
  \frac{1}{2}m_A\left(v_A^i\right)^2 + \frac{1}{2}m_B\left(v_B^i\right)^2 =
  \frac{1}{2}m_A\left(v_A^f\right)^2 +
  \frac{1}{2}m_B\left(v_B^f\right)^2
  \label{equ:momentum:elasticKinetic}
\end{equation}
Here we have one equation and six unknowns, $m_A$, $m_B$, $v_A^i$,
$v_B^i$, $v_A^f$, and $v_B^f$. Normally, however, we know the
conditions before the collision, and want to determine the result. In
this scenario we would then know $m_A$, $m_B$, $v_A^i$, and $v_B^i$,
but still have two unknowns, $v_A^f$ and $v_B^f$. With two unknowns we
need two equations to uniquely determine the solution, and so we use
conservation of momentum to impose our second relation.
\begin{equation}
  m_Av_A^i+m_Bv_B^i = m_Av_A^f+m_Bv_B^f
  \label{equ:momentum:elasticMomentum}
\end{equation}
We can simultaneously solve the system of equations above using
substitution to find values for $v_A^f$ and $v_B^f$.
\begin{equation}
  \begin{aligned}
    v_A^f &= \frac{m_Av_A^i+m_B\left(2v_B^i-v_A^i\right)}{m_B+m_A},
    ~~~v_A^f = v_A^i\\
    v_B^f &= \frac{m_Bv_B^i+m_A\left(2v_A^i-v_B^i\right)}{m_B+m_A},
    ~~~v_B^f = v_B^i\\
    \label{equ:momentum:elasticCollision}
  \end{aligned}
\end{equation}

Because of the quadratic terms in Equation \ref{equ:momentum:elasticKinetic} we
have two solutions for both $v_A^f$ and $v_B^f$. Physically the first
solution corresponds to when the two objects collide with each other
after some time period. The second set of trivial equations, where the
initial velocities of the objects match their final velocities,
corresponds to when the objects do not collide. This can occur when
object $A$ is fired at $B$ with a velocity slower than $B$, when $A$
is fired away from $B$ and $B$ is moving at a velocity slower than
$A$, or when $A$ is fired in the opposite direction of $B$.

Let us now consider a real world example of an elastic
collision. Elastic collisions are somewhat rare, but in pool, the
interactions between the pool balls are almost completely elastic. Of
course the balls roll, which adds another level of complexity (now
angular momentum and kinetic energy must be conserved as well) but for
now let us assume that the balls slide and do not experience
friction. Consider hitting the cue ball (object $A$) at the eight ball
(object $B$) in the final shot of a game. We know that the masses of
the two balls are around $0.10$ kg and that the eight ball begins at
rest. Let us also assume that we know the speed of the cue ball to be
$10\mbox{ m/s}$.\footnote{These numbers, while approximate, are
  relatively close to actual values that would be measured in a pool
  game.}

Plugging all these quantities into Equation
\ref{equ:momentum:elasticCollision}, we can determine the final velocities of
both the cue ball and the eight ball.
\begin{equation}
  \begin{aligned}
    v_A^f &= \frac{0.1\times10+0.1\left(2\times0-10\right)}{0.1+0.1} &&=
    \frac{1-1}{0.2} &&= 0\mbox{ m/s}\\
    v_B^f &= \frac{0.1\times0+0.1\left(2\times10-0\right)}{0.1+0.1} &&= 
    \frac{2}{0.2} &&= 10\mbox{ m/s}\\
  \end{aligned}
\end{equation}
From the results above, we see that after the collision the cue ball
is completely at rest while the eight ball is now moving with the
initial velocity of the cue ball, and has hopefully gone into the
pocket.

Let us return to the setup of Figure \ref{fig:momentum:collision} and assume a
completely inelastic collision now. From this we know that objects $A$
and $B$ are moving at the same final velocity.
\begin{equation}
  v_A^f = v_B^f
\end{equation}
Again we impose conservation of momentum from Equation
\ref{equ:momentum:elasticMomentum}, and solve for the final velocities in terms
of the initial velocities and the mass of the objects.
\begin{equation}
  v_A^f = v_B^f = \frac{m_Av_A^i+m_Bv_B^i}{m_A+m_B}
  \label{equ:momentum:inelasticCollision}
\end{equation}
We can use the pool example above to again explore these results, but
replace the eight ball with a lump of clay that sticks to the cue ball
after the collision.
\begin{equation}
  v_A^f = v_B^f = \frac{0.1\times10+0.1\times0}{0.1+0.1} = 5\mbox{ m/s}
\end{equation}
Both the clay and the cue ball are traveling at half the initial
velocity of the cue ball. Unlike the elastic collision, the cue ball
has a positive non-zero velocity, and so the shot would be a
scratch. If collisions in pool were inelastic and not elastic, the
game would be nearly impossible; every straight on shot would be a
scratch!

The methods and equations outlined above are valid for any type of
\term{two-body collision}. While the examples were given in one
dimension (along the $x$-axis only), they can be applied to three
dimensional problems as well. Now the velocities are given by vectors,
but these can be split into their components, $v_x$, $v_y$, and $v_z$,
and Equations \ref{equ:momentum:elasticCollision} and
\ref{equ:momentum:inelasticCollision} can be applied to each component
individually.

\section{Experiment}

In deriving Equations \ref{equ:momentum:elasticCollision} and
\ref{equ:momentum:inelasticCollision} we made the assumption that
momentum is conserved without any basis to do so. The goal of this
experiment associated with this lab is to verify the theory outlined
above by experimentally confirming conservation of momentum.

The apparatus used in this lab consists of two carts of equal mass
that move on a near frictionless track. The idea is that these carts
can be collided with velocities and masses measured before and after
the collision so that the two momentums can be compared. The
experiment is split into three investigations. In the first
investigation cart $A$ is fired at cart $B$ and the initial and final
velocities of the carts are measured. In the second investigation,
cart $A$ is still fired at $B$, but now the mass of $B$ is changed. In
the final investigation cart $A$ and $B$ are placed back to back, and
fired apart in an ``explosion''.

One important point to remember for this experiment is that the
collisions between the carts are neither elastic, nor fully
inelastic. This is because the carts have both velcro and magnets on
them which dampen the collisions. Consequently, the behaviors of the
carts will not resemble the pool ball examples above, but be a
combination of the two.

\begin{figure}[h]
  \begin{center}
    \executeiffilenewer{Code/Momentum/velocity.m}{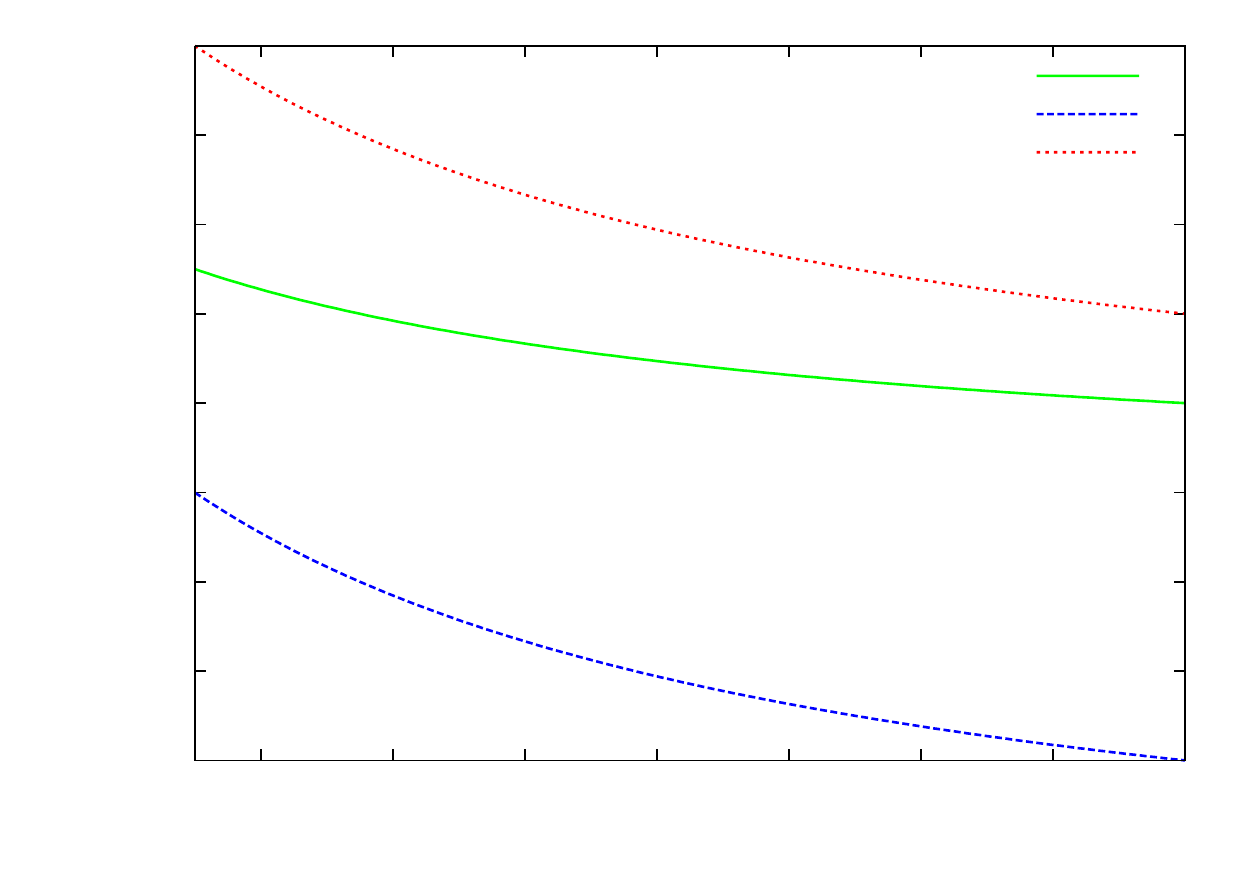}
  {octave --silent --eval "addpath([pwd(),'/Code']); 
    addpath([pwd(),'/Code/Momentum']); velocity(1);"} 
  \setlength{\unitlength}{\columnwidth*\real{0.00013}}
  \input{Figures/Momentum/velocityLabel.tex}
    \caption{Theoretical final velocities for carts $A$ and $B$ in
      elastic (red and blue curves) and fully inelastic (green curve)
      collisions. The initial conditions for the collision are $v_A^i
      = 0.5$ m/s, $v_B^i = 0$ m/s, and $m_A = 0.5$ kg which are
      similar to values obtained in the
      experiment.\label{fig:momentum:velocity}}
  \end{center}
\end{figure}

In the second investigation (where the mass of $B$ is changed), the
lab manual requests for plots to be be made of $v_A^f$ against $m_B$
and $v_B^f$ against $m_B$. In Figure \ref{fig:momentum:velocity} these
plots have been made for the elastic and fully inelastic cases using
Equations \ref{equ:momentum:elasticCollision} and
\ref{equ:momentum:inelasticCollision}. Here the solid green line gives
the velocity of both cart $A$ and $B$ after a fully inelastic
collision. This curve is proportional to $1/m_B$ as expected from
Equation \ref{equ:momentum:inelasticCollision}. For an elastic
collision the blue dashed line gives $v_A^f$ and the red dotted line
gives $v_B^f$. Notice that for any $m_B > m_A$, cart $A$ bounces
backwards off cart $B$.

Because the collisions between the carts in this lab are partially
inelastic, the plots from experiment will not match Figure
\ref{fig:momentum:velocity} but fall between the two curves (assuming same
$v_A^i$ which will not be the case, but should be close). The $v_A^f$
data points should fall below the green curve but above the blue
curve, while the $v_B^f$ data points should fall above the green curve
but below the red curve. It is important to understand that these
plots when made in the lab should not be fitted with any function
because we do not have a theoretical form for the fitting curve.
\graphicspath{{Figures/Rotation/}}

\chapter{Rotation}\label{chp:rotation}

Linear motion describes many of the interactions around us in the
physical world. Objects under the influence of gravity fall linearly,
cars normally accelerate in a straight line, and the collisions of
pool balls are most easily visualized and solved within a linear
system. But sometimes, linear motion simply is not sufficient. A
perfect example of this can be found in the example of compound
pendulum, given in Chapter \ref{chp:pendulum}. Here, attempting to
find the period of a compound pendulum using linear motion is
exceptionally complicated. Instead, we introduce a new force system,
rotational motion, to help solve the problem.

Rotational motion is everywhere. Every time a door is opened,
principals of rotational motion are exhibited. Cars accelerating
around turns, satellites orbiting the earth, wound clocks, weather
systems, all of these phenomena are more intuitively described in a
rotating system, rather than a linear system. More examples of systems
that are best solved in a rotating frame are readily available in
almost any introductory physics text book. Further details and
development of rotating systems can be found in
\href{http://www.amazon.com/Introduction-Mechanics-Daniel-Kleppner/dp/0070854238/ref=sr_1_1?ie=UTF8&s=books&qid=1258283716&sr=8-1-spell}{\it
  An Introduction to Mechanics} by Kleppner and Kolenkow as well as
the MIT OpenCourseWare materials for
\href{http://ocw.mit.edu/OcwWeb/Physics/8-012Fall-2008/CourseHome/index.htm}{8.012}
as taught by Adam Burgasser.

\section{Coordinate System}

But what exactly do we mean by a rotating system, or rotational
mechanics? In linear motion we describe the interactions of objects
through forces using the Cartesian coordinate system. Using this
coordinate system is convenient because forces, accelerations, and
velocities are oftentimes in straight lines. In rotational mechanics
we don't change any of the fundamental laws of physics, we just change
our coordinate system. We then transform the relations we have for
linear motion to this new rotational frame. Perhaps the easiest way to
understand this is to make a direct comparison between the quantities
used to describe a system in linear and rotational frames.

\begin{figure}
  \begin{center}
    \subfigure[Cartesian Coordinates]{
      \executeiffilenewer{Figures/Rotation/cartesian.svg}
  {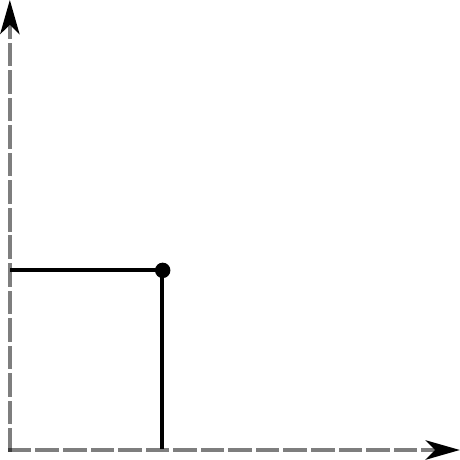}{inkscape-0.48pre1 -z -D --file=Figures/Rotation/cartesian.svg 
    --export-pdf=Figures/Rotation/cartesian.pdf --export-latex} \def\svgwidth{4cm}
  \input{Figures/Rotation/cartesianLabel.tex}
      \label{fig:rotation:cartesian}}
       \hspace{2cm}
    \subfigure[Polar Coordinates]{
      \executeiffilenewer{Figures/Rotation/polar.svg}
  {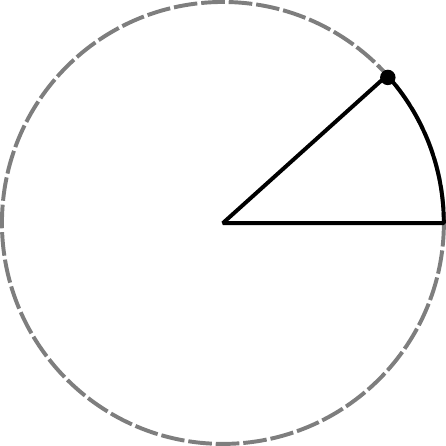}{inkscape-0.48pre1 -z -D --file=Figures/Rotation/polar.svg 
    --export-pdf=Figures/Rotation/polar.pdf --export-latex} \def\svgwidth{4cm}
  \input{Figures/Rotation/polarLabel.tex}
      \label{fig:rotation:polar}}
  \end{center}
  \caption{The Cartesian and polar coordinate systems used in linear
    and rotational motion.\label{fig:rotation:coordinates}}
\end{figure}

To begin, we must introduce the two coordinate systems. In linear
motion \term{Cartesian coordinates} are used; the location of an
object or the components of a vector are given by $x$, and $y$ as
shown in Figure \ref{fig:rotation:cartesian}. In the simplest
one-dimensional case (such as the examples in Chapter
\ref{chp:momentum}), \term{linear position} is given by the variable
$x$. In rotational motion \term{polar coordinates} are used (or in the
three dimensions, spherical coordinates) as shown in Figure
\ref{fig:rotation:polar}. Here the position of an object or the
direction of a vector is described by the variables $\theta$ (the
Greek letter theta) and $r$ where the units for $\theta$ are always
radians, and $r$ is a distance. In the one-dimensional case for
rotational motion, \term{angular position} is given by the variable
$\theta$. Consequently, we see that $x$ transforms into $\theta$ and
$y$ into $r$ for rotational motion.\footnote{Of course it is possible
  to also look at the three dimensional case, in which case spherical
  coordinates are used. However, as the variables used in spherical
  coordinates are not consistent across disciplines and can cause
  confusion, we stick to the two dimensional case for this
  discussion.}

In linear motion, the velocity or speed of an object is described by,
\begin{equation}
  \vec{v} = \frac{\Delta x}{\Delta t}
\end{equation}
where $\Delta x$ is change in position and $\Delta t$ is change in
time. In rotational motion, time remains the same, as changing the
coordinate system does not affect the passage of time. If we
substitute $\Delta \theta$ for $\Delta x$ we arrive at the formula for
\term{angular velocity} which is denoted by the Greek letter $\omega$
(spelled omega).
\begin{equation}
  \omega = \frac{\Delta \theta}{\Delta t}
\end{equation}
Angular velocity is given in units of radians per unit
time.\footnote{We drop the vector sign for angular velocity as we
  always know it is either in the plus or minus $\theta$ direction.}
Alternatively, angular velocity can also be thought of as number of
rotations per second, and is sometimes also called \term{angular
  frequency}.\footnote{Angular frequency is the magnitude of angular
  velocity.} For the purposes of this chapter, we will always refer to
$\omega$ as angular velocity.

Now that we have position and speed for an object in rotational
motion, the only remaining quantity we need to describe the motion of
an object is acceleration. In linear mechanics,
\begin{equation}
  \vec{a} = \frac{\Delta \vec{v}}{\Delta t}
\end{equation}
where $\Delta \vec{v}$ is change in velocity. Again, we just replace
linear quantities with rotational quantities.
\begin{equation}
  \alpha = \frac{\Delta \omega}{\Delta t}
\end{equation}
Here, \term{angular acceleration} is represented by the Greek letter
$\alpha$ (spelled alpha).\footnote{Again, we drop the associated
  vector sign as we can express direction with positive and negative
  values.} The units of $\alpha$ are just radians per second per
second. Now we have a full arsenal of quantities to describe motion in
a rotational system.

\section{Moment of Inertia}

Before we are able to describe physical laws with rotational motion,
we need to introduce a rotational quantity analogous to mass. In a
linear system mass is just given by the quantity $m$. In a rotational
system, the equivalent of mass is given by \term{moment of inertia}
oftentimes denoted by the letter $I$. The moment of inertia for an
object is defined as,
\begin{equation}
  I \equiv \int r^2\,dm
  \label{equ:rotation:inertia}
\end{equation}
where $dm$ is an infinitesimal mass, and $r$ the distance of that
infinitesimal mass from the axis around which the object is
rotating. Let us consider an example to illustrate this. A mass $m$ is
rotating around an axis $P$ at a distance $r$. If we assume the mass
is a point mass, the moment of inertia for this system is just $mr$.

\begin{figure}
  \begin{center}
    \subfigure[Center of Mass]{
      \executeiffilenewer{Figures/Rotation/horizontalRod.svg}
  {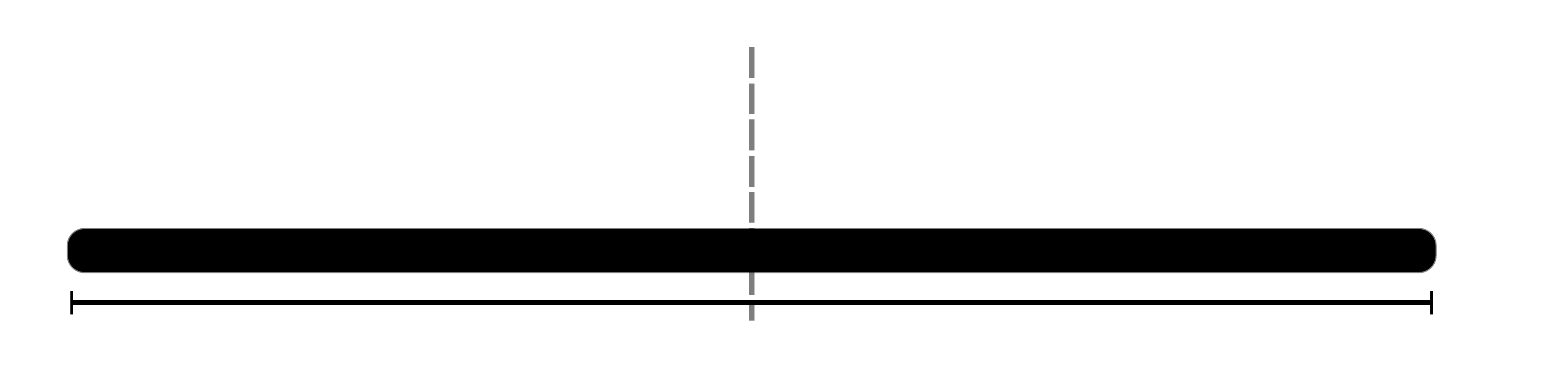}{inkscape-0.48pre1 -z -D --file=Figures/Rotation/horizontalRod.svg 
    --export-pdf=Figures/Rotation/horizontalRod.pdf --export-latex} \def\svgwidth{\columnwidth}
  \input{Figures/Rotation/horizontalRodLabel.tex}
      \label{fig:rotation:horizontalRod}}
    \subfigure[End]{
      \executeiffilenewer{Figures/Rotation/parallelAxis.svg}
  {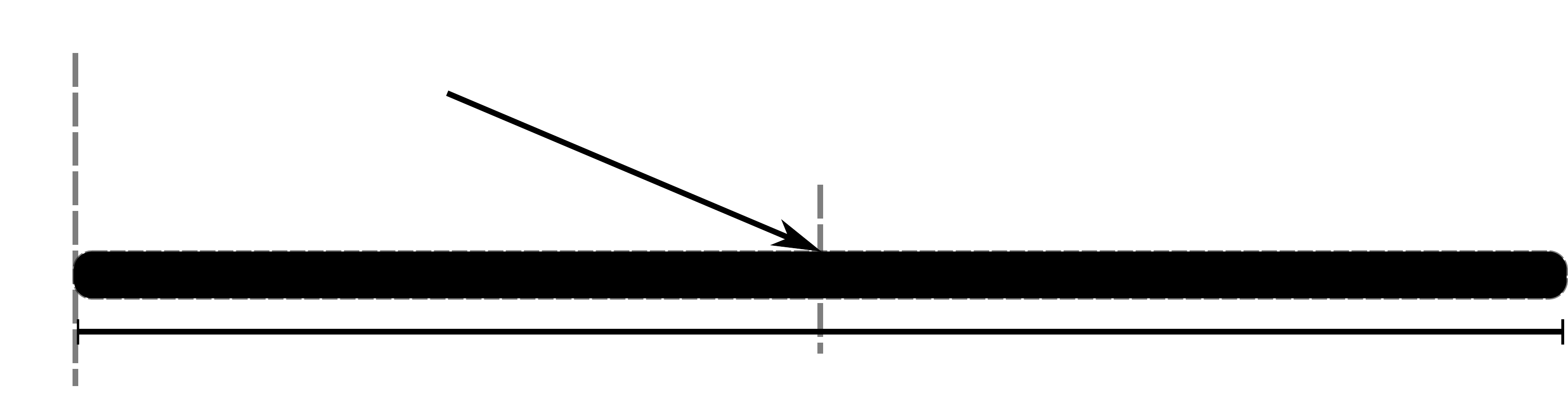}{inkscape-0.48pre1 -z -D --file=Figures/Rotation/parallelAxis.svg 
    --export-pdf=Figures/Rotation/parallelAxis.pdf --export-latex} \def\svgwidth{0.9\columnwidth}
  \input{Figures/Rotation/parallelAxisLabel.tex}
      \label{fig:rotation:parallelAxis}}
  \end{center}
  \caption{The moments of inertia for a rod rotating about its center
    of mass and rotating about its end.\label{fig:rotation:inertia}}
\end{figure}

Now let us consider a slightly more difficult example. In Figure
\ref{fig:rotation:horizontalRod} a rod with length $\ell$ is rotating about its
center of mass horizontally. We can then write,
\begin{equation}
  I = \int_{-\ell/2}^{\ell/2} r^2\,dm = \int_{-\ell/2}^{\ell/2} r^2\mu\,dr
\end{equation}
where we assume the rod has a uniform linear density of $\mu$ so $dm =
\mu\,dr$. Taking this integral we obtain,
\begin{equation}
  I = \left[\frac{\mu r^3}{3}\right]_{-\ell/2}^{\ell/2} =
  \frac{2\mu}{3}\left(\frac{\ell^3}{8}\right) = \frac{\mu \ell
    \ell^2}{12} = \frac{m\ell^2}{12}
  \label{equ:rotation:rodMiddle}
\end{equation}
where in the final step we realize that the mass of the rod, $m$, is
equal to $\mu\ell$.

There are two very important properties to remember when dealing with
moments of inertia. The first is that moments of inertia can be added
if the objects are rotating about the same axis. The second property
is the \term{parallel axis theorem}. This theorem states that if an
object has a moment of inertia $I_\mathrm{cm}$ around its center of
mass, then the moment of inertia for that object when rotating around
an axis distance $r$ from the center of mass is just its center of
mass moment of inertia plus the the objects mass times $r$ squared.
\begin{equation}
  I = I_\mathrm{cm}+mr^2
  \label{equ:rotation:parallelAxis}
\end{equation}

Figure \ref{fig:rotation:parallelAxis} demonstrates the use of the parallel
axis theorem for the same rod of Figure \ref{fig:rotation:horizontalRod}. Now
the rod is rotating about one of its ends, so $r = \ell/2$. This gives
us,
\begin{equation}
  I = \frac{m\ell^2}{12}+\frac{m\ell^2}{4} = \frac{m\ell^2}{3}
\end{equation}
where we have used Equations \ref{equ:rotation:rodMiddle} and
\ref{equ:rotation:parallelAxis}. Notice that in this case the moment of inertia
increases. This is because we have moved more of the mass of the rod
away from the axis. It is very important to realize that the moment of
inertia for an object can change without the mass of the object being
changed, merely the distance of the mass from the axis of rotation.

\section{Momentum, Energy, and Force}

In linear motion momentum is always conserved (as hopefully shown in
the momentum lab for this course) and so the same must apply to
rotational motion. In linear motion, momentum is just mass times
velocity.
\begin{equation}
  \vec{p} = m\vec{v}
\end{equation}
In rotational motion, we just replace mass with moment of inertia and
velocity with angular velocity.
\begin{equation}
  L = I\omega
\end{equation}
The resulting quantity, $L$, is called \term{angular momentum}. The
angular momentum for a system, just as the momentum for a system in
linear mechanics is always conserved. This is why a figure skater can
increase the speed of their spin by pulling in his or her arms. This
lowers the moment of inertia of the skater, and so to conserve angular
momentum, the angular velocity of the spin must increase.

Another important derived quantity in physics is kinetic
energy.\footnote{Kinetic energy is extremely useful for helping solve
  a variety of problems, especially in elastic collisions where it is
  conserved. See Chapter \ref{chp:momentum} for more detail.}
\begin{equation}
  K_t = \frac{1}{2}m\vec{v}^2
\end{equation}
The subscript $t$ on the $K$ stands for translational, which is just
normal linear kinetic energy. In rotational motion we have
\term{angular kinetic energy},
\begin{equation}
  K_r = \frac{1}{2}I\omega^2
\end{equation}
where the subscript $r$ stands for rotational. The total energy of a
system, whether translational kinetic energy, angular kinetic energy,
or potential energy is always conserved.

One final piece of the puzzle is still missing, and that is force. In
linear motion,
\begin{equation}
  \vec{F} = m\vec{a}
\end{equation}
by Newton's second law. In rotational motion a new quantity,
\term{torque} (represented by the Greek letter $\tau$), is
used.\footnote{Torque is a vector, but for the purpose of this lab it
  is presented just as a scalar.}
\begin{equation}
  \tau = I\alpha
  \label{equ:rotation:newtonRotational}
\end{equation}
If a force is applied to a rotating system the torque on the system
is,
\begin{equation}
  \tau = rF\sin\phi
  \label{equ:rotation:torque}
\end{equation}
where $\phi$ is the angle at which the force is applied, and $r$ is
the distance from the axis of rotation.\footnote{In full vector
  notation, $\vec{\tau} = \vec{r}\times\vec{F}$.}

\section{Comparison}

Quite a few new terms have been introduced in the sections above,
which can be very daunting for someone who has never seen rotational
motion before. The important point to remember is that for every
quantity and law in linear motion, the same exact law or quantity is
available in rotational motion. To help, Table \ref{tab:rotation:comparison}
summarizes all the relations discussed above between linear and
rotational motion.

\begin{table}
  \footnotesize
  \begin{center}
    \begin{tabular}{l|l|l|l|l}
      \toprule
      \multicolumn{2}{c|}{\bf Linear Motion} &  &
      \multicolumn{2}{c}{\bf Rotational Motion}\\
      {\bf Quantity} & {\bf Symbol} & \multicolumn{1}{c|}{\bf Relation} & {\bf
        Quantity} & {\bf Symbol} \\
      \midrule
      distance        & $x$ & & angular distance & $\theta$ \\
      velocity        & $\vec{v}$ & & angular
      velocity & $\omega$ \\
      acceleration    & $\vec{a}$ & & angular
      acceleration & $\alpha$ \\
      mass            & $m$ & $I = \int r^2\,dm$ & moment of inertia & $I$ \\
      momentum        & $\vec{p} = m\vec{v}$ & $L = rp\sin\phi$ & angular
      momentum & $L = I\omega$ \\ 
      kinetic energy  & $K_t = \frac{1}{2}m\vec{v}^2$ &  & angular kinetic
      energy & $K_r = \frac{1}{2}I\omega^2$ \\
      force           & $\vec{F} = m\vec{a}$ & $\tau = rF\sin\phi$ &
      torque & $\tau = I\alpha$ \\
      \bottomrule
    \end{tabular}
    \caption{A summary of useful physical quantities used to describe
      a system in linear motion and rotational motion. The middle
      column gives the relation between the linear and rotational
      quantities.\label{tab:rotation:comparison}}
  \end{center}
\end{table}

The equations of motion for an object moving under constant
acceleration in linear motion can also be recast into rotational
motion using the quantities discussed in the previous sections and
summarized in Table \ref{tab:rotation:comparison}. To do this, a simple
substitution is made between linear quantities and rotational
quantities with the results given in Table \ref{tab:rotation:motion}.

\begin{table}
  \begin{center}
    \begin{tabular}{l|l}
      \toprule
      {\bf Linear Motion} & {\bf Rotational Motion} \\
      \midrule
      $x = v_0t + \frac{at^2}{2}$ & $\theta = \omega_0t + \frac{\alpha
        t^2}{2}$ \\
      $x = vt-\frac{at^2}{2}$ & $\theta = \omega t-\frac{\alpha
        t^2}{2}$ \\
      $x = \frac{1}{2}\left(v_0+v\right)t$ & $\theta =
      \frac{1}{2}\left(\omega_0+\omega\right)t$ \\
      $v = v_0 + at$ & $\omega = \omega_0 + \alpha t$ \\
      $v = \sqrt{v_0^2+ax}$ & $\omega =
      \sqrt{\omega_0^2+\alpha\theta}$ \\
      $a = \frac{v-v_0}{t}$ & $\alpha = \frac{\omega-\omega_0}{t}$ \\
      \bottomrule
    \end{tabular}
    \caption{A summary of the equations of motion for an object moving
      under constant acceleration in both linear and rotational
      motion. Here, $v_0$ and $\omega_0$ indicate initial velocity and
      initial angular velocity respectively.\label{tab:rotation:motion}}
  \end{center}
\end{table}

\section{Periodic Harmonic Motion}

A defining characteristic of rotational motion is harmonic
motion. Consider a rod rotating slowly at a constant angular velocity,
but instead of viewed from above or below, viewed from profile. As the
rod rotates, from the experimenter's view, the length of the rod will
appear to grow and shrink periodically. When the rod is parallel with
the experimenter its apparent length will be its actual length,
$\ell$. When at an angle of $45^\circ$ with the experimenter the rod
will appear to have a length of $\ell/\sqrt{2}$ and when perpendicular
to the experimenter, the rod will appear to have a length of
zero. Plotting the relative length of the rod as observed by the
experimenter versus time will yield a sine wave, clearly periodic
harmonic motion.

Of course the pendulum is also another example of harmonic motion,
most naturally understood by using rotational
motion.\footnote{Understanding the basics of rotational motion is
  necessary for understanding the derivation of the period of a
  compound pendulum, as is done in Chapter \ref{chp:pendulum}.}
Another system, analogous to the pendulum (and to a simple linear
spring oscillator) is the torsion spring. With a simple linear
oscillator, force is related to displacement by Hooke's law.
\begin{equation}
  F = -kx
\end{equation}
Now we simply substitute the rotational quantities discussed earlier
into Hooke's law to obtain a rotational motion form.
\begin{equation}
  \tau = -\kappa x
\end{equation}
Here, $\kappa$ is the torque constant, the rotational version of the
linear force constant, $k$, with units traditionally given in
$\mathrm{Nm/radian}s$. The period of oscillation for a harmonic
oscillator governed by Hooke's law is given by\footnote{See Chapter
  \ref{chp:pendulum} for a derivation of the period for a simple
  harmonic oscillator.},
\begin{equation}
  T = 2\pi\sqrt{\frac{m}{k}}
\end{equation}
where $T$ is period given in units of time. Using the rotational
analogue of Hooke's law, we can find the period of a torsion spring by
substitution.
\begin{equation}
  T = 2\pi\sqrt{\frac{I}{\kappa}}
\end{equation}
This relation is very useful for finding the moment of inertia for
complex objects, where the moment of inertia can not be calculated
analytically. By placing the object on a torsion spring with a known
torque constant, the period of oscillation can be timed, and the
moment of inertia solved for.
\begin{equation}
  I = \frac{\kappa T^2}{4\pi^2}
\end{equation}

\section{Experiment}

One of the best ways to begin acquiring an intuitive feel for
rotational motion problems is to observe rotational motion in the lab
setting. The experiment associated with this chapter provides two
investigations, one to explore rotational motion under constant
acceleration, and the second to explore periodic harmonic oscillations
from a torsion spring.

In the first investigation, to obtain constant angular acceleration on
a rotating object, a constant torque must be applied. Looking back at
Table \ref{tab:rotation:comparison} we can see that by applying a
constant force at a constant radius, we obtain a constant torque. Of
course one of the best ways to apply a constant known force is to take
advantage of the force due to gravity on an object.\footnote{We used
  this exact same method in the experiment associated with Chapter
  \ref{chp:momentum}.}

By applying a constant angular acceleration to an object we can verify
that our system for rotational motion is consistent. First we can
calculate out a value for the torque being applied to the system by
using Equation \ref{equ:rotation:torque} and the moment of inertia for the
system using Equation \ref{equ:rotation:inertia}. From these quantities we can
calculate a value for $\alpha$ using Equation
\ref{equ:rotation:newtonRotational}. Next we record angular position $\theta$
versus time. This relation should be governed by the first equation of
motion from Table \ref{tab:rotation:motion} if theory is correct. As we can
start the experiment with $\omega_0 = 0$ we can plot $\theta$ versus
$t^2$ and calculate $\alpha$ from the slope of the graph.

We know that angular velocity is just change in angular position over
change in time, and so from the $\theta$ versus time data we can
calculate out $\omega$.\footnote{What we are doing here is numerical
  differentiation using a method known as the ``finite difference''
  method. The specific method we use in the experiment associated with
  this chapter is equivalent to the ``midpoint rule'' oftentimes used
  for numerical integration.} We associate the calculated $\omega$
with the average time used to calculate the time difference. If the
fourth equation of Table \ref{tab:rotation:motion} holds, then we can
now plot $\omega$ versus $t$ and again calculate out $\alpha$ from the
slope of the plot. If our theory is correct, than the values for
$\alpha$ calculated using the three different methods should match.

The second investigation allows us to experimentally measure the
moment of inertia for the rod used in the previous investigation, but
more importantly validates the theory behind moment of inertia. This
portion of the lab consists of measuring the period for a rod with
masses on it in different configurations. From the period, the moment
of inertia for the rod and masses can be calculated and compared with
the theoretical moment of inertia for a rod calculated earlier in this
chapter.
\graphicspath{{Figures/Pendulum/}}

\chapter{Pendulum}\label{chp:pendulum}

Pendulums in today's modern age may seem insignificant, but less than
a century ago the simple pendulum was still the most accurate method
for keeping time on the planet.\footnote{Marrison, Warren. "The
  Evolution of the Quartz Crystal Clock". Bell System Technical
  Journal {\bf 27} 1948. pp. 510-588.} With the development of the
crystal quartz clock and now the use of nuclear sources, the pendulum
has been rendered obsolete as a time keeping mechanism, yet remains as
an integral part of most physics curriculums. This begs the question,
why is the pendulum so important in physics?

There are many answers to this question, but one answer dominates. The
simple pendulum is an experimentally demonstrable, yet theoretically
solvable example of simple harmonic motion accessible to most students
without requiring an advanced knowledge of differential equations but
can also be studied to a very advanced level of physics.\footnote{An
  excellent paper demonstrating the incredible depth of the pendulum
  problem is available from
  \href{http://fy.chalmers.se/~f7xiz/TIF080/pendulum.pdf}{Nelson and
    Olson}.} The combination of an analytically solvable theory with
experiment is very rare in physics. Most real world scenarios do not
have a theoretical solution as simple and beautiful as the solution to
the pendulum.

\section{Simple Harmonic Motion}

\begin{figure}[h]
  \begin{center}
    \executeiffilenewer{Figures/Pendulum/springDiagram.svg}
  {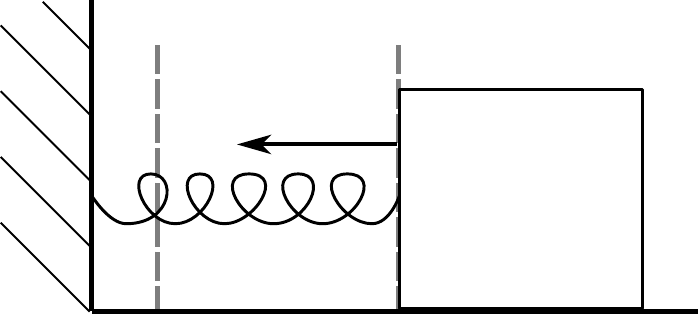}{inkscape-0.48pre1 -z -D --file=Figures/Pendulum/springDiagram.svg 
    --export-pdf=Figures/Pendulum/springDiagram.pdf --export-latex} \def\svgwidth{0.8\columnwidth}
  \input{Figures/Pendulum/springDiagramLabel.tex}
    \caption{Force diagram for simple harmonic motion for a block of
      mass $m$ attached to a spring with force constant
      $k$.\label{fig:pendulum:springDiagram}}
    \end{center}
\end{figure}

Understanding the theory behind the physics of the pendulum begins
with understanding \term{simple harmonic motion} (SHO). The mathematics
behind SHO can at times become a little complex, but the end result is
well worth the wait. To introduce SHO, let us consider the classic
example used, a spring attached to a mass $m$ at one end and an
immoveable wall at the other end as shown in Figure
\ref{fig:pendulum:springDiagram}.

The block of mass $m$ is displaced by a distance $x$ from
\term{equilibrium}, $x_0$. Here, equilibrium is defined as the
location where the potential energy of the system is minimized, or in
this case, zero. Physically, this is where the block is placed such
that the spring is not exerting a force on the block. The force on the
block is then given by \term{Hooke's law},
\begin{equation}
  F = -kx 
  \label{equ:pendulum:hooke}
\end{equation}
where $k$ is the \term{spring constant} and indicates how stiff the
spring is. A large $k$ means that the spring is very stiff, while a
small $k$ means the spring is easily compressed. Notice the negative
sign in the equation above, this is intentional. When the block is
pulled away from the wall, the spring exerts a \term{restoring force}
drawing it back towards the wall. Because we are working with a
one-dimensional example, we have dropped the vector symbols for both
force and displacement.

Using Newton's second law we can relate the mass and acceleration of
the block to the force exerted on the block by Hooke's law.
\begin{equation}
F = ma = m\frac{d^2x}{dt^2} = -kx
\label{equ:pendulum:sode}
\end{equation}
This is a second order differential equation of degree one, for which
the solution is well known. As this chapter is not about differential
equations we will assume we know the solution, but for those curioius
about the details there is an excellent book on differential
equations,
\href{http://www.amazon.com/exec/obidos/ASIN/013145773X/ref=nosim/mitopencourse-20}{\it
  Elementary Differential Equations} by Arthur Mattuck.\footnote{This
  is the textbook used for the MIT OpenCourseWare materials for
  \href{http://ocw.mit.edu/OcwWeb/Mathematics/18-03Spring-2006/CourseHome/index.htm}{18.03}
  also taught by Arthur Mattuck. This course is the most popular
  introductory math course at MIT, and with good reason, some of his
  more hilarious quotes are
  \href{http://web.mit.edu/marissav/www/mattuck_quotes.html}{here}.}
\begin{equation}
x(t) = A\cos\left(\omega t\right)
\label{equ:pendulum:position}
\end{equation}

The equation above is a particular solution to Equation
\ref{equ:pendulum:sode}, but suits our needs perfectly. We see that the value
$A$ represents the amplitude of the oscillations (or the maximum
displacement of the block), as the cosine function reaches a maximum
value of one, and can be rewritten as the initial displacement of the
block, or $x$. The value $\omega$ is the \term{angular frequency} at
which the block is oscillating. We can see that for every $2\pi$
seconds the block will have returned to its initial position.

Taking the first derivative of Equation \ref{equ:pendulum:position} with
respect to time we can find the velocity of the block as a function of
time, and by taking the second derivative we can find the acceleration
of the block also as a function of time.
\begin{equation}
  \begin{aligned}
    v(t) &= -\omega A\sin(\omega t)\\
    a(t) &= -\omega^2A\cos(\omega t)\\
  \end{aligned}
\end{equation}
Plugging acceleration and position back into Equation \ref{equ:pendulum:sode} we can
find $\omega$ in terms of $m$ and $k$.
\begin{equation}
  -m\omega^2A\cos(\omega t) = -kA\cos\left(\omega t\right)
  ~~\Rightarrow~~ m\omega^2 = k ~~\Rightarrow~~ \omega =
  \sqrt{\frac{k}{m}}
  \label{equ:pendulum:springConstant}
\end{equation}
From the angular frequency we can then find the \term{period} of
oscillation, or the time it takes the block to return to its initial
displacement distance, $x$.
\begin{equation}
  T = \frac{2\pi}{\omega} = 2\pi\sqrt{\frac{m}{k}}
  \label{equ:pendulum:period}
\end{equation}
This result has one very important result, the period is not dependent
on how far the block is initially displaced from equilibrium! This is
a very important result for pendulums as will be shown shortly.

\section{Simple Pendulum}

\begin{figure}[h]
  \begin{center}
    \subfigure[Simple Pendulum]{
      \executeiffilenewer{Figures/Pendulum/simplePendulum.svg}
  {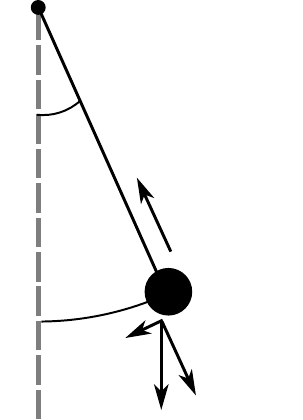}{inkscape-0.48pre1 -z -D --file=Figures/Pendulum/simplePendulum.svg 
    --export-pdf=Figures/Pendulum/simplePendulum.pdf --export-latex} \def\svgwidth{0.39\columnwidth}
  \input{Figures/Pendulum/simplePendulumLabel.tex}
      \label{fig:pendulum:simplePendulum}}
    \subfigure[Compound Pendulum]{
      \executeiffilenewer{Figures/Pendulum/compoundPendulum.svg}
  {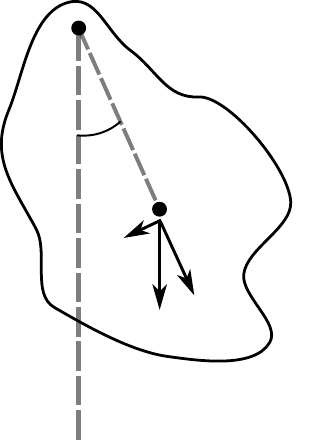}{inkscape-0.48pre1 -z -D --file=Figures/Pendulum/compoundPendulum.svg 
    --export-pdf=Figures/Pendulum/compoundPendulum.pdf --export-latex} \def\svgwidth{0.4\columnwidth}
  \input{Figures/Pendulum/compoundPendulumLabel.tex}
      \label{fig:pendulum:compoundPendulum}}
  \end{center}
  \caption{Force diagrams for a simple pendulum and general compound
    pendulum.\label{fig:pendulum:pendulums}}
\end{figure}

The \term{simple pendulum} consists of a pendulum bob of mass $m$
attached to a string of length $\ell$ pivoting about a pivot $P$. The
string is assumed to have no mass, and the pendulum bob is assumed to
be a point mass (all of its mass is within an infinitely small
point). The force diagram for a simple pendulum is given in Figure
\ref{fig:pendulum:simplePendulum}. The pendulum bob is experiencing two forces,
the tension of the string $T$ from centripetal force, and
gravitational force $mg$. The entire pendulum is displaced from
equilibrium by an angle $\theta$ and or by an arc distance $s$.

We can further split the gravitational force into two components,
centripetal and tangential. The centripetal component must be cancelled by
the tension of the string otherwise the pendulum bob would go flying
through the air. The tangential force however is not counter-acted and
so the pendulum will accelerate towards the equilibrium; this force
acts as the restoring force for the system just as the spring provided
the restoring force in the previous example.

We can now write the equivalent of Hooke's law from Equation
\ref{equ:pendulum:hooke}, but now for the pendulum.
\begin{equation}
  ma = -mg\sin\theta = -ks = -k\ell\theta ~~\Rightarrow~~ mg\sin\theta
  = k\ell\theta
  \label{equ:pendulum:linearNewton}
\end{equation}
In the first step we replace $a$ with the tangential acceleration
$-mg\sin\theta$. In the second step we replace the displaced distance
$x$ with the displaced arc length $s$. In the final step we write arc
length in terms of $\theta$ and length $\ell$ using simple
trigonometry.

We must now take one final step, and that is the \term{small angle
  approximation}. This approximation states that if the angle $\theta$
is sufficiently small (usually under $10^\circ$) then $\sin\theta =
\theta$ or $\cos\theta = 1$. By looking at a unit circle you can
convince yourself of the validity of this approximation. By making
this approximation we can now solve $k$ in terms of $\ell$, $m$, and
$g$.
\begin{equation}
  k = \frac{mg}{\ell}
\end{equation}
Furthermore we can place this $k$ back into the formula we derived
for period, Equation \ref{equ:pendulum:period}, and find the period of the
pendulum.
\begin{equation}
T = 2\pi\sqrt{\frac{m}{\frac{mg}{\ell}}} = 2\pi\sqrt{{\ell}{g}}
\label{equ:pendulum:periodSimple}
\end{equation}
We see that the masses cancel and the period is dependent only on $g$
(a relative global constant) and the length of the pendulum $\ell$,
not the mass or the initial displacement! This result,
\term{isochronism}, demonstrates why pendulums are so useful as time
keepers, and also shows how we can measure the value of $g$ if we know
the length of the pendulum. It is, however, important to remember that
we did make use of the small angle approximation. This means that this
equation is only valid when we displace the pendulum by small angles
$\theta$. Later on we will see the difference this approximation makes
in the period of a pendulum at large angles.

\section{Compound Pendulum}

A \term{compound pendulum} is any \term{rigid body} which rotates
around a pivot, as shown in Figure
\ref{fig:pendulum:compoundPendulum}. The method outlined above for
determining the period of the simple pendulum works well, but for the
compound pendulum this method no longer works as gravity is now acting
on the entire pendulum, not just the pendulum bob. To circumvent this
problem we introduce a new method to determine the period of the
pendulum, using rotational motion. For those readers not familiar with
rotational motion, reread Chapter \ref{chp:rotation}. The general idea
however is that for every quantity used in \term{linear motion} to
describe the motion of an object, there is an equivalent quantity in
\term{rotational motion}. A summary of the relationship between linear
and rotational motion is given in Table \ref{tab:rotation:comparison}
of Chapter \ref{chp:rotation}.

Returning to the force diagram of Figure
\ref{fig:pendulum:compoundPendulum} we see that a gravitational force
of $mg$ is still being exerted on the center of mass of the pendulum
at an angle $\theta$. From this, the \term{torque} on the system is
just $-\ell mg\sin\theta$. We can use Newton's second law in
rotational motion to set torque equal to \term{moment of inertia}
times \term{angular acceleration}, with the equivalent relation in
linear motion given previously by Equation
\ref{equ:pendulum:linearNewton}. The similarities are striking.
\begin{equation}
  I\alpha = -\ell mg\sin\theta = -\kappa\theta ~~\Rightarrow~~
  mg\ell\sin\theta = \kappa\theta
\end{equation}
We again make the small angle approximation and solve for $\kappa$
(our rotational motion equivalent of $k$).
\begin{equation}
  \kappa = mg\ell
\end{equation}
Substituting $\kappa$ for $k$, and $I$ for $m$ in Equation
\ref{equ:pendulum:periodSimple} we now have the solution for the period of any
pendulum with moment of inertia $I$ and distance $\ell$ between the
pivot of the pendulum and the center of mass of the pendulum.
\begin{equation}
  T = 2\pi\sqrt{\frac{I}{mg\ell}}
  \label{equ:pendulum:periodCompound}
\end{equation}
For the case of the simple pendulum the moment of inertia is just
$\ell^2m$ which when plugged back into Equation
\ref{equ:pendulum:periodCompound} returns the same period as Equation
\ref{equ:pendulum:periodSimple}.

\section{Small Angle Approximation}

One final issue to discuss is the validity of the small angle
approximation made for both the simple and compound pendulums. As
stated earlier, this approximation is only valid for $\theta \ll 1$,
usually $\theta < 10^\circ$. However, it is important to understand
how this approximation affects the results of our theory. A closed
analytic solution to Equation \ref{equ:pendulum:sode} when replacing $kx$ with
$mg\sin x$ is not currently known, but can be expressed by a
perturbation series in $\theta$.\footnote{Nelson, Robert and
  M. G. Olsson. "The pendulum - Rich physics from a simple
  system". American Journal of Physics {\bf 54} (2). February
  1986. pp. 112-121. }
\begin{equation}
  T = 2\pi\sqrt{\frac{\ell}{g}} \left[1 + \sum_{n=1}^\infty
    \left(\frac{(2n)!}{2^{2n}(n!)^2}\right)\sin^{2n}\left(\theta/2\right)
  \right]
  \label{equ:pendulum:periodCorrect}
\end{equation}

\begin{figure}[h]
  \scriptsize
  \setlength{\unitlength}{0.0108mm}
  \begin{center}
    \subfigure[]{
      \executeiffilenewer{Code/Pendulum/periodCorrect.m}{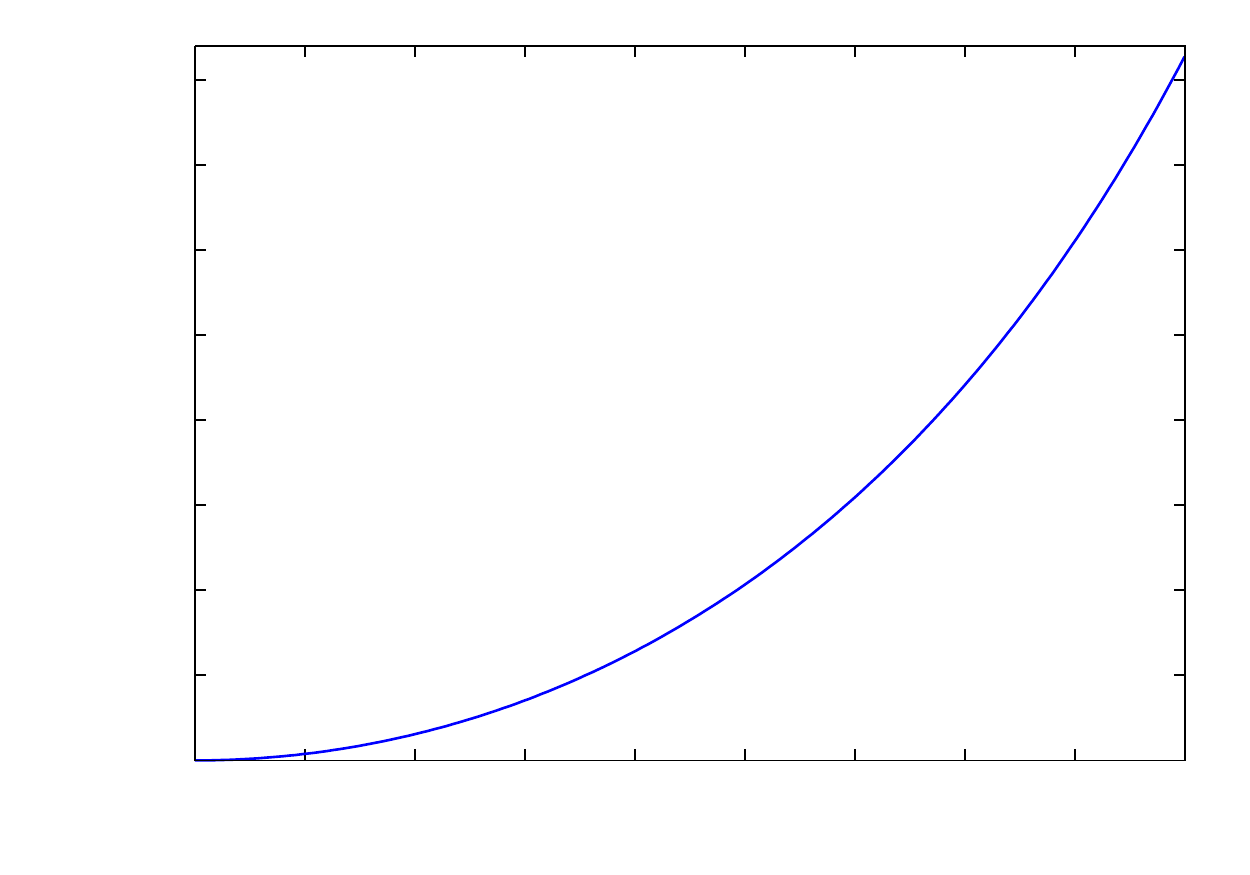}
  {octave --silent --eval "addpath([pwd(),'/Code']); 
    addpath([pwd(),'/Code/Pendulum']); periodCorrect(1);"} 
  \setlength{\unitlength}{\columnwidth*\real{0.00013}}
  \input{Figures/Pendulum/periodCorrectLabel.tex}
      \label{fig:pendulum:periodCorrect}}
    \subfigure[]{
      \executeiffilenewer{Code/Pendulum/periodCompare.m}{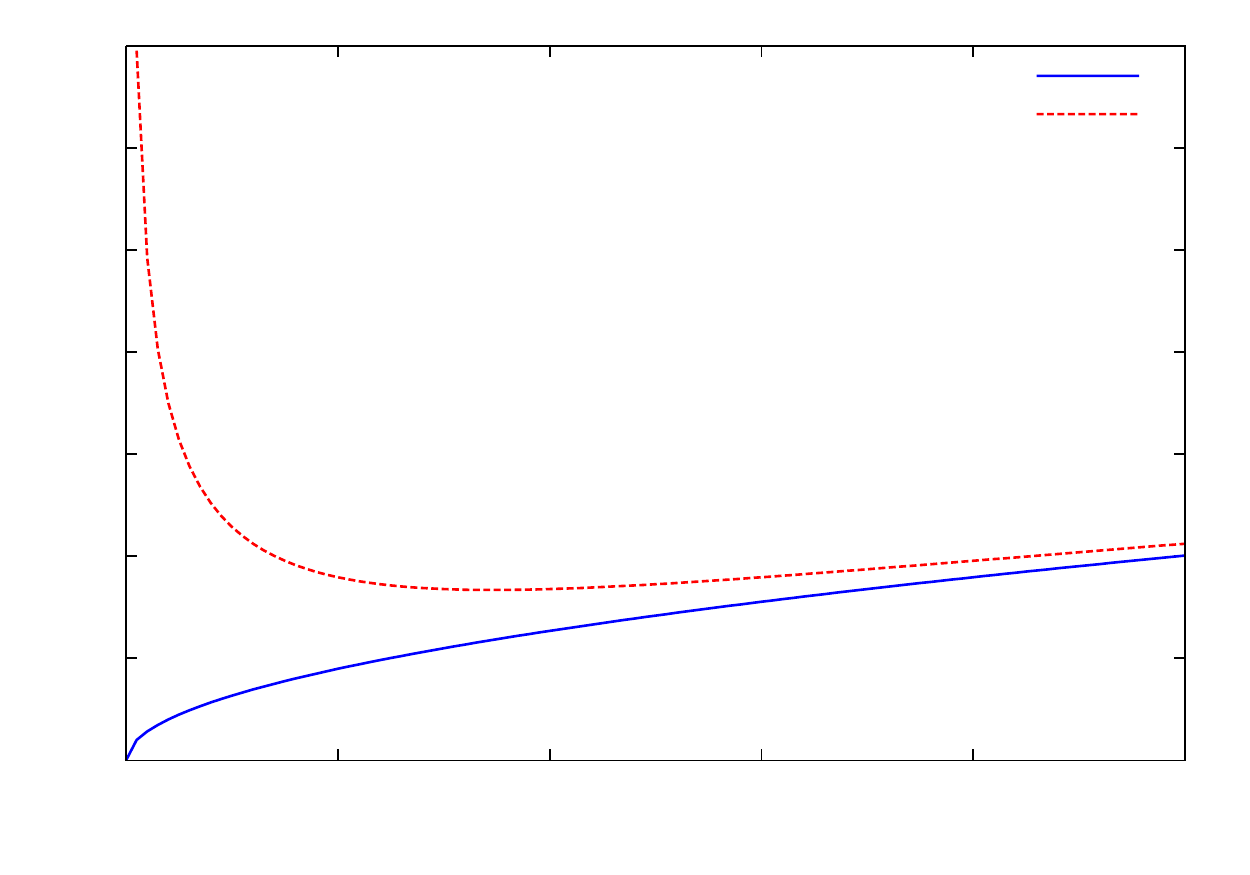}
  {octave --silent --eval "addpath([pwd(),'/Code']); 
    addpath([pwd(),'/Code/Pendulum']); periodCompare(1);"} 
  \setlength{\unitlength}{\columnwidth*\real{0.00013}}
  \input{Figures/Pendulum/periodCompareLabel.tex}
      \label{fig:pendulum:periodCompare}}
  \end{center}
  \caption{The period of a simple pendulum with
    $\ell=\frac{g}{4\pi^2}$ versus initial displacement angle $\theta$
    without making the small angle approximation is given on in the
    top plot. A comparison between the periods of a rod pendulum with
    $L=1.2$ m and the simple pendulum is given in the bottom
    plot.\label{fig:pendulum:period}}
\end{figure}

It is not important to necessarily understand the above equation, but
is important to understand the results given in Figure
\ref{fig:pendulum:periodCorrect}. Here a convenient length for the
simple pendulum has been chosen, $\ell = \frac{g}{4\pi^2}$, such that
the period for the pendulum as given by Equation
\ref{equ:pendulum:periodSimple} is just one second. For initial
displacement angles less than $16^\circ$ the percentage error caused
by making the small angle approximation is less than $1\%$. As
$\theta$ grows, the error from the small angle approximation grows as
well, with an initial displacement angle of $90^\circ$ causing nearly
$50\%$ error.

\section{Experiment}

For the experiment associated with this chapter a compound pendulum is
used to validate Equation \ref{equ:pendulum:periodCompound} and to
experimentally measure a value for $g$. For the purposes of this
experiment we assume the width of the pendulum is much less than that
of its length $L$, and so we can approximate the moment of inertia for
the pendulum as that of a rod. The moment of inertia for a rod
rotating around its center of mass is,
\begin{equation}
  I = \frac{mL^2}{12}
\end{equation}
as derived in Chapter \ref{chp:rotation}. Using the parallel axis
theorem we see that a rod rotating about an axis at distance $\ell$
from its center of mass has the following moment of inertia.
\begin{equation}
  I = \frac{mL^2}{12} +m\ell^2
\end{equation}
Plugging this back into Equation \ref{equ:pendulum:periodCompound} we
can theoretically predict the period for the compound pendulum used in
this experiment.
\begin{equation}
  T = 2\pi\sqrt{\frac{\frac{mL^2}{12} +m\ell^2}{mg\ell}} =
  2\pi\sqrt{\frac{mL^2}{12g\ell^2}+\frac{\ell}{g}} =
  2\pi\sqrt{\frac{\ell}{g}}\sqrt{1+\frac{L^2}{12\ell^2}}
  \label{equ:pendulum:periodRod}
\end{equation}
Notice that this formula is dependent on both $\ell$, the distance of
the pivot from the center of mass of the rod (in this case just the
center of the rod), and $L$, the entire length of the rod. It is
important to not confuse these two quantities.

Using Equations \ref{equ:pendulum:periodSimple} and
\ref{equ:pendulum:periodRod} we can compare the periods of oscillation
for a simple pendulum and a compound pendulum made from a rod versus
the variable length of either pendulum, $\ell$. For the simple
pendulum we expect a simple parabola from the square root term. The
simple pendulum behavior is shown by the solid line in blue in Figure
\ref{fig:pendulum:periodCompare}. The change in period with respect to
$\ell$ for the rod compound pendulum is more complicated as is
apparent from Equation \ref{equ:pendulum:periodRod}. For a rod of $L =
1.2$ m (close to the length of the compound pendulum used for the
experiment associated with this chapter), the behavior of the period
with respect to $\ell$ is plotted in dashed red in Figure
\ref{fig:pendulum:periodCompare} as well.

Perhaps one of the most noticeable aspects of the comparison between
the two periods is that for the simple pendulum, the period approaches
zero as $\ell$ approaches zero. For the compound pendulum however, the
period approaches $\infty$ as $\ell$ approaches zero.
\begin{equation}
  \lim_{\ell\rightarrow +0}
  2\pi\sqrt{\frac{\ell}{g}} = 0,~~~~
  \lim_{\ell\rightarrow +0}
  2\pi\sqrt{\frac{\ell}{g}}\sqrt{1+\frac{L^2}{12\ell^2}} =
  \lim_{\ell\rightarrow +0} 2\pi\sqrt{\frac{L^2}{12g\ell}} = +\infty
\end{equation}
For very large $\ell$ the periods of the simple and rod compound
pendulum match. This can be seen mathematically by taking the limit of
Equation \ref{equ:pendulum:periodRod},
\begin{equation}
  \lim_{\ell\rightarrow +\infty}
  2\pi\sqrt{\frac{\ell}{g}}\sqrt{1+\frac{L^2}{12\ell^2}} =
  \lim_{\ell\rightarrow +\infty}
  2\pi\sqrt{\frac{\ell}{g}}\sqrt{1+0} = \lim_{\ell\rightarrow +\infty}
  2\pi\sqrt{\frac{\ell}{g}}
\end{equation}
which we see is exactly the same as the period for the simple
pendulum. Physically this makes sense as well. For $\ell \gg L$ we see
that the physical set up approaches that of the simple pendulum. The
rod is no longer rotating about itself, but must be attached to the
pivot by some massless connector. As $\ell$ becomes larger and larger,
the rod becomes more and more like a point mass.

There is another important observation to make about Figure
\ref{fig:pendulum:periodCompare}. The period of the rod compound pendulum is not
monotonic (always increasing or decreasing) like that of the simple
pendulum but begins large, decreases rapidly, and then begins to
increase again. By taking the derivative of Equation
\ref{equ:pendulum:periodRod} and setting this to zero we can find the $\ell$
which provides the minimum period.
\begin{equation}
  \begin{aligned}
    \frac{d}{d\ell}
    \left[2\pi\sqrt{\frac{\ell}{g}}\sqrt{1+\frac{L^2}{12\ell^2}}\right]
    &= \frac{\pi}{2\ell\sqrt{3g}}
    \left(\frac{\left(12\ell^2-L^2\right)}
    {\sqrt{\ell^3\left(L^2+12\ell^2\right)}}\right)\\
    \frac{\pi}{2\ell\sqrt{3g}}
    \left(\frac{\left(12\ell^2-L^2\right)}
    {\sqrt{\ell^3\left(L^2+12\ell^2\right)}}\right) &= 0\\
    \ell &= \frac{L}{\sqrt{12}}\\
    T_\mathrm{min} &= \sqrt{8}\pi\sqrt{\frac{L}{g\sqrt{12}}}\\
  \end{aligned}
\end{equation}
Calculating this out with a value of $g = 9.81~\mathrm{ms^{-2}}$ and
$L = 1.2$ m gives a minimum period of $T_\mathrm{min} = 1.7$ s at
$\ell = 0.35$ m for the rod compound pendulum used in the experiment
associated with this chapter.

\graphicspath{{Figures/Gas/}}

\chapter{Gas}\label{chp:gas}

Understanding how gasses interact with their environment has many
connections to every day life, ranging from important matters such as
how the earth's atmosphere is contained (so that we can breath), to
more mundane matters (yet still very important) such as how quickly it
takes flatulence to spread through a closed room.

Before diving directly into the theory of gasses, a little history is
necessary to understand the beginning of the fields of {\bf
  thermodynamics} and \term{statistical mechanics}, both of which are
closely tied with modeling the behavior of gasses. In $1660$ Robert
Boyle\footnote{Robert Boyle was a born a native of Ireland to the
  $1\mathrm{st}$ Earl of Cork and amongst his contemporaries was
  regarded as one of the world's leading physicists.} published a book
with the rather long name of {\it New Experiments Physico-Mechanicall,
  Touching the Spring of the Air, and its Effects}. In this book Boyle
described a series of experiments that he had done, with the help of
his colleague Robert Hooke, which are the first known rigorous
experiments regarding the behavior of gasses. Two years later, after
some of his colleagues suggested he rewrite his book, he formulated
\term{Boyle's law} purely from experimental observation.
\begin{equation}
  pV = k_B
  \label{equ:gas:boyle}
\end{equation}
Here $p$ is the pressure of a gas, $V$ the volume of the gas, and
$k_B$ some constant that is dependent upon the experimental
setup. Physically speaking, the law states that the pressure of a gas
is inversely proportional to the volume of the gas.\footnote{J Appl
  Physiol 98:31-39,
  2005. \url{http://jap.physiology.org/cgi/reprint/98/1/31}}

Despite this rather important breakthrough, the field of
thermodynamics languished for another hundred years until the arrival
of Carnot and others, primarily because the physicists of the time
were too distracted trying to build steam engines.\footnote{There may
  be other reasons as well.} The next major breakthrough was made by
Gay-Lussac who postulated that the volume of a gas is directly
proportional to its temperature.\footnote{{\it The Expansion of Gasses
    through Heat}. Annales de Chimie {\bf 43}, 137
  (1802). \url{http://web.lemoyne.edu/~giunta/gaygas.html}}
\begin{equation}
  V = k_CT
  \label{equ:gas:charles}
\end{equation}
This law is called \term{Charles' Law} as Gay-Lussac claimed the law
was experimentally discovered previous to his own
discovery.\footnote{Gay-Lussac, while having a rather long and
  difficult to pronounce name, certainly seemed to be an all around
  good guy.} Additionally Gay-Lussac also notice that the pressure of
a gas is proportional to its temperature as well, and so he also
postulated what is known as \term{Gay-Lussac's Law}.
\begin{equation}
  p = k_GT
  \label{equ:gas:gay}
\end{equation}
Notice that the constant of proportionality here, $k_G$, is not the
same as the constant of proportionality in Boyle's Law, $k_B$ nor the
same as in Charles' Law, $k_C$, and that each of these constants are
entirely dependent upon the experimental setup.

\begin{figure}[h]
  \begin{center}
    \executeiffilenewer{Code/Gas/ideal.m}{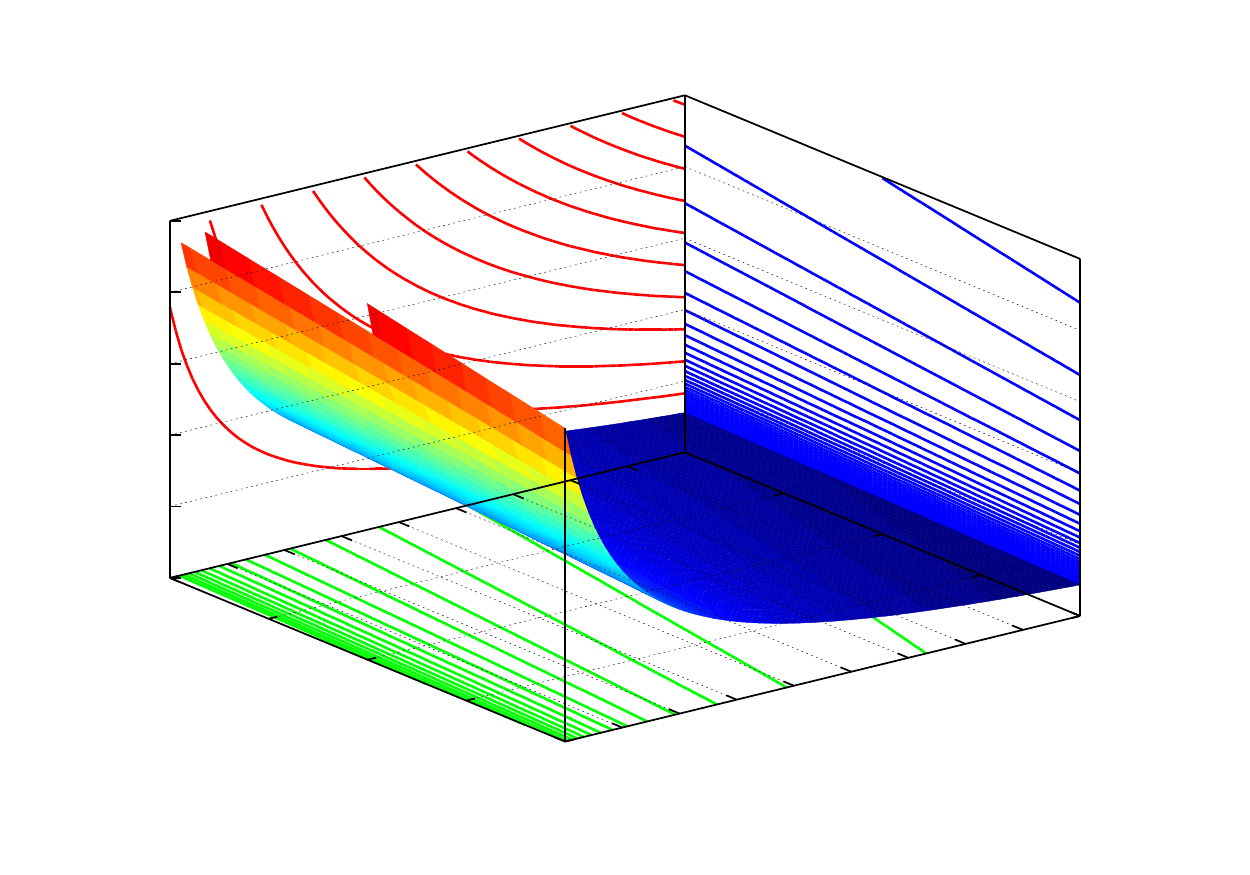}
  {octave --silent --eval "addpath([pwd(),'/Code']); 
    addpath([pwd(),'/Code/Gas']); ideal(1);"} 
  \setlength{\unitlength}{\columnwidth*\real{0.00013}}
  \input{Figures/Gas/idealLabel.tex}
    \caption{The ideal gas law given in Equation \ref{equ:gas:ideal} with
      isoline profiles for Boyle's, Charles', and Gay-Lussac's laws
      given in red, green, and blue respectively. The isolines for
      Boyle's law are given for a temperature every $300$ degrees
      Celsius, the isolines for Charles' law are given for a pressure
      every $0.3$ bar, and the isolines for Gay-Lussac's law are given
      for a volume every $0.3$ cm$^3$.\label{fig:gas:ideal}}
  \end{center}
\end{figure}

With these three laws the three macroscopic \term{state variables} of a
gas, pressure, volume, and temperature, are connected. Of course it
would be nice to have one equation instead of three, and this was
eventually both experimentally and theoretically arrived at in the
\term{ideal gas law}.
\begin{equation}
  pV = nRT
  \label{equ:gas:ideal}
\end{equation}
Here $n$ is the number of moles of gas and $R$ is the \term{ideal gas
  constant}. Figure \ref{fig:gas:ideal} shows a three dimensional plot of
the ideal gas law with the $z$-axis corresponding to pressure, the
$x$-axis to volume, and the $y$-axis to temperature. The isolines in
red demonstrate Boyle's law, the isolines in blue Gay-Lussac's law,
and the isolines in green Charles' law.

The ideal gas law was originally arrived at experimentally, but
eventually was derived using kinetic gas theory. Later, as
thermodynamics developed into statistical mechanics, a statistical
derivation was also discovered. The statistical derivation can be a
bit daunting without the proper background in statistical mechanics,
and so the following section attempts to give an intuitive feel for
the the theoretical origin of the ideal gas law through kinetic gas
theory.

\section{Kinetic Theory}

In the previous section we briefly mentioned the pressure, volume, and
temperature of a gas, but what exactly do these quantities mean? The
volume of a gas is just the volume of the container in which the gas
is confined. The pressure of a gas is the amount of force per unit
area that the gas is exerting on its container. The temperature of a
gas is technically defined as the partial derivative of the energy of
the gas taken with respect to the \term{entropy} of the gas. This
definition is not very helpful without more theory, so for the
purposes of this chapter, we can also write temperature of an ideal
gas as,
\begin{equation}
  T = \frac{2\overline{K}}{3R}
  \label{equ:gas:temp}
\end{equation}
where $\overline{K}$ is the average kinetic energy of the
gas.\footnote{For those who are curious this relation can be found
  from the \term{Boltzmann distribution} and by assuming that an ideal
  gas has three translational degrees of freedom, corresponding to the
  three physical dimensions.}

There is one final definition that needs to be given, and that is for
an ideal gas. An \term{ideal gas} must satisfy the following
assumptions.
\begin{enumerate}
\item The atoms are point-like with the same mass and do not interact
  except through collisions.
\item All collisions involving the atoms must be elastic\footnote{For
    more information on elastic collisions read over Chapter
    \ref{chp:momentum}. In elastic collisions two quantities are
    conserved, momentum and kinetic energy.} (with either the
  container of the wall, or between individual gas atoms).
\item The atoms obey Newton's laws.
\item There are a large number of atoms, moving at random speeds which
  do not change over time.
\end{enumerate}
With these assumptions for an ideal gas, and the definitions for
pressure, volume, and temperature above, we can now begin the
derivation of the ideal gas law.

\begin{figure}[h]
  \begin{center}
    \executeiffilenewer{Figures/Gas/cube.svg}
  {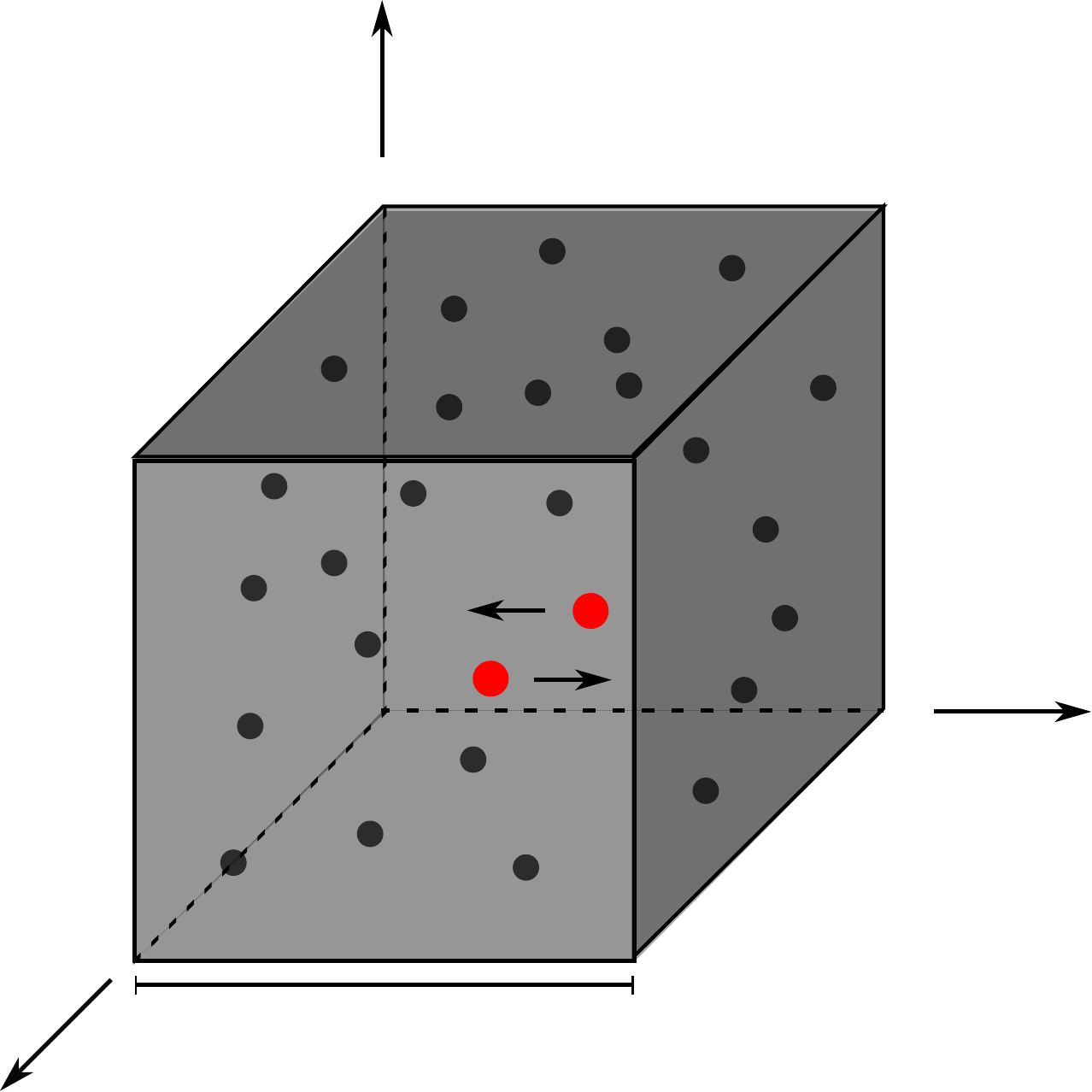}{inkscape-0.48pre1 -z -D --file=Figures/Gas/cube.svg 
    --export-pdf=Figures/Gas/cube.pdf --export-latex} \def\svgwidth{0.5\columnwidth}
  \input{Figures/Gas/cubeLabel.tex}
    \caption{A cube with width, length, and height $L$ filled with an
      ideal gas of $N$ particles.\label{fig:gas:cube}}
  \end{center}
\end{figure}

Consider a cube filled with gas as shown in Figure
\ref{fig:gas:cube}. Each side of the cube is of length $L$ and so the
volume of the cube is just $L^3$. The cube is filled with $N$ gas
atoms each with a velocity $\vec{v}_i$. The magnitude of the velocity,
or the speed of each gas atom, does not change over time because all
the collisions within the cube are elastic (we used assumptions $1$,
$2$, and $3$ to arrive at this conclusion). The average of the squared
velocities for all the gas atoms is just,
\begin{equation}
  \overline{v^2} =
  \frac{\vec{v}_1^2 + \vec{v}_2^2 + \cdots +
    \vec{v}_N^2}{N}
  \label{equ:gas:averageSquare}
\end{equation}
where $\overline{v^2}$ is the average of the squared velocities. In
general, a straight line segment over a symbol indicates an
average. Notice we have not written $\overline{v}^2$ as this would
indicate taking the average of the velocity and then squaring rather
than squaring the velocities and then averaging. Using the value for
the average of the squared velocities in Equation
\ref{equ:gas:averageSquare}, we are able to write the average kinetic
energy of the gas.
\begin{equation}
  \overline{K} = \frac{1}{2}m\overline{v^2}
  \label{equ:gas:kineticEnergy}
\end{equation}
Here, $m$ is the mass of one gas atom. Notice from assumption $1$ all
the particles have the same mass, and so the average mass is just the
mass of one gas atom.

We can also break the average of the squared velocities into the
components of the average velocity.\footnote{For readers unfamiliar
  with vectors, the magnitude squared of a vector is equal to the sum
  of the squares of the components. This is exactly the same as
  the Pythagorean theorem which states the square of the hypotenuse is
  equal to the sum of the square of each side of a triangle.}
\begin{equation}
  \overline{v^2} = \overline{v_x^2} + \overline{v_y^2} +
  \overline{v_z^2}
  \label{equ:gas:squareComponent}
\end{equation}
By assumption $4$ we know there are a large number of randomly moving
gas atoms in the cube, and so if the cube was rotated $90^\circ$ to
the left with the $y$-axis now where the $z$-axis was, no one could
tell the difference. This tells us that the average component
velocities must be equal and so the average of each component velocity
squared must also be equal.
\begin{equation}
  \overline{v_x^2} = \overline{v_y^2} = \overline{v_z^2}
  ~~~\Rightarrow~~~ \overline{v^2} = 3\overline{v_y^2}
  ~~~\Rightarrow~~~ \overline{K} =
  \frac{3}{2}m\overline{v_y^2}
  \label{equ:gas:components}
\end{equation}
In the second step of the equation above we have rewritten Equation
\ref{equ:gas:squareComponent}, but now with only the $y$-component of
velocity as all the average components must be equal. In the third
step we have just plugged this value for $\overline{v^2}$ into
Equation \ref{equ:gas:kineticEnergy}. Now we can take the final result for
the average kinetic energy in Equation \ref{equ:gas:components} and plug
this into our equation for temperature, given in Equation
\ref{equ:gas:temp}.
\begin{equation}
  \overline{v_y^2} = \frac{m}{RT}
  \label{equ:gas:yTemp}
\end{equation}

Returning to Figure \ref{fig:gas:cube}, we can consider a single gas atom
with a velocity $\vec{v} = v_y$. The atom could be (and probably is)
also moving in the $x$ and $z$ directions, but we ignore that for
now. From Newton's laws (assumption number $3$) we know that,
\begin{equation}
  \overline{F} = m\overline{a} = m\frac{\Delta v}{\Delta t} =
  \frac{\Delta P}{\Delta t}
  \label{equ:gas:newtonsSecond}
\end{equation}
where in the second step we have just written out the definition for
acceleration, and in the third step substituted in momentum for
$m\Delta v$. We are using a capital $P$ to denote momentum to avoid
confusion with pressure, which is represented by a lower case
$p$.\footnote{In every other chapter, $p$ is used to notate momentum.}
Now consider what happens if the atom bounces of the wall of the
cube. Because the collision is elastic (by assumption $2$) and we
assume the wall of the cube is very massive with respect to the gas
atom, the velocity of the gas atom after the collision with the wall
is just $-v_y$.\footnote{Check this with Equation $6$ from Chapter
  \ref{chp:momentum}. Let the mass of the gas atom approach $0$ and
  check the result.}

The initial momentum of the gas particle was $mv_y$ and the final
momentum of the gas particle was $-mv_y$ so the change in momentum of
the gas particle was $\Delta P = 2mv_y$. Every time the atom bounces
of the wall, there is a momentum change of $2mv_y$. If the atom
travels between the two walls of the cube, bouncing off each time, we
know that the time between bounces is the distance traveled, $2L$,
divided by the velocity and so $\Delta t = 2L/v_y$. Plugging $\Delta
P$ and $\Delta t$ into Equation \ref{equ:gas:newtonsSecond} we arrive at a
value for the average force exerted by a single gas atom.
\begin{equation}
  \overline{F} = \Delta P \left(\frac{1}{\Delta t}\right) =
  2mv_y\left(\frac{v_y}{2L}\right) = \frac{mv_y^2}{L}
  \label{equ:gas:force}
\end{equation}

Now we can think of the total average force of $N$ particles with
average velocity $\overline{v_y}$.
\begin{equation}
  \overline{F}_N = \frac{mN\overline{v_y^2}}{L}
\end{equation}
We can set the average total force of the particles equal to the
pressure, as pressure is just force over area. The area here is just
the area of the wall from the cube, or $L^2$.
\begin{equation}
  \overline{F}_N = pA = pL^2 ~~~\Rightarrow~~~
  \frac{mN\overline{v_y^2}}{L} =  pL^2 ~~~\Rightarrow~~~ p =
  \frac{mN\overline{v_y^2}}{L^3} = \frac{mN\overline{v_y^2}}{V}
\end{equation}
If we plug in $\overline{v_y^2}$ from Equation \ref{equ:gas:yTemp} into
the result above, we obtain the ideal gas law!
\begin{equation}
  p = \left(\frac{mN}{V}\right)\left(\frac{RT}{m}\right)
  ~~~\Rightarrow~~~ pV = NRT
\end{equation}
Notice we have a capital $N$ (number of atoms) rather than lower case
$n$ (number of moles). This is just a matter of notation. Typically
the ideal gas constant, $R$, is given in Joules per mole per Kelvin,
in which case small $n$ (moles) should be used instead of large $N$
(atoms).

\section{Experiment}

The experiment associated with this chapter consists of experimentally
verifying both Boyle's law from Equation \ref{equ:gas:boyle} and
Gay-Lussac's law from Equation \ref{equ:gas:gay}. This first
experiment is performed by pressurizing a column of air with a bike
pump and taking volume measurements as more and more pressure is
added. As the pressure increases, the volume must decrease according
to Boyle's law. The raw data from this part of the experiment should
look similar to the red lines in Figure \ref{fig:gas:ideal}, and be
nearly identical to the lowest isoline (corresponding to a temperature
of $20^\circ$ Celsius). By making a linear plot of pressure on the
$x$-axis and $1/V$ on the $y$-axis, it is possible to verify Boyle's
law with a linear fit.

Gay-Lussac's law can be verified by keeping a constant volume of gas
and changing the temperature while measuring the pressure. In this
experiment a small container of gas is heated, and a pressure gauge
allows the temperature to be read. According to Equation \ref{equ:gas:gay}
the temperature and pressure should rise linearly, and if a plot is
made it should closely resemble the blue isolines of Figure
\ref{fig:gas:ideal}. The volume of the apparatus is near $10$ cm$^3$, and
so the raw data from this part of the experiment should closely follow
the lower isolines. Because of this we expect that a large change in
temperature will yield a relatively small change in pressure.

While verifying Gay-Lussac's law, it is possible to experimentally
determine absolute zero ($0$ Kelvin) in degrees Celsius! In all the
formulas involving temperature in this chapter, the temperature must
be given in Kelvin. Because the Kelvin scale is the same as the
Celsius scale, except with a constant term added on, we can rewrite
Equation \ref{equ:gas:gay}.
\begin{equation}
  p = k_G(T_C-T_0)
\end{equation}
Here $T_C$ is temperature in degrees Celsius, and $T_0$ absolute zero
in degrees Celsius. By formatting this linear relationship in
slope-intercept form we see that the intercept of the plot is just,
\begin{equation}
  b = k_gT_0 = mT_0
\end{equation}
where we already know $k_g$ from the slope of the plot. Plugging in
values for $m$ and $b$ as obtained from the best fit of the plot, we
can solve for $T_0$ and find absolute zero!

\graphicspath{{Figures/Resistance/}}

\chapter{Resistance}\label{chp:resistance}

Electricity has captivated the imagination of humans since the
earliest recordings of civilizations, yet it was not until the
$17^\mathrm{th}$ century that we truly began to understand the
fundamental nature of electricity. Now, electromagnetism is described
by the most precise theory ever developed by the physics
community. What began with experiments ranging from flying kites in
thunderstorms to catching electric eels has now culminated in
providing the base for all modern technology. The study of
electromagnetism today can be broadly broken down into applied
electrical engineering and theoretical electrodynamics.

Despite being united by electricity, these two fields of study are
completely different, yet equally important. The engineering aspect
allows for new technologies to be developed, while the the theoretical
side develops new ideas that can be implemented in practice. However,
both of these fields require an understanding of the fundamentals of
electrodynamics. There are a large number of books on the subject;
some are good while others are terrible. Two particularly excellent
resources are MIT's OpenCourseWare materials for
\href{http://ocw.mit.edu/OcwWeb/Physics/8-02Electricity-and-MagnetismSpring2002/CourseHome/index.htm}{8.02}
as taught by Walter Lewin, and an
\href{http://www.amazon.com/Introduction-Electrodynamics-3rd-David-Griffiths/dp/013805326X}{\it
  Introduction to Electrodynamics} by David Griffiths. More in depth
discussions of certain parts of electrodynamics are given in Chapters
\ref{chp:induction} and \ref{chp:planck}.

\section{Circuits}

A quick review of the quantities used in electric circuits is given in
Table \ref{tab:resistance:quantities}. Here $Q$ indicates charge, $t$ time, $U$
energy, and $\Phi$ magnetic flux. The two most commonly measured
quantities of any circuit are \term{current} and \term{voltage}. Current
is the amount of charge crossing a point in a circuit per unit
time. For the purposes of this chapter we will assume that current
always remains constant, hence writing $Q/t$ rather than $dQ/dt$. Both
inductors and capacitors have a time dependence, unless at
equilibrium, and subsequently will only be touched on briefly in this
chapter.

Voltage is the electric potential at a point in a circuit per unit
charge. Essentially, current can be equated to the width of a river
while voltage can be compared to the velocity at which the river is
flowing. This comparison does not quite match, but sometimes can be
useful for thinking intuitively about circuits, as electricity behaves
like water in many ways. The current of a circuit is measured with an
ammeter\footnote{A galvanometer is a specific type of ammeter, a
  significant step up from the original ammeter which was the
  scientist shocking him or herself and trying to gauge the power of
  the shock by how much it hurt.} while voltage is measured with a
voltmeter. The symbol for both of these devices in an electric circuit
is given in Table \ref{tab:resistance:quantities}.

There are three more important basic components of circuits (ignoring
transistors) which are \term{resistance}, \term{capacitance}, and
\term{inductance}. Resistance is the amount of energy dissipated by a
component of a circuit per unit time. Oftentimes resistors dissipate
their energy through heat, but can also emit through light. Capacitors
and inductors provide the exact opposite purpose of resistors in
circuits; rather than dissipating energy, they store
energy. Capacitors store energy by creating an electric field and so
capacitance is given by the charge stored on the capacitor divided by
the voltage gap across the capacitor.

Inductors store potential energy in the form of a magnetic field,
oftentimes created by electric current flowing. The inductance of an
inductor is given by the magnetic flux (change in magnetic field)
divided by the current flowing around the magnetic field. Inductors
are usually small solenoids which consist of many turns of wires
wrapped around a cylindrical core. In a case like the solenoid, the
current must be divided by the number of times it circles the magnetic
field and so the inductance for a solenoid is usually given as
$N\Phi/I$ where $N$ is the number of turns within the solenoid. The
idea of inductance is explored in more detail in Chapter
\ref{chp:induction}.

\begin{table}
  \small
  \vspace{1.1cm}
  \begin{center}
    \begin{tabular}{cm{2cm}m{1.2cm}cm{4cm}}
      \midrule
      \vspace{-1.1cm}\\
      \vspace{-0.55cm}{\bf Quantity} & {\bf Definition} & {\bf Symbol}
      & {\bf Units} & {\bf~~~~~Base Units}\\
      \toprule
      \vspace{-1cm}current &
      $$I = \frac{Q}{t}$$ &
      \executeiffilenewer{Figures/Resistance/ammeter.svg}
  {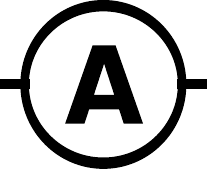}{inkscape-0.48pre1 -z -D --file=Figures/Resistance/ammeter.svg 
    --export-pdf=Figures/Resistance/ammeter.pdf --export-latex} \def\svgwidth{0.8cm}

\begingroup
  \makeatletter
  \providecommand\color[2][]{%
    \errmessage{(Inkscape) Color is used for the text in Inkscape, but the package 'color.sty' is not loaded}
    \renewcommand\color[2][]{}%
  }
  \providecommand\transparent[1]{%
    \errmessage{(Inkscape) Transparency is used (non-zero) for the text in Inkscape, but the package 'transparent.sty' is not loaded}
    \renewcommand\transparent[1]{}%
  }
  \providecommand\rotatebox[2]{#2}
  \ifx\svgwidth\undefined
    \setlength{\unitlength}{59.525pt}
  \else
    \setlength{\unitlength}{\svgwidth}
  \fi
  \global\let\svgwidth\undefined
  \makeatother
  \begin{picture}(1,0.81646347)%
    \put(0,0){\includegraphics[width=\unitlength]{ammeter}}%
  \end{picture}%
\endgroup
 &
      amperes [A] &
      $$\mathrm{\frac{coulombs}{seconds}}$$ \\
      \vspace{-1cm}voltage &
      $$V = \frac{U}{Q}$$ &
      \executeiffilenewer{Figures/Resistance/voltmeter.svg}
  {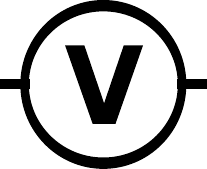}{inkscape-0.48pre1 -z -D --file=Figures/Resistance/voltmeter.svg 
    --export-pdf=Figures/Resistance/voltmeter.pdf --export-latex} \def\svgwidth{0.8cm}

\begingroup
  \makeatletter
  \providecommand\color[2][]{%
    \errmessage{(Inkscape) Color is used for the text in Inkscape, but the package 'color.sty' is not loaded}
    \renewcommand\color[2][]{}%
  }
  \providecommand\transparent[1]{%
    \errmessage{(Inkscape) Transparency is used (non-zero) for the text in Inkscape, but the package 'transparent.sty' is not loaded}
    \renewcommand\transparent[1]{}%
  }
  \providecommand\rotatebox[2]{#2}
  \ifx\svgwidth\undefined
    \setlength{\unitlength}{59.525pt}
  \else
    \setlength{\unitlength}{\svgwidth}
  \fi
  \global\let\svgwidth\undefined
  \makeatother
  \begin{picture}(1,0.81646347)%
    \put(0,0){\includegraphics[width=\unitlength]{voltmeter}}%
  \end{picture}%
\endgroup
 &
      volts [V] &
      $$\mathrm{\frac{joules}{coulombs}}$$ \\
      \vspace{-1cm}resistance &
      $$ R = \frac{U}{t}$$ &
      \executeiffilenewer{Figures/Resistance/resistor.svg}
  {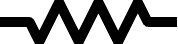}{inkscape-0.48pre1 -z -D --file=Figures/Resistance/resistor.svg 
    --export-pdf=Figures/Resistance/resistor.pdf --export-latex} \def\svgwidth{0.9cm}

\begingroup
  \makeatletter
  \providecommand\color[2][]{%
    \errmessage{(Inkscape) Color is used for the text in Inkscape, but the package 'color.sty' is not loaded}
    \renewcommand\color[2][]{}%
  }
  \providecommand\transparent[1]{%
    \errmessage{(Inkscape) Transparency is used (non-zero) for the text in Inkscape, but the package 'transparent.sty' is not loaded}
    \renewcommand\transparent[1]{}%
  }
  \providecommand\rotatebox[2]{#2}
  \ifx\svgwidth\undefined
    \setlength{\unitlength}{51.05pt}
  \else
    \setlength{\unitlength}{\svgwidth}
  \fi
  \global\let\svgwidth\undefined
  \makeatother
  \begin{picture}(1,0.24779628)%
    \put(0,0){\includegraphics[width=\unitlength]{resistor}}%
  \end{picture}%
\endgroup
 &
      ohms [$\Omega$] &
      $$\mathrm{\frac{joules}{seconds}}$$ \\
      \vspace{-1cm}capacitance &
      $$ C = \frac{Q}{V}$$ &
      \executeiffilenewer{Figures/Resistance/capacitor.svg}
  {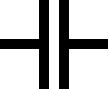}{inkscape-0.48pre1 -z -D --file=Figures/Resistance/capacitor.svg 
    --export-pdf=Figures/Resistance/capacitor.pdf --export-latex} \def\svgwidth{0.5cm}

\begingroup
  \makeatletter
  \providecommand\color[2][]{%
    \errmessage{(Inkscape) Color is used for the text in Inkscape, but the package 'color.sty' is not loaded}
    \renewcommand\color[2][]{}%
  }
  \providecommand\transparent[1]{%
    \errmessage{(Inkscape) Transparency is used (non-zero) for the text in Inkscape, but the package 'transparent.sty' is not loaded}
    \renewcommand\transparent[1]{}%
  }
  \providecommand\rotatebox[2]{#2}
  \ifx\svgwidth\undefined
    \setlength{\unitlength}{31.2pt}
  \else
    \setlength{\unitlength}{\svgwidth}
  \fi
  \global\let\svgwidth\undefined
  \makeatother
  \begin{picture}(1,0.81891026)%
    \put(0,0){\includegraphics[width=\unitlength]{capacitor}}%
  \end{picture}%
\endgroup
 &
      farads [F] &
      $$\mathrm{\frac{coulombs^2}{joules}}$$ \\
      \vspace{-1cm}inductance &
      $$ L = \frac{\Phi}{I}$$ &
      \executeiffilenewer{Figures/Resistance/inductor.svg}
  {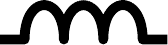}{inkscape-0.48pre1 -z -D --file=Figures/Resistance/inductor.svg 
    --export-pdf=Figures/Resistance/inductor.pdf --export-latex} \def\svgwidth{0.9cm}

\begingroup
  \makeatletter
  \providecommand\color[2][]{%
    \errmessage{(Inkscape) Color is used for the text in Inkscape, but the package 'color.sty' is not loaded}
    \renewcommand\color[2][]{}%
  }
  \providecommand\transparent[1]{%
    \errmessage{(Inkscape) Transparency is used (non-zero) for the text in Inkscape, but the package 'transparent.sty' is not loaded}
    \renewcommand\transparent[1]{}%
  }
  \providecommand\rotatebox[2]{#2}
  \ifx\svgwidth\undefined
    \setlength{\unitlength}{48.175pt}
  \else
    \setlength{\unitlength}{\svgwidth}
  \fi
  \global\let\svgwidth\undefined
  \makeatother
  \begin{picture}(1,0.26543986)%
    \put(0,0){\includegraphics[width=\unitlength]{inductor}}%
  \end{picture}%
\endgroup
 &
      henries [H] &
      $$\mathrm{\frac{webers\cdot seconds}{coulombs}}$$ \\
      & & \\
      \bottomrule
    \end{tabular}
    \caption{A review of basic quantities used to describe
      circuits.\label{tab:resistance:quantities}}
  \end{center}
\end{table}

Current, voltage, and resistance are all connected through \term{Ohm's
  law} which states that voltage is just current times resistance.
\begin{equation}
  V = IR
  \label{equ:resistance:ohm}
\end{equation}
Additionally, the power dissipated by a resistor is equal to the
square of the current running through the resistor times its
resistance.
\begin{equation}
  P = I^2R
  \label{equ:resistance:power}
\end{equation}
While Ohm's law seems extraordinarily simple, it is the basis for
simple circuits, and the underlying theory is very involved on a
microscopic level.

\begin{figure}
  \begin{center}
    \subfigure[]{
      \executeiffilenewer{Figures/Resistance/series.svg}
  {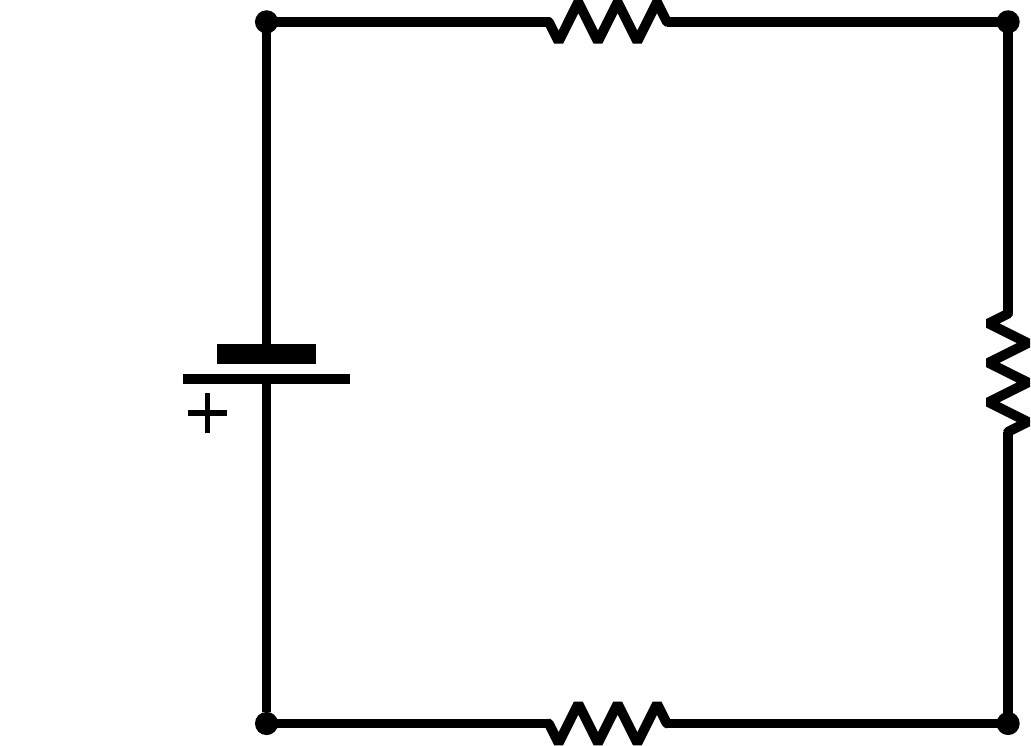}{inkscape-0.48pre1 -z -D --file=Figures/Resistance/series.svg 
    --export-pdf=Figures/Resistance/series.pdf --export-latex} \def\svgwidth{0.5\columnwidth}
  \input{Figures/Resistance/seriesLabel.tex}
      \label{fig:resistance:series}
    }
    \hspace{2cm}
    \subfigure[]{
      \executeiffilenewer{Figures/Resistance/parallel.svg}
  {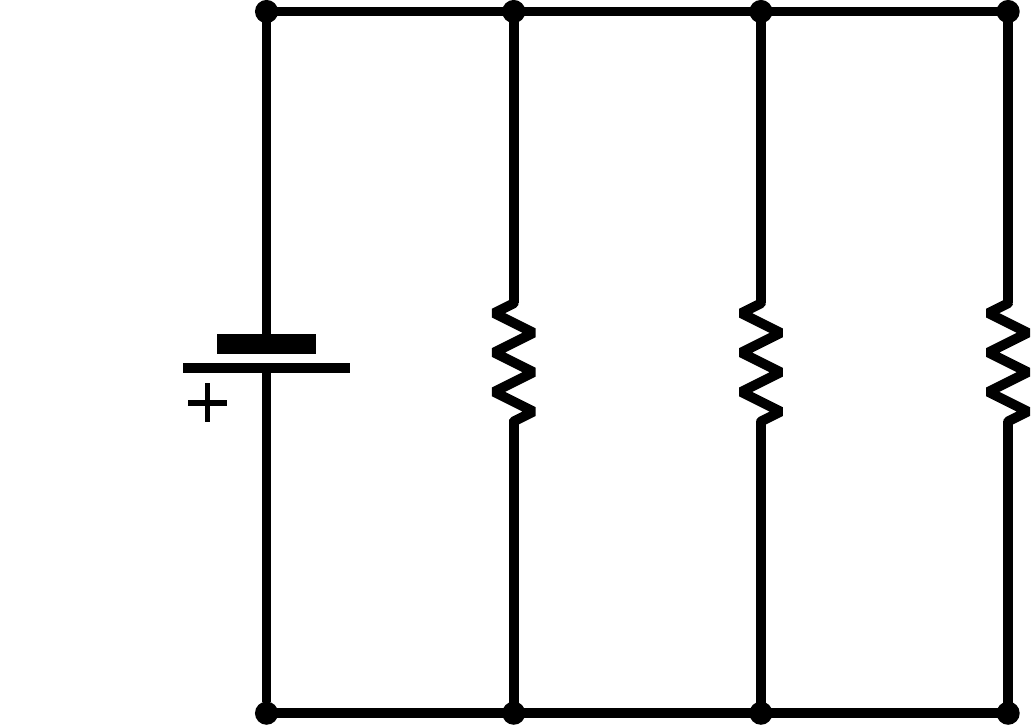}{inkscape-0.48pre1 -z -D --file=Figures/Resistance/parallel.svg 
    --export-pdf=Figures/Resistance/parallel.pdf --export-latex} \def\svgwidth{0.5\columnwidth}
  \input{Figures/Resistance/parallelLabel.tex}
      \label{fig:resistance:parallel}
    }
    \caption{Examples of a series and parallel circuit given in
      Figures \ref{fig:resistance:series} and \ref{fig:resistance:parallel} respectively.}
  \end{center}
\end{figure}

Ohm's law in the form of Equation \ref{equ:resistance:ohm} is only useful for
determining the current, voltage, or resistance of a simple circuit
consisting of a power source and a resistor. However, more complex
circuits can be broken into two categories, \term{series circuits} and
\term{parallel circuits}. In a series circuit all the electrical
components are placed one after another on a single electrical path as
shown in Figure \ref{fig:resistance:series}. For this type of circuit,
resistances can be added together to find a total resistance,
\begin{equation}
  R_\mathrm{total} = R_1+R_2+R_3+\cdots
  \label{equ:resistance:seriesResistance}
\end{equation}
whereas the inverse of capacitances must be added.
\begin{equation}
  \frac{1}{C_\mathrm{total}} =
  \frac{1}{C_1}+\frac{1}{C_2}+\frac{1}{C_3}+\cdots
  \label{equ:resistance:seriesCapacitance}
\end{equation}
Inductors in a series circuit are added just like resistors.

In a parallel circuit a single electrical path breaks into multiple
electrical paths, and then combines back into a single electrical
path, like the diagram given in Figure \ref{fig:resistance:parallel}. For
parallel circuits the inverse of the resistances and inductances must
be added to find the total resistance or inductance,
\begin{equation}
  \frac{1}{R_\mathrm{total}} = \frac{1}{R_1}+\frac{1}{R_2}+\frac{1}{R_3}+\cdots
  \label{equ:resistance:parallelResistance}
\end{equation}
while the
capacitances may just be added.
\begin{equation}
  C_\mathrm{total} = C_1+C_2+C_3+\cdots
  \label{equ:resistance:parallelCapacitance}
\end{equation}

\section{Kirchhoff's Laws}

Equations \ref{equ:resistance:seriesResistance} through
\ref{equ:resistance:parallelCapacitance} are not fundamental laws
themselves, but rather, can be derived from what are known as
\term{Kirchhoff's laws} which are given below.
\begin{enumerate}
\item Conservation of charge. All the current flowing
  into a junction must equal the current flowing out of the junction.
\item Conservation of energy. The net voltage of any loop within a
  circuit must be zero.
\end{enumerate}
Both of these laws have an even more fundamental basis from
\term{Maxwell's equations}, but for the purposes of this chapter, let
us accept the two laws above without further derivation. In Chapter
\ref{chp:planck} Maxwell's equations are introduced.

\begin{figure}
  \begin{center}
    \subfigure[]{
      \executeiffilenewer{Figures/Resistance/example1.svg}
  {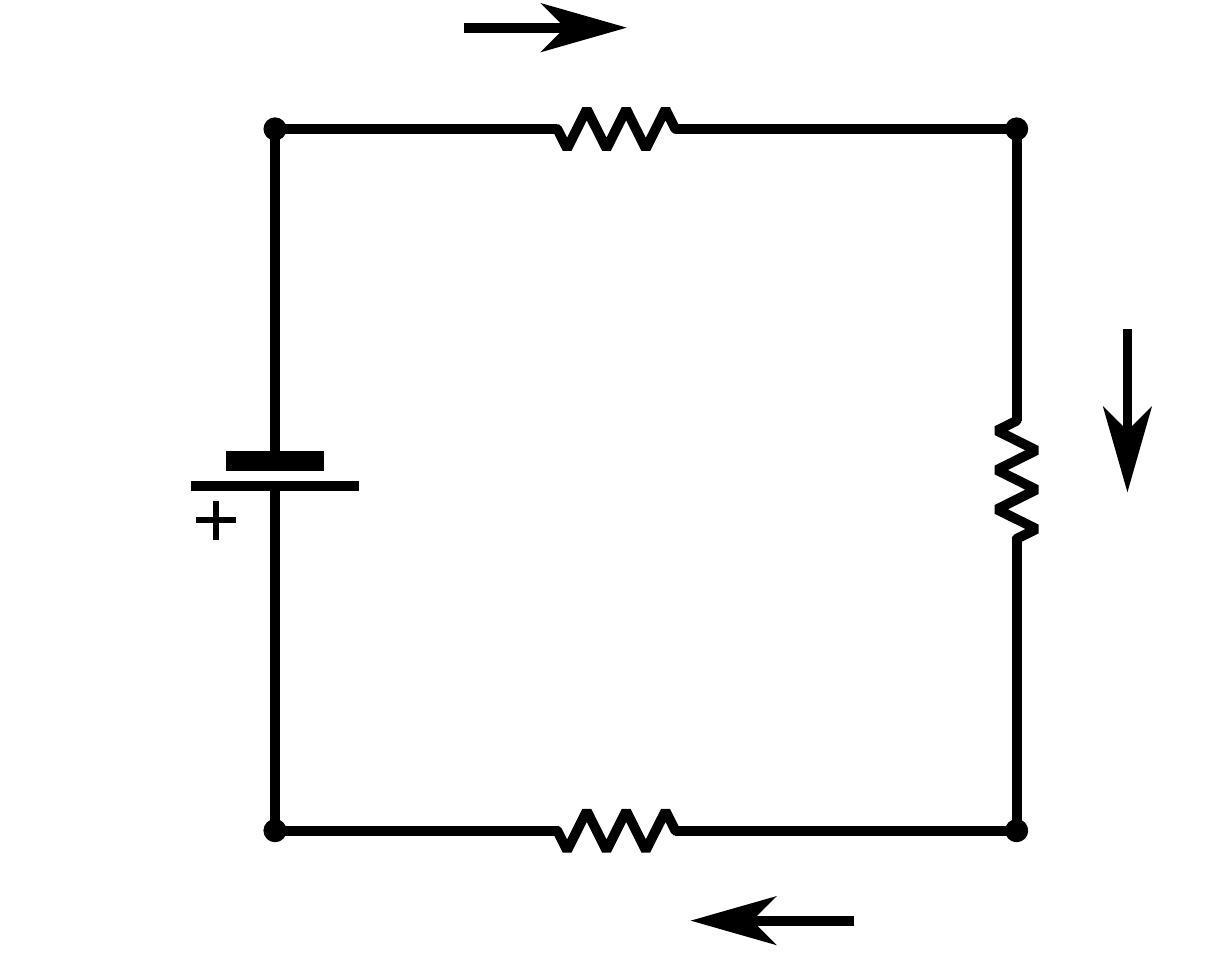}{inkscape-0.48pre1 -z -D --file=Figures/Resistance/example1.svg 
    --export-pdf=Figures/Resistance/example1.pdf --export-latex} \def\svgwidth{0.6\columnwidth}
  \input{Figures/Resistance/example1Label.tex}
      \label{fig:resistance:example1}
    }
    \subfigure[]{
      \executeiffilenewer{Figures/Resistance/example2.svg}
  {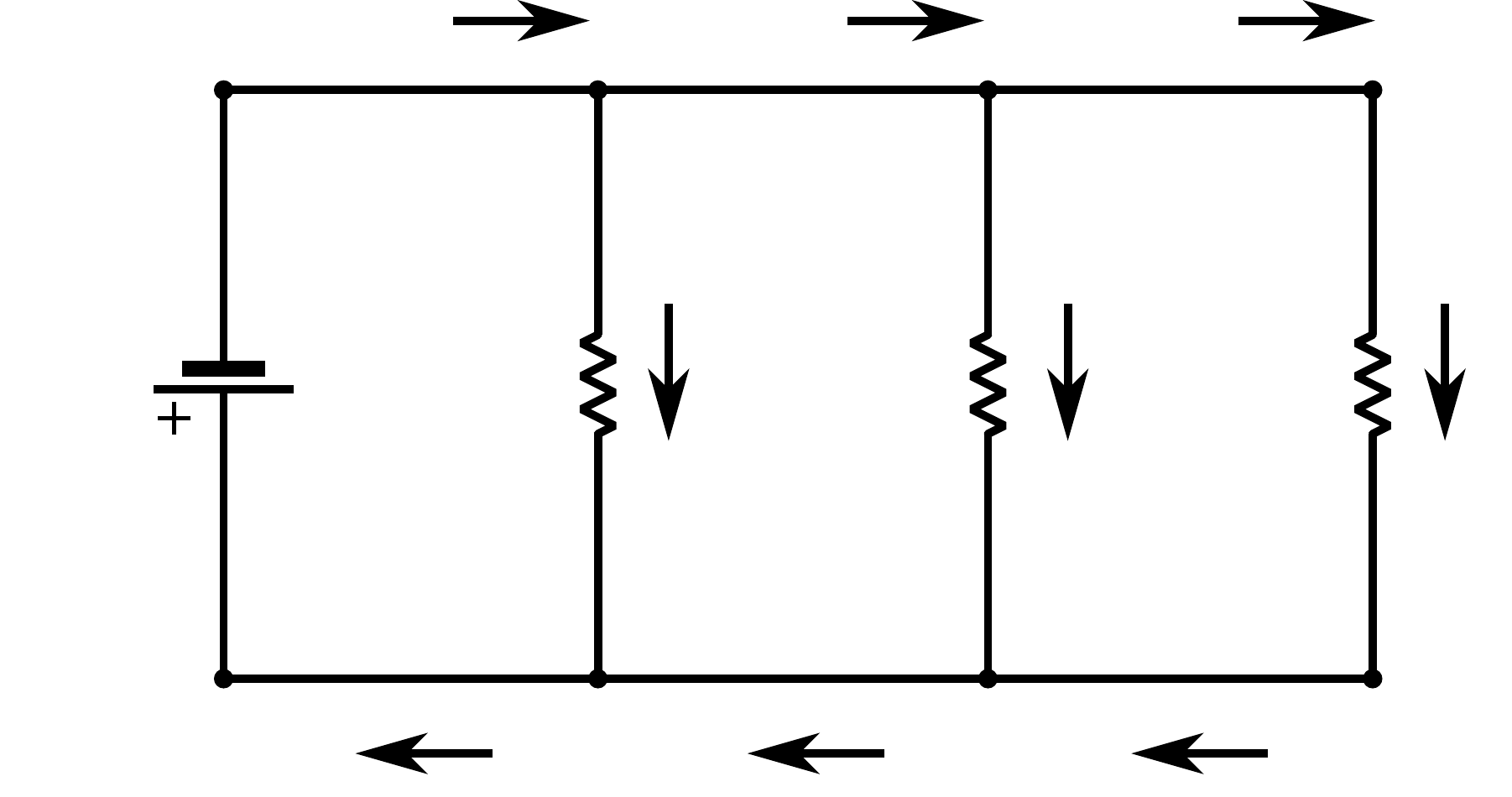}{inkscape-0.48pre1 -z -D --file=Figures/Resistance/example2.svg 
    --export-pdf=Figures/Resistance/example2.pdf --export-latex} \def\svgwidth{0.8\columnwidth}
  \input{Figures/Resistance/example2Label.tex}
      \label{fig:resistance:example2}
    }
    \caption{Figures \ref{fig:resistance:series} and
      \ref{fig:resistance:parallel} with current labels and arrows
      added.}
  \end{center}
\end{figure}

But how do we apply these laws to circuits? Let us first take the
series circuit of Figure \ref{fig:resistance:series} as an example. The first
step when using Kirchhoff's laws is to ensure that all currents and
resistances are labeled. The currents must all be labeled with arrows;
the direction of the arrow does not matter, as long as it is
maintained consistently throughout the application of the laws. In
Figure \ref{fig:resistance:example1} all the resistances and currents have been
labeled and all currents have been given a direction.

The next step is to apply Kirchhoff's first law to every junction in
the circuit. The first junction in this example occurs at the upper
right hand corner of the circuit between currents $I_1$ and $I_2$. The
current $I_1$ is going into the junction (the arrow is pointing to the
junction) while $I_2$ is leaving the junction (the arrow is pointing
away from the junction). Using the first law we have $I_1 =
I_2$. Similarly we can apply the same logic to obtain $I_2 = I_3$ and
consequently $I_1 = I_2 = I_3$.

Now Kirchhoff's second law can be applied to the circuit. To apply the
second law, locate all the loops within the circuit and apply the law
to each loop individually. In this example there is only one loop and
so our job here is simplified. For each loop start at any point on the
loop and trace around the loop. For each resistor crossed going in the
direction of the current subtract the current times the resistance at
that point. For every resistor crossed going in the opposite direction
of the current, add on the current times resistance. For every power
supply crossed going in the direction of the current, add on the power
supply voltage. For every power supply crossed in the opposite
direction of the current, subtract that voltage.

After the loop is finished, equate all of these values with zero. For
the example of the series circuit we begin in the upper left hand
corner and first cross $R_1$ with current $I_1$ in the direction we
are moving and so we must subtract $I_1R_1$. Next we across $R_2$ with
current $I_2$ so we must subtract $I_2R_2$. Crossing $R_3$ we must
again subtract $I_3R_3$ and finally we cross the power supply in the
direction of the current so we add on $V_s$. Putting this all together
and equating to zero gives us the following,
\begin{equation}
  V_s-I_1R_1-I_2R_2-I_3R_3 ~~~\rightarrow~~~ V_s = I_1\left(R_1+R_2+R_3\right)
\end{equation}
where in the second step the relation between the currents obtained
using Kirchhoff's first law was applied. From the second step we see
that we have arrived at Equation \ref{equ:resistance:seriesResistance}!

The above example is somewhat trivial, but it is important to be
careful about skipping steps when using Kirchhoff's laws. Oftentimes
shortcuts can seriously reduce the time needed to complete a problem,
but mistakes can easily creep in without notice. By following the laws
down to even the trivial steps, these mistakes can oftentimes be
avoided.

Now let us consider the slightly more difficult case of the parallel
circuit in Figure \ref{fig:resistance:parallel}. Again we begin by drawing
current labels with associated directions as is done in Figure
\ref{fig:resistance:example2}. Applying Kirchhoff's first law we obtain,
\begin{equation}
  \begin{array}{lcl}
    I_4 = I_5+I_1  &&  I_8 = I_7+I_2 \\
    I_5 = I_6+I_2  &&  I_9 = I_8+I_1 \\
    I_6 = I_3      &&  I_4 = I_9     \\
    I_3 = I_7      &&                \\
  \end{array}
\end{equation}
where the currents flowing in and out of each junction have been
equated, beginning in the upper left hand corner and proceeding
clockwise.

After a little manipulation we see that
\begin{equation}
  I_4 = I_1+I_2+I_3
  \label{equ:resistance:parallel1}
\end{equation}
and that $I_5 = I_8$, etc. We could have come to this conclusion much
more quickly be simplifying the diagram so that $I_1$, $I_2$, and
$I_3$ are all joined at the same points. However, the law was applied
in full to demonstrate the method.

Next we apply the second law. Starting in the upper left hand corner
of the diagram and proceeding clockwise there are three possible
loops, with each loop crossing $V_s$ and $R_1$, $R_2$, or $R_3$, all
in the direction of the current. This yields the following three
equations.

\begin{equation}
  V_s = I_1R_1,~~~V_s = I_2R_2,~~~V_s = I_3R_3
  \label{equ:resistance:parallel2}
\end{equation}

Using the last relation of Equation \ref{equ:resistance:parallel2} and
substituting in $I_3$ from Equation \ref{equ:resistance:parallel1} we obtain,
\begin{equation}
  V_s = \left(I_4-I_2-I_1\right)R_3
\end{equation}
which can be simplified further by using the first two relations of
Equation \ref{equ:resistance:parallel2} to replace $I_1$ and $I_2$. 
\begin{equation}
  \begin{aligned}
    V_s &= \left(I_4 -\frac{V_s}{R_1} - \frac{V_s}{R_2}\right)R_3 \\
    I_4 &= \frac{V_s}{R_1} + \frac{V_s}{R_2}+ \frac{V_s}{R_3} \\
    V_s &= I_4\left(\frac{1}{\frac{V_s}{R_1} + \frac{V_s}{R_2}+
        \frac{V_s}{R_3}}\right) \\
  \end{aligned}
\end{equation}
From this we have derived Equation \ref{equ:resistance:parallelResistance}. The
methods for finding how to add capacitance and inductance together is
similar, and can be explored further by the reader.

\section{Experiment}

\begin{figure}
  \begin{center}
    \executeiffilenewer{Code/Resistance/battery.m}{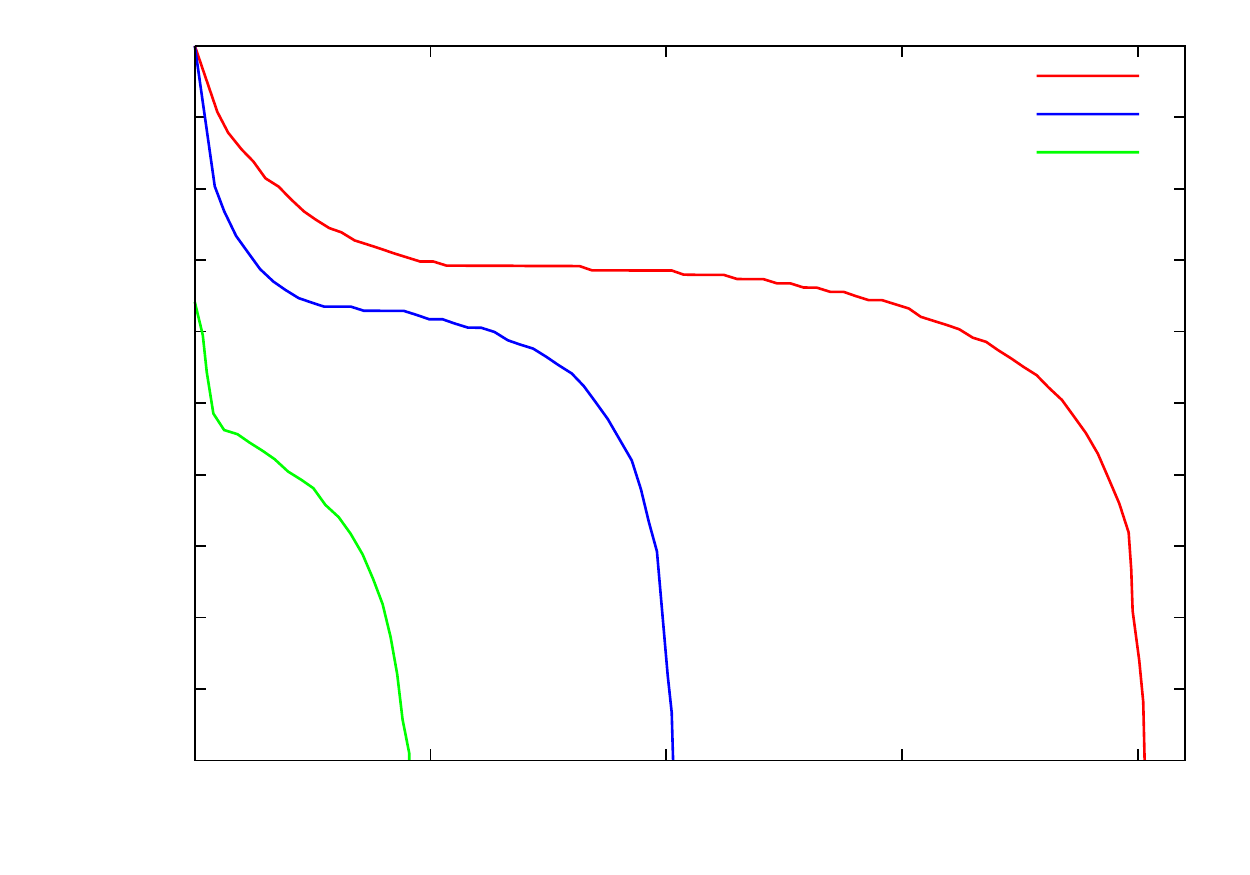}
  {octave --silent --eval "addpath([pwd(),'/Code']); 
    addpath([pwd(),'/Code/Resistance']); battery(1);"} 
  \setlength{\unitlength}{\columnwidth*\real{0.00013}}
  \input{Figures/Resistance/batteryLabel.tex}
    \caption{Voltage versus time (in hours) for a typical AA battery
      at varying levels of constant current. The data from this plot
      was taken from the
      \href{http://data.energizer.com/PDFs/nh15-2500.pdf}{product data
        sheet} for the Energizer NH15-2500 rechargeable AA
      battery.\label{fig:resistance:battery}}
  \end{center}
\end{figure}

In the two examples given above, the resistances have been assumed to
be known, while the currents and consequently voltages, were
unknown. This is because in general, the resistance of components used
within a circuit are known to very high precision, and at standard
temperature and pressure remain very stable. Current and voltage on
the other hand can change greatly, depending upon the power
supply. Power supplies typically are constant voltage; they always
provide the exact same voltage, and change the supplied current as
necessary. Of course, most power supplies can only supply up to a
maximum current after which the voltage is compromised. As an example,
batteries, such as the AA battery supply a constant voltage of between
$1.2$ and $1.5$ V and whatever current is required. These batteries
can only supply a very limited amount of current and over time the
voltage degrades quickly.

Figure \ref{fig:resistance:battery} shows the discharge
characteristics of a rechargeable AA battery. When a very low current
of $1,250$ mA is required the voltage remains relatively constant over
a period of about $2$ hours. When a very large current such as $5,000$
mA is required the battery lasts only about half an hour and the
voltage does not plateau. Oftentimes electrical devices such as cell
phones provide an estimate of the battery life on screen by digitally
measuring the voltage of the battery. However, as can be seen from
Figure \ref{fig:resistance:battery}, the voltage does not provide a
very good indicator of remaining battery life, and so usually the
battery estimators on cell phones are not very accurate.

\begin{figure}
  \begin{center}
    \subfigure[]{
      \executeiffilenewer{Figures/Resistance/resistance.svg}
  {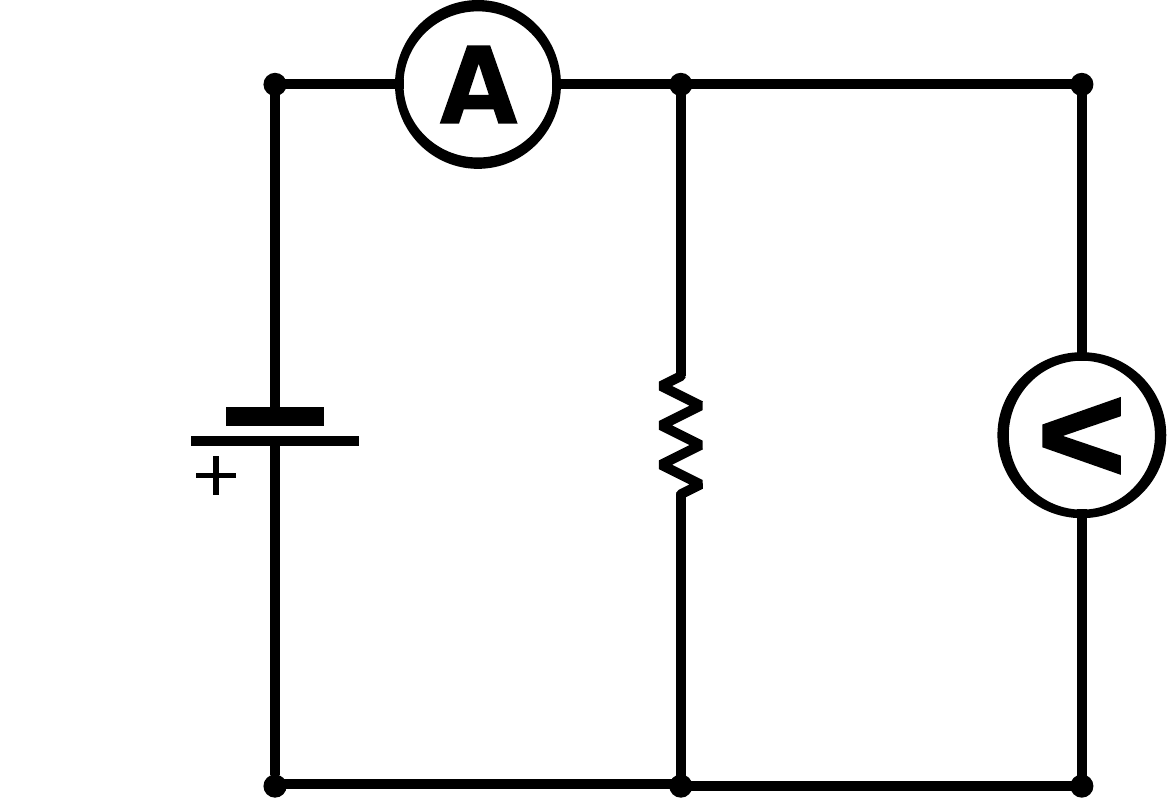}{inkscape-0.48pre1 -z -D --file=Figures/Resistance/resistance.svg 
    --export-pdf=Figures/Resistance/resistance.pdf --export-latex} \def\svgwidth{0.6\columnwidth}
  \input{Figures/Resistance/resistanceLabel.tex}
      \label{fig:resistance:resistance}
    }
    \hspace{2cm}
    \subfigure[]{
      \executeiffilenewer{Figures/Resistance/bridge.svg}
  {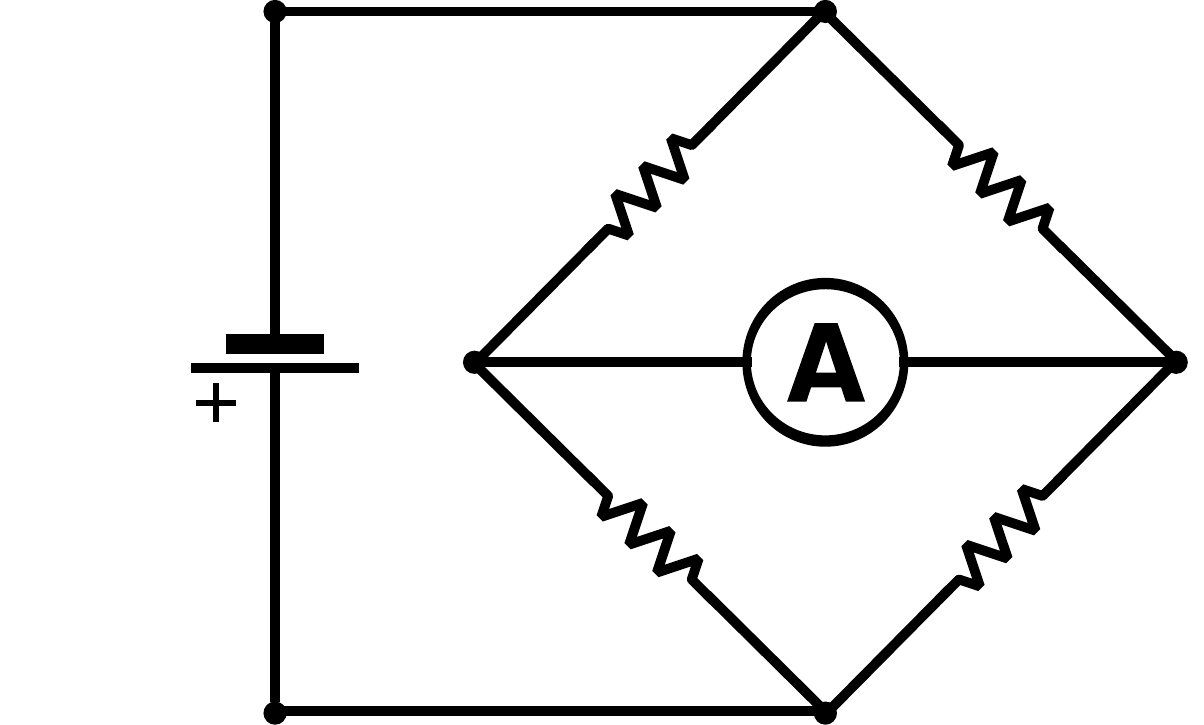}{inkscape-0.48pre1 -z -D --file=Figures/Resistance/bridge.svg 
    --export-pdf=Figures/Resistance/bridge.pdf --export-latex} \def\svgwidth{0.6\columnwidth}
  \input{Figures/Resistance/bridgeLabel.tex}
      \label{fig:resistance:bridge}
    }
    \caption{Two methods for measuring resistance. In Figure
      \ref{fig:resistance:resistance} the current and voltage are
      measured and the resistance is determined using Ohm's law. In
      Figure \ref{fig:resistance:bridge} a Wheatstone bridge is used.}
  \end{center}
\end{figure}

The point of the discussion above is that usually the current and
voltage of a power supply are not known to a very high precision, so
how then are we able to measure resistance to such high precision? One
possibility is given in the circuit diagram of
\ref{fig:resistance:resistance}. Here a power supply is connected to
an ammeter and resistor in series, with a voltmeter in parallel,
measuring the voltage drop. The first problem with this setup is that
two readings are being taken, current and voltage. The second problem
is that the leads to and from the voltmeter, along with the leads from
the ammeter provide resistance that is not taken into consideration.

To counter these effects, a device known as the Wheatstone bridge was
developed, which utilizes what is known as a \term{difference
  measurement}. The circuit for a typical Wheatstone bridge is given
in Figure \ref{fig:resistance:bridge}. In this diagram it is assumed
that the values for $R_1$ and $R_3$ are known to high precision. The
resistor $R_2$ is a variable resistor and $R_x$ is the unknown
resistor being measured. But how does this setup help us determine the
value for $R_x$? Let us apply Kirchhoff's laws to find out.

First, we label all the currents in Figure \ref{fig:resistance:bridge}
and apply Kirchhoff's first law. For resistor $R_1$ through $R_x$
consider currents $I_1$ through $I_x$ all flowing downwards. Again, we
could have chosen to have the currents flowing upwards, or
counterclockwise, or with whatever configuration we would like, but
this configuration is the most intuitive physically. Now let us define
$I_A$ as the current flowing across the ammeter in the center from
right to left. We actually don't know which direction this current is
flowing as this is dependent upon the values of the resistors, but we
can arbitrarily decide it is flowing from right to left. Finally, we
define $I_s$ to be the current flowing from the power supply to the
top of the bridge, and the current flowing from the bottom of the
bridge to the power supply.

From Kirchhoff's first law we have,
\begin{equation}
  \begin{aligned}
    &I_s = I_1+I_3  \\  
    &I_3 = I_A+I_x  \\
    &I_x = I_2+I_s  \\
    &I_1 = I_2+I_A  \\ 
  \end{aligned}
  \label{equ:resistance:bridge1}
\end{equation}
from starting at the top junction of the bridge and moving clockwise
about the diagram. Now we can apply Kirchhoff's second law to the
diagram. There are two loops of interest. The first loop is the top
triangle of the bridge and starts at the top of the bridge and moves
across $R_3$, then the ammeter, and back to the top across $R_1$. The
second loop is the bottom triangle of the bridge and starts at the far
right point and crosses $R_x$, then $R_2$, and returns across the
ammeter to the far right point. Applying Kirchhoff's second law to
these loops we arrive at,
\begin{equation}
  \begin{aligned}
    0 &= -I_3R_3-I_AR_A+I_1R_1 \\
    0 &= -I_xR_x+I_2R_2+I_AR_A \\
  \end{aligned}
  \label{equ:resistance:bridge2}
\end{equation}
where $R_A$ is the resistance of the ammeter.

After moving the negative terms in Equation
\ref{equ:resistance:bridge2} to the left side of the relations we have
the following.
\begin{equation}
  \begin{aligned}
    I_3R_3+I_AR_A &= I_1R_1 \\
    I_xR_x &= I_2R_2+I_AR_A \\
  \end{aligned}
  \label{equ:resistance:bridge3}
\end{equation}
Dividing the bottom relation by the top relation of Equation
\ref{equ:resistance:bridge3},
\begin{equation}
  \frac{I_xR_x}{I_3R_3+I_AR_A} = \frac{I_2R_2+I_AR_A}{I_1R_1}
  \label{equ:resistance:bridge4}
\end{equation}
gives us a ratio between the resistance and currents.

Taking the second and last relation of Equation
\ref{equ:resistance:bridge1} we can now substitute out $I_x$ and $I_1$
from Equation \ref{equ:resistance:bridge4}.
\begin{equation}
  \frac{\left(I_3-I_A\right)R_x}{I_3R_3+I_AR_A} =
  \frac{I_2R_2+I_AR_A}{\left(I_2+I_A\right)R_1} 
  \label{equ:resistance:bridge5}
\end{equation}
From this we now see something interesting, although it may not be
obvious at first glance. If the current flowing across the ammeter is
zero, Equation \ref{equ:resistance:bridge5} becomes,
\begin{equation}
  \frac{I_3R_x}{I_3R_3} = \frac{I_2R_2}{I_2R_1} ~~~\rightarrow~~~
  \frac{R_x}{R_3} = \frac{R_2}{R_1} ~~~\rightarrow ~~~ R_x = \frac{R_3R_2}{R_1}
  \label{equ:resistance:bridge6}
\end{equation}
and we can find the unknown resistance $R_x$ if we know all the other
resistances! This is how the Wheatstone bridge works. The variable
resistor is changed until the current flowing across the center of the
bridge becomes zero. Once this occurs, the unknown resistance can be
determined using the other three known resistances.

In the experiment associated with this chapter the resistance for a
piece of wire is determined along with the resistivity constant,
$\rho$, for the metal the wire is made from. For a conductor through
which direct current is running, the resistance is,
\begin{equation}
  R = \frac{\rho\ell}{A}
  \label{equ:resistance:rho}
\end{equation}
where $\rho$ is the resistivity constant of the resistive material,
$\ell$ is the length of the material, and $A$ is the cross-sectional
area of the material.\footnote{For wires we assume they are circular
  and so $A$ becomes $\pi r^2$ where $r$ is the radius of the wire.}
Subsequently, if we measure the total resistance for a length of wire,
along with the length of the wire, and the width of the wire, we can
determine $\rho$.

The apparatus used to determine the resistance of the length of wire
is the same as that of Figure \ref{fig:resistance:bridge} but instead
of using a variable resistor for $R_2$ and a known resistor for $R_1$
we take advantage of Equation \ref{equ:resistance:rho} and replace
both with a single length of wire. Consequently, the resistances of
$R_1$ and $R_2$ become,
\begin{equation}
  R_1 = \frac{\rho\ell_1}{A},~~~R_2 = \frac{\rho\ell_2}{A}
\end{equation}
which can be substituted back into Equation \ref{equ:resistance:bridge6}.
\begin{equation}
  R_x = \frac{R_3\ell_2}{\ell_1}
  \label{equ:resistance:bridge7}
\end{equation}
The left side of the bridge becomes a wire with a fixed length, and
the ammeter is connected to it with a moveable connection. The
connection is moved up and down the wire until the ammeter indicates
zero current flow, and the lengths $\ell_1$ and $\ell_2$ are then
measured. From these lengths $R_x$ is determined using Equation
\ref{equ:resistance:bridge7}!
\graphicspath{{Figures/Induction/}}

\chapter{Induction}\label{chp:induction}

Whenever electricity is discussed, such as in Chapter
\ref{chp:resistance}, the basic components of an electric circuit are
usually discussed: resistors, capacitors, measuring tools, and power
supplies. But oftentimes another basic electrical component, the
inductor, is ignored, or at least given only a brief explanation. This
is not because \term{inductance} is unimportant, but because the
theory behind inductance can be more complicated than resistance or
capacitance.

Explaining inductance is not simple, so let us begin with the
explanation of inductors, given in Chapter \ref{chp:resistance}. A
resistor is a component in a circuit that dissipates energy, whether
through heat, light, or some other energy transfer
mechanism. Capacitors and inductors are the opposites of resistors,
and store energy, rather than dissipate it. In the case of a
capacitor, the energy is stored in an electrical field, while in the
case of the inductor, the energy is stored in a magnetic field. This
is the core idea behind inductance, it connects electricity with
magnetism into \term{electromagnetism}.

Inductance plays an important role in everyday life. Without
inductance, radios and televisions, AC power transformers, electrical
motors, all would not work. Perhaps most importantly, electricity
would no longer be available, as the concept by which all electrical
generators operate is inductance. To begin understanding the theory
behind inductance we first need to understand the fundamental
interaction between charged particles and electromagnetic fields

\section{Lorentz Force}

The \term{Lorentz force} describes the force felt on a charged particle,
such as an electron, as it passes through electric and magnetic
fields. From intuition, we know that placing two objects with the same
charge next to each other causes the objects to be repelled. This
force due to the electric fields of the two objects is
called the \term{Coulomb force}. But what happens if a charged object
is placed into a magnetic field? Does the object feel a force? The
answer is, it depends.

If the object is not moving, it does not feel a force. However, if the
object is moving, it feels a force proportional to the charge of the
object, the strength of the magnetic field, and the velocity at which
the object is moving in the magnetic field. At this point, perhaps the
obvious question to ask is, why does the force on the object depend
upon its velocity? The answer can be very complicated but the
following explanation, while greatly simplified, will hopefully shed
some light.

The idea in electromagnetism is that electric fields and magnetic
fields are actually the same thing, it just depends upon which
\term{reference frame} the electric or magnetic field is observed
in. This is a consequence of \term{relativity}, which while an
incredibly beautiful theory, will not be discussed here!\footnote{For
  intrepid readers who would like to read more, I would suggest
  A.P. French's book
  \href{http://www.amazon.com/Special-Relativity-M-I-T-Introductory-Physics/dp/0393097935}{Special
    Relativity}. } For example, take a common bar magnet. If we set it
on a table, we will just observe a pure magnetic field. However, if we
run by the counter top, we will begin to observe an electric field,
and less of a magnetic field. This is why the force on a charged
particle from a magnetic field is due to the velocity of the
particle. As soon as the particle begins to move, it starts to see an
electric field and begins to experience Coulomb's force.

So enough qualitative discussion and time for an equation. The
Lorentz force is expressed mathematically as,
\begin{equation}
  \vec{F} = q\vec{E} + q\vec{v}\times\vec{B}
  \label{equ:induction:lorentzForce}
\end{equation}
where $q$ is the charge of the object, $\vec{v}$ is the velocity of
the object, $\vec{E}$ is the electric field, and $\vec{B}$ is the
magnetic field. Notice that everything here is a vector (except for
charge)! That is because all of the quantities above, force, velocity,
electric field, and magnetic field have direction.\footnote{Perhaps at
  this point the astute reader will then ask, why isn't the first term
  also dependent upon velocity? If magnetic fields transform into
  electric fields, shouldn't magnetic fields transform into electric
  fields? The answer is yes, and the explanation given above really is
  not correct, but gives the general idea of what is happening without
  spending an entire book on it.}

It is important to realize that the second term in Equation
\ref{equ:induction:lorentzForce} is the cross product of the velocity of
the charged object with the magnetic field, or $vB\sin\theta$ where
$\theta$ is the angle between the velocity vector and the magnetic
field vector. This means that if the charged particle is moving in the
same direction that the magnetic field is pointing, it experiences no
force, whereas if the charged particle is moving perpendicular to the
magnetic field, it experiences a force of $qvB$ where $v$ and $B$ are
the magnitudes of $\vec{v}$ and $\vec{B}$.

\begin{figure}
  \begin{center}
    \executeiffilenewer{Figures/Induction/lorentzForce.svg}
  {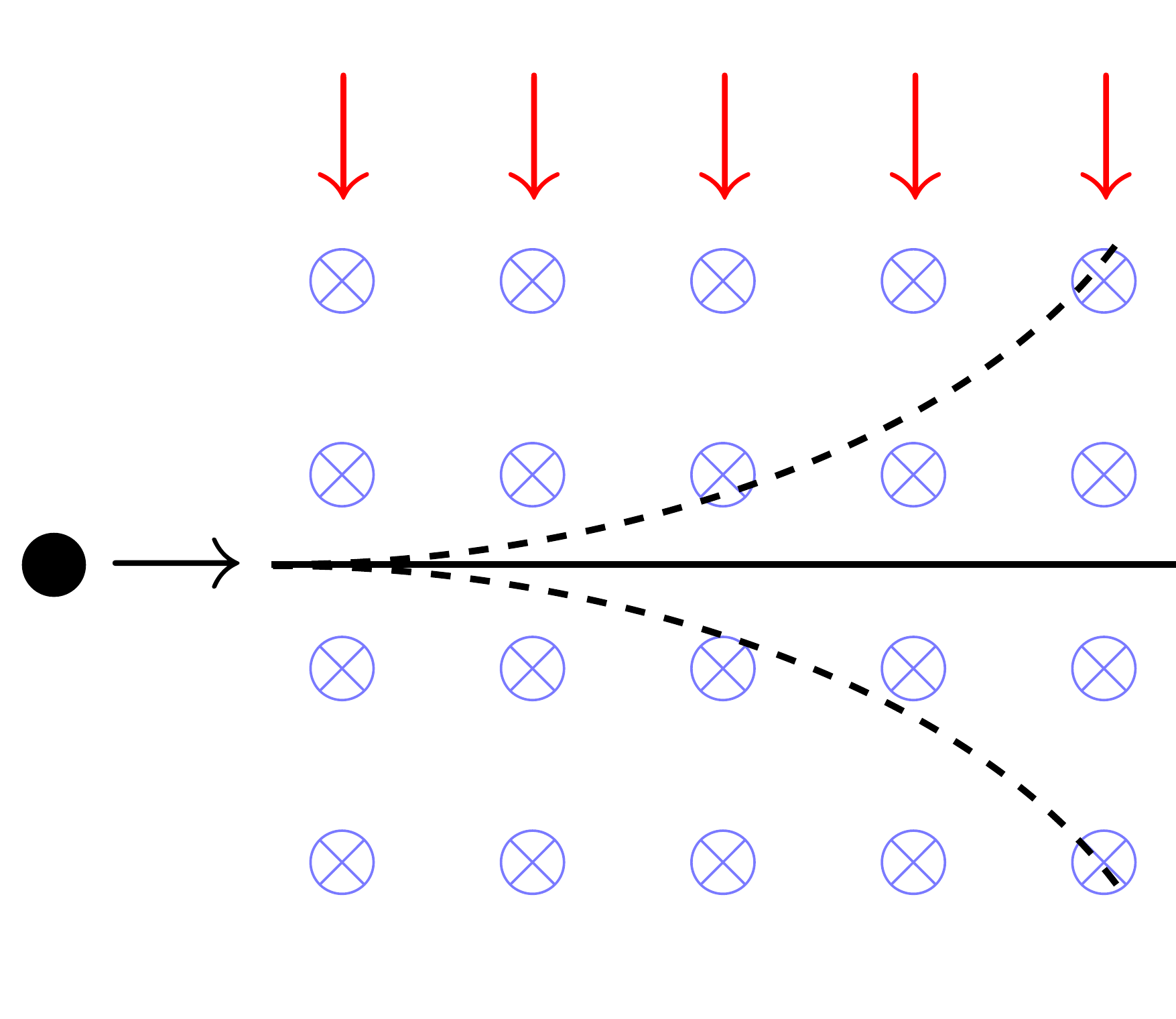}{inkscape-0.48pre1 -z -D --file=Figures/Induction/lorentzForce.svg 
    --export-pdf=Figures/Induction/lorentzForce.pdf --export-latex} \def\svgwidth{6cm}
  \input{Figures/Induction/lorentzForceLabel.tex}
    \caption{An example of the Lorentz force where a charged particle
      is traveling through both an electric and magnetic field. If
      $B = E/v$ then the particle travels in a straight line. If $B >
      E/v$ the particle curves upwards and if $B < E/v$ the particle
      curves downwards.\label{fig:induction:lorentzForce}}
  \end{center}
\end{figure}

Let us consider a simple example that requires the application of
Equation \ref{equ:induction:lorentzForce} and the Lorentz
force. Consider a particle with positive charge $q$ moving through
both a constant electric field and magnetic field, as depicted in
Figure \ref{fig:induction:lorentzForce}. The directions of the fields are
important, and in this diagram the electric field is pointing down the
page, while the magnetic field is pointing into the page. The particle
is traveling from left to right across the page with velocity
$\vec{v}$. If the electric field has a magnitude of $E$, what is the
magnitude of the magnetic field, $B$, required so that the particle
moves in a straight line?

If the particle moves in a straight line, we know from Chapter
\ref{chp:newton} and Newton's first law that the external force
acting on the electron must be zero.
\begin{equation}
  \vec{F} = 0 = q\vec{E} + q\vec{v}\times\vec{B} 
\end{equation}
Next, because the particle has a positive charge of $+q$ we know that
the electric field is exerting a force downward of $qE$, where $E$ is
the magnitude of the electric field. The magnetic field is
perpendicular to the velocity of the particle, and so we know that the
force being exerted on the particle from the magnetic field is $qvB$.

But what is the direction of the force from the magnetic field? From
the \term{right hand rule}, we know that the direction of the force
from the magnetic field must be upward.\footnote{The right hand rule
  is a useful tool for determining the direction of cross products
  like in Equation \ref{equ:induction:lorentzForce}. Using your right
  hand, point your fingers along the direction of the first vector (in
  this case $\vec{v}$). Next bend your fingers at the knuckles so that
  they point in the direction of the second vector, or $\vec{B}$. Now
  look at the direction that your thumb is pointing. This is the
  direction of the cross product of the two vectors. The right hand
  rule is easier to see in action than have described on paper.} This
means that the force from the electric field and the magnetic field
are in opposite directions and so,
\begin{equation}
  qE = qvB
\end{equation}
which gives a value of $B = E/v$ for the magnetic field.

What happens if $B \neq E/v$? Now the particle does experience a net
force, and so it will no longer travel in a straight line. The upper
dotted path in Figure \ref{fig:induction:lorentzForce} illustrates the
trajectory of the particle if $B > E/v$ and the lower dotted path
illustrates the trajectory of the particle if $B < E/v$. This
technique of controlling the trajectory of a charged particle with a
certain velocity using electric and magnetic fields is common in
particle physics (and even devices such as a mass spectrometer) and is
called a \term{velocity selector} as only a particle with the proper
velocity can pass through without being deflected.

\section{Biot-Savart Law}

We now know how a magnetic field can effect an electrically charged
object through the Lorentz force, but can the opposite happen? Can an
electrically charged object effect a magnetic field? The answer is
yes, and the phenomena is described by the \term{Biot-Savart law}. The
Biot-Savart law (conveniently abbreviated the B.S. law) states that any
charge flowing in a loop will produce a magnetic field. While the
Biot-Savart law might sound new, it is a physics phenomena that is
encountered daily. A prime example of the law in action is in the
operation of electromagnets which are used in everything from electric
motors to audio speakers.

An electromagnet usually consists of a \term{ferromagnetic} material,
wrapped in coils of wire. A ferromagnetic material is just a material
that can be magnetized by an external magnetic field, but after the
magnetic field is removed, it slowly loses its magnetization. For
example, a needle can be magnetized using a bar magnet, but slowly
over time the needle demagnetizes. This magnetization of the needle
occurs because the magnetic dipoles of the molecules line up in the
magnetic field, causing a net magnetic field, but slowly, over time,
come out of alignment due to random movements of the molecules in the
needle.

Back to the principle driving an electromagnet. Charge runs through
the coils of wire surrounding the ferromagnet, and because the charge
is flowing in a loop, a magnetic field is created. This magnetic field
aligns the magnetic moments of the molecules in the ferromagnet, and
the ferromagnet produces a net magnetic field. As soon as charge stops
flowing through the coils of the electromagnet, the ferromagnet is no
longer subjected to an external magnetic field, and so it loses its
magnetization.

\begin{figure}
  \begin{center}
    \executeiffilenewer{Figures/Induction/biotSavart.svg}
  {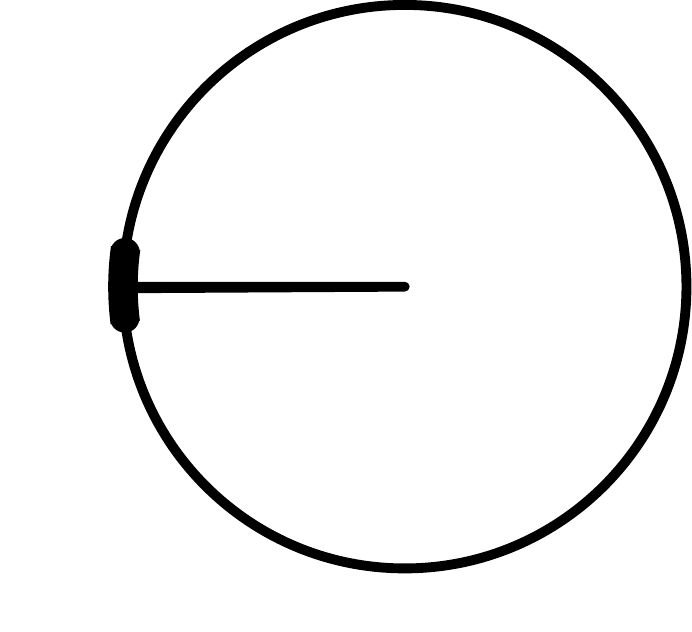}{inkscape-0.48pre1 -z -D --file=Figures/Induction/biotSavart.svg 
    --export-pdf=Figures/Induction/biotSavart.pdf --export-latex} \def\svgwidth{4cm}
  \input{Figures/Induction/biotSavartLabel.tex}
    \caption{An example of the Biot-Savart law. Current is flowing
      clockwise through the loop, and consequently creating a magnetic
      field. \label{fig:induction:biotSavart}}
  \end{center}
\end{figure}

But what is the magnetic field given by a single loop of an electromagnet?
The Biot-Savart law is given by,
\begin{equation}
  \vec{B} = \oint \frac{\mu_0}{4\pi}\frac{Id\vec{l}\times\vec{r}}{|r|^3}
  \label{equ:induction:biotSavart}
\end{equation}
where $\vec{B}$ is the magnetic field created by the current loop,
$\mu_0$ is the \term{magnetic constant}, $I$ is the current flowing in
the loop, $\vec{r}$ is the vector from the current loop to the point
where the magnetic field is being calculated, and $d\vec{l}$ is an
infinitesimal length along the current loop.\footnote{The $\oint$
  symbol is the mathematical notation for a \term{path integral}. For
  example, if we are finding the magnetic field from a square with
  current running through it, the integral of Equation
  \ref{equ:induction:biotSavart} is performed over the path traced out
  by current passing through the square.} Describing Equation
\ref{equ:induction:biotSavart} with just words is not that useful, so
we turn to the diagram of Figure \ref{fig:induction:biotSavart} to
hopefully make things a little clearer.

The current, $I$, is flowing around the circular loop in a clockwise
direction from above. We wish to calculate the magnetic field at the
center of the loop $(x,y) = (0,0)$.\footnote{Of course this problem
  could be done in three dimensions, but this makes the example
  slightly easier to follow.} We split the loop into infinitesimal
pieces of size $dl$, with one of these pieces indicated in bold on the
diagram. Each of these pieces has a direction which points along the
direction of the current. In the example here, the direction of
$d\vec{l}$ is straight upwards in the $y$ direction (tangential to the
circular loop). We need to take the cross product of $d\vec{l}$ with
the vector $\vec{r}$, which points from $d\vec{l}$ to the center of
the circle.

Because $d\vec{l}$ is perpendicular to $\vec{r}$ (a nice property of
circles), the cross product of the two vectors is just $rdl$ and
points into the page by the right hand rule. From the infinitesimal
portion of the current loop, a magnetic field of strength,
\begin{equation}
  d\vec{B} = \frac{\mu_0Idl}{4\pi r^2}(-\khat)
  \label{equ:induction:contributeBiotSavart}
\end{equation}
is created, where $-\khat$ indicates the vector is pointing along the
negative $z$ direction, or in this case, into the page. \footnote{For
  those unfamiliar with \term{hat} notation, the \term{unit vectors}
  $\ihat$, $\jhat$, and $\khat$ point in the $x$, $y$, and $z$
  directions and have a length of $1$.} To find the total magnetic
field at the center of the loop we now need to add up all the magnetic
fields contributed from the infinitesimal pieces, $d\vec{l}$.

Luckily, it turns out that wherever $d\vec{l}$ is located on the
circle, it always provides the same contribution to the magnetic
field, Equation \ref{equ:induction:contributeBiotSavart}. This means
that we can just multiply Equation
\ref{equ:induction:contributeBiotSavart} by $2\pi r$ (the
circumference of the circle) to perform the integration of Equation
\ref{equ:induction:biotSavart}. This gives us a total magnetic field
of,
\begin{equation}
  \vec{B} = \frac{\mu_0 I }{2r}(-\khat)
\end{equation}
again, pointing into the page.

Actually, it turns out that it was not luck that all the contributions
from $d\vec{l}$ were the same. This example was given because the
answer is relatively simple to calculate. If we had tried to find the
magnetic field at the point $(x,y) = (0,r/2)$ the problem would have
become much more complicated. Now the magnetic field contributions
from $d\vec{l}$ are all different, and even more importantly, they
don't all point in the same direction. Calculating out the magnetic
field this way is possible, but very tedious.

\section{Lenz's and Faraday's Laws}

Up to this point we have looked at constant electric and magnetic
fields. But what happens if we look at a changing magnetic field? This
requires the introduction of both \term{Lenz's law} and
\term{Faraday's law}. Lenz's law states that if a magnetic field is
passing through a loop made of some conductor, and the magnetic field
changes, a current will be produced in the conductor that creates a
magnetic field (through the Biot-Savart law) which tries to keep the
magnetic field the same. If this concept sounds somewhat familiar, it
should. Looking all the way back to Chapter \ref{chp:newton},
Newton's second law essentially states that objects like to stay in
the state that they currently are in. A block at rest does not want to
move, and a block that is moving does not want to stop moving.

Lenz's law is exactly the same idea, but now dealing with magnetic
fields. A conductor loop in a magnetic field wants the magnetic field
to stay the same, and if the field changes the conductor tries to
compensate by creating its own magnetic field. The idea is very much
like inertia. Just as it is difficult to bring a block to a stop, it
is difficult to change a magnetic field surrounded by a conducting
loop. The process by which a current is created within the conductor
is called \term{induction}.

Faraday's law takes Lenz's law and adds some math behind the concept
by stating that the induced voltage in the conducting loop is equal to
the derivative of the magnetic field passing through the loop with
respect to time, or,
\begin{equation}
  V = \left|\frac{d (\vec{B}\cdot\vec{A})}{dt}\right|
  \label{equ:induction:faraday}
\end{equation}
where $V$ is the voltage induced in the conducting loop, $\vec{B}$ is
the magnetic field passing through the loop, and $\vec{A}$ is the area
surrounded by the loop through which the magnetic field passes. It is
important to notice that the area of the loop is a vector, not just a
number. The magnitude of $\vec{A}$ is $A$, but the direction is
important as well, because the dot product between $\vec{B}$ and
$\vec{A}$ needs to be taken. The direction of $A$ is the direction of
the \term{normal} to the surface $A$. What this means is that
$\vec{A}$ points in a direction perpendicular to the surface which it
represents.

Again, it may be simpler to explain Equation
\ref{equ:induction:faraday} with an example. Consider a circular loop
of some conductor with a setup similar to Figure
\ref{fig:induction:biotSavart}, but now a magnetic field is passing
through the loop. First, what happens if we consider a magnetic field
that does not change over time? What is the current induced in the
conducting loop? The answer is zero, because the derivative of a
constant is zero. The magnetic field is not changing, and so the
conducting loop does not need to compensate to keep the magnetic field
constant. Looking at a slightly more complicated example, let us
consider a magnetic field given by,
\begin{equation}
  \vec{B}(t) = \frac{B_0}{t^2}\khat
\end{equation}
where $B_0$ is some constant, and $\vec{B}$ is pointing out of the
page, while decreasing over time.

To find out what the voltage induced in the circuit is, we must take
the dot product of $\vec{B}$ with $\vec{A}$. The magnetic field is
pointing out of the page, as is $\vec{A}$, and so the dot product of
the two is $B_0A/t^2$. Taking the time derivative of this yields the
induced voltage,
\begin{equation}
  V = \frac{2 \pi r^2 B_0}{t}
  \label{equ:induction:voltage}
\end{equation}
which will create a magnetic field through the Biot-Savart law. By
Lenz's law we know that the magnetic field will compensate for the
loss of $\vec{B}(t)$ and so we know that the current in the conducting
loop must be flowing counterclockwise. If the resistance $R$ of the
conductor was known, we could calculate the \term{self-induced}
magnetic field using Ohm's law (Equation \ref{equ:resistance:ohm}) to
determine $I$ from Equation \ref{equ:induction:voltage}, and plug this
into Equation \ref{equ:induction:biotSavart}.

\section{Dipoles}

Equation \ref{equ:induction:lorentzForce} allows us to calculate the
force felt on a charged object, but requires knowledge of the electric
field, $\vec{E}$, the magnetic field, $\vec{B}$, the charge of the
object, and the velocity of the object. These last two quantities,
charge and velocity should already be familiar, but how do we
determine the electric and magnetic fields? The electric field from a
single charge, such as an electron is given by,
\begin{equation}
  \vec{E} = \frac{q}{4\pi\epsilon_0}\frac{\rhat}{r^2}
  \label{equ:induction:electricFieldPolar}
\end{equation}
where $q$ is charge, $\epsilon_0$ is the electric constant, and $r$,
just as before, is the distance from the charge. Notice that the
electric field, $\vec{E}$, is a vector quantity, and has a
direction. The direction of $\vec{E}$ is given by the only vector
quantity on the right hand side of the equation, $\rhat$, which has
a length of one, and points from the electric charge to the point
where $\vec{E}$ is being measured. This type of vector which has a
length of one is called a \term{unit vector}.

There are two important points to notice about Equation
\ref{equ:induction:electricFieldPolar}. The first is that if an
observer is measuring the electric field, the direction of the field
will always be pointing directly towards or away (depending upon the
electric charge) from the charged object. The second point is that the
electric field falls off quadratically as the distance $r$ between the
observer and the object increases. For example, if an observer
measures the electric field from an object at one meter, and then at
two meters, the electric field will have been reduced by a factor of
four. In the remainder of this section, we explore slightly more
complicated electric fields from multiple charged objects, and then
extend this to magnetic fields. The math can be very involved, so even
if the following part of this section seems incomprehensible, remember
the two points above.

Equation \ref{equ:induction:electricFieldPolar} gives the electric
field in polar coordinates, but what if we would like the electric
field in Cartesian coordinates?\footnote{For those readers unfamiliar
  with polar and Cartesian coordinates, briefly read over Chapter
  \ref{chp:rotation}.} We can simply write $r$ and $\rhat$ in
Cartesian coordinates. The distance $r$ is given by the Pythagorean
theorem, $r^2 = x^2 +y^2$. The unit vector $\rhat$, however, is a bit
trickier. We must split $\rhat$ into the unit vector in the $x$
direction, $\ihat$, and the unit vector in the $y$ direction,
$\jhat$. This gives us the vector $x\ihat+y\jhat$, which is in the
same direction as $\rhat$, but this is not a unit vector, because it
has length $r$. If we divide by $r$, then we obtain a vector with
length one, which gives us $\rhat =
(x\ihat+y\jhat)/\sqrt{x^2+y^2}$. Plugging in the values we have found
for $r$ and $\rhat$ yields Equation
\ref{equ:induction:electricFieldCartesian}, which is the electric
field from a single charge in Cartesian coordinates.

\begin{equation}
  \vec{E} = 
  \frac{q}{4\pi\epsilon_0}\frac{x\ihat+y\jhat}{\left(x^2+y^2\right)^{3/2}}
  \label{equ:induction:electricFieldCartesian}
\end{equation}

This is convenient, because now we can write the electric field in
terms of its $y$ component, $E_y$, and its $x$ component,
$E_x$. Notice that these are no longer vector quantities, because we
already know which direction they point. If we place an object with
charge $q$ at the coordinates $(0,0)$, then the electric field along
the $y$ direction at the point $(x,y)$ is given by,
\begin{equation}
  E_y = \frac{q}{4\pi\epsilon_0}\frac{y}{\left(x^2+y^2\right)^{3/2}}
  \label{equ:induction:electricFieldY}
\end{equation}
which was obtained by taking only the terms of Equation
\ref{equ:induction:electricFieldCartesian} which were multiplied by
$\jhat$. We could perform the exact same step for $E_x$, but let us
stick with the $E_y$ for now.

\begin{figure}
  \begin{center}
    \executeiffilenewer{Figures/Induction/dipole.svg}
  {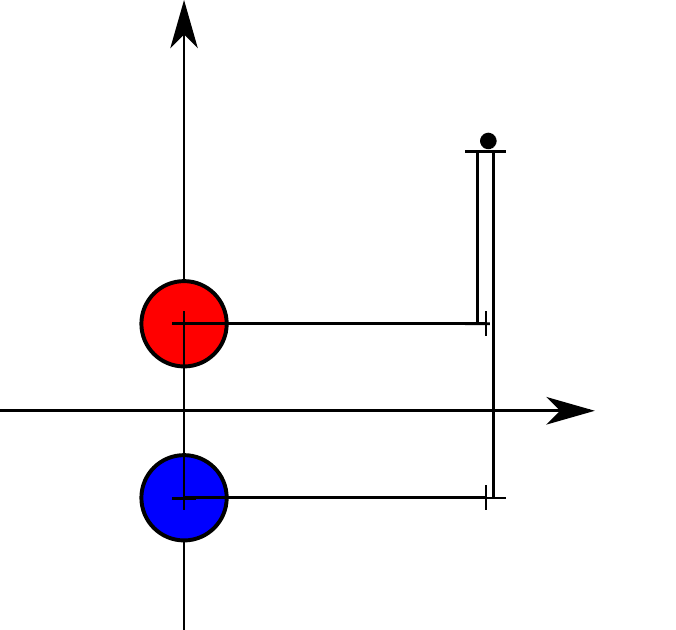}{inkscape-0.48pre1 -z -D --file=Figures/Induction/dipole.svg 
    --export-pdf=Figures/Induction/dipole.pdf --export-latex} \def\svgwidth{0.6\columnwidth}
  \input{Figures/Induction/dipoleLabel.tex}
    \caption{A diagram of a dipole (either electric of magnetic),
      where the dipole is aligned along the $y$-axis, the dipole
      charges are separated by a distance of $\ell$, and the field is
      being measured at point $P$. \label{fig:induction:dipole}}
  \end{center}
\end{figure}

Consider Figure \ref{fig:induction:dipole}, where there are two
charged objects, one with charge $+q$ and one with charge $-q$, placed
along the $y$-axis. We call this configuration of electric charges an
\term{electric dipole}.\footnote{Each charge is considered a ``pole''
  and their are two, so it is called a dipole.} How do we calculate
the electric field from this configuration? For this we need to use
the idea of \term{linear superposition}, which states that the total
electric field from a collection of charged objects can be found by
just adding all the individual electric fields for each charged
object, given by Equation
\ref{equ:induction:electricFieldCartesian}. The term superposition
just means that the fields can be added, and linear means that they
can be added without having to perform some mathematical operation on
each field.\footnote{A good example of something that does not
  normally obey linear superposition is Gaussian uncertainty, as
  discussed in Chapter \ref{chp:uncertainty}.}

Returning to Figure \ref{fig:induction:dipole}, at the point $P =
(x,y)$, the electric fields in the $y$ direction for the positive and
negative charges are given by,
\begin{equation}
  \begin{aligned}
    E_y^+ &= \frac{+q}{4\pi\epsilon_0} \frac{y-\frac{\ell}{2}}
    {\left(x^2+\left(y-\frac{\ell}{2}\right)^2\right)^{3/2}} \\
    E_y^- &= \frac{-q}{4\pi\epsilon_0} \frac{y+\frac{\ell}{2}}
    {\left(x^2+\left(y+\frac{\ell}{2}\right)^2\right)^{3/2}}\\
    \label{equ:induction:electricFieldDipoleComponents}
  \end{aligned}
\end{equation}
where $E_y^+$ is the electric field from the positive charge and
$E_y^-$ is the electric field from the negative charge. Adding these
two electric fields together,
\begin{equation}
  E_y^\mathrm{dipole} = E_y^++E_y^-
  \label{equ:induction:superposition}
\end{equation}
yields the $y$ component of the electric field for both
charges. Unfortunately, this is where the math starts to get a little
more complicated.

We would like to add $E_y^+$ and $E_y^-$ together, but their
denomitators are not the same, differing by a $y-\ell/2$ in $E_y^+$
and a $y+\ell/2$ in $E_y^-$. To get around this problem we use what is
called a \term{Taylor expansion} which allows us to approximate a
function about a given value for a variable.\footnote{Taylor
  expansions were used earlier in Chapter \ref{chp:uncertainty} to
  determine how to combine normal uncertainty.} The Taylor expansion
for the function $f(x)$ is given by,
\begin{equation}
  f(x) \approx f(x_0)+\frac{df(x_0)}{dx}\frac{x-x_0}{1!} +
  \frac{d^2f(x_0)}{dx^2}\frac{(x-x_0)^2}{2!} + \cdots 
  \label{equ:induction:taylor}
\end{equation}
where $x_0$ is the value for the variable $x$ about which we are
expanding.\footnote{There is a lot more to Taylor expansions then just
  giving the definition, but unfortunately understanding them more
  fully requires more detail which is not relevant to this
  discussion.}

Looking at Figure \ref{fig:induction:dipole} again, we can imagine
looking at the charge configuration from very far away. When we do
this, the distance $\ell$ approaches zero, at least from our
viewpoint. Of course the absolute distance between the two charges
stays the same, but to us, the distance looks much smaller. What this
means is that we can approximate the denominators of $E_y^+$ and
$E_y^-$ by performing a Taylor expansion around the point $\ell =
0$. Applying Equation \ref{equ:induction:taylor} to the denominators
of Equation \ref{equ:induction:electricFieldDipoleComponents} yields,
\begin{equation}
  \begin{aligned}
    \frac{1}{\left(x^2+\left(y\pm\frac{\ell}{2}\right)^2\right)^{3/2}}
    \approx \,
    & \frac{1}{\left(x^2+y^2\right)^{3/2}} \pm \frac{3 y \ell}{2
      \left(x^2+y^2\right)^{5/2}}\\
    &+ \frac{3 \left(4 y^2 - x^2\right)
      \ell^2}{8 \left(x^2+y^2\right)^{7/2}} + \cdots\\
    \label{equ:induction:denominatorExpansion}
  \end{aligned}
\end{equation}
expanded out to a term quadratic in $\ell$, or $\ell^2$. We can
actually ignore the term with $\ell^2$, because $\ell \approx 0$ and
so this number will be very small.

Using the first two terms of \ref{equ:induction:denominatorExpansion}
to approximate the denominators of Equation
\ref{equ:induction:electricFieldDipoleComponents} and plugging this
back into Equation \ref{equ:induction:superposition} yields,
\begin{equation}
  \begin{aligned}
    E_y^\mathrm{dipole}
    &= \frac{q}{4\pi\epsilon_0} \left(\frac{3 \ell
        y^2 } {\left(x^2+y^2\right)^{5/2}} - \frac{\ell}
      {\left(x^2+y^2\right)^{3/2}}\right) \\
    &= \frac{q}{4\pi\epsilon_0} \left( \frac{\ell\left(3 y^2 - x^2 -
          y^2\right)}
      {\left(x^2+y^2\right)^{5/2}}\right) \\
    &= \frac{p}{4\pi\epsilon_0}
    \left(\frac{2y^2-x^2}{(x^2+y^2)^{5/2}}\right) \\
    \label{equ:induction:electricDipole}
  \end{aligned}
\end{equation}
which is the $y$ component of the electric field from a dipole. In the
final step we have made the substitution $p = q\ell$ which we call the
\term{electric dipole moment}.

While the result above is interesting, one might wonder, why did we go
to all this trouble? The reason is that magnets cannot be split into
individual ``poles''. They always consist of a north pole and a south
pole.\footnote{If you can show that a magnet can be split, or that a
  \term{magnetic monopole} actually exists, you would win the Nobel
  prize.} What this means is that we cannot use
Equation \ref{equ:induction:electricFieldPolar} to represent a
magnetic field from a typical magnet, but rather an adjusted form of
Equation \ref{equ:induction:electricDipole}. Of course we would also
like to know $E_x$ and $E_z$ if we were working in three dimensional
space, but these can be determined using the exact same method above.

The magnetic field from a magnetic dipole is given by,
\begin{equation}
    B_y^\mathrm{dipole}
    = \frac{\mu_0\mu}{4\pi}
    \left(\frac{2y^2-x^2}{(x^2+y^2)^{5/2}}\right)
    \label{equ:induction:magneticDipole}
\end{equation}
where $\mu_0$ is the magnetic constant and $\mu$ is the \term{magnetic
  dipole moment} for a specific dipole. Notice that only the
substitutions $\epsilon_0 \rightarrow 1/\mu_0$ and $p \rightarrow \mu$
have been made. Remember that the dipole must be aligned along the
$y$-axis for both Equations \ref{equ:induction:electricDipole} and
\ref{equ:induction:magneticDipole} to be valid.

\begin{figure}
  \begin{center}
    \executeiffilenewer{Code/Induction/dipoleFieldLines.m}{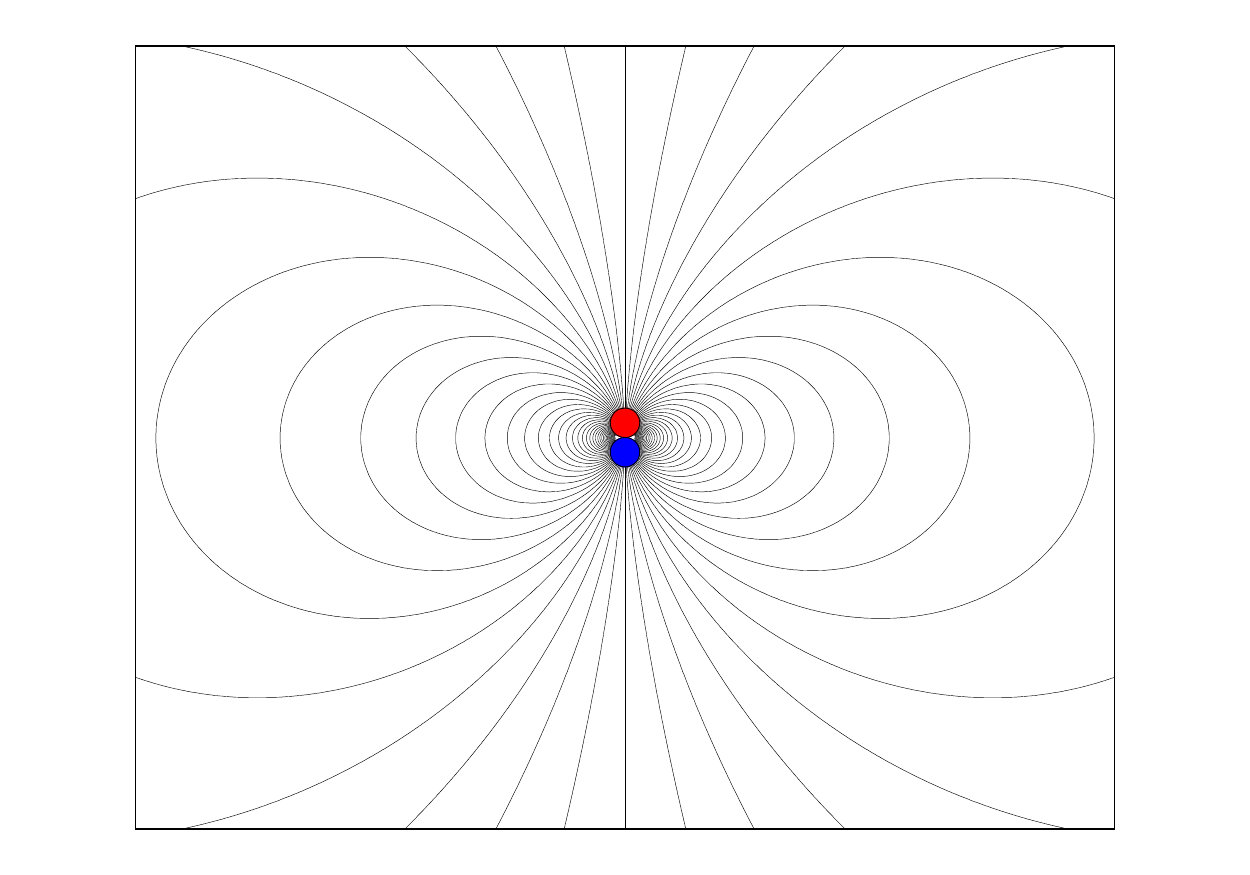}
  {octave --silent --eval "addpath([pwd(),'/Code']); 
    addpath([pwd(),'/Code/Induction']); dipoleFieldLines(1);"} 
  \setlength{\unitlength}{\columnwidth*\real{0.00013}}
  \input{Figures/Induction/dipoleFieldLinesLabel.tex}
    \caption{The field lines of a dipole drawn around the dipole setup
      of Figure \ref{fig:induction:dipole}. If a charge is placed on
      one of the field lines it will begin to move along the
      line.\label{fig:induction:dipoleFieldLines}}
  \end{center}
\end{figure}

Now we have an equation for the electric field from a single charge, a
dipole, and know how to calculate the electric field for multiple
charges using linear superposition. We also have an equation for the
magnetic field from a typical magnetic dipole. But oftentimes
equations, while useful for calculations, do not give a physical
intuition for the problem. Consider what would happen if another free
electric charge was placed into the electric field from an electric
dipole. The charge would feel a force from the electric field, given
by the Lorentz force, and would begin to move.

In Figure \ref{fig:induction:dipoleFieldLines} the dipole is drawn in
red and blue, and drawn around the dipole are \term{field
  lines}. These lines do not represent the strength of the electric
field!\footnote{It is important to remember this, and even many
  experienced physicists forget. The field lines of a magnetic dipole
  are almost always shown instead of the actual magnitude of the force
  because it helps readers understand the direction in which the field
  is pointing.} Instead, the field lines represent the direction the
electric field is pointing. Physically, if the free charge was placed
on a field line, it would begin to move along the field line. The
exact same idea can be applied to a magnetic dipole, but now we can no
longer use a single free charge to map out the field lines, and
instead must use another dipole. If a piece of paper filled with iron
filings is placed over a common bar magnet, the filings will align
along the field lines and create the dipole field line pattern seen in
Figure \ref{fig:induction:dipoleFieldLines}.

\section{Experiment}

A large amount of material was presented above, and while the details
may have escaped, hopefully the general concepts remain. Specifically,
electric charges feel a force from electric fields and magnetic fields
described by the Lorentz force, loops of current create magnetic
fields described by the Biot-Savart law, current loops are created by
changing magnetic fields described by Faraday's law, and most magnetic
fields can be described by a dipole. The experiment for this chapter
manages to connect all the ideas above into a rather simple to perform
but theoretically complex experiment.

\begin{figure}
  \begin{center}
    \executeiffilenewer{Figures/Induction/experiment.svg}
  {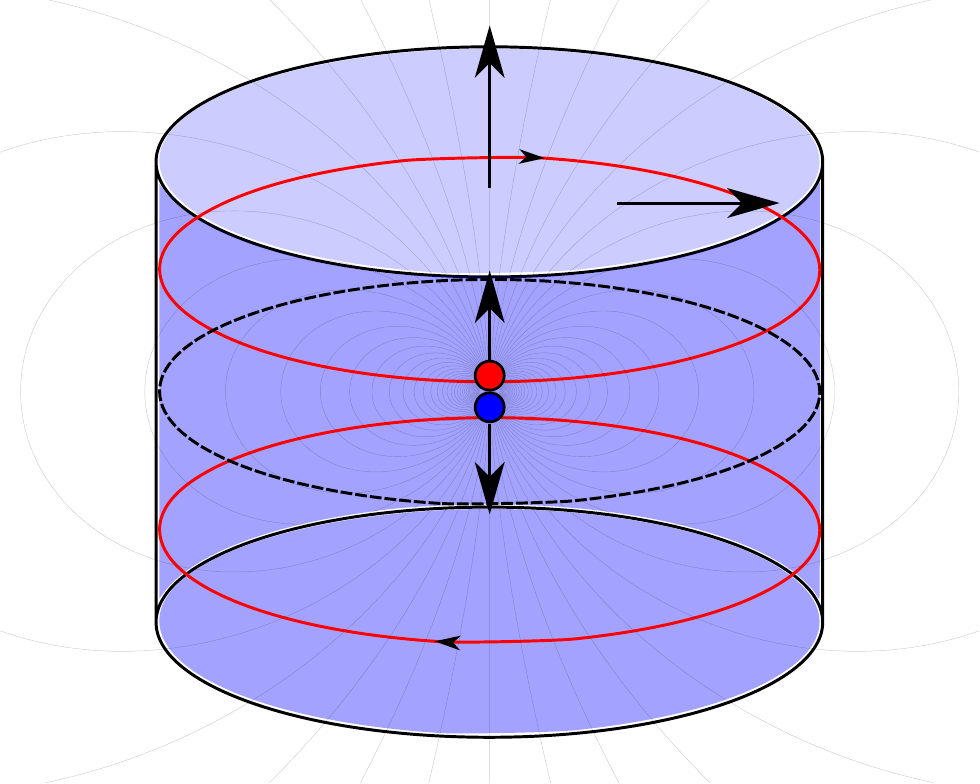}{inkscape-0.48pre1 -z -D --file=Figures/Induction/experiment.svg 
    --export-pdf=Figures/Induction/experiment.pdf --export-latex} \def\svgwidth{\columnwidth}
  \input{Figures/Induction/experimentLabel.tex}
    \caption{The experimental setup for this chapter. A magnetic
      dipole, indicated by the red and blue circles, is falling
      through a copper tube, due to a gravitational force $mg$. The
      changing magnetic fields from the dipole induce eddy currents
      above and below the magnetic field (shown in red), which exert a
      magnetic force, $F_B$, on the magnetic dipole, counteracting the
      gravitational force.\label{fig:induction:experiment}}
  \end{center}
\end{figure}

Consider a conducting pipe (for this experiment we use copper) which
has some resistance $R$, and some radius $r_0$. What happens if we
take a magnet with a dipole moment $\mu$ and drop it down the copper
tube as shown in Figure \ref{fig:induction:experiment}? Without taking
into account any of the theory above we would say quite simply that
the magnet will fall with an acceleration of $g \approx 9.81$ m/s$^2$
by Newton's second law. But this is not the whole story! A changing
magnetic field is passing through a conducting loop (the copper tube),
so an electric current, called an \term{eddy current}, is induced
within the copper tube and can be described by Faraday's law. The
current loop within the pipe creates a magnetic field described by the
Biot-Savart law, which produces a force on the falling magnet,
counteracting the force of gravity. If the magnet falls fast enough,
the induced magnetic field in the pipe will be large enough to
completely counteract the force of gravity and the magnet will reach a
terminal velocity.

The idea of the experiment associated with this chapter is to measure
the terminal velocity at which the magnet falls through the copper
tube. While we have a qualitative prediction (the magnet will fall
more slowly than just through air) it would be be even better to make
a quantitative prediction for the terminal velocity of the
magnet. While the modeling of eddy currents is very complex, we can
make a reasonable estimate of the terminal velocity using only the
theory described above and a few simple assumptions. To begin, let us
first think of the magnet falling through a single copper loop, rather
than a pipe, and calculate the current induced within the copper loop
using Faraday's law.

We can approximate the magnetic field of the magnet using the dipole
field of Equation \ref{equ:induction:magneticDipole} with just a few
adjustments. First, we have changed from two dimensions to three
dimensions but this actually is not a problem. The magnetic dipole is
aligned along the $z$ direction, so we simply substitute $z$ for
$y$. Next, because the pipe is symmetrical in $x$ and $y$, we can
substitute $r$ for $x$. Making these substitutions yields,
\begin{equation}
  B_z =
  \frac{\mu_0\mu}{4\pi}\left(\frac{2z^2-r^2}{(z^2+r^2)^{\frac{5}{2}}}\right)
\end{equation}
which tells us the magnetic field along the $z$ direction due to the
magnetic dipole. For Faraday's law, the magnetic field component along
the $r$ direction, $B_r$, does not matter because this is parallel to
the area of the copper loop, and so $B_r \cdot A = 0$.

Now, to calculate Faraday's law using Equation
\ref{equ:induction:faraday} we must take the dot product of the
magnetic field, $B_z$, with the area, $A$, through which the magnetic
field passes. Unlike the previous example for Faraday's law, the
magnetic field is not constant with respect to $r$, so we must
integrate over the magnetic field.
\begin{equation}
  B_z \cdot A = \int_0^{r_0} \frac{\mu_0\mu}{4\pi}
  \left(\frac{2z^2-r^2}{(z^2+r^2)^{\frac{5}{2}}}\right) 2\pi r\,dr = 
  \frac{\mu_0\mu r_0^2}{2(z^2+r_0^2)^{\frac{3}{2}}}
\end{equation}
Taking the derivative of this with respect to time is a bit tricky,
because there is no time dependence! To get around this we must apply
the chain rule, and first differentiate the magnetic field with
respect to $z$.
\begin{equation}
  V = \left|\frac{d(B_z\cdot A)}{dt}\right| =  \left|\frac{d(B_z\cdot
      A)}{dz} \frac{dz}{dt}\right| =
  \left| \frac{\mu_0\mu r_0^2z}{2(z^2+r_0^2)^{\frac{5}{2}}}
    \frac{dz}{dt} \right|
\end{equation}
Something quite interesting just occurred. We now have a $dz/dt$ term,
which is just the velocity at which the magnet is falling away from
the current loop, $v = dz/dt$!

The induced current within the copper loop, $I$, can be found using
Ohm's law\footnote{If Ohm's law is a little hazy go back and take a
  look at Equation \ref{equ:resistance:ohm} in Chapter \ref{chp:resistance}.},
\begin{equation}
  I = \frac{V}{R} = \frac{v\mu_0\mu
    r_0^2z}{2R(z^2+r_0^2)^{\frac{5}{2}}}
  \label{equ:induction:loopCurrent}
\end{equation}
assuming the copper loop has a resistance $R$. We could use the
Biot-Savart law to calculate the induced magnetic field from this
current loop and then determine the force exerted on the falling
magnet, but it is simpler to use the Lorentz force and calculate the
force exerted on the flowing charge in the copper loop by the falling
magnet. Because there must always be an equal and opposite force, the
same force must be applied to the magnet, but in the opposite
direction.

The definition of current, given in Chapter \ref{chp:resistance}, is
the amount of charge passing per unit time. If the current is
multiplied by the length of the loop through which it is circulating,
this can be multiplied by the magnetic field from the falling magnet
to find the $q\vec{v}\times \vec{B}$ force term of Equation
\ref{equ:induction:lorentzForce}.\footnote{This step might require a
  bit of thought.} If we calculate the force exerted on the loop from
the $B_z$ component of the magnetic field, the direction of the force
is pointing radially outward, and will not counteract
gravity. However, if we calculate the Lorentz force from the radial
$B_r$ component of the magnetic field, the direction will counteract
gravity. The radial component from a magnetic dipole is\footnote{This
  was not derived, but the exact same method used in the previous
  section to determine $E_y^\mathrm{dipole}$ can be used, but now for
  $E_x^\mathrm{dipole}$.},
\begin{equation}
  B_r = \frac{3\mu_0\mu r z}{4\pi(z^2+r^2)^{\frac{5}{2}}}
\end{equation}
and so the force exerted on the magnet is,
\begin{equation}
  F_B = 2\pi r_0 B_r I = \frac{9\mu_0^2\mu^2}{4 R} \frac{r_0^4
      z^2}{(z^2+r_0^2)^5} v
    \label{equ:induction:loopForce}
\end{equation}
where $B_r$ was evaluated at the radius of the copper loop, $r_0$. We
now know the force exerted on the falling magnet, given the distance
$z$ which the magnet is from the copper loop, and the velocity $v$ at
which the magnet is falling.

Let us now return to the more complicated idea of the magnet falling
through a copper pipe. First, assume that the eddy currents within the
pipe caused by the falling magnet can be approximated by an eddy
current flowing directly above the magnet and another eddy current
flowing in the opposite direction, directly below the magnet. As the
magnet falls, these eddy current follow above and below the
magnet. Next, assume that the current in these loops is described
reasonably well by Equation \ref{equ:induction:loopCurrent}. In this
case we don't know $z$, the distance of the eddy currents from the
magnet, but we can write $z$ in terms of $r_0$,
\begin{equation}
  z = Cr_0
\end{equation}
where $C$ is some unknown constant. Additionally, if we assume that
the magnet has reached its terminal velocity, than $v$ is constant and
no longer dependent upon $z$. Plugging this into Equation
\ref{equ:induction:loopForce} we obtain,
\begin{equation}
  F_B = 2\frac{9\mu_0^2\mu^2 C^2}{4R(1+C^2)^5r_0^4} v
  \label{equ:induction:pipeCurrent}
\end{equation}
which is just in terms of the resistance of the copper tube, the
radius of the copper tube, the terminal velocity of the magnet, and
the dipole moment of the magnet. Notice the factor of two in front of
Equation \ref{equ:induction:pipeCurrent}. This is because two eddy
currents are contributing to the force on the magnet.

Because the magnet is at terminal velocity it is no longer
accelerating, and so the forces acting on it must balance. The
gravitational force on the magnet is just $mg$, where $m$ is mass of
the magnet, and $g$ is the acceleration due to gravity. Setting the
gravitational force equal to the magnetic force gives,
\begin{equation}
  \begin{aligned}
    mg &= 2\frac{9\mu_0^2\mu^2 C^2}{4R(1+C^2)^5r_0^4} v\\
    v &= \frac{2 mg  R(1+C^2)^5r_0^4}{9\mu_0^2\mu^2 C^2}
    \label{equ:induction:velocity}
  \end{aligned}
\end{equation}
where in the second line $v$ has been solved for. It turns out from
experiment that $C \approx 1.37$, which corresponds to an eddy current
one third the maximum possible current of Equation
\ref{equ:induction:loopCurrent}. Physically, this means that we can
think of the eddy currents being created above and below the magnet at
a distance of $1.37r_0$, or $1.37$ times the radius of the pipe. Now,
with Equation \ref{equ:induction:velocity}, we have a velocity that we
can verify with experiment.
\graphicspath{{Figures/Waves/}}

\chapter{Waves}\label{chp:waves}

Waves are a critical part of everyday life, but are oftentimes
overlooked. Without waves we could not see or hear; television, radio,
internet, and cell phones all would not be possible without an
extensive understanding of waves and how they behave. Fully
understanding waves is not a simple task, but a large body of
literature is available to those eager to learn more. A good
introduction to wave phenomena can be found in
\href{http://www.amazon.com/Vibrations-Waves-Mit-Introductory-Physics/dp/0748744479/ref=sr_1_1?ie=UTF8&s=books&qid=1258036503&sr=8-1}{\it
  Vibrations and Waves} by A.P. French, while a more detailed analysis
is given in
\href{http://www.amazon.com/Electromagnetic-Vibrations-Radiation-George-Bekefi/dp/0262520478/ref=pd_bxgy_b_img_b}{\it
  Electromagnetic Vibrations, Waves, and Radiation} by Bekefi and
Barrett. A large body of course notes and examples is available
through the MIT OpenCourseWare materials for
\href{http://ocw.mit.edu/OcwWeb/Physics/8-03Fall-2004/CourseHome/index.htm}{8.03}
as taught by Walter Lewin.

\section{Types of Waves}

But what exactly is a wave? The following definition is given by the
Oxford English Dictionary.

\begin{quote}
\small
``Each of those rhythmic alternations of disturbance and recovery of
configuration in successively contiguous portions of a fluid or solid
mass, by which a state of motion travels in some direction without
corresponding progressive movement of the particles successively
affected. Examples are the waves in the surface of water (sense 1),
the waves of the air which convey sound, and the (hypothetical) waves
of the ether which are concerned in the transmission of light, heat,
and electricity.''
\end{quote}

This definition is hardly satisfying and rather long-winded, but
illustrates some of the difficulties of rigorously defining exactly
what a wave is. Perhaps a simpler definition would be ``a periodic
variation traveling within a medium''. Of course this simpler
definition is somewhat inadequate as well; an electromagnetic wave
such as a radio wave does not need to travel through any medium, yet
it is still a wave\footnote{A consequence of the wave/particle duality
  of electromagnetic radiation.}. Despite the difficulty in defining a
wave, most people have a decent intuition for waves from everyday
interactions with everything from water waves to sound and light
waves.

\begin{figure}
  \begin{center}
    \subfigure[Transverse Wave]{
      \executeiffilenewer{Figures/Waves/transverseWave.svg}
  {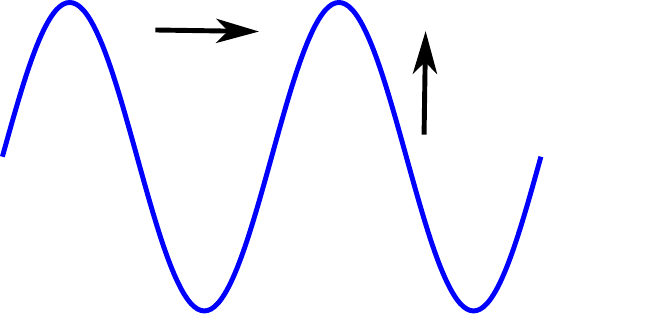}{inkscape-0.48pre1 -z -D --file=Figures/Waves/transverseWave.svg 
    --export-pdf=Figures/Waves/transverseWave.pdf --export-latex} \def\svgwidth{0.6\columnwidth}
  \input{Figures/Waves/transverseWaveLabel.tex}
      \label{fig:waves:transverseWave}}
    \subfigure[Longitudinal Wave]{
      \executeiffilenewer{Figures/Waves/longitudinalWave.svg}
  {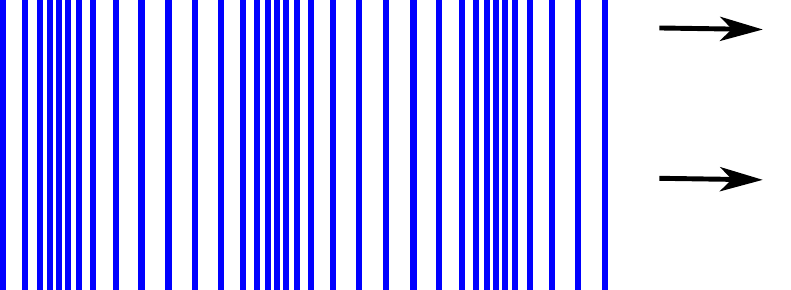}{inkscape-0.48pre1 -z -D --file=Figures/Waves/longitudinalWave.svg 
    --export-pdf=Figures/Waves/longitudinalWave.pdf --export-latex} \def\svgwidth{0.7\columnwidth}
  \input{Figures/Waves/longitudinalWaveLabel.tex}
      \label{fig:waves:longitudinalWave}}
  \end{center}
  \caption{The two different types of waves.\label{fig:waves:typesWave}}
\end{figure}

There exist two general types of waves, \term{transverse} and {\bf
  longitudinal}. In a transverse wave the ``periodic variation'' is
perpendicular to the direction of propagation of the wave as shown in
Figure \ref{fig:waves:transverseWave}. A classic example of this is shaking
the end of a rope. The wave produced by the shaking of the rope
travels horizontally away from the experimenter along the rope while
the actual displacement of the rope is vertical, either up or
down. Similarly, when a guitar string is plucked the string is
displaced vertically, but the wave travels horizontally. Other common
types of transverse waves are light waves (as are all electromagnetic
waves) and water waves.

For a longitudinal wave, shown in Figure \ref{fig:waves:longitudinalWave}
the periodic variation is parallel to the direction of
propagation. Perhaps the most commonly experienced longitudinal waves
are sound waves. Vibrations from an object, whether an instrument or
voice, compress the surrounding air and create a pocket of high
pressure. This pocket of compressed air travels in the direction of
the sound wave until it impacts another object, such as the human ear.

\section{Properties of Waves}

\begin{figure}
  \begin{center}
    \executeiffilenewer{Figures/Waves/propertiesWave.svg}
  {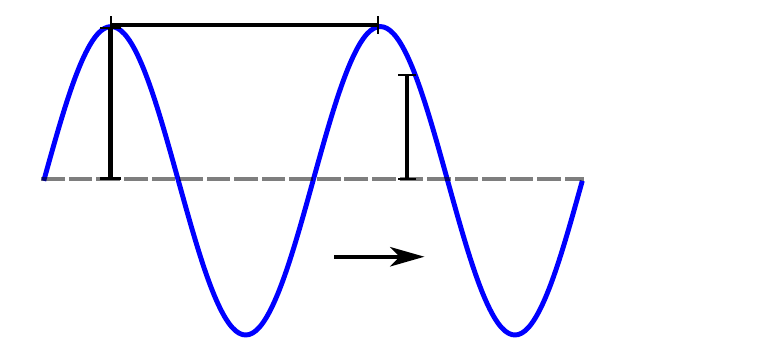}{inkscape-0.48pre1 -z -D --file=Figures/Waves/propertiesWave.svg 
    --export-pdf=Figures/Waves/propertiesWave.pdf --export-latex} \def\svgwidth{0.8\columnwidth}
  \input{Figures/Waves/propertiesWaveLabel.tex}
  \end{center}
  \caption{The fundamental properties used to describe a
    wave.\label{fig:waves:propertiesWave}}
\end{figure}

While the shape of waves can vary greatly, three fundamental
properties can be used to describe all waves. These three properties
are \term{wavelength}, \term{velocity}, and \term{amplitude}. Further
quantities derived from the three quantities above but also useful to
describe waves are frequency, angular frequency, and period. Of the
two types of waves, transverse and longitudinal, it is oftentimes
simpler to visualize transverse waves and so here, the properties of
waves are described using a transverse wave in the form of a sine wave
in Figure \ref{fig:waves:propertiesWave}. The exact same properties describe
longitudinal waves and other types of transverse waves as well, but
are not as simple to depict.

The \term{wavelength} for a wave is usually denoted by the Greek letter
$\lambda$ (spelled lambda) and is given in units of distance,
traditionally meters. In Figure \ref{fig:waves:propertiesWave} the
wavelength is defined as the distance between identical points on two
consecutive oscillations of the wave. Here, the wavelength is measured
from the crest of the first oscillation to the crest of the second
oscillation. It is also possible to measure the wavelength from the
equilibrium of the wave, but this leads to the common problem of
accidentally measuring only half a wavelength. The wavelength of
visible light ranges from $400$ nm (violet) up to about $600$ nm
(red). Audible sound waves have wavelengths ranging from around $20$ m
(very low) to $2$ cm (very high).

The \term{velocity} of a wave, $\vec{v}$, is a vector, and consequently has
an associated direction with units of distance over time. The velocity
for a wave can be found by picking a fixed point on the wave and
measuring the distance traveled by the point over a certain time. The
speed of light within a vacuum is around $3.0\times10^8$ meters per
second, while the speed of sound at standard temperature and pressure
is only $340$ meters per second. The speed of sound is highly
dependent upon the density of the medium through which the sound wave
is traveling and is given by,
\begin{equation}
v_\mathrm{sound} = \sqrt{\frac{C}{\rho}}
\end{equation}
where $C$ is the elasticity of the medium, and $\rho$ the density. As
can be seen, the lower the density of the medium, the faster the
velocity of the sound wave (echos from the top of Mount Kilimanjaro
return faster than echos here in Dublin).

The frequency for a wave is obtained by dividing the velocity of the
wave by the wavelength of the wave.
\begin{equation}
  f = \frac{\left|\vec{v}\right|}{\lambda}
\end{equation}
Frequency is usually denoted by the letter $f$ but is sometimes
denoted using the Greek letter $\nu$ (spelled nu). With dimensional
analysis and the equation above it is possible to see that the units
of frequency are one over time. A more intuitive explanation of
frequency is to imagine a wave passing in front of an observer. The
frequency is the number of oscillations that pass in front of the
observer per unit time. Traditionally, frequency is given in Hertz
(Hz) or oscillations per second. Using the velocities and wavelengths
given above for both light and sound, the frequency of visible light
is between the range of $7.5\times10^{14}$ Hz (violet light) and
$5\times10^{14}$ Hz (red light) while the frequency of audible sound
is between $20$ and $20,000$ Hz.

Another method of expressing frequency is \term{angular frequency} or
\term{angular velocity}, given by the Greek letter $\omega$, and
introduced earlier in Chapter \ref{chp:rotation}. As implied by the
name, angular frequency is oftentimes used for problems involving
rotational motion, and is just the velocity in radians per second at
which an object is rotating. It is possible to write angular frequency
in terms of normal frequency,
\begin{equation}
\omega = 2\pi f
\end{equation}
with $f$ given in Hz and $\omega$ in radians per second.

The \term{period} of a wave is the time it takes one oscillation to
pass an observer, or just the inverse of the frequency. Period is
denoted by the letter $T$ and given in units of time.
\begin{equation}
  T = \frac{1}{f}
\end{equation}

The \term{amplitude} of a wave, as shown in Figure
\ref{fig:waves:propertiesWave}, is denoted by the letter $A$ and is a
measure of the distance of the peak of an oscillation to the
equilibrium of the wave. The unit for amplitude is generally distance,
but depends on the type of wave. Sometimes waves are not symmetric
like the wave shown in Figure \ref{fig:waves:propertiesWave}, and so the
type of amplitude defined above is renamed \term{peak amplitude} and a
new amplitude called \term{root mean square (RMS) amplitude} is used
instead. The RMS amplitude for a wave is found first by squaring the
distance of the wave from equilibrium. The mean of this squared value
is found, and the square root taken to return the RMS amplitude. More
mathematically,
\begin{equation}
A_{RMS} = \sqrt{\frac{\int x^2\,dt}{\int\,dt}}
\end{equation}
where $x$ is the distance of the wave at time $t$ from equilibrium. The
RMS amplitude is important for sound and light waves because the
average power of the wave is proportional to the square of the RMS
amplitude. For example, a very bright light source, such as a halogen
bulb, emits light waves with very large RMS amplitudes.

\section{Standing Waves}

\begin{figure}
  \begin{center}
    \executeiffilenewer{Figures/Waves/harmonics.svg}
  {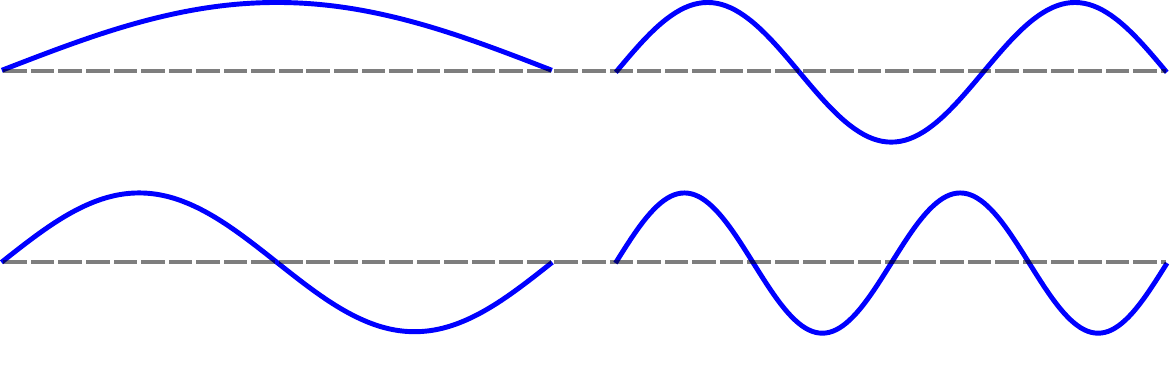}{inkscape-0.48pre1 -z -D --file=Figures/Waves/harmonics.svg 
    --export-pdf=Figures/Waves/harmonics.pdf --export-latex} \def\svgwidth{\columnwidth}
  \input{Figures/Waves/harmonicsLabel.tex}
  \end{center}
  \caption{The first four harmonics for a standing wave fixed at both
    ends.\label{fig:waves:harmonics}}
\end{figure}

One of the most interesting and important phenomena of waves is the
\term{standing wave} which is what makes music possible. In a standing
wave, places along the vibrating medium (take for example a guitar
string) do not move. These non-moving points are called \term{nodes}. Using the guitar string example, it is possible to see that
by necessity there must be at least two nodes, one for each fixed end
of the string. When there are only nodes at the end points of the
string, no nodes in the middle, the sound emitted is called the \term{fundamental} or \term{first harmonic}. The standing wave produced
when the frequency of the wave is increased sufficiently to produce a
node in the middle of the guitar string is called the \term{first
  overtone} or \term{second harmonic}. Whenever the frequency of the
wave is increased enough to produce another node, the next harmonic or
overtone is reached. Figure \ref{fig:waves:harmonics} gives the first four
harmonics of a guitar string.\footnote{Here we use a sine wave for
  simplicity, but technically this can be any wave that satisfies the
  wave equation.}

Looking again at the example above we can derive a relationship for
the fundamental frequency of a vibrating string. To begin, we notice
that the wavelength is just twice the length of the string and so we
can write frequency in terms of velocity of the wave on the string,
$v$, and length of the string, $L$.
\begin{equation}
f = \frac{v}{2L}
\label{equ:waves:fundementalFrequency}
\end{equation}
Unfortunately we do not usually know the velocity at which a wave
travels through a string. However, we can find the velocity by
examining the force diagram for a infinitely small portion of the
string. The next portion of this chapter becomes a little
mathematically involved, but the end result is beautiful.

For a wave traveling in two dimensions , $x$ and $y$, along the
$x$-axis, we can write that the second derivative of $y$ with respect
to $x$ is equal to some coefficient $C$ times the acceleration of the
wave in the $y$ direction.
\begin{equation}
\frac{\partial^2 y}{\partial x^2} = C\frac{\partial^2 y}{\partial t^2}
\label{equ:waves:waveEquation}
\end{equation}
This comes directly from the two-dimensional case of the wave
equation.\footnote{If you want to read more about the wave equation,
  consult the resources given at the start of this chapter.} If we
substitute in some function for our wave, in this case,
\begin{equation}
y = A\sin(x-v t)
\end{equation}
where $v$ is the velocity of the wave, we can find a value for $C$.
\begin{equation}
\begin{aligned}
\frac{\partial^2 y}{\partial x^2} = A\sin(v t-x),~~
\frac{\partial^2 y}{\partial t^2} = v^2A\sin(v t-x)\\
A\sin(v t-x) = Cv^2A\sin(v t-x) \Rightarrow C = \frac{1}{v^2}\\
\label{equ:waves:waveConstant}
\end{aligned}
\end{equation}

\begin{figure}
  \begin{center}
    \executeiffilenewer{Figures/Waves/forceDiagram.svg}
  {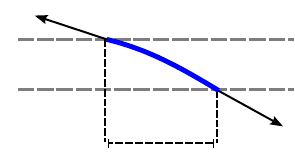}{inkscape-0.48pre1 -z -D --file=Figures/Waves/forceDiagram.svg 
    --export-pdf=Figures/Waves/forceDiagram.pdf --export-latex} \def\svgwidth{6cm}
  \input{Figures/Waves/forceDiagramLabel.tex}
  \end{center}
  \caption{Force diagram for an infinitesimally small portion of a
    string vibrating at its fundamental frequency.\label{fig:waves:forceDiagram}}
\end{figure}

Now we use a force diagram to obtain a relationship similar to
Equation \ref{equ:waves:waveEquation} but for this particular situation. The
force diagram for an infinitesimally small portion of the string is
shown in Figure \ref{fig:waves:forceDiagram}. We begin by finding the net
force in the $y$ direction.
\begin{equation}
F_y = T_1\sin\theta_1-T_2\sin\theta_2
\end{equation}
We can now use Newton's second law to equate this to $ma$,
\begin{equation}
T_1\sin\theta_1-T_2\sin\theta_2 = \mu\Delta x \frac{\partial^2 y}{\partial t^2}
\label{equ:waves:newtonSecond}
\end{equation}
where $\mu\Delta x$ is the mass of the piece of string (density $\mu$
times distance), and ${\partial^2 y}/{\partial t^2}$ the acceleration
of the string along the $y$ direction.

Because the string is not moving in the $x$ direction and because the
string is not deforming much from equilibrium, we can assume that the
net force in the $x$ direction is $0$ and in either direction is
approximately $T$, the tension on the entire string.
\begin{equation}
  F_x = T_1\cos\theta_1 - T_2\cos\theta_2 \approx T - T = 0
\end{equation}
Dividing Equation \ref{equ:waves:newtonSecond} by $T$ and using $T \approx
T_{1,2}\cos\theta_{1,2}$ from above we obtain the following.
\begin{equation}
  \frac{T_1\sin\theta_1}{T}-\frac{T_2\sin\theta_2}{T} =
  \frac{T_1\sin\theta_1}{T_1\cos\theta_1}-\frac{T_2\sin\theta_2}{T_2\cos\theta_2}
  = \tan\theta_1-\tan\theta_2 = \frac{\mu\Delta x}{T} \frac{\partial^2
    y}{\partial t^2}
\end{equation}
As $\Delta x$ approaches zero by definition $\tan\theta_1-\tan\theta_2
\rightarrow \partial y/\partial x$. Dividing both sides by $\Delta x$
then gives,
\begin{equation}
  \frac{\partial y}{\Delta x\partial x} = \frac{\partial^2 y}{\partial
    x^2} = \frac{\mu}{T} \frac{\partial^2 
    y}{\partial t^2}
\end{equation}
which looks almost exactly like Equation \ref{equ:waves:waveEquation}!
Setting,
\begin{equation}
\frac{1}{v^2} = \frac{\mu}{T}
\end{equation}
we can solve for $v$ in terms of $\mu$, the linear density of the
string, and $T$, the tension of the string.
\begin{equation}
v = \sqrt{\frac{T}{\mu}}
\end{equation}
Placing this value for $v$ back into Equation
\ref{equ:waves:fundementalFrequency}, we can now write the fundamental
frequency for a string in terms of length of the string $L$, density
of the string $\mu$, and tension of the string $T$.
\begin{equation}
  f = \frac{1}{2L}\sqrt{\frac{T}{\mu}}
\end{equation}
Intuitively this equation makes sense. Shorter strings produce higher
frequencies; a bass produces very low notes, while a violin produces
very high notes. Similarly, very taught strings, like those on a
violin produce higher sounds than loose strings. Anyone who has tuned
a string instrument knows that by tightening the string the instrument
becomes sharp, and by loosening the string, the instrument becomes
flat. Finally, the denser the string, the lower sound. This is the
reason most stringed instruments do not use gold or lead strings, but
rather tin, steel, or plastic.

\section{Experiment}

Despite some of the mathematical intricacies of the argument above we
have the end result that the fundamental frequency for a vibrating
string is given by,
\begin{equation}
f = \frac{1}{2L}\sqrt{\frac{T}{\mu}}
\end{equation}
where $\mu$ is the linear density of the string, and $T$ the tension
on the string. The equation above can be rewritten as,
\begin{equation}
  f = \frac{1}{2}L^{-1}\left(\frac{T}{\mu}\right)^{1/2} =
  kL^n\left(\frac{T}{\mu}\right)^r
  \label{equ:waves:theory}
\end{equation}
which matches the general formula given in the lab manual when
$k=1/2$, $n=-1$, and $r=1/2$.

The goal of this experiment is to verify the theory above by
determining values for $n$ and $r$ experimentally. This is done by
oscillating a string and varying the length and tension of the string
to determine values for $n$ and $r$ respectively.

\begin{figure}
  \begin{center}
    \executeiffilenewer{Figures/Waves/experiment.svg}
  {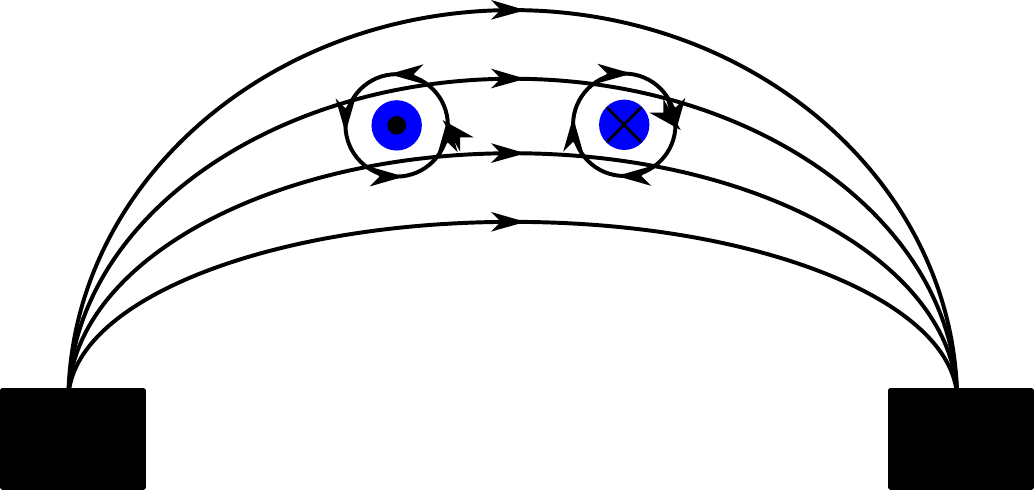}{inkscape-0.48pre1 -z -D --file=Figures/Waves/experiment.svg 
    --export-pdf=Figures/Waves/experiment.pdf --export-latex} \def\svgwidth{8cm}
  \input{Figures/Waves/experimentLabel.tex}
  \end{center}
  \caption{The magnetic field lines of the experiment. The blue
    circles represent wires running perpendicular to the plane of the
    page. The wire with a $\times$ symbol in the middle has current
    running into the page, whereas the wire with $\cdot$ symbol has
    current running out of the page.The black boxes represent the
    poles of the horse-shoe magnet.\label{fig:waves:experiment}}
\end{figure}

The experimental apparatus consists of a metal wire (acting as a our
string) through which an alternating current is driven. The changing
current causes the magnetic field lines surrounding the wire (due to
moving charge) to flip between clockwise and counter-clockwise as
shown in Figure \ref{fig:waves:experiment}. The horse-shoe magnetic provides
a constant magnetic field so that the string is pulled back and forth,
causing the wire to oscillate. The weight at the end of the wire
provides a known tension on the wire.

Because it is oftentimes easier to visualize linear relationships, we
use a common trick in physics and take the natural log of both sides
of Equation \ref{equ:waves:theory}.
\begin{equation}
  \ln f = \ln\left[kL^n\left(\frac{T}{\mu}\right)^r\right]
\end{equation}
Next we use some of the algebraic properties of logarithms,
\begin{equation}
\begin{aligned}
  \ln \left(ab\right) &= \ln a + \ln b\\
  \ln \left(\frac{a}{b}\right) &= \ln a - \ln b\\
  \ln \left(a^b\right) &= b\ln a\\
\end{aligned}
\end{equation}
to write a linear relationship between $\ln f$ and $\ln L$, and $\ln f$
and $\ln T$.
\begin{equation}
\begin{aligned}
  \ln f &= n\ln L + \ln \left[k\left(\frac{T}{\mu}\right)^r\right]\\
  \ln f &= rln T + \ln \left[kL\left(\frac{1}{\mu}\right)^r\right]\\
\end{aligned}
\end{equation}
From here we can determine values for $n$ when we vary $L$ and measure
$f$, and similarly we can find $r$ when we vary $T$ and measure
$f$. Finally, if we assume that $k=1/2$ we can determine the linear
density of the resonating wire, $\mu$.
\graphicspath{{Figures/Optics/}}

\chapter{Optics}\label{chp:optics}

Optics is a subject oftentimes used to study other phenomena in
physics such as quantum mechanics or general relativity, but is rarely
studied as its own subject. Unfortunately, this means that many
physicists have learned optics as a patchwork of examples rather than
a complete field. This chapter attempts to give a brief cohesive
introduction into geometrical optics, but only touches the tip of the
iceberg. Fermat's principle is introduced to explain the phenomena of
reflection and refraction (and subsequently diffraction) in the first
section and these principles are applied to thin lenses in the second
section. A slightly more advanced look at geometrical optics using
matrix methods is introduced in the fourth section, and these methods
are then applied to the thick lens in the fifth section to derive the
lensmaker's formula. Finally, the Fresnel equations are discussed in
the sixth section.

To begin, we must first differentiate between \term{geometrical optics}
and \term{physical optics}. In geometrical optics the wavelength of the
light passing through the optical setup is much smaller than the size
of the apparatus. For example, visible light passing through a pair of
glasses would be accurately modeled by geometrical optics as the
wavelength of the light is $\approx 5\times10^{-5}$ cm while the size
of the glasses is on the order of $1$ cm. Physical optics models light
when the wavelength of the light is of the same order as the optical
apparatus. One can think of geometric optics as the limit of physical
optics for $\lambda \ll x$ where $x$ is the size of the optical
apparatus. Additionally, because light is treated as rays, geometrical
optics cannot account for the polarization and interference of light.

In geometrical optics light can be \term{translated}, \term{refracted},
\term{diffracted}, and \term{reflected}. The translation of light is
just light traveling in a straight line through some medium such as
glass or water. The refraction of light is when light passes from one
medium to another medium, where the index of refraction for the first
medium is different from the second. When this process occurs, the
light is bent by an angle which is governed by Snell's law. The
diffraction of light occurs when light is separated by
wavelength. Diffraction occurs in a prism through the process of
refraction, and so diffraction will not be discussed further in this
chapter. The final phenomena, reflection, is when light bounces off an
object governed by the law of reflection.

\section{Fermat's Principle}

All three principle actions that can be performed on light,
translation, reflection, and refraction can be derived from
\term{Fermat's principle}. This principle states that light is always
in a hurry; it tries to get from point $A$ to point $B$ in the
shortest amount of time.\footnote{Alternatively, because light must
  always go the same speed in a specific medium, we can think of light
  trying to take the shortest possible path, and subsequently light is
  not in a hurry, but just lazy. On another note, this statement of
  Fermat's principle is the original form, but also is not quite
  correct mathematically speaking; sometimes light will take the
  longest path.} The first consequence of this principle is that light
translates or travels in a straight line. Ironically, while the idea
that a straight line is the fastest route between two points seems
painfully obvious, the actual mathematics behind it are
non-trivial.\footnote{For those readers who are interested in learning
  more, try reading
  \href{http://severian.mit.edu/philten/math/geodesics.pdf}{\it
    Geodesics: Analytical and Numerical Solutions} by Coblenz, Ilten,
  and Mooney.} Proving the laws behind reflection and refraction,
however, are much simpler.

\begin{figure}
  \begin{center}
    \subfigure[]{
      \executeiffilenewer{Figures/Optics/reflectProof.svg}
  {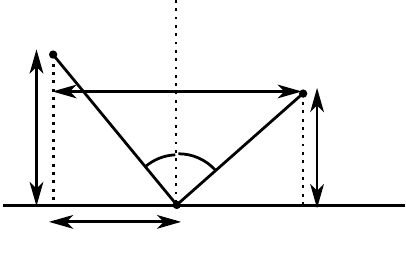}{inkscape-0.48pre1 -z -D --file=Figures/Optics/reflectProof.svg 
    --export-pdf=Figures/Optics/reflectProof.pdf --export-latex} \def\svgwidth{0.4\columnwidth}
  \input{Figures/Optics/reflectProofLabel.tex}
      \label{fig:optics:reflectProof}
    }
    \subfigure[]{
      \executeiffilenewer{Figures/Optics/refractProof.svg}
  {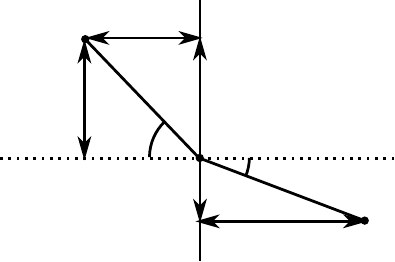}{inkscape-0.48pre1 -z -D --file=Figures/Optics/refractProof.svg 
    --export-pdf=Figures/Optics/refractProof.pdf --export-latex} \def\svgwidth{0.4\columnwidth}
  \input{Figures/Optics/refractProofLabel.tex}
      \label{fig:optics:refractProof}
    }
    \caption{Diagrams used to prove both the law of reflection, Figure
      \ref{fig:optics:reflectProof}, and Snell's law,
      \ref{fig:optics:refractProof}, using Fermat's
      principle.\label{fig:optics:fermatProof}}
  \end{center}
\end{figure}

Consider the diagram of Figure \ref{fig:optics:reflectProof} where a light
ray begins at point $A$, hits a surface at point $R$ at distance $x$
from $A$, and reflects to point $B$ at distance $d$ from $A$. We
already know that the light must travel in a straight line, but what
is the fastest path between $A$ and $B$? The first step is to write
out the total time the path of the light takes. This is just the
distance traveled divided by the speed of the light as it travels
through a medium with \term{refractive index} $n_0$ where,
\begin{equation}
  n \equiv \frac{c}{v}
  \label{equ:optics:refractiveIndex}
\end{equation}
or in words, the refractive index is the speed of light in a vacuum
divided by the speed of light in the medium through which it is
traveling.

The total time of the path is then just $t = n_0L/c$ where $L$ is the
path length. Using the Pythagorean theorem the path length is,
\begin{equation}
  L = \sqrt{a^2+x^2}+\sqrt{b^2+(d-x)^2}
\end{equation}
where $a$ is the distance of point $A$ above the surface, and $b$ the
distance of $B$ above the surface. Now we wish to find the value for
$x$ that provides the minimum total time for the path. First we take
the derivative of the path time $t$ with respect to the distance $x$.
\begin{equation}
  \frac{dt}{dx} = \frac{n_0}{c}\left(\frac{x}{\sqrt{a^2+x^2}} -
    \frac{d-x}{\sqrt{b^2+(d-x)^2}}\right)
  \label{equ:optics:reflectDerivative}
\end{equation}
This is just slope of $t(x)$, which when equal to zero yields either a
local minimum or maximum of the function $t(x)$.\footnote{We know we
  have reached the bottom of a valley or the top of a hill when the
  ground is no longer at a slope, but is flat.}

Setting Equation \ref{equ:optics:reflectDerivative} to zero yields an
extremum for the path time $t$. Without looking at a graph or taking
the second derivative of $t(x)$ it is impossible to tell if this
extremum is a maximum or a minimum, but to save time, let us proceed
with the knowledge that this extremum is a minima. This means that the
path time is minimized when the following relation is true.
\begin{equation}
  \frac{x}{\sqrt{a^2+x^2}} = \frac{d-x}{\sqrt{b^2+(d-x)^2}}
  \label{equ:optics:reflectMinima}
\end{equation}
But the left side of the equation is the angle of incidence,
$\sin\theta_0$, and the right side of the equation is the angle of
reflection, $\sin\theta_1$. Note that for both reflection and
refraction the angle of incidence and reflection is defined as the
angle between the light ray and the \term{normal} to the surface of the
reflecting or refracting object. The normal to a surface is the line
perpendicular to the surface at that point; for example, the normal to
the surface of a sphere is always the radius. Plugging $\sin\theta_0$
and $\sin\theta_1$ back into Equation \ref{equ:optics:reflectMinima} just
gives us the \term{law of reflection},
\begin{equation}
  \theta_0 = \theta_1
  \label{equ:optics:lawOfReflection}
\end{equation}
which tells us the incident angle is equal to the reflected angle!

For refraction we can proceed with the exact same procedure that we
used for reflection, but now the velocity of the light from point $A$
to $R$, shown in Figure \ref{fig:optics:refractProof}, is different from the
velocity of the light traveling from point $R$ to $B$. This yields the
total time traveled as,
\begin{equation}
  t = \frac{n_0}{c}\sqrt{a^2+x^2}+\frac{n_1}{c}\sqrt{b^2+(d-x)^2}
  \label{equ:optics:refractTime}
\end{equation}
where $n_0$ is the index of refraction for the first medium and $n_1$
is the index of refraction for the second medium. Finding the minima
using the method above results in,
\begin{equation}
  \frac{n_0}{c}\frac{x}{\sqrt{a^2+x^2}} =
  \frac{n_1}{c}\frac{d-x}{\sqrt{b^2+(d-x)^2}}
\end{equation}
which can be further reduced to \term{Snell's law},
\begin{equation}
  n_0\sin\theta_0 = n_1\sin\theta_1
  \label{equ:optics:snell}
\end{equation}
which dictates how light is refracted! Another way to think of Snell's
law is a lifeguard trying to save someone drowning in the ocean. For
the lifeguard to get to the victim as quickly as possible they will
first run along the beach (low index of refraction) and then swim to
the victim (high index of refraction).

\section{Thin Lenses}

\begin{figure}
  \begin{center}
    \subfigure[]{
      \executeiffilenewer{Figures/Optics/convexOutside.svg}
  {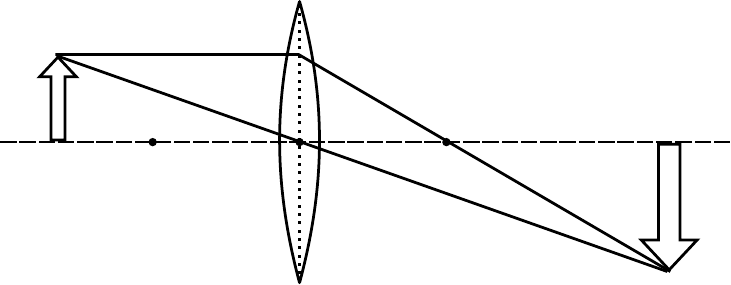}{inkscape-0.48pre1 -z -D --file=Figures/Optics/convexOutside.svg 
    --export-pdf=Figures/Optics/convexOutside.pdf --export-latex} \def\svgwidth{8cm}
  \input{Figures/Optics/convexOutsideLabel.tex}
      \label{fig:optics:convexOutside}
    }
    \subfigure[]{
      \executeiffilenewer{Figures/Optics/convexInside.svg}
  {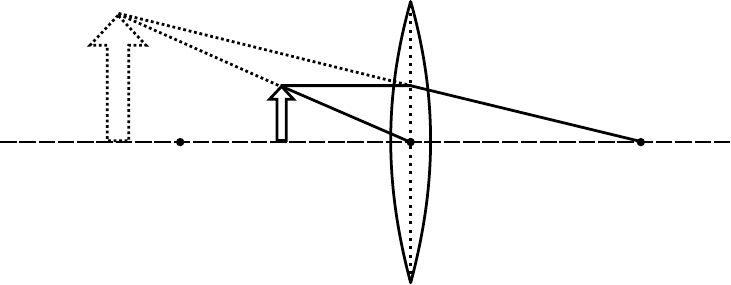}{inkscape-0.48pre1 -z -D --file=Figures/Optics/convexInside.svg 
    --export-pdf=Figures/Optics/convexInside.pdf --export-latex} \def\svgwidth{8cm}
  \input{Figures/Optics/convexInsideLabel.tex}
      \label{fig:optics:convexInside}
    }
    \subfigure[]{
      \executeiffilenewer{Figures/Optics/concaveOutside.svg}
  {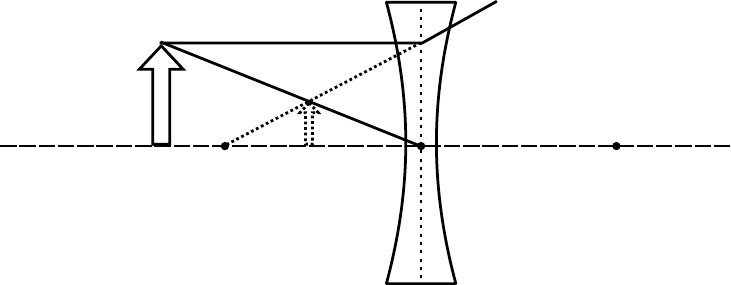}{inkscape-0.48pre1 -z -D --file=Figures/Optics/concaveOutside.svg 
    --export-pdf=Figures/Optics/concaveOutside.pdf --export-latex} \def\svgwidth{8cm}
  \input{Figures/Optics/concaveOutsideLabel.tex}
      \label{fig:optics:concaveOutside}
    }
    \subfigure[]{
      \executeiffilenewer{Figures/Optics/concaveInside.svg}
  {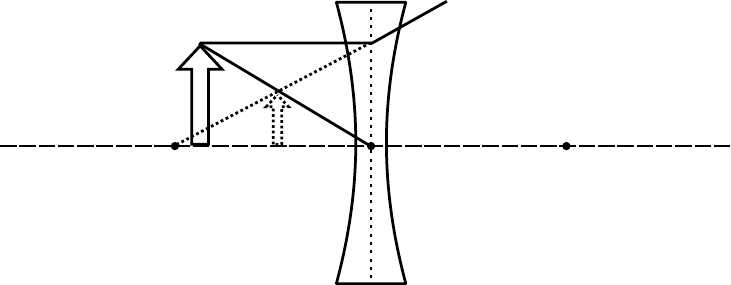}{inkscape-0.48pre1 -z -D --file=Figures/Optics/concaveInside.svg 
    --export-pdf=Figures/Optics/concaveInside.pdf --export-latex} \def\svgwidth{8cm}
  \input{Figures/Optics/concaveInsideLabel.tex}
      \label{fig:optics:concaveInside}
    }
    \caption{Ray tracings for convex and concave thin lenses. Figures
      \ref{fig:optics:convexOutside} and \ref{fig:optics:convexInside}
      give diagrams for a convex lens with the object outside and
      inside the focal length. Figures \ref{fig:optics:concaveOutside}
      and \ref{fig:optics:concaveOutside} give diagrams for a concave
      lens with the object outside and inside the focal length as
      well.\label{fig:optics:rays}}
  \end{center}
\end{figure}

Now that we know how light translates, reflects, and refracts, we can
apply \term{ray diagrams} to \term{thin lenses}. A thin lens is a
spherical lens where the thickness of the lens is much less than the
\term{focal length} of the lens. Before the definition of focal length
is given, let us consider the lens of Figure
\ref{fig:optics:convexOutside}. We can draw light rays from some object,
given by the arrow in the diagram, up to the lens as straight
lines. However, once they pass through the lens they must bend because
of refraction.

It turns out that every light ray parallel to the \term{optical axis},
the dotted line passing through the center of the lens, converges to a
single \term{focal point} after passing through the lens. We will not
prove this now, but by using the matrix method of the following
section it can be seen that focal points must exist for spherical
lenses. The focal length of a lens is just the distance from the
center of the lens to the focal point and is given by the \term{thin
  lens approximation},
\begin{equation}
  \frac{1}{f_0} =
  \frac{n_1-n_0}{n_0}\left(\frac{1}{R_2}-\frac{1}{R_1}\right)
  \label{equ:optics:thinFocal}  
\end{equation}
where $R_1$ is the radius of curvature for the front of the lens,
$R_2$ the radius of curvature for the back of the lens, $n_0$ the
index of refraction for the medium surrounding the lens, and $n_1$ the
index of refraction for the lens.

A ray diagram then is a geometrical method to determine how light rays
will propagate through an optical set up. An optical system can be
fully visualized by drawing the two following rays.
\begin{enumerate}
\item A ray from the top of the object, parallel to the optical axis,
  is drawn to the centerline of the lens, and then refracted by the
  lens to the focal point.
\item A ray from the top of the object is drawn through the center of
  lens and passes through unrefracted.
\end{enumerate}
In Figure \ref{fig:optics:convexOutside} these two rays are traced for a
\term{convex lens} with the object outside of the focal length of the
lens.

This brings about an important point on notation. The radii of
curvature and focal lengths for lenses have signs associated with
them. Oftentimes different conventions for the signs are used between
sources. For the purposes of this chapter the radius for a sphere
starts at the surface and proceeds to the center of the sphere. The
focal length begins at the center of the lens and proceeds to the
focal point. Positive values are assigned to focal lengths or radii
moving from left to right, while negative values are assigned for the
opposite movement. A convex lens has $R_1 > 0$ and $R_2 < 0$ while a
\term{concave lens} has $R_1 < 0$ and $R_2 > 0$.

The diagrams of Figure \ref{fig:optics:rays} demonstrate the four basic
configurations possible with thin lenses: object outside focal length
of convex lens, object inside focal length of convex lens, object
outside focal length of concave lens, and object inside focal length
of concave lens. In the diagrams, a dotted arrow indicates a
\term{virtual image} while a solid arrow indicates a \term{real image}
or object. A real image is an image that can be projected onto a
surface, while a virtual image cannot. Notice the only optical
configuration that produces a real image is with an object outside the
focal length of a convex lens.

\section{The Matrix Method}

But where does the thin lens approximation of Equation
\ref{equ:optics:thinFocal} come from? How do we know the assumptions
of the previous section are true? One method for deriving Equation
\ref{equ:optics:thinFocal} involves a large amount of rather tedious
geometry along with the use of Snell's law. Unfortunately, this
derivation is just more of the ray tracing diagrams above, and cannot
be easily adapted to other optical systems, for example a system with
a concave lens followed by a convex lens. What happens if we want a
method for determining the focal length for any optical system? Such a
method does exist and is called the \term{matrix method}, but as its
name implies can be mathematically challenging at times for readers
unfamiliar with matrix operations.\footnote{For those who do not know
  how to use matrices, I suggest just quickly reading up on basic
  matrix operations because the method outlined below is well worth
  the work to understand it. All that is needed is a basic
  understanding of how to multiply $2\times2$ matrices along with
  creating $2\times2$ matrices from a system of linear equations.}

The idea behind the matrix method is to split any optical system into
building blocks of the three basic actions: translation, refraction,
and reflection. Putting these three actions together in an order
dictated by the set up allows us to build any optical system

Let us consider the example of a lens. To begin, an incident light ray
is diffracted when entering the front of the lens, and so the first
action is diffraction. Next the ray must pass through the lens, and so
the second action is translation. Finally, when the ray exits the lens
it is diffracted again, and so the third action is diffraction. If we
placed a mirror some distance behind the lens, then the light would
translate to the mirror, reflect off the mirror, and then translate
back to the lens followed by diffraction, translation, and diffraction
from the lens.

Now that we understand how to use these optical building blocks to
create optical systems, we must define the blocks more mathematically
using linear algebra.\footnote{Matrices are oftentimes used to solve
  linear systems of equations algebraically, hence linear algebra
  refers to the branch of mathematics dealing with matrices.}  The
idea is to describe a single light ray as a vector and an optical
building block as a matrix that modifies that vector. Luckily, a light
ray in two dimensions can be fully described by its angle to the
optical axis, $\theta$, and its distance above the optical axis,
$y$. This means that a light ray can be fully described by a vector of
two components, $(\theta,y)$, and subsequently each optical building
block is represented by a $2\times2$ matrix.

\begin{figure}
  \begin{center}
    \subfigure[Translation]{
      \executeiffilenewer{Figures/Optics/translate.svg}
  {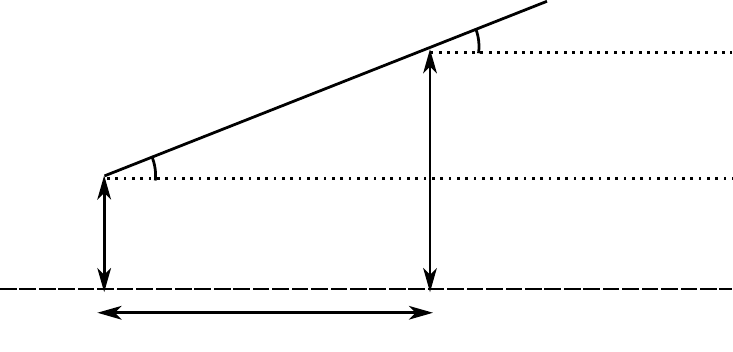}{inkscape-0.48pre1 -z -D --file=Figures/Optics/translate.svg 
    --export-pdf=Figures/Optics/translate.pdf --export-latex} \def\svgwidth{0.7\columnwidth}
  \input{Figures/Optics/translateLabel.tex}
      \label{fig:optics:translate}
    }
    \subfigure[Refraction]{
      \executeiffilenewer{Figures/Optics/refract.svg}
  {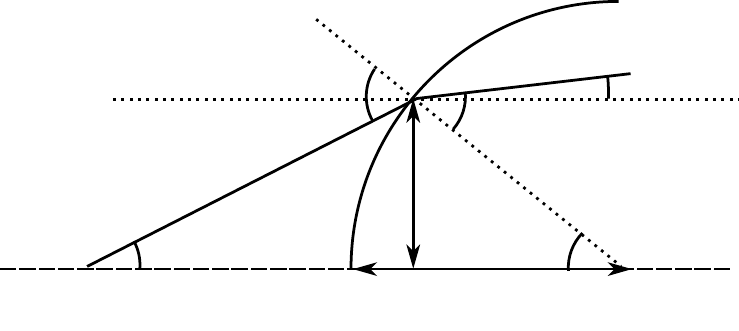}{inkscape-0.48pre1 -z -D --file=Figures/Optics/refract.svg 
    --export-pdf=Figures/Optics/refract.pdf --export-latex} \def\svgwidth{0.7\columnwidth}
  \input{Figures/Optics/refractLabel.tex}
      \label{fig:optics:refract}
    } \subfigure[Reflection]{
      \executeiffilenewer{Figures/Optics/reflect.svg}
  {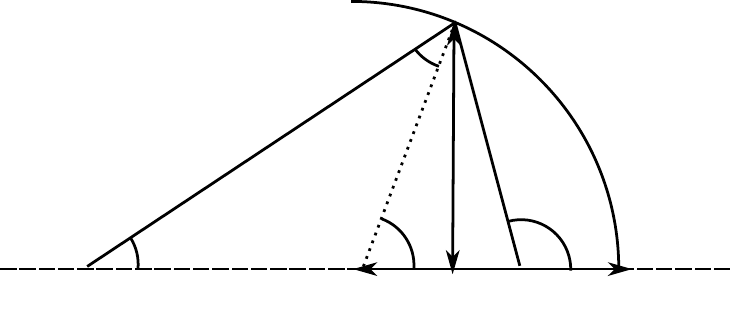}{inkscape-0.48pre1 -z -D --file=Figures/Optics/reflect.svg 
    --export-pdf=Figures/Optics/reflect.pdf --export-latex} \def\svgwidth{0.7\columnwidth}
  \input{Figures/Optics/reflectLabel.tex}
      \label{fig:optics:reflect}
    }
    \caption{Diagrams of the three possible actions performed on a
      light ray which can be represented as matrices.\label{fig:optics:matrices}}
  \end{center}
\end{figure}

Using Figure \ref{fig:optics:translate}, we can determine the matrix that
represents the translation action for a light ray. First we write out
a general equation,
\begin{equation}
  \left[\begin{array}{c}
      \theta_1 \\
      y_1 \\
    \end{array}\right] = 
  \left[\begin{array}{cc}
      T_1 & T_2 \\
      T_3 & T_4 \\
    \end{array}\right]
  \left[\begin{array}{c}
      \theta_0 \\
      y_0 \\
    \end{array}\right]
  \label{equ:optics:translate1}
\end{equation}
which states that the outgoing ray represented by the vector
$(\theta_1,y_1)$ is equal to the incoming light ray, $(\theta_0,y_0)$,
multiplied by the translation matrix $\mathbf{T}$. Performing matrix
multiplication on Equation \ref{equ:optics:translate1} yields,
\begin{equation}
  \begin{aligned}
    \theta_1 &= T_1\theta_0+T_2y_0 \\
    y_0 &= T_3\theta_0+T_4y_0 \\
  \end{aligned}
  \label{equ:optics:translate2}
\end{equation}
but by using the diagram of Figure \ref{fig:optics:translate}, we can also
write the following system of equations.
\begin{equation}
  \begin{aligned}
    \theta_1 &= \theta_0 \\
    y_0 &= x\tan\theta_0 + y_0 \\
  \end{aligned}
  \label{equ:optics:translate3}
\end{equation}

Unfortunately, Equation \ref{equ:optics:translate3} is not linear, it has a
tangent of $\theta_0$! To avoid this problem we make the assumption
that all light rays passing through the optical system are
\term{paraxial}; the angle $\theta$ between the ray and the optical
axis is very small and so the small angle approximations $\sin\theta
\approx \theta$, $\cos\theta \approx 1$, and $\tan\theta \approx
\theta$ hold. Using the small angle approximation on Equation
\ref{equ:optics:translate3} yields,
\begin{equation}
  \begin{aligned}
    \theta_1 &= \theta_0 \\
    y_0 &=  x\theta_0 + y_0 \\
  \end{aligned}
  \label{equ:optics:translate4}
\end{equation}
which when combined with Equation \ref{equ:optics:translate2} allows us to
determine the matrix elements of $\mathbf{T}$.

\begin{equation}
  \mathbf{T} = 
  \left[\begin{array}{cc}
      1 & 0 \\
      x & 1 \\
    \end{array}\right]
  \label{equ:optics:translate}
\end{equation}

With the matrix for translation, we can move on to the matrices for
refraction and reflection. The method for determining these matrices
is exactly the same as for the translation matrix, but unlike the
translation matrix, these matrices are dependent upon the geometry of
the optical component being modeled for either diffraction or
reflection. Spherical lenses and mirrors are the most common optical
components used, and so we will find the refraction and reflection
matrices for spherical components.

The diagram of Figure \ref{fig:optics:refract}, shows a light ray refracting
through a spherical lens with radius $R$. By geometry and the small
angle approximation we know,
\begin{equation}
  \begin{aligned}
    \alpha_0 &= \theta_0+\phi \\
    \alpha_1 &= \theta_1+\phi \\
    \sin\phi &= \frac{y_0}{R} ~~~\rightarrow~~~ \phi \approx
    \frac{y_0}{R} \\
    y_0 &= y_1 \\
  \end{aligned}
  \label{equ:optics:refract1}
\end{equation}
while by Snell's law (and again the small angle approximation) we know,
\begin{equation}
  n_0\sin\alpha_0 = n_1\sin\alpha_1 ~~~\rightarrow~~~ n_0\alpha_0
  \approx n_1\alpha_1
  \label{equ:optics:refract2}
\end{equation}
where $n_0$ is the index of refraction outside the lens, and $n_1$ is
the index of refraction within the lens. Plugging in Equation
\ref{equ:optics:refract1} into Equation \ref{equ:optics:refract2} yields,
\begin{equation}
  \theta_1 = 
  \left(\frac{n_0}{n_1}\right)\theta_0 +
  \frac{1}R{}\left(\frac{n_0}{n_1}-1\right)y_0
\end{equation}
with a little algebraic manipulation. Using the relation for
$\theta_1$ above and $y_0 = y_1$, the elements of the refraction
matrix $\mathbf{R}$ can be found and are given below.

\begin{equation}
  \mathbf{R} = 
  \left[\begin{array}{cc}
      \frac{n_0}{n_1} & \frac{1}{R}\left(\frac{n_0}{n_1}-1\right) \\
      0 & 1 \\
    \end{array}\right]
  \label{equ:optics:refract}
\end{equation}

Finally, we need to determine the elements for the reflection matrix,
$\mathbf{M}$, of a spherical mirror with radius $-R$, shown in the
diagram of Figure \ref{fig:optics:reflect}. First we know that $y_0 = y_1$. Next
by the law of reflection we see that the light ray must reflect with
angle $\alpha$ about the radius perpendicular to the point of
reflection. By requiring the angles of triangles to sum to $\pi$, we
arrive at,
\begin{equation}
  \begin{aligned}
    \theta_0 &= \phi-\alpha \\
    \theta_1 &= \phi+\alpha \\
  \end{aligned}
  \label{equ:optics:reflect1}
\end{equation}
into which $\phi \approx y_0/R$ (using the small angle approximation
and geometry) can be substituted. This results in,
\begin{equation}
  \theta_1 = \theta_0+\frac{2}{R}y_0
  \label{equ:optics:reflect2}
\end{equation}
which, along with $y_1 = y_0$ and a little manipulation, can again be
used to determine the elements of the reflection matrix given below in
Equation \ref{equ:optics:reflect}.

\begin{equation}
  \mathbf{M} = 
  \left[\begin{array}{cc}
      1 & \frac{2}{R} \\
      0 & 1 \\
    \end{array}\right]
  \label{equ:optics:reflect}
\end{equation}

\section{Lensmaker's Equation}

\begin{figure}
  \begin{center}
    \executeiffilenewer{Figures/Optics/thickLens.svg}
  {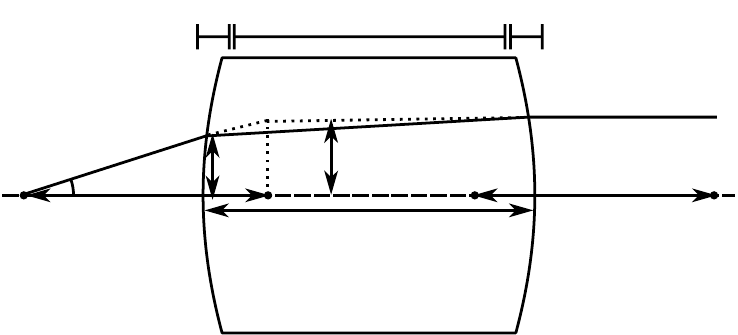}{inkscape-0.48pre1 -z -D --file=Figures/Optics/thickLens.svg 
    --export-pdf=Figures/Optics/thickLens.pdf --export-latex} \def\svgwidth{0.7\columnwidth}
  \input{Figures/Optics/thickLensLabel.tex}
    \caption{Diagram of a thick lens with the cardinal points,
      excluding nodal points $N_0$ and $N_1$. Here the medium
      surrounding the lens has index of refraction $n_0$ while the
      lens has index of refraction $n_1$.\label{fig:optics:thickLens}}
  \end{center}
\end{figure}

Finding the three matrices for translation, refraction, and reflection
above is rather involved, and the advantage of using the matrix method
may not be readily apparent, so let us apply it to a thick lens, as
shown in Figure \ref{fig:optics:thickLens}. However, before we can dive into
the matrix method, we must define the \term{cardinal points} of the
thick lens. The focal points $F_0$ and $F_1$ are defined as points
through which light rays parallel to the optical axis will pass after
refracting through the lens. The \term{principal points} $P_0$ and
$P_1$ are the points where the parallel rays intersect the rays from
the focal points when neglecting refraction. Finally, the \term{nodal
  points} $N_0$ and $N_1$ are the points through which a ray enters
the lens, refracts, and exits on a parallel trajectory. The nodal
points are equivalent to the center of a thin lens, and are not shown
in Figure \ref{fig:optics:thickLens} because they are not relevant for the
discussion below.

Consider the light ray leaving the focal point $F_0$ and exiting the
thick lens parallel to the optical axis. The incoming ray has some
initial angle, $\theta_0$, and some initial height, $y_0$. After
exiting the lens, the ray has some final angle $\theta_1$ and some
final height $y_1$. For this specific scenario we know the final angle
must be $0$ by the definition of the focal point. Just like with
Equation \ref{equ:optics:translate1} we can write,
\begin{equation}
  \left[\begin{array}{c}
      \theta_1 \\
      y_1 \\
    \end{array}\right] = 
  \mathbf{S}
  \left[\begin{array}{c}
      \theta_0 \\
      y_0 \\
    \end{array}\right] =
  \left[\begin{array}{cc}
      S_1 & S_2 \\
      S_3 & S_4 \\
    \end{array}\right]
  \left[\begin{array}{c}
      \theta_0 \\
      y_0 \\
    \end{array}\right]
  \label{equ:optics:smatrix2}
\end{equation}
where $\mathbf{S}$ is the transforming matrix for the optical system,
or in this case, the thick lens.

We don't know what $\mathbf{S}$ is, but by looking at the diagram and
using the translation, refraction, and reflection matrices, we can
build $\mathbf{S}$. First, the ray refracts, and so we must multiply
the initial ray vector by the refraction matrix. Next, the ray is
translated over a distance $x$ and so we must multiply the initial
ray vector and refraction matrix with a translation matrix. Finally,
the ray refracts again, and so we must multiply the previous matrices
and ray vector with another refraction matrix.
\begin{equation}
  \left[\begin{array}{c}
      \theta_1 \\
      y_1 \\
    \end{array}\right] = 
  \mathbf{R}
  \mathbf{T}
  \mathbf{R}
  \left[\begin{array}{c}
      \theta_0 \\
      y_0 \\
    \end{array}\right] ~~~\rightarrow~~~ 
  \mathbf{S} =
  \mathbf{R}
  \mathbf{T}
  \mathbf{R}
  \label{equ:optics:smatrix1}
\end{equation}
It is important to note that the order matters. In Equation
\ref{equ:optics:smatrix1} the $\mathbf{R}$ farthest to the right represents
the first refraction, while the $\mathbf{R}$ on the left is the second
refraction.

Now, we can use Equations \ref{equ:optics:translate} and \ref{equ:optics:refract}
for $\mathbf{T}$ and $\mathbf{R}$ to explicitly calculate out
$\mathbf{S}$.
\begin{equation}
  \begin{aligned}
    \left[\begin{array}{cc}
        S_1 & S_2 \\
        S_3 & S_4 \\
      \end{array}\right] &= 
    \left[\begin{array}{cc}
        \frac{n_1}{n_0} & \frac{1}{R_2}\left(\frac{n_1}{n_0}-1\right) \\
        0 & 1 \\
      \end{array}\right]
    \left[\begin{array}{cc}
        1 & 0 \\
        x & 1 \\
      \end{array}\right]
    \left[\begin{array}{cc}
        \frac{n_0}{n_1} & \frac{1}{R_1}\left(\frac{n_0}{n_1}-1\right) \\
        0 & 1 \\
      \end{array}\right] \\
    &= 
    \left[\begin{array}{cc}
        \frac{n_1}{n_0} & \frac{1}{R_2}\left(\frac{n_1}{n_0}-1\right) \\
        0 & 1 \\
      \end{array}\right]
    \left[\begin{array}{cc}
        \frac{n_0}{n_1} & \frac{1}{R_1}\left(\frac{n_0}{n_1}-1\right) \\
        \frac{xn_0}{n_1} & \frac{x}{R_1}\left(\frac{n_0}{n_1}-1\right)+1 \\
      \end{array}\right] \\
    &= 
    \left[\begin{array}{cc}
        1 +
        \frac{x}{R_2}\left(1-\frac{n_0}{n_1}\right) & 
        \frac{1}{R_1}\left(1-\frac{n_1}{n_0}\right) +
        \frac{1}{R_2}\left(\frac{n_1}{n_0}-1\right)
        \left[\frac{x}{R_1}\left(\frac{n_0}{n_1}-1\right)+1\right] \\
        \frac{xn_0}{n_1} & \frac{x}{R_1}\left(\frac{n_0}{n_1}-1\right)+1 \\
      \end{array}\right] \\
  \end{aligned}
  \label{equ:optics:smatrix3}
\end{equation}
Returning to Equation \ref{equ:optics:smatrix2} we can write out relations
for $\theta_1$ and $y_1$,
\begin{equation}
  \begin{aligned}
    \theta_1 &= S_1\theta_0+S_2y_0 ~~~\rightarrow~~~ y_0 =
    \frac{-S_1\theta_0}{S_2} \\
    y_1 &= S_3\theta_0+S_4y_0 \\
  \end{aligned}
  \label{equ:optics:smatrix4}
\end{equation}
where in the first relation we have utilized the fact that the
outgoing ray is parallel to the optical axis, i.e. $\theta_1 =
0$. Looking back at Figure \ref{fig:optics:thickLens} we see from the
geometry that,
\begin{equation}
  \tan\theta_0 = \frac{y_1}{-f_0} ~~~\rightarrow~~~ f_0 \approx
  -\frac{y_1}{\theta_0}
  \label{equ:optics:smatrix5}
\end{equation}
where the small angle approximation was used in the second
step. Plugging in Equation \ref{equ:optics:smatrix4} into Equation
\ref{equ:optics:smatrix5} yields,
\begin{equation}
  \begin{aligned}
    f_0 &= -\frac{y_1}{\theta_0}\\
    &= -\frac{S_3\theta_0+S_4y_0}{\theta_0}\\
    &= \frac{S_4S_1}{S_2} - S_3\\
    &= \frac{S_4S_1-S_3S_2}{S_2}\\
    &= \frac{1}{S_2}
  \end{aligned}
  \label{equ:optics:focal1}
\end{equation}
which gives the focal point $f_0$, in terms of the elements of
$\mathbf{S}$. The algebra in the second to last step is found by
plugging in values for the elements of $S$ in the numerator using
Equation \ref{equ:optics:smatrix3} and is rather tedious, but the result
$S_1S_4-S_3S_2 = 1$ is well worth it.

If we take the final result of Equation \ref{equ:optics:focal1} and plug in
$S_2$ from Equation \ref{equ:optics:smatrix3}, we obtain,
\begin{equation}
  \frac{1}{f_0} = S_2 =
  \frac{n_1-n_0}{n_0}\left[\frac{1}{R_2}-\frac{1}{R_1} -
    \left(\frac{n_1-n_0}{n_1}\right)\left(\frac{x}{R_1R_2}\right)\right]
  \label{equ:optics:lensmaker}
\end{equation}
which is known as
the \term{lensmaker's equation}. As $x$ becomes very small, the lens
approaches the thin lens approximation and Equation
\ref{equ:optics:lensmaker} becomes,
\begin{equation}
  \frac{1}{f_0} =
  \frac{n_1-n_0}{n_0}\left(\frac{1}{R_2}-\frac{1}{R_1}\right)
  \label{equ:optics:lensmakerLimit}  
\end{equation}
which is the thin lens approximation for focal length given in
Equation \ref{equ:optics:thinFocal}!

\section{Fresnel Equations}

The previous sections have exclusively dealt with geometrical optics,
which while important, cannot describe all light phenomena,
specifically polarization. To understand polarization we must leave
the realm of geometrical optics and enter the world of physical
optics. The first thing we must do is understand the details of what
exactly light is. Light can be thought of as the combination of two
fields, an electric field oscillating up and down, and a magnetic
field, perpendicular to the electric field, also oscillating. The
cross product of the electric field with the magnetic field,
$\vec{E}\times\vec{B}$, is always in the direction that the light wave
is traveling.

If an observer were to see a light wave directly approaching them,
they would see the electric field as a vector, always pointing in the
same direction, but growing and shrinking. Similarly, they would see
another vector representing the magnetic field, perpendicular to the
electric field vector, also growing and shrinking as the light wave
approached. The vectors that the observer sees are two dimensional,
and so like any two dimensional vector, they can be broken into $x$
and $y$ components.

So what happens if the observer is in jail and the light must pass
through parallel vertical jail bars that block electric fields? Any
bit of the electric field that is not vertical will bounce off the
bars, while any part of the electric field that is vertical will pass
through. In other words, the $x$ component of the electric field will
not survive, but the $y$ component will. Additionally, if the the
electric field is blocked, then so is the associated magnetic field,
and so the $y$ component of the magnetic field associated with the $x$
component of the electric field will be blocked as well. Now the
observer sees the electric field vector oscillating vertically up and
down, while the magnetic field vector is oscillating left and right.

The thought experiment above is the general idea behind \term{plane
  polarization}. Incoming parallel light waves have a myriad of
different electric and magnetic field directions, but after passing
through a vertical polarizer, only the vertical components of the
electric field, and the horizontal components of the magnetic field
survive. Another type of polarization, \term{circular polarization} is
also possible and is the same idea as plane polarized light, but is
more difficult to visualize.

\begin{figure}
  \begin{center}
    \subfigure[]{
      \executeiffilenewer{Figures/Optics/perpendicular.svg}
  {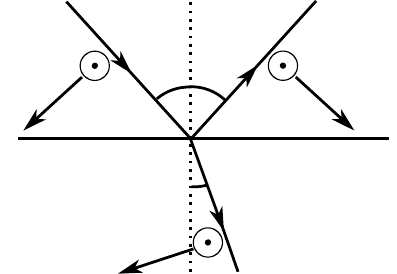}{inkscape-0.48pre1 -z -D --file=Figures/Optics/perpendicular.svg 
    --export-pdf=Figures/Optics/perpendicular.pdf --export-latex} \def\svgwidth{0.4\columnwidth}
  \input{Figures/Optics/perpendicularLabel.tex}
      \label{fig:optics:perpendicular}
    }
    \subfigure[]{
      \executeiffilenewer{Figures/Optics/parallel.svg}
  {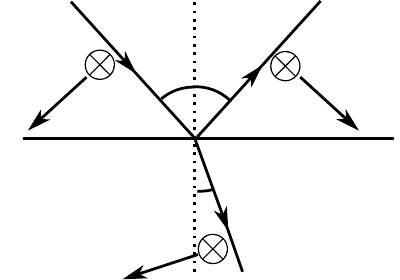}{inkscape-0.48pre1 -z -D --file=Figures/Optics/parallel.svg 
    --export-pdf=Figures/Optics/parallel.pdf --export-latex} \def\svgwidth{0.4\columnwidth}
  \input{Figures/Optics/parallelLabel.tex}
      \label{fig:optics:parallel}
    }
  \end{center}
\end{figure}

Now what happens if we consider light bouncing off a piece of glass?
The light incident on the glass can be either vertically or
horizontally polarized, again because the vectors can be broken into
their $x$ and $y$ components. The composite of the two polarizations
is just normal unpolarized light, but by looking at the two components
individually we can see what happens to the light reflected off the
glass. In Figure \ref{fig:optics:perpendicular} the light is polarized so
that the $\vec{E}$ field is coming out of the page and perpendicular
to the plane of incidence.\footnote{This notation can be a bit
  confusing but is the standard. Perpendicular to the plane of
  incidence means that the electric field is parallel with the surface
  of the reflector if the reflector is a plane.} The light wave is
traveling in the direction $\vec{E}\times\vec{B}$ and so we can then
draw the direction of the magnetic field.

When the light waves hits the glass surface, part of the wave is
reflected, while part of the wave passes through the glass. Because
the electric field is parallel to the surface of the glass, the
electric field will remain pointing in the same direction as the
incident light wave for both the reflected and transmitted waves. This
means that,
\begin{equation}
  E_{i,\perp}+E_{r\perp} = E_{t,\perp}
\end{equation}
or that the incident and reflected electric field amplitudes must
equal the transmitted electric field amplitude.

Next, we see that the vertical components of the magnetic fields in
the incident and transmitted waves should are in the same direction,
while the horizontal component of the magnetic field in the reflected
wave has been flipped by the reflection. This gives us,
\begin{equation}
  B_{i,\parallel}\cos\theta_i-B_{r,\parallel}\cos\theta_r =
  B_{t,\parallel}\cos\theta_t
  \label{equ:optics:bCondition}
\end{equation}
which states that the incident horizontal component of the magnetic
field less the reflected field must equal the transmitted magnetic
field. Using the relation that $E = \frac{c}{n}B$, Equation
\ref{equ:optics:bCondition} can be rewritten in terms of the electric field,
\begin{equation}
  n_0E_{i,\perp}\cos\theta_i-n_1E_{r,\perp}\cos\theta_r =
  n_1E_{t,\perp}\cos\theta_t
\end{equation}
where $n_0$ is the index of refraction for the incident and reflected
medium, and $n_1$ the index of refraction for the transmitted
medium. Note that the speed of light cancels out of the equation
because both sides are divided by $c$. We now have three unknowns and
only two equations. This is not a problem however, because we only
want to know the percentage of light reflected back, and so we can
eliminate $E_{t,\perp}$, let $\theta_r = \theta_i$ by the law of
reflection, and solve for the ratio of the reflected electric field to
the incident electric field.

\begin{equation}
  \frac{E_{r,\perp}}{E_{i,\perp}} = \frac{\cos\theta_i -
    \frac{n_1}{n_0}\cos\theta_t}{\cos\theta_i +
    \frac{n_1}{n_0}\cos\theta_t}
  \label{equ:optics:ratioPerp}
\end{equation}

What happens if we look at the light with magnetic fields parallel to
the plane of incidence rather than perpendicular as is shown in Figure
\ref{fig:optics:parallel}?\footnote{Again, to clear up the confusion, this
  means the electric field is now perpendicular to the surface of the
  reflector.} We can repeat the exact same process except now all the
magnetic fields stay the same direction,
\begin{equation}
  B_{i,\perp}+B_{r\perp} = B_{t,\perp}
\end{equation}
and the horizontal component of the electric field is flipped by
reflection.
\begin{equation}
  E_{i,\parallel}\cos\theta_i-E_{r,\parallel}\cos\theta_r =
  E_{t,\parallel}\cos\theta_t
  \label{equ:optics:eCondition}
\end{equation}
The substitution for $E=\frac{c}{n}B$ can be made again and we can
obtain the ratio of the reflected electric field to the incident
electric field.
\begin{equation}
  \frac{E_{r,\parallel}}{E_{i,\parallel}} = \frac{\frac{n_1}{n_0}\cos\theta_i -
    \cos\theta_t}{\frac{n_1}{n_0}\cos\theta_i +
    \cos\theta_t}
  \label{equ:optics:ratioParallel}
\end{equation}

Knowing the ratio of reflected to incident electric fields is nice,
but it does not tell us anything that we can easily measure. However,
the intensity of light is just the square of the electric field, so if
we square Equations \ref{equ:optics:ratioPerp} and \ref{equ:optics:ratioParallel} we
can find the percentage of light reflected for an incidence angle
$\theta_i$ and transmitted angle $\theta_t$ for both perpendicular and
parallel polarized light! Taking this one step further, we can
eliminate $\theta_t$ with Snell's law, $n_0\sin\theta_i =
n_1\sin\theta_t$. Finally, trigonometric properties can be used to
reduce the equations into an even more compact form. These equations
are known as the Fresnel equations.\footnote{Pronounced {\it
    fray-nell}, the {\it s} is silent.}

\begin{equation}
  \begin{aligned}
    \frac{I_{r,\perp}}{I_{i,\perp}}
    &= \left(\frac{E_{r,\perp}}{E_{i,\perp}}\right)^2 
    = \left(\frac{\cos\theta_i -
        \frac{n_1}{n_0}\cos\theta_t}{\cos\theta_i +
        \frac{n_1}{n_0}\cos\theta_t}\right)^2\\
    &= \left(\frac{\cos\theta_i -
        \sqrt{\left(\frac{n_1}{n_0}\right)^2-\sin^2\theta_i}}{\cos\theta_i +
        \sqrt{\left(\frac{n_1}{n_0}\right)^2-\sin^2\theta_i}}\right)^2\\
    &= \frac{\sin^2\left(\theta_i-\theta_t\right)}
    {\sin^2\left(\theta_i+\theta_t\right)} \\
  \end{aligned}
  \label{equ:optics:fresnel1}
\end{equation}

\begin{equation}
  \begin{aligned}
    \frac{I_{r,\parallel}}{I_{i,\parallel}}
    &= \left(\frac{E_{r,\parallel}}{E_{i,\parallel}}\right)^2 
    = \left(\frac{\frac{n_1}{n_0}\cos\theta_i -
        \cos\theta_t}{\frac{n_1}{n_0}\cos\theta_i +
        \cos\theta_t}\right)^2\\
    &= \left(\frac{\frac{n_1}{n_0}\cos\theta_i -
        \sqrt{\left(\frac{n_1}{n_0}\right)^2-\sin^2\theta_i}}
      {\frac{n_1}{n_0}\cos\theta_i +
        \sqrt{\left(\frac{n_1}{n_0}\right)^2-\sin^2\theta_i}}\right)^2\\
    &= \frac{\tan^2\left(\theta_i-\theta_t\right)}
    {\tan^2\left(\theta_i+\theta_t\right)} \\
  \end{aligned}
  \label{equ:optics:fresnel2}
\end{equation}

\begin{figure}
  \begin{center}
    \executeiffilenewer{Code/Optics/fresnel.m}{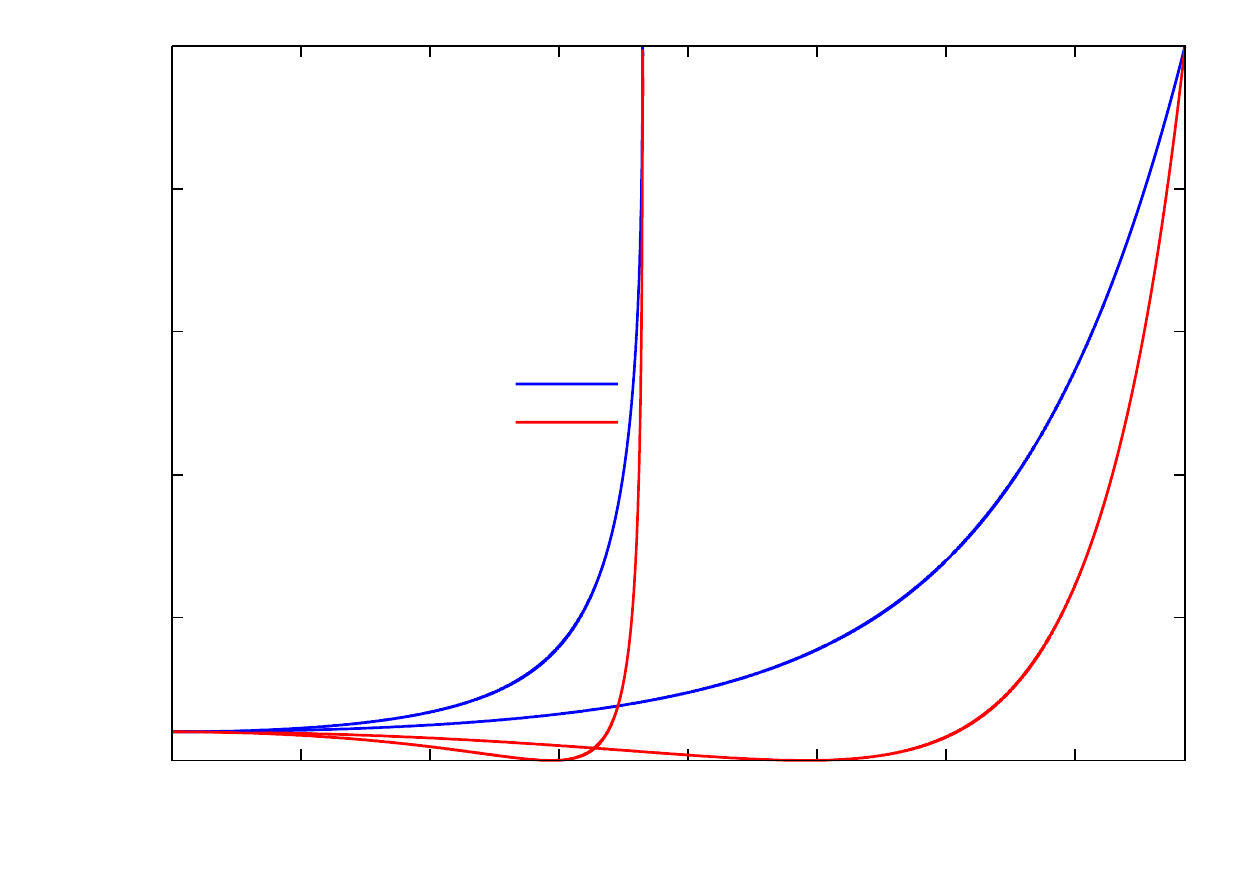}
  {octave --silent --eval "addpath([pwd(),'/Code']); 
    addpath([pwd(),'/Code/Optics']); fresnel(1);"} 
  \setlength{\unitlength}{\columnwidth*\real{0.00013}}
  \input{Figures/Optics/fresnelLabel.tex}
    \caption{Physical interpretation of the Fresnel equations given in
      Equations \ref{equ:optics:fresnel1} and \ref{equ:optics:fresnel2} for light
      transitioning from glass to air and air to
      glass.\label{fig:optics:fresnel}}
  \end{center}
\end{figure}

So where does all of this math get us? Figure \ref{fig:optics:fresnel} plots
the percentage light reflected for perpendicular and parallel
polarized light with respect to the plane of incidence for the
scenario of light passing from air to glass, $n_0 = 1, n_1 = 1.5$, and
glass to air, $n_0 = 1.5, n_1 = 1$. For the scenario of light passing
from air to glass, there is a point on the plot at $\theta_i = 0.98$
where no parallel polarized light, only perpendicular polarized light
is reflected. This angle is called \term{Brewster's angle} and is the
angle at which the light reflecting off a surface is completely
polarized! For the second scenario, light passing from glass to
air, the light is completely reflected for values greater than
$\theta_i = 0.73$. This is called \term{total internal reflection}. By
taking into account the polarization of light we are able to see two
very real light phenomena!

\section{Experiment}

The experiment for this chapter explores both geometrical and physical
optics and consists of three parts. In the first part of the
experiment, the focal length of a convex lens is measured by using the
relation,
\begin{equation}
  \frac{1}{x_i}+\frac{1}{x_o} = \frac{1}{f}
  \label{equ:optics:measureFocal}
\end{equation}
where $x_i$ is the distance of the image from the lens and $x_o$ is
the distance of the object from the lens. This relationship can be
derived by tracing an additional ray in Figure \ref{fig:optics:convexOutside}
and equating ratios from the similar triangles formed.

The second part of the experiment measures the focal length of a
concave lens. Because concave lenses do not project real images, a
concave lens must be placed before the convex lens. The focal point
for the system is just,
\begin{equation}
  \frac{1}{f} = \frac{1}{f_\mathrm{concave}}+\frac{1}{f_\mathrm{convex}}
\end{equation}
where $f_\mathrm{concave}$ and $f_\mathrm{convex}$ are the focal
points for the concave and convex lenses. The focal length for the
compound system can be found by using Equation \ref{equ:optics:measureFocal},
and so after plugging in the focal length of the convex lens from the
first part of the experiment, the focal length of the concave lens can
be found.

The final part of the experiment verifies the Fresnel equations by
producing a plot very similar to Figure \ref{fig:optics:fresnel}. A laser is
polarized using a filter and bounced off a glass plate. The reflected
laser is directed to a photo-multiplier tube which creates a voltage
proportional to the intensity of the light. Brewster's angle can then
be determined from the plots made. It is important in this part of the
experiment to not mix up the polarizations of the light. The line on
the polarizing filter indicates the direction the electric field is
polarized so for perpendicular polarized light with respect to the
plane of incidence the line should be vertical.
\graphicspath{{Figures/Diffraction/}}

\chapter{Diffraction and Interference}\label{chp:diffraction}

At the beginning of the $19^\mathrm{th}$ century a great debate had
been raging within the physics community for over $100$ years, sparked
by the diametrically opposing theories of Isaac Newton and Christiaan
Huygens regarding the nature of light. Newton, in his book
\href{http://books.google.com/books?id=GnAFAAAAQAAJ&vq=corpuscle&output=text&source=gbs_navlinks_s}{\it
  Optiks}\footnote{He based this publication on his first series of
  lectures at Trinity College, Cambridge in $1704$.}, outlined a
theory where light was made up of small particles or {\bf
  corpuscles}. Geometrical optics, as introduced in Chapter
\ref{chp:optics}, is modeled well by rays or straight lines, which can
be thought of as the paths traced out by individual light particles,
and so Newton's theory described the optics of the day well. Huygens,
however, in his
\href{http://www.gutenberg.org/files/14725/14725-h/14725-h.htm}{\it
  Treatise on Light}, written in $1690$, $14$ years earlier than
Newton's {\it Optiks}, proposed that light was not made up of
particles, but rather waves, similar to ocean waves. This theory had
its own merits, but was generally ignored in favor of Newton's
particle theory.

The debate between the theories of Huygens and Newton came to its
first\footnote{Why this is just the first resolution, and not the
  final resolution of the debate will be explained in the double slit
  diffraction section of this chapter.} resolution in $1802$ when
Thomas Young performed his famous \term{double slit} experiment which
emphatically demonstrated the wave nature of light. In the experiment,
Young set up a light source which passed through a cover with two
slits. The light then passed from these two slits onto a screen, where
the pattern of the light could be observed. According to Newton's
theory, the light should project two slits onto the screen. What Young
found however, was a complicated \term{interference} pattern.

\section{Interference}

Before diving into the details of Young's experiment, we first need to
understand exactly what is meant by an interference pattern. Looking
back at both Chapters \ref{chp:waves} and \ref{chp:optics} we know
that a light wave consists of an electric field perpendicular to a
magnetic field. Because the magnetic field is directly related to the
electric field (and vice versa), to fully describe a light wave it is
only necessary to specify either the electric or magnetic field. By
convention, the electric field is typically used to describe the wave,
and is given by the general wave function (in one dimension),
\begin{equation}
  E = E_0\sin\left(kx-\omega t\right)
  \label{equ:diffraction:waveEquation}
\end{equation}
where $x$ is the position at which the wave is measured, $t$ is the
time at which the wave is measured, $E_0$ is the amplitude of the
wave, $\omega$ is the angular frequency of the wave\footnote{See the
  Chapter \ref{chp:waves} for more detail; the angular frequency is
  related to the frequency of a wave by $\omega = 2\pi f$.}, and $k$
is the wave number, defined as,
\begin{equation}
  k \equiv \frac{2\pi}{\lambda}
  \label{equ:diffraction:waveNumber}
\end{equation}
where $\lambda$ is wavelength.

So what happens if we place two light waves on top of each other? In
electrodynamics, electric fields can be combined by {\bf
  superposition}, which just means that the electric fields are added
together. If the electric field of Equation \ref{equ:diffraction:waveEquation} was
added to another electric field described by the exact same equation
(same $k$ and $\omega$), the result would be Equation
\ref{equ:diffraction:waveEquation} but now with an amplitude of $2E_0$ rather than
$E_0$. This is called \term{complete constructive interference}, where
two fields are added together with the same \term{phase}.

\begin{figure}
  \begin{center}
    \subfigure[]{
      \executeiffilenewer{Code/Diffraction/interference.m}{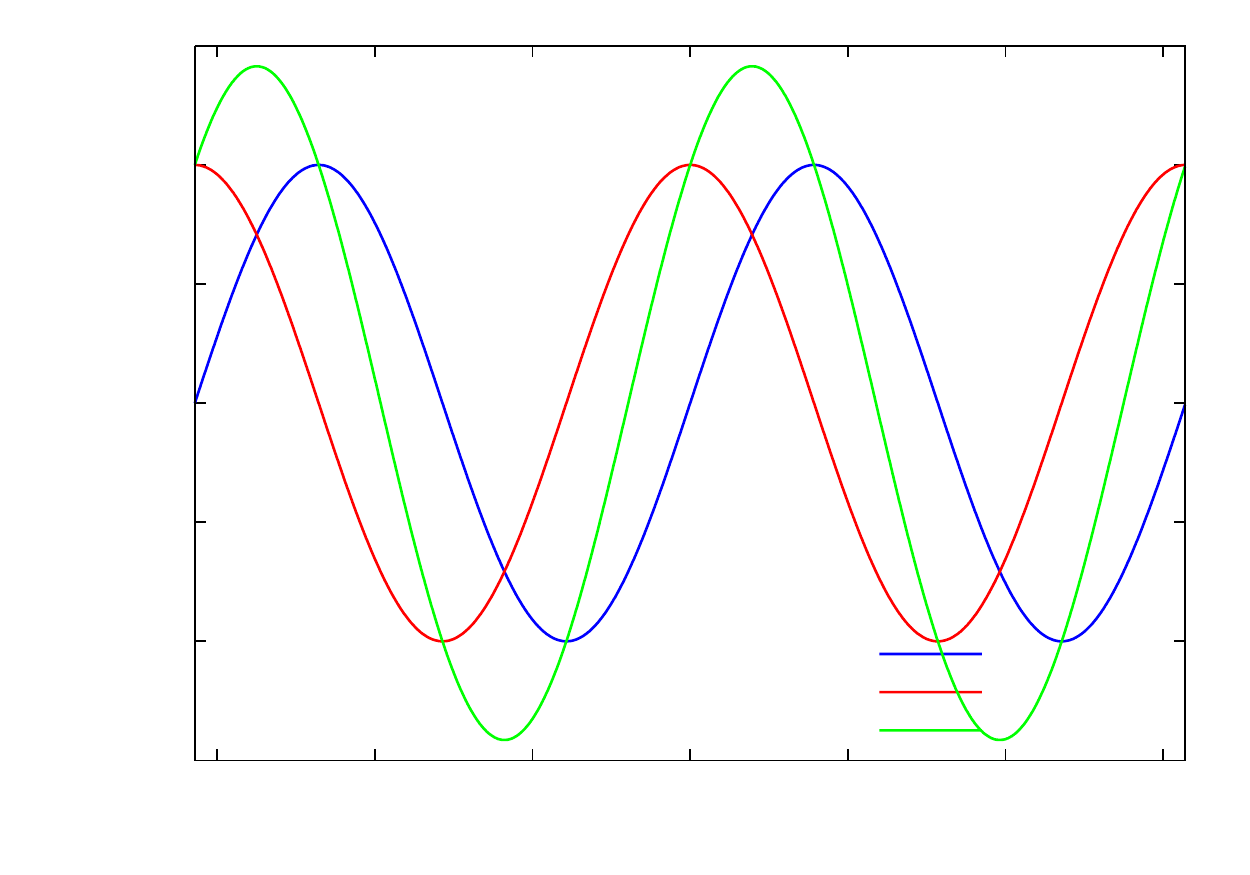}
  {octave --silent --eval "addpath([pwd(),'/Code']); 
    addpath([pwd(),'/Code/Diffraction']); interference(1);"} 
  \setlength{\unitlength}{\columnwidth*\real{0.00013}}
  \input{Figures/Diffraction/interferenceLabel.tex}
      \label{fig:diffraction:interference}
    }

    \subfigure[]{
      \executeiffilenewer{Figures/Diffraction/huygen.svg}
  {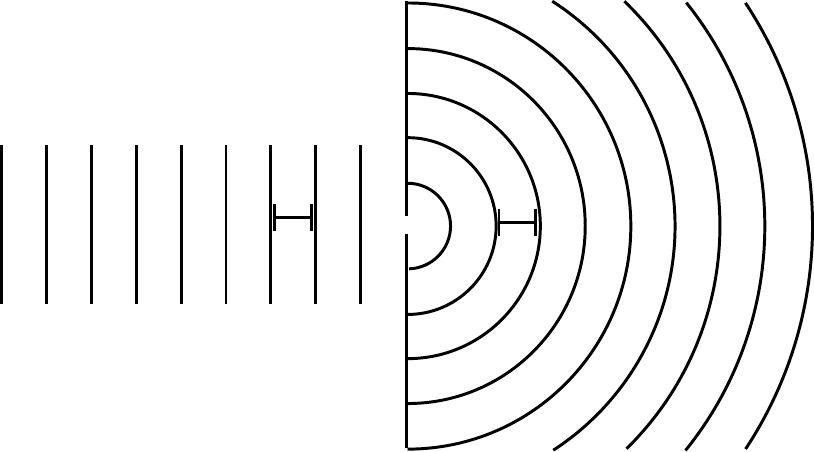}{inkscape-0.48pre1 -z -D --file=Figures/Diffraction/huygen.svg 
    --export-pdf=Figures/Diffraction/huygen.pdf --export-latex} \def\svgwidth{0.7\columnwidth}
  \input{Figures/Diffraction/huygenLabel.tex}
      \label{fig:diffraction:huygen}
    }
    \caption{Diagram of interference between two waves is given in
      Figure \ref{fig:diffraction:interference} and Huygens principle is
      demonstrated in Figure \ref{fig:diffraction:huygen}.\label{fig:diffraction:intro}}
  \end{center}
\end{figure}

But what is the phase of a wave? We can rewrite Equation
\ref{equ:diffraction:waveEquation} as,
\begin{equation}
  E = E_0\sin\left(kx-\omega t+\delta\right)
  \label{equ:diffraction:phaseEquation}
\end{equation}
where $\delta$ is the phase of the wave. From the equation, we see
that the phase of the wave just shifts the wave to the right by
$\delta$ if the phase is negative, and to the left by $\delta$ if the
phase is positive. Now, if we add two waves together, one with a field
described by Equation \ref{equ:diffraction:phaseEquation} with $\delta = 0$, and
another with $\delta = \pi$, the peaks of the first wave match with
the valleys of the second wave, and so when the two waves are added
together, the net result is zero! This is called \term{complete
  destructive interference}. There are of course combinations between
destructive and constructive interference, as shown in Figure
\ref{fig:diffraction:interference} where the wave $\sin(x)$ is added onto the wave
$\sin(x+\pi/2)$, but in general, noticeable interference is either
complete constructive or destructive.

Now let us take a slightly more mathematical approach to the idea of
interference between two waves,
\begin{equation}
  E_1 = E_0\sin\left(kx-\omega t+\frac{\delta}{2}\right), ~~~ E_2 =
  E_0\sin\left(kx-\omega t-\frac{\delta}{2}\right)
  \label{equ:diffraction:twoWaves}
\end{equation}
where the wavenumber and angular frequencies are the same but with a
phase difference between the two waves of $\delta/2-(-\delta/2) =
\delta$. We can again just add the two waves together to find the
total electric field wave, but physically this is not very
interesting, as the human eye cannot observe the pattern of an
electric wave. However, what is of interest is the intensity (or
brightness) of the wave, which is proportional to the electric field
squared. Adding the two electric fields of Equation \ref{equ:diffraction:twoWaves}
together and squaring then gives us a quantity proportional to the
intensity.

\begin{equation}
  \begin{aligned}
    I \propto E^2 &=& &\left(E_1+E_2\right)^2\\
    &=& &E_1^2+E_2^2+2E_1E_2\\
    &=& &E_0^2\sin^2\left(kx-\omega t + \frac{\delta}{2}\right) +
    E_0^2\sin^2\left(kx-\omega t - \frac{\delta}{2}\right)\\
    & & &+
    2E_0^2\sin\left(kx-\omega t + \frac{\delta}{2}\right)
    \sin\left(kx-\omega t -
      \frac{\delta}{2}\right)\\
  \end{aligned}
  \label{equ:diffraction:interferenceIntensity}
\end{equation}

The final expanded step of the intensity above looks rather nasty, but
luckily we can simplify it. We note that it is impossible to look
everywhere (i.e. at all $x$) at the same time (i.e. for a specific
$t$). Humans can however, look at a specific point over a period of
time. Subsequently, we want to look at the final step of Equation
\ref{equ:diffraction:interferenceIntensity} at a specific point over a period of
time. This means $x$ becomes a constant, and we need to find the time
average of the trigonometric functions. Without giving a rigorous
mathematical argument, the time average, indicated by angled brackets,
for $\sin^2$ and $\cos^2$ is just $1/2$ and for sine and cosine is
just $0$.
\begin{equation}
  \left\langle\sin^2 t\right\rangle = \frac{1}{2}, ~~~
  \left\langle\sin t\right\rangle = 0, ~~~
  \left\langle\cos^2 t\right\rangle = \frac{1}{2}, ~~~
  \left\langle\cos t\right\rangle = 0
  \label{equ:diffraction:averageTrig}
\end{equation}

Using the time average of the trigonometric functions, the first two
terms in the final step of Equation \ref{equ:diffraction:interferenceIntensity}
become $E_0^2/2$.
\begin{equation}
  E^2 =  E_0^2 + 2E_0^2\sin\left(kx-\omega t + \frac{\delta}{2}\right)
  \sin\left(kx-\omega t -
    \frac{\delta}{2}\right)
  \label{equ:diffraction:averageFirstTerms}
\end{equation}
By using the trigonometric identity,
\begin{equation}
  \sin\theta\sin\phi \equiv \frac{\cos\left(\theta-\phi\right) +
    \cos\left(\theta+\phi\right)}{2}
\end{equation}
the final term can be recast in terms of cosines and the terms
dependent upon $t$ can be time averaged.
\begin{equation}
  \begin{aligned}
    2E_0^2\sin^2\left(kx-\omega t + \frac{\delta}{2}\right)
    \sin^2\left(kx-\omega t -
      \frac{\delta}{2}\right) &=
    E_0^2\cos\left(\delta\right) + E_0^2\cos\left(2kx-2\omega
        t\right) \\
    &= E_0^2\cos\delta \\
  \end{aligned}
\end{equation}
Combining this result for the final term with the first two terms of
Equation \ref{equ:diffraction:averageFirstTerms} results in,
\begin{equation}
  E^2 = \frac{1}{2}E_0^2+\frac{1}{2}E_0^2+E_0^2\cos\delta =
  E_0^2\left(1+\cos\delta\right)
\end{equation}
which can be further simplified using the trigonometric identity,
\begin{equation}
  1+\cos\delta \equiv 2\cos^2\left(\frac{\delta}{2}\right)
\end{equation}
for the cosine function.
\begin{equation}
  E^2 = 2E_0^2\cos^2\left(\frac{\delta}{2}\right)
\end{equation}

Putting this all back together, we have found the time averaged
intensity at any point of two overlapping light waves of the same
wavelength, but with a phase separation of $\delta$.
\begin{equation}
  I \propto \cos^2\left(\frac{\delta}{2}\right)
  \label{equ:diffraction:twoWaveIntensity}
\end{equation}
We can now check that this mathematical description for the
interference between two waves makes sense. If the two waves are
completely in phase, or $\delta = 0$ we should have complete
constructive interference and Equation \ref{equ:diffraction:twoWaveIntensity}
should be at a maximum, which we see it is. Similarly, Equation
\ref{equ:diffraction:twoWaveIntensity} is at a minimum of $0$ when $\delta = \pi$,
or the waves have complete destructive interference.

\section{Double Slit Interference}

So how do we create an experiment to check Equation
\ref{equ:diffraction:twoWaveIntensity}? The first item needed is a light source
that produces waves that have the same phase and wavelength or a {\bf
  coherent} and \term{monochromatic} light source. Next we need to
combine two waves from the light source with a known phase difference
between the two, and observe the intensity pattern. This can be
accomplished by taking advantage of what is known as \term{Huygens
  principle}, illustrated in Figure \ref{fig:diffraction:huygen}. Here, plane
waves of wavelength $\lambda$ are incident on some surface with a
slit. The waves pass through the slit and propagate out in a circular
manner but still with wavelength $\lambda$.

\begin{figure}
  \begin{center}
    \subfigure[]{
      \executeiffilenewer{Figures/Diffraction/doubleSetup.svg}
  {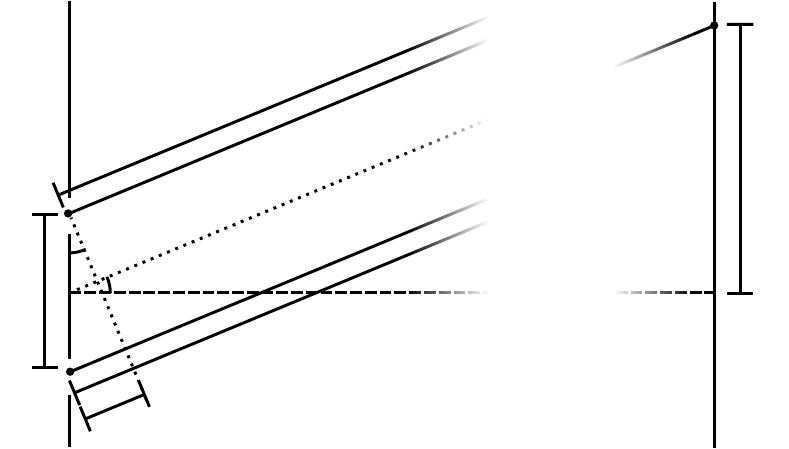}{inkscape-0.48pre1 -z -D --file=Figures/Diffraction/doubleSetup.svg 
    --export-pdf=Figures/Diffraction/doubleSetup.pdf --export-latex} \def\svgwidth{0.7\columnwidth}
  \input{Figures/Diffraction/doubleSetupLabel.tex}
      \label{fig:diffraction:doubleSetup}
    }
    \subfigure[]{
      \executeiffilenewer{Code/Diffraction/doublePower.m}{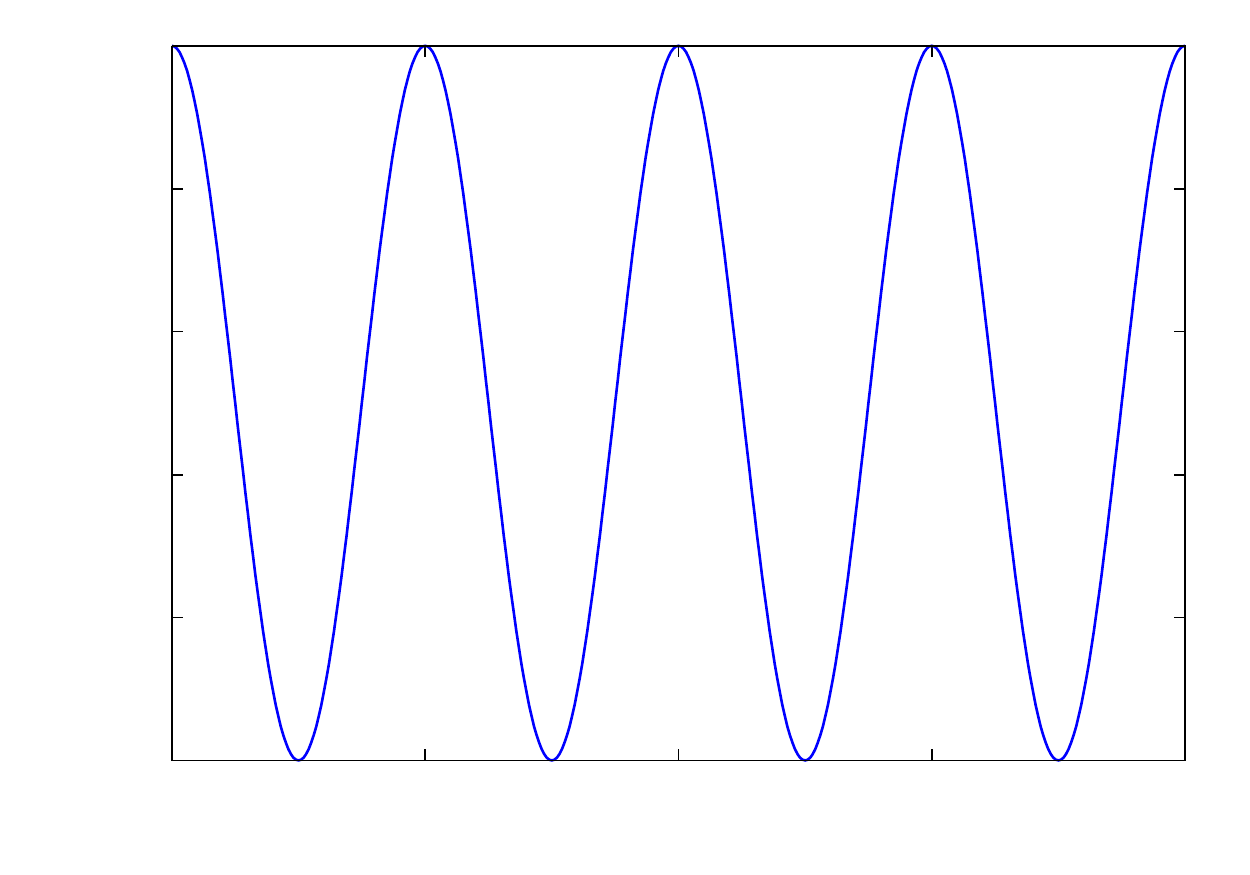}
  {octave --silent --eval "addpath([pwd(),'/Code']); 
    addpath([pwd(),'/Code/Diffraction']); doublePower(1);"} 
  \setlength{\unitlength}{\columnwidth*\real{0.00013}}
  \input{Figures/Diffraction/doublePowerLabel.tex}
      \label{fig:diffraction:doublePower}
    }
    \caption{Diagram of the double slit experiment is given in Figure
      \ref{fig:diffraction:doubleSetup} and the intensity pattern is given in
      Figure \ref{fig:diffraction:doublePower}.\label{fig:diffraction:double}}
  \end{center}
\end{figure}

Using this principle, a setup can be made with a coherent
monochromatic light source emitting plane waves which pass through two
slits separated by a distance $d$, shown in Figure
\ref{fig:diffraction:doubleSetup}. The plane waves then become spherical and
propagate outwards from the slit, interfering with the wave from the
adjacent slit. A screen is placed at a very far distance $L$ from the
double slit, such that $L\gg d$. This setup is called a {\bf
  Fraunhofer} or \term{far-field} interference experiment because $L$
is so large in comparison to the distance between the slits.

Consider now looking at a point $P$ on the projection screen at a
distance $y$ from the centerline between the two slits, forming an
angle $\theta$ with the centerline. We can trace a line along the
spherical waves emitted from each slit to this point, and label the
distance of the path from the upper slit as $r_1$ and the lower slit
as $r_2$. Because the screen is so far away, we can approximate $r_1$
and $r_2$ as being parallel, as shown in Figure
\ref{fig:diffraction:doubleSetup}. Drawing a line perpendicular to $r_1$ from the
upper slit to $r_2$ creates a small right triangle, of which we label
the base length as $\Delta r$. The upper angle of this triangle is
$\theta$ by geometry, and so $\Delta r$ is just $d\sin\theta$. Because
$r_1$ and $r_2$ are approximately parallel, we can write $r_2$ as $r_2
= r_1+\Delta r$.

The waves emitted from the slits are spherical, but we can see that
the profile of the wave as it travels along either $r_1$ or $r_2$ is
described by the general wave equation of Equation
\ref{equ:diffraction:waveEquation}, $r$ replacing $x$ as the position variable. We
can then write the equation of both electric fields, $E_1$ and $E_2$
at the point where they overlap on the projection screen.
\begin{equation}
  \begin{aligned}
    E_1 &= E_0\sin\left(kr_1-\omega t\right) \\
    E_2 &= E_0\sin\left(kr_2-\omega t\right) \\
    &= E_0\sin\left(k(r_1+\Delta r)-\omega t\right) \\
    &= E_0\sin\left(kr_1-\omega t+
      \frac{2\pi}{\lambda}a\sin\theta\right) \\
  \end{aligned}
  \label{equ:diffraction:doubleWaves}
\end{equation}
In the second to last step, $r_1+\Delta R$ is substituted for $r_2$, and
in the final step $d\sin\theta$ is substituted for $\Delta r$.

From the first and final line of Equation \ref{equ:diffraction:doubleWaves}, we
see that the phase difference between the two waves at point $P$ on
the projection screen is just $k\Delta r$.
\begin{equation}
  \delta = \frac{2\pi d\sin\theta}{\lambda}
  \label{equ:diffraction:doubleDeltaTheta}
\end{equation}
If we assume that the angle $\theta$ is small, we can then make the
small angle approximation between sine and tangent,
\begin{equation}
  \sin\theta \approx \tan\theta = \frac{y}{L}
  \label{equ:diffraction:smallAngle}
\end{equation}
and plug this back into the phase difference of Equation
\ref{equ:diffraction:doubleDeltaTheta} for $\sin\theta$.
\begin{equation}
  \delta = \frac{2\pi d y}{\lambda L}
\end{equation}
We now have the phase difference between the two different waves at
point $P$ in terms of $y$, the distance $y$ from the centerline of the
setup, $L$, the distance of the projection screen from the slits, $d$,
the distance between the two slits, and $\lambda$, the wavelength of
the monochromatic coherent light being used.

Because we have the phase difference between the two waves, and the
two waves have the same wavelength and frequency, we can use Equation
\ref{equ:diffraction:twoWaveIntensity} to determine the intensity pattern on the
screen as a function of the distance $y$.
\begin{equation}
  I(y) \propto 2E_0^2\cos^2\left(\frac{\pi d y}{\lambda L}\right)
\end{equation}
As this is just a proportionality, we can absorb all the constant
coefficients and replace them with some maximal intensity, $I_0$, to
make an equality,
\begin{equation}
  I(y) = I_0 cos^2\left(\frac{\pi d y}{\lambda L}\right)
  \label{equ:diffraction:doubleIntensity}
\end{equation}
which is plotted in Figure \ref{fig:diffraction:doublePower}. From this plot we
can see that the intensity pattern will be at a maximum when,
\begin{equation}
  y_\mathrm{max} = \frac{m\lambda L}{d},~~~m=0,1,2,\dots
  \label{equ:diffraction:doubleMax}
\end{equation}
where $m$ is the order of the maxima found.

The intensity pattern of Figure \ref{fig:diffraction:doublePower} is an incredible
result that allows us to see whether light is a wave or particle using
a simple experimental setup. However, it turns out that in the
derivation above we have assumed that a single spherical wave is
emitted from each slit, which is not a good approximation unless the
width of the slits is much smaller than the separation between the
slits. We will now take this into account, first with single slit
diffraction and then with double slit diffraction.

\section{Single Slit Diffraction}

While reading through this chapter, one may have noticed that the word
interference has been used to describe the previous two sections. So
what then is \term{diffraction}? The definition for diffraction can be
rather tricky, but in general, diffraction is a phenomena that occurs
from the interference of a continuous set of waves, rather than a
discrete number of waves, such as two in the double slit interference
example above. In the following example of a single slit, we must now
consider an infinite number of light waves interfering, rather than a
set of two waves.

\begin{figure}
  \begin{center}
    \subfigure[]{
      \executeiffilenewer{Figures/Diffraction/singleSetup.svg}
  {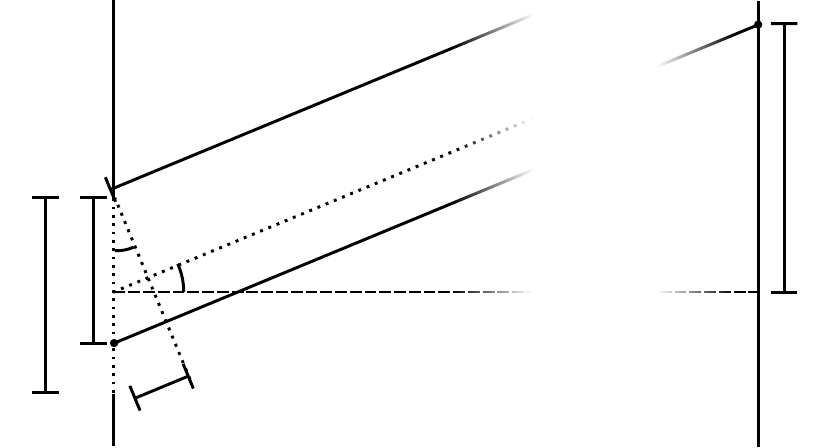}{inkscape-0.48pre1 -z -D --file=Figures/Diffraction/singleSetup.svg 
    --export-pdf=Figures/Diffraction/singleSetup.pdf --export-latex} \def\svgwidth{0.7\columnwidth}
  \input{Figures/Diffraction/singleSetupLabel.tex}
      \label{fig:diffraction:singleSetup}
    }
    \subfigure[]{
      \executeiffilenewer{Code/Diffraction/singlePower.m}{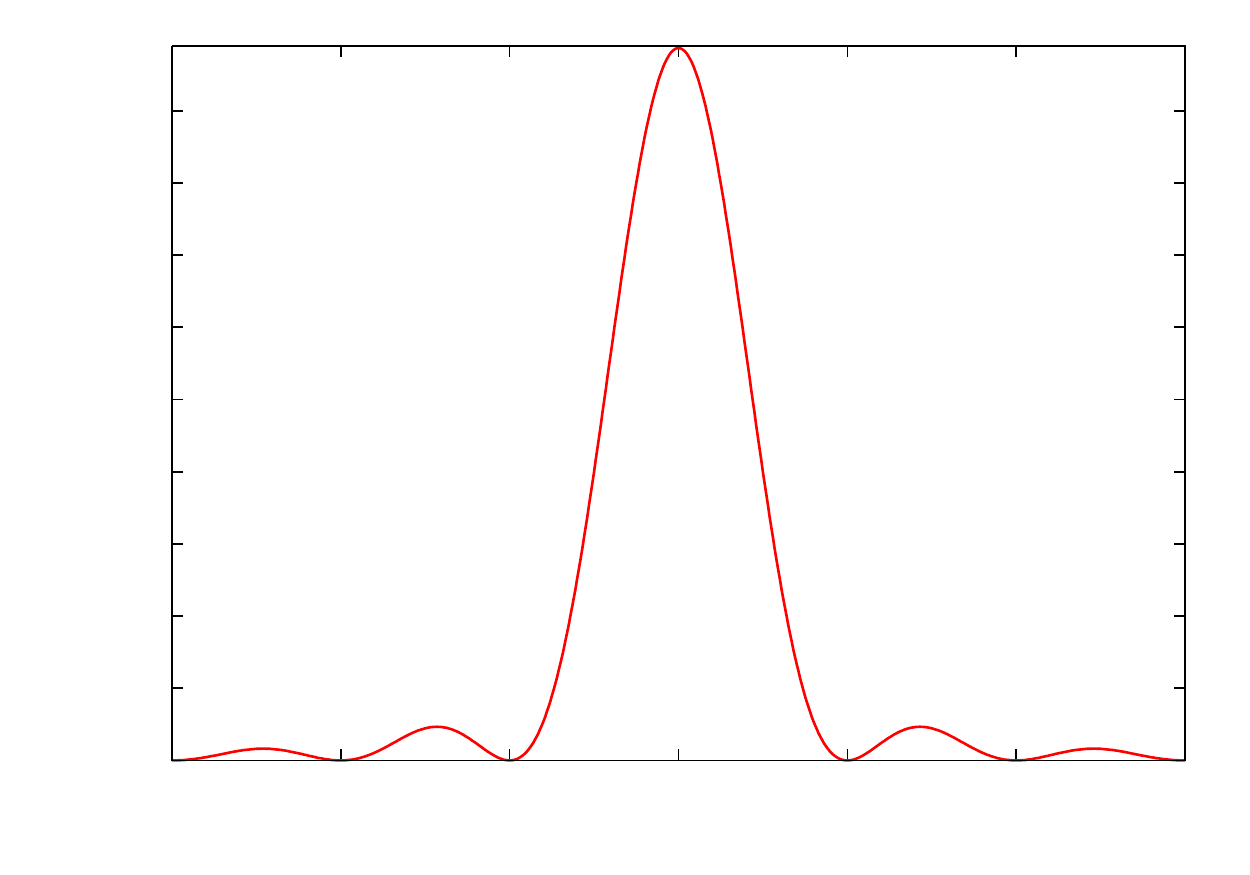}
  {octave --silent --eval "addpath([pwd(),'/Code']); 
    addpath([pwd(),'/Code/Diffraction']); singlePower(1);"} 
  \setlength{\unitlength}{\columnwidth*\real{0.00013}}
  \input{Figures/Diffraction/singlePowerLabel.tex}
      \label{fig:diffraction:singlePower}
    }
    \caption{Diagram of the double slit experiment.\label{fig:diffraction:single}}
  \end{center}
\end{figure}

Figure \ref{fig:diffraction:singleSetup} shows a single slit experiment setup
where a coherent monochromatic light source passes through a slit of
width $a$ and is projected onto a screen at a distance $L$ from the
slit. Again, we assume that $L \gg a$, or Fraunhofer diffraction, and
that we observe the interference pattern at point $P$ a distance $y$
above the centerline of the slit. In this experimental setup we must
consider adding together an infinite number of electric fields,
$E_1+E_2+E_3+\cdots+E_\infty$, each with a slightly different path
length $r$, and squaring the result to determine the intensity of the
light on the projection screen. Of course, adding together an infinite
number of electric fields by hand is not fun, and so instead we will
use an integral.

First, we notice that Figure \ref{fig:diffraction:singleSetup} is very
similar to Figure \ref{fig:diffraction:doubleSetup}; the only
difference, as mentioned before, is that we now must consider a
continuous source of electric fields rather than two discrete electric
fields. We can write the first infinitesimally small contribution to
the total electric field as,
\begin{equation}
  dE = \frac{dE_0}{r_0}\sin\left(kr_0-\omega t\right)
\end{equation}
where $r_0$ is the distance from the top of the slit to point $P$. The
value $dE_0$ is the infinitesimal amplitude of the electric field.

We can move down the slit a distance $x$ and consider an infinitesimal
contribution being emitted from this point. The path distance, just as
in the double slit setup, is now increased by an amount $\Delta r$
from $r_0$.
\begin{equation}
  dE = \frac{dE_0}{r_0+\Delta r}\sin\left(k(r_0+\Delta r)-\omega
    t\right)
  \label{equ:diffraction:dE}
\end{equation}
Now we can write $dE_0$ as the electric field amplitude density,
$\rho$, times the length over which the field is being emitted,
$dx$. Additionally, we can write $\Delta r$ in terms of $x$ as
$x\sin\theta$. Finally, we want to add together all the infinitesimal
electric fields, so we integrate from $x = 0$ to $x = a$, the width of
the slit.

Making these substitutions into Equation \ref{equ:diffraction:dE}, and integrating
yields the following.
\begin{equation}
  \begin{aligned}
    E &= \int_0^a\frac{\rho}{r_0+x\sin\theta}\sin\left(kr_0-\omega
      t+kx\sin\theta\right)\,dx \\
    &\approx \frac{\rho}{r_0}\int_0^a\sin\left(kr_0-\omega
      t+kx\sin\theta\right)\,dx \\
    &=
    \frac{2\rho}{kr_0}\left(\frac{\sin\left(\frac{ka\sin\theta}{2}\right)
        \sin\left(kr_0-\omega t+\frac{ka\sin\theta}{2}\right)}
      {\sin\theta}\right) \\
  \end{aligned}
  \label{equ:diffraction:singleE}
\end{equation}
In the first step, we have just made the substitutions and set up the
integral. Performing this integral is very messy, so we make the
approximation that $\rho/(r_0+x\sin\theta)$ is approximately
$\rho/r_0$ in the second step. This is a valid approximation because
$r_0 \gg \Delta r$. Notice that we cannot make the same approximation
in the sine term. This is because the phase difference of the waves is
entirely determined by $\Delta r$. In the third and final step we
perform the definite integral. Now we have the total electric field
from the infinitesimal contributions along the slit!

Just as with the double slit experiment, we are not very interested in
the electric field amplitude at point $P$ but rather the average
intensity. To find this, we square the electric field given in
Equation \ref{equ:diffraction:singleE}, and time average the trigonometric
functions dependent on time.
\begin{equation}
  \begin{aligned}
    I \propto E^2 &=
    \frac{4\rho^2}{k^2r_0^2}\left(\frac{\sin^2\left(\frac{ka\sin\theta}{2}
        \right) \sin^2\left(kr_0-\omega
          t+\frac{ka\sin\theta}{2}\right)}
      {\sin^2\theta}\right) \\
    &=
    \frac{4\rho^2}{k^2r_0^2}\left(\frac{\sin^2\left(\frac{ka\sin\theta}{2}
        \right)}
      {2\sin^2\theta}\right) \\
    &= \frac{\lambda^2\rho^2}{2\pi^2r_0^2}
    \left(\frac{\sin^2\left(\frac{\pi a\sin\theta}{\lambda} \right)}
      {\sin^2\theta}\right) \\
    &= \frac{\lambda^2\rho^2}{2\pi^2r_0^2}
    \left(\frac{L^2\sin^2\left(\frac{\pi a y}{\lambda L} \right)}
      {y^2}\right) \\
  \end{aligned}
  \label{equ:diffraction:singlePower}
\end{equation}
For the first step, we have just squared the electric field to find
the intensity. In the second step we have replaced
$\sin^2\left(kr_0-\omega t+ka\sin\theta/2\right)$ with a time averaged
value of $1/2$. In the third step we have replaced the wavenumber $k$
with its definition given in Equation \ref{equ:diffraction:waveNumber}. In the
final step we have made the small angle approximation of Equation
\ref{equ:diffraction:smallAngle}.

The final result of Equation \ref{equ:diffraction:singlePower} is proportional to
the intensity of the diffraction pattern at $y$ on the
projection screen, and so we can absorb all the constants into some
maximal intensity $I_0$ and equate Equation \ref{equ:diffraction:singlePower} with
intensity.
\begin{equation}
  I(y) = I_0 \frac{\sin^2\left(\frac{\pi a y}{\lambda L} \right)}
  {y^2}
  \label{equ:diffraction:singleIntensity}
\end{equation}
This intensity pattern is nearly identical to that of the double slit
given in Equation \ref{equ:diffraction:doubleIntensity}, but sine squared has been
replaced with cosine squared, and the whole quantity is divided by
$y^2$. This makes all the difference in the world, as the shape of the
intensity pattern for the single slit, shown in Figure
\ref{fig:diffraction:singlePower} is significantly different from that of the
double slit.

Looking at Figure \ref{fig:diffraction:singlePower}, we see that we can write a
relation similar to Equation \ref{equ:diffraction:doubleMax}, but instead of
locating the maxima, we locate the minima.
\begin{equation}
  y_\mathrm{min} = \frac{m\lambda L}{a},~~~m=0,1,2,\dots
  \label{equ:diffraction:singleMin}
\end{equation}
Again, we have derived a theoretical result which allows for a simple
experimental verification of the wave nature of light!

\section{Double Slit Diffraction}

As mentioned earlier, the interference pattern from two discrete slits
given in Figure \ref{fig:diffraction:doublePower}, is difficult to observe in most
experimental slits as oftentimes the slit widths is comparable to the
slit separation. This means that what was an interference problem now
becomes a diffraction problem and the exact same method used for the
single slit can be used but with different limits of integration. The
derivation does not introduce any new concepts, and drudging through
the math is not useful to the discussion here, so the results of
double slit diffraction will be presented without
derivation.\footnote{For readers who do not trust me, do the
  derivation yourself. Use the exact same method as the single slit,
  but now perform two integrals. Define $a$ as the slit width and $d$
  as the distance between the middle of both slits.  For the first
  slit the limits of integration will be from $0$ to $a$. For the
  second integral the limits will be from $d-a/2$ to $d+a/2$. Then
  enjoy slogging through all the math!}

The intensity for double slit diffraction is given by,
\begin{equation}
  I(y) = I_0\cos^2\left(\frac{\pi d y}{\lambda L}\right)
  \left(\frac{\sin^2\left(\frac{\pi a y}{\lambda L}
      \right)}{y^2}\right)
  \label{equ:diffraction:bothIntensity}
\end{equation}
where $a$ is the width of the two slits, and $d$ is the separation
between the the middle of the two slits. But wait just one moment,
this equation looks very familiar! That's because it is; the intensity
pattern for double slit diffraction is just the intensity pattern for
double slit interference, given in Equation \ref{equ:diffraction:doubleIntensity}
multiplied by the intensity pattern for single slit diffraction,
given in Equation \ref{equ:diffraction:singleIntensity}.

\begin{figure}
  \begin{center}
    \executeiffilenewer{Code/Diffraction/bothPower.m}{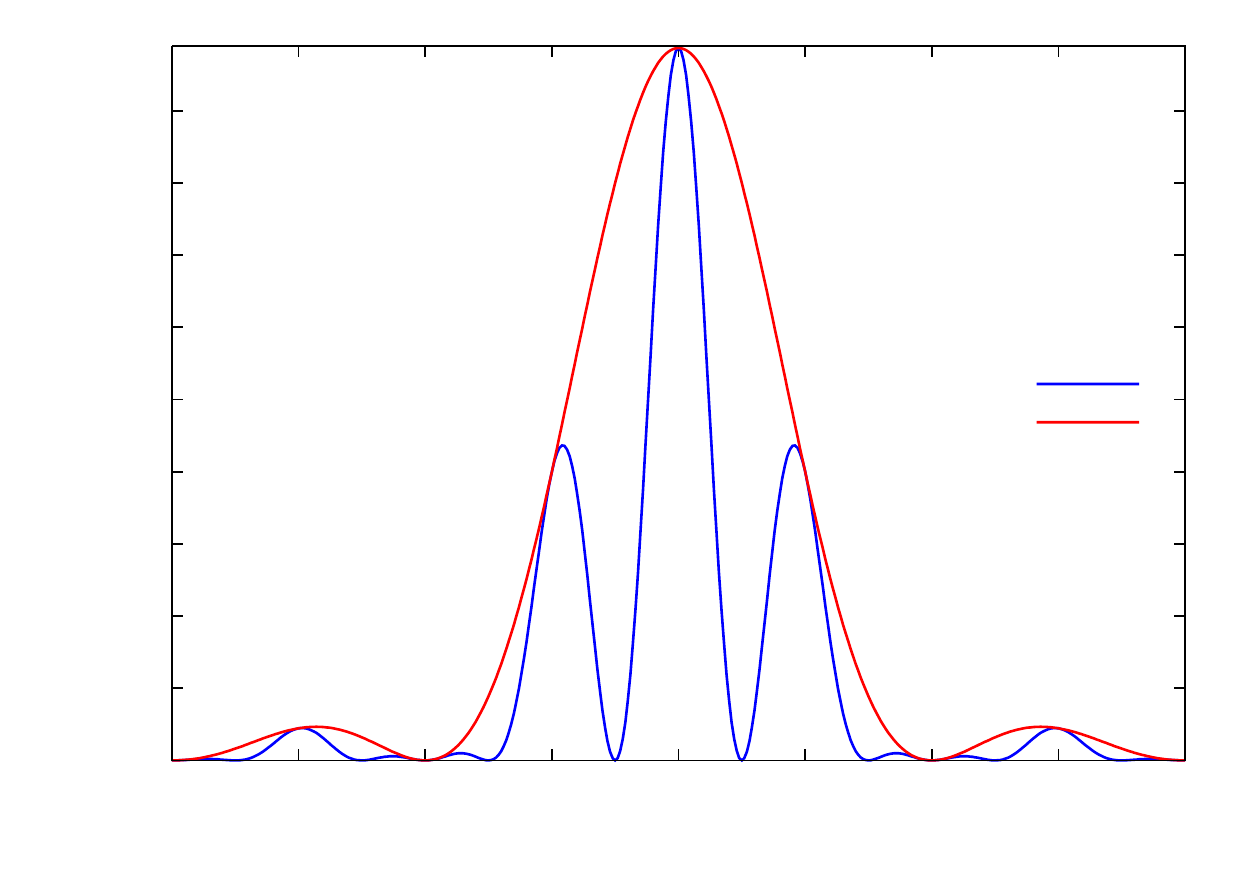}
  {octave --silent --eval "addpath([pwd(),'/Code']); 
    addpath([pwd(),'/Code/Diffraction']); bothPower(1);"} 
  \setlength{\unitlength}{\columnwidth*\real{0.00013}}
  \input{Figures/Diffraction/bothPowerLabel.tex}
    \caption{Intensity pattern for double slit diffraction. Here,
      $a=2d$.\label{fig:diffraction:bothPower}}
  \end{center}
\end{figure}

This means we know what Equation \ref{equ:diffraction:bothIntensity} should look
like, just Figure \ref{fig:diffraction:doublePower} multiplied by Figure
\ref{fig:diffraction:singlePower}. The intensity plot as a function of $y$ for
double slit diffraction is shown in Figure \ref{fig:diffraction:bothPower}, where
the complete intensity pattern is given in blue, and the overlying
single slit intensity pattern is given in red. Notice that when the
distance between the slits is much less than the width of the slits,
$d \ll a$, the intensity pattern becomes similar to that of a single
slit. Additionally, when $d$ is much larger than $a$, the double slit
interference separation becomes very small and is difficult to resolve
from the single slit diffraction. In Figure \ref{fig:diffraction:bothPower} the
width of the slits is twice the distance between the slits, $a = 2d$.

There is one final comment that needs to be made about double slit
interference and diffraction, and single slit diffraction. Throughout
the entirety of this chapter we have assumed that light is purely a
wave, and that Huygens was correct. It turns out that this is not the
case as shown by Einstein's famous \term{photoelectric
  experiment}. With this experiment, Einstein demonstrated that
discrete quanta of light, \term{photons}, carry an energy $hf$ where
$f$ is the frequency of the light and $h$ is Planck's constant. The
argument that physicists had thought resolved since $1802$ was yet
again revived, and the idea that light is both a particle and a wave
was developed. This idea is now known as the \term{wave-particle
  duality} of light, and applies not only to light, but everything.

The investigation of the wave-particle duality of light has lead to
many breakthroughs in the field of quantum mechanics. One of the most
interesting results is that every particle, and consequently all
matter, is made up of waves. The wavelength of these waves is given by
the \term{de Broglie wavelength}, $\lambda = h/p$ where $p$ is the
momentum of the object. We are not able to observe these waves in
everyday objects such as cars because the momentum of these objects is
so much larger than that of electrons or photons that the wavelength
is tiny in comparison to elementary particles.

Because the waves of particles are governed by quantum mechanics, the
results of some double slit examples are rather mind-bending. For
example, electrons were shot through a double slit experiment, and a
time lapse photograph of the results was made. The interference
patterns of Figure \ref{fig:diffraction:bothPower} were observed, along with the
individual impacts of the electrons! The same experiment was performed
again, but it was ensured that only one single electron passed through
a slit at a time. The interference pattern remained, despite the fact
that there were no other electrons to interfere with!  Finally,
electrons were passed through the slits one at a time, and one of the
two slits was covered so the experimenters knew from which slit the
electron emerged. When this was done, the interference pattern
disappeared! This incredible result provides direct evidence of the
hypothesis in quantum mechanics that the mere action of observing a
particle changes the particle, even if the measurement was
non-destructive.

\section{Experiment}

The experiment associated with this chapter consists of three parts:
in the first part the width of a single slit is measured by measuring
the distance between first and second order minima. The intensity
pattern should look very similar to the pattern given in Figure
\ref{fig:diffraction:singlePower}, and so the slit width $a$ can be determined
using Equation \ref{equ:diffraction:singleMin}. In the second part of the
experiment, the width of a human hair is measured. This is
accomplished by what is known as \term{Babinet's principle}.

Babinet's principle states that the diffraction pattern for any
combination of slits (or more generally shapes) is the same as the
diffraction pattern for the exact opposite setup, i.e. each opening is
replaced with a blocking, and each blocking is replaced with an
opening. For the case of a human hair, the exact opposite setup is a
single slit with the width of a human hair. But how exactly does this
work?

Consider first the electric field present from just shining a laser
onto the projection screen, let us call this $E_u$. Then consider
blocking the light with a human hair and call this electric field
$E_h$. Finally, we block the light with the exact opposite of the
human hair, a single slit the width of the human hair, and call this
electric field $E_s$. We can write the equation,
\begin{equation}
  E_u = E_h+E_s
\end{equation}
or the total unobstructed electric field is equal to the electric
field not blocked by the hair, plus the electric field not blocked by
the slit. Now we make a rather bold assumption and say that $E_u
\approx 0$. If this is the case, then $E_h = -E_s$ and since intensity
is proportional to the square of the electric field, $I_h = I_s$!

But by now, alarm bells should be going off. How can we possible
justify $E_u \approx 0$? If this were the case we would not be able to
see any interference pattern! The justification behind this assumption
is a bit unsatisfying, but the end result can be experimentally
verified. The idea is that for Fraunhofer or far-field diffraction and
interference the projection screen is so far away from the initial
light source that the light has spread sufficiently to yield an
electric field of $\approx 0$. Because the light is spreading out in
spherical waves, the electric field decreases by a factor of $1/r$,
and so for a very large $r$ the electric field is virtually
non-existent. This assumption, and consequently Babinet's principle
does not apply to \term{near-field} or \term{Fresnel diffraction}.

The final part of the experiment consists of measuring the separation
between two slits in a double slit diffraction (not interference)
experiment similar to that of Figure \ref{fig:diffraction:bothPower}. This means
that $a\sim d$ and so the single slit diffraction and double slit
interference patterns will both be visible. Make sure not to confuse
the two patterns; oftentimes the double slit interference pattern is
very small, and must be observed using a magnifying glass.
\graphicspath{{Figures/Planck/}}

\chapter{Planck's Constant}\label{chp:planck}

While exploring the phenomena of diffraction in Chapter
\ref{chp:diffraction}, we stated that the ``first'' resolution of the
light wave-particle duality was arrived at by the experimental
confirmation of Young's slit experiment that light was a wave. But
what about the ``second'' resolution? This chapter first looks at the
classical theory of light as a wave, first verified by Young's
experiment, and then given a theoretical groundwork by James Clerk
Maxwell. The theoretical framework for light and electromagnetism,
Maxwell's equations, provide an incredibly accurate theory, yet at the
turn of the $20^\mathrm{th}$ century problems began to
arise. Specifically, Albert Einstein began to question the purely wave
nature of light, and hypothesized his famous photoelectric effect
theory which stated that light was made up of particles called
photons.

With the photoelectric effect, Einstein postulated a famous
relationship between the energy of a photon and its frequency, which
requires the use of what is known as \term{Planck's constant}. This
constant had been experimentally determined by Max Planck a few years
earlier through his successful attempt to model what is known as
black-body radiation. In his model, Planck required the quantization
of light, yet he did not truly recognize the implications of this
quantization until Einstein's photoelectric effect. The final section
of this chapter briefly explores the theory of black-body radiation in
an attempt to put the history of the constant into context.

\section{Maxwell's Equations}

In Chapter \ref{chp:optics}, while deriving the Fresnel equations we
learned that not only is light a wave, but that it is an
electromagnetic wave, or an electric wave, $\vec{E}$, perpendicular to
a magnetic wave, $\vec{B}$ travelling in the direction
$\vec{E}\times\vec{B}$. Furthermore, we learned that the intensity of
light is proportional to the amplitude of the electric wave squared,
$E_0^2$. From Young's slit experiment know that light must be a wave,
but how do we know that light is made up of a magnetic and electric
field, and how do we know that the intensity of light is the amplitude
squared of the electric field? To answer these questions we must turn
to what are known as \term{Maxwell's equations}.

During the mid $1800$'s physicists such as Gauss, Faraday, Amp\'{e}re,
and Maxwell had become increasingly interested by both electrical and
magnetic phenomena, and through experimental trial and error,
determined the two phenomena were governed by the same force, or the
\term{electromagnetic force}. Maxwell went even further, and united
four laws that governed this force, Maxwell's equations, which are
given in a \term{differential form} by Equation
\ref{equ:planck:maxwellDerivative} and an \term{integral form} by
Equation \ref{equ:planck:maxwellIntegral}.

\begin{center}
\begin{minipage}[h]{6cm}
  \begin{subequations}
    \begin{equation}
      \nabla\cdot\vec{E} = \frac{\rho}{\epsilon_0}
      \label{equ:planck:maxwellDerivative1}
    \end{equation}
    \begin{equation}
      \nabla\times\vec{E} = -\frac{\partial\vec{B}}{\partial t}
      \label{equ:planck:maxwellDerivative2}
    \end{equation}
    \begin{equation}
      \nabla\cdot\vec{B} = 0
      \label{equ:planck:maxwellDerivative3}
    \end{equation}
    \begin{equation}
      \nabla\times\vec{B} = \mu_0\vec{J} +
      \mu_0\epsilon_0\frac{\partial\vec{E}}{\partial t}
      \label{equ:planck:maxwellDerivative4}
    \end{equation}
    \label{equ:planck:maxwellDerivative}
  \end{subequations}
\end{minipage}
\begin{minipage}[h]{6cm}
  \begin{subequations}
    \begin{equation}
      \ointint{V}{} \vec{E}\cdot d\vec{A} = \frac{Q}{\epsilon_0}
      \label{equ:planck:maxwellIntegral1}
    \end{equation}
    \begin{equation}
      \oint_A \vec{E}\cdot d\vec{l} = -\frac{\partial\Phi_B}{\partial t} 
      \label{equ:planck:maxwellIntegral2}
    \end{equation}
    \begin{equation}
      \ointint{V}{} \vec{B}\cdot d\vec{A} = 0
      \label{equ:planck:maxwellIntegral3}
    \end{equation}
    \begin{equation}
      \oint_A \vec{B}\cdot d\vec{l} =
      \mu_0I-\mu_0\epsilon_0\frac{\partial\Phi_E}{\partial t} 
      \label{equ:planck:maxwellIntegral4}
    \end{equation}
    \label{equ:planck:maxwellIntegral}
  \end{subequations}
\end{minipage}
\end{center}

At first glance these equations are very intimidating, but after
getting past all the symbols, their meaning is quite simple and
elegant. To begin, we must first explain the difference between the
differential and integral form of the equations. Both sets of
equations, Equations \ref{equ:planck:maxwellDerivative} and
\ref{equ:planck:maxwellIntegral}, represent the exact same physical
laws, but are different mathematical and physical interpretations of
the laws. Oftentimes one form is taught in class, and the other form
is ignored, but both forms of Maxwell's equations provide important
physical insights. More importantly, when solving electromagnetism
problems, choosing the appropriate form of the law can greatly simplify
the math behind the problem.

Looking at Equation \ref{equ:planck:maxwellDerivative1} we see that
the \term{divergence} of the electric field, $\nabla\cdot\vec{E}$, is
equal to the \term{charge density}, $\rho$, divided by the
\term{electric constant}. The divergence of a vector field, $\vec{F}$,
is given mathematically by $\partial F_x/\partial x + \partial
F_y/\partial y + \partial F_z/\partial z$ and is a scalar quantity (a
number not a vector). Visually, the divergence of a field is the
magnitude per unit area of all the vectors passing in and out of a
surface drawn around a single point in the field. For example, the
divergence of the electric field from a single electron would be the
magnitude per unit area of all the electric field vectors passing
through a sphere drawn around the electron. In this example, because
electrons have a negative charge, the divergence of $\vec{E}$ would be
negative. This makes sense, as $\rho$ should be negative as well.

The definition of divergence relates directly to the integral form of
Equation \ref{equ:planck:maxwellDerivative1} given by Equation
\ref{equ:planck:maxwellIntegral1}. This equation states that if a
volume of any shape, $V$, is drawn around some amount of enclosed
charge $Q$, the surface integral of the electric field is equal to the
enclosed charge, divided by the electric constant. The surface
integral is given by integrating over the dot product of the electric
field with the infinitesimally small piece of area, $d\vec{A}$, through
which the electric field is passing for the entire volume. Notice that
$d\vec{A}$ is a vector quantity, and the direction of $d\vec{A}$ is
perpendicular or \term{normal} to the surface of the volume at that
point. Oftentimes a surface can be chosen so that $\vec{E}$ and
$d\vec{A}$ are perpendicular and consequently $\vec{E}\cdot d\vec{A}$
is just $|\vec{E}||d\vec{A}|$.

Moving onto Equation \ref{equ:planck:maxwellDerivative2}, we see that
\term{curl} of the electric field, $\nabla \times \vec{E}$, is equal
to the negative of the partial derivative of the magnetic field with
respect to time. For readers not familiar with the curl of a field,
look at Chapter \ref{chp:superconductivity}. Briefly, the curl of a
vector field is how much the vector field is curling (like a whirlpool)
about a certain point. To understand what this means physically, we
turn to the integral form given by Equation
\ref{equ:planck:maxwellIntegral2}, which states that the line integral
of the electric field is equal to the opposite of the change in
magnetic flux over time through the area, $A$, of the line
integral. As an example, consider an electromagnet that is becoming
more powerful over time. If we draw a circle around the electromagnet,
the dot product of the electric field going through the circumference
of the circle is equal to the change in the magnetic field going
through the area of the circle. In this case the magnetic field is
growing, and so this means the electric field is pointing inwards
towards the electromagnet.

If we now look at Equations \ref{equ:planck:maxwellDerivative3} and
\ref{equ:planck:maxwellDerivative4} we see that their left hand sides
look very similar to Equations \ref{equ:planck:maxwellDerivative1} and
\ref{equ:planck:maxwellDerivative2} but with $\vec{B}$ swapped in for
$\vec{E}$. The same observation applies to the integral forms as
well. But, the right hand sides of these equations do not match. This
is because electric fields are caused by electric charges, such as the
electron or positron, which can be either negative or
positive. Magnetic fields, however, are caused by magnets which always
consist of a north and south pole. The idea of a magnet that is only a
north pole, or only a south pole is called a \term{magnetic monopole}
and has not yet been experimentally observed. Equations
\ref{equ:planck:maxwellDerivative3} and
\ref{equ:planck:maxwellIntegral3} essentially state this fact:
whenever a magnetic field is enclosed, the total ``magnetic charge''
has to be zero because there are no magnetic monopoles.

The same idea applies to Equation
\ref{equ:planck:maxwellDerivative4}. Here, electric charge can flow
(think electrons) and so the term $\mu_0\vec{J}$, the \term{magnetic
  constant} times the \term{current density}, must be included. The
analogous quantity for magnetism does not exist, and so this term is
missing from Equation \ref{equ:planck:maxwellDerivative2}. In the
integral form, Equation \ref{equ:planck:maxwellIntegral4}, current
density is just replaced with total current flowing through some line
integral.

\section{Electromagnetic Waves}

Entire books have been written just to explain the application of
Maxwell's equations, so it is not expected that the brief outline
above is anywhere close to a full explanation. However, hopefully it
provides a general idea of how the equations work, and their
inter-relations. For readers curious in learning more about Maxwell's
equations and how to use them to solve electromagnetism problems, the
book \href{http://www.amazon.com/dp/013805326X}{Introduction to
  Electrodynamics} by David J. Griffiths provides an excellent
introduction.

But back to our original questions, how do we know light is an
electromagnetic wave, and why is the intensity of light proportional
to amplitude squared? If we combine Equation
\ref{equ:planck:maxwellDerivative1} with Equation
\ref{equ:planck:maxwellDerivative2} we  can arrive at,
\begin{equation}
  \nabla^2\vec{E} = \mu_0\epsilon_0\frac{\partial^2\vec{E}}{\partial
    t^2}
  \label{equ:planck:waveElectric}
\end{equation}
where a few steps have been skipped in the process. Hopefully this
equation is vaguely familiar. Looking back all the way to
Chapter \ref{chp:waves}, we see that Equation
\ref{equ:planck:waveElectric} satisfies a three dimensional form of
the wave equation, given by Equation \ref{equ:waves:waveEquation}!
Now here's the exciting part. We know that the coefficient on the
right hand side of Equation \ref{equ:planck:waveElectric} is equal to
one over the velocity of the wave squared, by using Equation
\ref{equ:waves:waveConstant}. This gives us,
\begin{equation}
  C = \mu_0\epsilon_0 = \frac{1}{v^2} \Rightarrow v =
  \frac{1}{\sqrt{\mu_0\epsilon_0}} = 3.0\times10^8~\mathrm{\frac{m}{s}}
\end{equation}
which is the speed of light. We can also repeat the same proces using
the equations governing the magnetic field, Equations
\ref{equ:planck:maxwellDerivative3} and
\ref{equ:planck:maxwellDerivative4}, and again arrive at a velocity
equal to the speed of light. Because of these relations, we know that
light waves are made up of electric fields and magnetic fields and can
predict the speed of light because Maxwell's equations satisfy the
wave equation!

So that answers the first question, but how do we know that the
intensity of light (or the energy of light) is proportional to the
amplitude of the electric field squared? We now know that a light wave
consists of a electric and magnetic field, and it turns out that the
energy density flux of an electromagnetic field is given by the
\term{Poynting vector},
\begin{equation}
  \vec{S} = \frac{\vec{E}\times\vec{B}}{\mu_0} = c\epsilon_0|\vec{E}|^2
\end{equation}
where we are able to perform the second step because we know $\vec{E}$
is perpendicular to $\vec{B}$ and $\vec{E} = c\vec{B}$ where $c$ is
the speed of light. The Poynting vector tells us the energy, or
intensity of light, is proportional to the amplitude of the electric
field squared. Of course where the Poynting vector comes from has not
been explained, but that requires quite a bit more theory, and is left
to the more intrepid readers to find out on their own.

\section{The Photoelectric Effect}

By the turn of the $20^\mathrm{th}$ century, problems with the
interpretation of Maxwell's equations had begun to arise, and the
question of whether light was merely a wave arose yet again. One of
the major problems observed was with an experiment known as the
\term{photoelectric effect}. In the photoelectric effect, a strong
source of monochromatic light is shown onto a metallic surface. The
light strikes electrons within the metallic surface, and some
electrons are ejected. These ejected electrons can be measured, as
they create a voltage which can be read by a voltmeter.

The electrons ejected from the surface have the kinetic energy,
\begin{equation}
  KE_\mathrm{max} = E_\gamma-\phi
  \label{equ:planck:electronEnergy}
\end{equation}
where $E_\gamma$ is the energy from the incident light striking the
electron, and $\phi$ is the \term{work function}, or energy required
to eject the electron from the surface. The work function of the
surface is dependent upon the material the surface is made from. While
reading through Chapter \ref{chp:hydrogen} it is possible to estimate
the order of magnitude for a typical work function by determining how
much energy is required to remove the most tightly bound electron from
a hydrogen atom.

Work function aside, the kinetic energy of the ejected electrons can
be measured by passing the electron through an electric field. Within
an electric field, the electrons feel a force governed by Coulomb's
force (see Equation \ref{equ:hydrogen:coulombForce}) from the electric
field. By Newton's first law, we know that if a force is applied to
the electron, the electron must decelerate, and if the electric field
is large enough, the electron will come to a stop. The electric field,
when it brings all ejected electrons to a stop, is called a
\term{stopping potential}.

By using this information, it is possible to determine the maximum
kinetic energy of the electrons being ejected from the surface. When
the electric field is increased until no electrons are able to pass
through the field then,
\begin{equation}
  KE_\mathrm{max} = V_se
  \label{equ:planck:stoppingPotential}
\end{equation}
or the maximum kinetic energy of the ejected electrons,
$KE_\mathrm{max}$, is equal to the voltage of the stopping potential,
$V_s$, times the charge of an electron, $e$.

Now, by Maxwell's equations, we know that if we increase the amplitude
of the incident light, we increase the intensity of the light, and we
also increase the energy of the light. This means that classically,
the kinetic energy of the escaping electrons should increase if we
increase the intensity of the light. If we change the frequency or
color of the light, nothing should happen, the escaping electrons
should have the same maximum kinetic energy. Unfortunately, this
effect of an increase in electron kinetic energy with an increase in
incident light intensity was not observed when the experiment was
performed. Rather, three disturbing phenomena were observed.

\begin{enumerate}
  \item An increase in light intensity increased the number of ejected
    electrons, but did not increase the maximum kinetic energy of the
    ejected electrons.
  \item An increase in the frequency of the light increased the
    maximum kinetic energy of the ejected electrons, but did not
    increase the number of ejected electrons.
  \item If the frequency of the light was too low, no electrons were
    ejected, no matter how bright a light was used.
\end{enumerate}

This baffled physicists, as these results did not match with Maxwell's
equations. Albert Einstein, however, looked at the results, and
decided that perhaps light was not just a wave, despite the rather
overwhelming evidence of Young's experiment and the incredible success
of Maxwell's equations. Instead, he postulated that light is made up
of particles or \term{quanta} called \term{photons}. More importantly,
he theorized that the energy of a photon is proportional to the
frequency of the photon, or,
\begin{equation}
  E_\gamma = hf
  \label{equ:planck:photonEnergy}
\end{equation}
where $E_\gamma$ is the energy of the photon, $h$ is Planck's
constant, and $f$ is the frequency of the photon.

Using Equation \ref{equ:planck:photonEnergy}, we can now explain the
odd results of the photoelectric experiment. Rather than thinking of
an incident wave of light striking an electron, we can think of a
single photon striking an electron. If we increase the number of
incident photons, we increase the intensity of the light, and
consequently the number of ejected electrons increases. The kinetic
energy of the ejected electrons does not change however, because the
energy of the incident light striking an electron is just the energy
of a single photon and does not change without changing the frequency
of the light. We have just explained the first observation!

But what about the second observation? The same logic applies. Now we
do not increase the intensity of the light, so the number of photons
stays the same, but we do increase the frequency and so by Equation
\ref{equ:planck:photonEnergy}, the energy of each photon is
increased. Going back to Equation \ref{equ:planck:electronEnergy},
$E_\gamma$ increases, and so the maximum kinetic energy of the
electrons must also increase. If however, the frequency of the
incoming photons is not high enough, the energy of the photon will be
less than the binding energy, $E_\gamma < \phi$, and so the electron
will have negative kinetic energy! This of course does not make sense,
as kinetic energy cannot be negative; it must be zero or positive.

What this does mean though, is that the electron absorbs less energy
than is needed to eject it from the surface. The electron will be less
strongly bound to the surface, but it cannot escape, so no ejected
electrons will be observed if $E_\gamma < \phi$, and so we now have an
explanation for the third observation. Of course, it is interesting to
know what happens when the electron is not ejected from the surface,
but its binding energy is decreased. This is covered more in the
chapter on fluorescence, Chapter \ref{chp:fluorescence}.

In Chapter \ref{chp:diffraction} it was mentioned that Einstein did
not suggest that light was just a particle, but rather he suggested
that it was both a particle and a wave. The implications of this
\term{wave-particle duality} were already discussed in Chapter
\ref{chp:diffraction}, but it is important to remember how Einstein
developed this new idea. The experimental evidence for light as a wave
was overwhelming, with Maxwell's equations providing a strong
theoretical groundwork, yet the photoelectric effect demonstrated
light must be made of particles. Einstein saw what no other physicist
at the time could see; if light exhibits the behavior of a wave, and
the behavior of a particle, it must be both a particle and a wave!

\section{Blackbody Radiation}

As stated in the introduction to this chapter, Einstein demonstrated
the quantum nature of light, yet Planck inadvertently required the
quantization of light through the \term{black-body radiation}
experiment and measured his constant, $h$, before realizing the full
implications of the constant. So what is a black-body? A black-body is
an object that absorbs all electromagnetic radiation, both visible and
invisible. Hence, it is given the name black-body because it cannot be
seen from reflected light. However, after absorbing electromagnetic
radiation, the black-body re-emits light on a variety of frequencies,
dependent upon the temperature of the black-body. Perfect black-bodies
do not exist in nature, but both the sun, and the human body act like
black-bodies for certain frequencies of light!

The theory behind black-body radiation is based on statistical
mechanics, and a variety of approaches can be taken. The theory behind
these approaches will not be explained here, as the math can be very
involved, but the resulting equations will be given. The first
equation describing black-body radiation was first proposed by Wein,
and later derived by Planck and is,
\begin{equation}
  I(f,t) = \frac{2hf^3}{c^2}e^{-\frac{hf}{kT}}
  \label{equ:planck:wein}
\end{equation}
where $I(f,t)$ is the intensity of the light at a frequency $f$
emitted from a black-body with a temperature $T$. The constant $k$ is
\term{Boltzmann's constant}, $c$ is the speed of light, and $h$ is
Planck's constant. The units of $I$ are Joules per square meter.

It was found that this equation matches black-body radiation well for
high frequencies, $f$, but deviates significantly from experiment for
low frequencies. Planck went back to the drawing board and postulated
\term{Planck's law},
\begin{equation}
  I(f,T) = \frac{2hf^3}{c^2\left(e^{\frac{hf}{kT}}-1\right)}
  \label{equ:planck:planck}
\end{equation}
which is very similar to Equation \ref{equ:planck:wein}, but matches
experiment for both high and low frequencies. In his derivation of
Equation \ref{equ:planck:planck}, Planck required that light obeyed
\ref{equ:planck:photonEnergy} but without realizing that he had
quantized light!

Finally, a third model based entirely on classical mechanics by
Jeans and Raleigh stated,
\begin{equation}
  I(f,T) = \frac{2f^2kT}{c^2}
  \label{equ:planck:jeans}
\end{equation}
but experimentally, this formula only matches the intensity of lower
frequencies of light. The formulation of this theory allowed physicist
to understand the importance of using quantum mechanics to explain
black-body radiation. Without the Jeans-Raleigh equation, Planck would
not have realized the implications of his quantization of light.

\begin{figure}
  \begin{center}
    \executeiffilenewer{Code/Planck/blackbody.m}{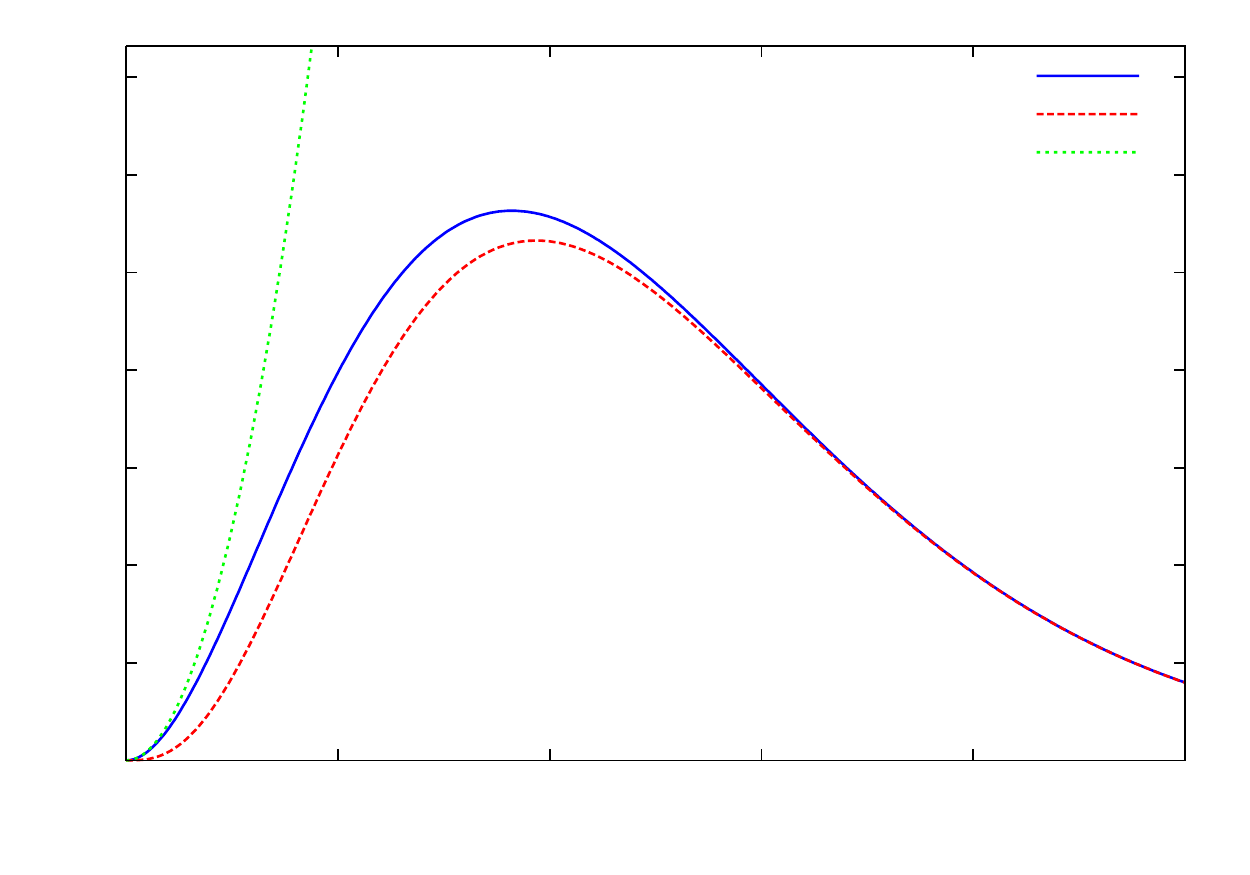}
  {octave --silent --eval "addpath([pwd(),'/Code']); 
    addpath([pwd(),'/Code/Planck']); blackbody(1);"} 
  \setlength{\unitlength}{\columnwidth*\real{0.00013}}
  \input{Figures/Planck/blackbodyLabel.tex}
    \caption{An example of the black-body radiation given off by the
      human body at $310$ K. The green curve gives the incorrect
      classical theory of Equation \ref{equ:planck:jeans}, the red
      curve the incorrect theory of \ref{equ:planck:wein}, and the
      blue curve the correct theory derived by Planck of Equation
      \ref{equ:planck:planck}.\label{fig:planck:blackbody}}
  \end{center}
\end{figure}

A comparison between Equations \ref{equ:planck:wein},
\ref{equ:planck:planck}, and \ref{equ:planck:jeans} is given in Figure
\ref{fig:planck:blackbody}. Equation \ref{equ:planck:planck} correctly
describes the spectrum of light from black-body radiation, while
Equation \ref{equ:planck:wein} describes the radiation well for high
frequencies, and \ref{equ:planck:jeans} for low frequencies. The
black-body radiation spectrum given in Figure
\ref{fig:planck:blackbody} is for an object with a temperature of
$310$ K, the average temperature of the human body. The light emitted
from the human body, due to black-body radiation, is on the infrared
scale. This is how many types of night vision systems work; they
detect the black-body radiation emitted from the human body in the
infrared spectrum.

\section{Experiment}

The experiment for this chapter demonstrates the photoelectric effect,
using the method discussed in the third section of this
chapter. Luckily, all of the complicated apparatus required during the
early $1900$'s to create a stopping potential and measure the maximum
kinetic energy of the ejected electrons can be condensed into a small
black box with the magic of circuits and a few other clever ideas. The
entire apparatus for the experiment consists of this black box, a
bright ultraviolet lamp, and various colored filters that can be used
to change the frequency of the light incident on the box.

Within the black box is a photodiode, essentially a surface that
ejects electrons when struck by photons. This surface has some work
function, $\phi$, which is unknown, and must be determined from the
data taken during the experiment. The ejected photons are then driven
into a capacitor which creates an electric field. Over time the
electric field within the capacitor increases until it reaches the
stopping potential, and then stabilizes at this value. A voltmeter is
connected to this capacitor and measures the voltage drop across the
capacitor, giving the stopping potential.

Using the voltmeter and various filters, a variety of data points can
be taken of light frequency and measured stopping potential. By
combining Equations \ref{equ:planck:electronEnergy},
\ref{equ:planck:stoppingPotential}, and \ref{equ:planck:photonEnergy}
we can write,
\begin{equation}
  V_s = \frac{h}{e}f - \frac{\phi}{e}
  \label{equ:planck:data}
\end{equation}
or the stopping potential is equal to the Planck's constant times
frequency less work function, all over fundamental electric charge. If
the theory above is correct, then we can plot the data taken with
$V_s$ on the $y$-axis and $f$ on the $x$-axis to obtain a relationship
governed by Equation \ref{equ:planck:data}. The slope of the plot will
yield $\frac{h}{e}$ while the intercept of the graph will give
$\frac{\phi}{e}$, and so if $e$ is known, it possible to calculate
both Planck's constant, and the work function of the apparatus from
the data!

\graphicspath{{Figures/Hydrogen/}}

\chapter{Hydrogen}\label{chp:hydrogen}

Over a century ago Lord Kelvin, while addressing a room full of the
world's leading physicists, stated, ``There is nothing new to be
discovered in physics now. All that remains is more and more precise
measurement.''\footnote{Weisstein,
  Eric. \href{http://scienceworld.wolfram.com/biography/Kelvin.html}{``
    Kelvin, Lord William Thomson (1824-1907)''}.} As it so happens,
Kelvin was wrong, as Rutherford and Bohr would so spectacularly
demonstrate in the following years in the field of
spectroscopy.\footnote{Lord Kelvin was wrong about quite a few
  things. A quick Google search will reveal pages and pages of rather
  humorous quotes from him. Despite his incredible failures in
  prediction, he pioneered many techniques in thermodynamics.}

At the beginning of the twentieth century most physicists would
describe the structure of an atom with what is known as the
\term{plum-pudding model}. In this model the atom is represented by a
fluidic ``pudding'' of positive charge with ``plums'' of electrons
floating around inside. A scientist of the time, Ernest Rutherford,
questioned this explanation and decided to investigate the inner
structure of the atom with his revolutionary \term{scattering}
experiment. In this experiment alpha particles were fired at a thin
gold foil. If the plum-pudding model was accurate most of these alpha
particles would be reflected back, but Rutherford found the exact
opposite. Nearly all the alpha particles passed through the foil
undeflected, indicating that the gold atoms consisted mainly of \ldots
nothing.

To explain this astonishing result, Rutherford introduced a new model
for the structure of the atom called the \term{Rutherford model}. In
this model the atom consists of a very hard and dense core (the
nucleus) about which electrons orbited. This model, while describing
the phenomena which Rutherford observed in his scattering experiment,
still suffered from a variety of flaws. In an attempt to correct these
flaws a scientist by the name of Neils Bohr introduced the \term{Bohr
  model} in 1913. This model enjoyed great success in experimental
observation and is taught to this day as an introduction to the
atom. The model suffers from a variety of drawbacks and has been
superseded by the \term{atomic orbital model} of the atom, but is still
useful in conceptualizing spectral emissions of simple atoms.

\section{Bohr Model}

\begin{figure}
  \begin{center}
    \executeiffilenewer{Figures/Hydrogen/bohrModel.svg}
  {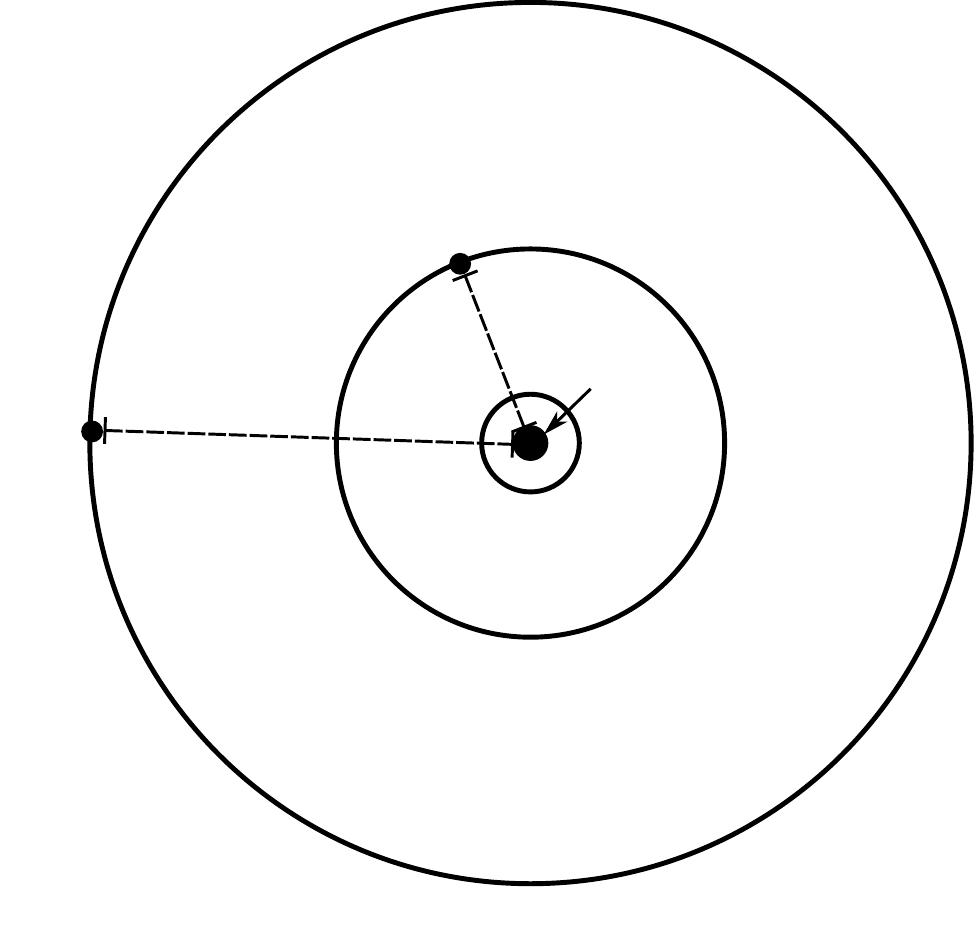}{inkscape-0.48pre1 -z -D --file=Figures/Hydrogen/bohrModel.svg 
    --export-pdf=Figures/Hydrogen/bohrModel.pdf --export-latex} \def\svgwidth{0.5\columnwidth}
  \input{Figures/Hydrogen/bohrModelLabel.tex}
    \caption{The Bohr model for a hydrogen atom with electrons at
      principal quantum numbers of $n=2$ and $n=3$. The symbol $e^-$
      denotes an electron and $p^+$ a proton.\label{fig:hydrogen:bohrModel}}
  \end{center}
\end{figure}

So what exactly is the Bohr model? The Bohr model is a planetary model
that describes the movement of electrons about the nucleus of an atom
as is shown in Figure \ref{fig:hydrogen:bohrModel}. Here, electrons (analogous
to planets) orbit about the charged, dense nucleus of the atom (the
star of the planetary system). More importantly, the Bohr model is a
combination of classical and quantum theory with surprisingly accurate
results for the hydrogen atom. Don't be disconcerted by the word
quantum here, it will be explained shortly. But first, we will begin
with the classical portion of the theory and apply it to the hydrogen
atom which consists of a single negatively charged electron, and a
single positively charged proton (plus a neutron but we don't care
about that).

Any object in regular circular motion experiences a force stopping it
from flying away, whether that force is gravity, tension, or
electromagnetism. This \term{centripetal force} can be written in terms
of mass of the orbiting object $m$, the radius of the orbit $r$, and
the tangential velocity of the object $v$.\footnote{There are a
  variety of methods to derive this formula, one being geometric. Try
  writing the period for one orbit of the object in terms of velocity
  and distance traveled, but also in terms of velocity and
  acceleration and equate the two.}
\begin{equation}
  F = \frac{mv^2}{r}
  \label{equ:hydrogen:centripitalForce}
\end{equation}
In the case of a negatively charged electron orbiting a positively
charged nucleus, the centripetal force is due to electromagnetism and
is described by the \term{Coulomb force}.
\begin{equation}
  F = -\frac{q_1q_2}{4\pi\epsilon_0r^2}
  \label{equ:hydrogen:coulombForce}
\end{equation}
Here $\epsilon_0$ is the \term{electric constant}\footnote{This
  constant is known by many names, permittivity of a vacuum, as used
  in the lab manual, permittivity of free space, etc. The bottom line
  is that this constant allows us to convert from units of charge
  squared over distance squared into force.}, $q_1$ the charge of the
electron, and $q_2$ the charge of the proton, while $r$ is the
distance between the two. Setting Equation \ref{equ:hydrogen:centripitalForce}
equal to Equation \ref{equ:hydrogen:coulombForce} we can find a value for the
velocity of the electron in terms of $r$, $\epsilon_0$, the charge of
the electron $e$, and the mass of the electron $m_e$.
\begin{equation}
  \frac{m_ev^2}{r} = \frac{e^2}{4\pi\epsilon_0r^2} ~~~\Rightarrow~~~ v =
  \sqrt{\frac{e^2}{m_e4\pi\epsilon_0r}}
  \label{equ:hydrogen:velocity}
\end{equation}

This is where we now need to use a very basic form of quantum
mechanics. Previously in the Rutherford model, Rutherford allowed
electrons to take on any possible velocity. The only problem with this
is that when electrons accelerate they lose energy by releasing energy
in the form of a photon.\footnote{For those of you who are interested,
  the idea of light being \term{quantized} in the form of a photon was
  first proven by Einstein in his photoelectric effect experiment.}
An electron orbiting a nucleus in circular motion is constantly
experiencing an acceleration towards the nucleus. Hence, the electrons
orbiting a nucleus must be radiating energy, and so all atoms must be
continuously emitting light!\footnote{For more details on electrons
  radiating energy under acceleration look up \term{synchrotron
    radiation} and the \term{Larmor formula}.} As everything around us
does not usually glow, this is clearly not the case. To provide a
solution to this dilemma, Bohr suggested that the the electrons could
only take on discrete energies, and that when at these energy levels
the electrons would not lose their energy to radiation unless forced
into a different energy level.

The idea that the energy of the electrons is \term{quantized}, or only
allowed at specific levels, is a very rudimentary form of quantum
mechanics. The reason Bohr suggested this rule was not because he had
incredible foresight into the intricacies of fully developed quantum
mechanics, but because such a rule would provide a theory that would
match experiment. Specifically, atoms in gasses had been observed
emitting light at discrete energies, rather than a continuum of
energies.

Despite the reasoning behind Bohr's rule of quantizing the energy, the
result is that the electrons can only take on quantized values of
angular momentum.
\begin{equation}
  L \equiv rp\sin\theta = rm_ev = \frac{nh}{2\pi}
  \label{equ:hydrogen:quantum}
\end{equation}
Here the letter $n$, called the \term{principal quantum number}, is a
positive integer greater than zero, i.e. $1$, $2$, $3$, etc., and $h$
is the \term{Planck constant}. Because momentum is quantized we see
that both the velocity and the radius at which the electron orbit the
nucleus must also be quantized. We can plug in the value that we
obtained for the velocity of the electron (using classical mechanics)
into Equation \ref{equ:hydrogen:quantum} and solve for the radius in terms of
the principal quantum number $n$, $h$, the mass of the electron $m_e$,
the electric constant $\epsilon_0$, and the fundamental unit of charge
$e$.
\begin{equation}
  rm_e\sqrt{\frac{e^2}{m_e4\pi\epsilon_0r}} = \frac{nh}{2\pi}
  ~~~\Rightarrow~~~
  r_n = \frac{\epsilon_0n^2h^2}{\pi m_ee^2}
  \label{equ:hydrogen:bohrRadius}
\end{equation}
In the second step we have replaced $r$ with $r_n$ to indicate that
the value of $r$ is entirely dependent upon the principal quantum number
$n$.

Looking at Equation \ref{equ:hydrogen:bohrRadius} we see an incredible
result. The radius at which an electron orbits a nucleus is only
dependent on $n$! What is even more exciting is that the radius at
which the electron orbits the nucleus gives us the effective size of
the atom. We can see what the smallest size of the atom is by letting
$n = 1$. We call this the \term{Bohr radius}, or $r_1$.
\begin{equation}
  r_1 =  \frac{\epsilon_0h^2}{\pi m_ee^2} \approx 0.5\times10^{-10}\mbox{ m}
\end{equation}
From first principals in classical mechanics and a little help from
quantum mechanics, we have derived the size of a ground state hydrogen
atom!

\section{Spectroscopy}

The idea that we can calculate the fundamental size of the hydrogen
atom from Equation \ref{equ:hydrogen:bohrRadius} is exciting, but is rather
difficult to experimentally verify; we can't just grab a ruler and go
measure the distance of an electron from the nucleus. Thankfully,
there is another way that we can experimentally verify our theory and
that is through \term{spectroscopy}.\footnote{Spectroscopy is literally
  the observation of a spectrum, in this case, a spectrum of light.}

Many people know Einstein for his work on general and special
relativity, but less commonly known is that Einstein never won the
Nobel prize for this work. Instead, he won the Nobel prize for what is
known as the \term{photoelectric effect}. What Einstein demonstrated is
that light is made up of tiny massless particles, called photons, and
that the energy of each photon is directly proportional to its
frequency.\footnote{This equation is more commonly written as $E =
  h\nu$ but for notational consistency we have stayed with designating
  frequency with the letter $f$ instead of $\nu$.}
\begin{equation}
  E_\gamma = hf
  \label{equ:hydrogen:photonEnergy}
\end{equation}
Here we have used the subscript $\gamma$ (the Greek letter gamma) to
denote the photon. What this means is that light with a high frequency
is more energetic than light with a low frequency. For example, a beam
of blue light would contain almost nearly one and a half times as much
energy as a beam of red light.

But how does this help us with experimentally verifying Bohr's model?
As the electrons orbit the nucleus, they have a potential energy
dependent upon their orbital radius and a kinetic energy dependent
upon their velocity.  The total energy of each electron is just its
kinetic energy plus its potential energy. The kinetic energy for the
electron is $\frac{1}{2}mv^2$, while the potential energy is the
electric potential energy between the proton of the nucleus and the
electron. The electric potential energy for two charges (exactly like
this case) is given by,
\begin{equation}
  V = \frac{q_1q_2}{4\pi\epsilon_0 r}
  \label{equ:hydrogen:electricPotential}
\end{equation}
where $q_1$ and $q_2$ are the charges, $\epsilon_0$ the electric
constant, and $r$ the separation between the two charges. Notice that
this relation is nearly identical to the Coulomb force given in
Equation \ref{equ:hydrogen:coulombForce}. This is because the integral of force
over distance is just energy and so Equation
\ref{equ:hydrogen:electricPotential} can be found by integrating Equation
\ref{equ:hydrogen:coulombForce} with respect to $r$.

Putting this all together, we can add our standard relation for
kinetic energy to our potential energy $V$ to find the total energy of
the electron as it orbits the nucleus.

\begin{equation}
  \begin{aligned}
    E_n &= \frac{1}{2}m_ev^2 + V\\
    &=
    \left(\frac{m_e}{2}\right)\left(\frac{e^2}{m_e4\pi\epsilon_0r_n}\right)
    - \frac{e^2}{4\pi\epsilon_0r_n}\\
    &=
    \left(\frac{m_e}{2}\right)\left(\frac{e^2}{m_e4\pi\epsilon_0}\right)
    \left(\frac{\pi m_ee^2}{\epsilon_0n^2h^2}\right) -
    \left(\frac{e^2}{4\pi\epsilon_0}\right) \left(\frac{\pi
        m_ee^2}{\epsilon_0n^2h^2}\right)\\
    &=
    \frac{m_ee^4}{8\epsilon_0^2n^2h^2}-\frac{m_ee^4}{4\epsilon_0^2n^2h^2}\\
    &= -\frac{m_ee^4}{8\epsilon_0^2n^2h^2} \approx
    \frac{-2.18\times10^{-18}\mbox{ J}}{n^2} = \frac{-13.6\mbox{ eV}}{n^2}\\
  \end{aligned}
  \label{equ:hydrogen:bohrEnergy}
\end{equation}
In the first step, we have just written the formula for the total
energy, kinetic energy plus potential energy. In the second step we
have replaced $v$ with Equation \ref{equ:hydrogen:velocity} and $V$ with
Equation \ref{equ:hydrogen:electricPotential}. In the third step we have
replaced $r_n$ with Equation \ref{equ:hydrogen:bohrRadius}. In the final steps
we have just simplified and then plugged in the values for all the
constants. Oftentimes when dealing with energies on the atomic (and
sub-atomic) scale we will express energy in terms of \term{electron
  volts} (eV) instead of joules. An electron volt is the energy an
electron gains when it passes through an electric potential of one
volt, hence the name electron volt. Equation \ref{equ:hydrogen:bohrEnergy} is
what we call the \term{binding energy}.

The first important result to notice about Equation
\ref{equ:hydrogen:bohrEnergy} is that the total energy of the electron is
negative! This means that once an electron becomes bound to a hydrogen
nucleus, it can't escape without external energy. The second result is
that for very large values of $n$ the binding energy is very close to
zero. This means that electrons very far away from the nucleus are
essentially free electrons. They do not need much help to escape from
the nucleus.

The question still remains, how can we use the results above to verify
our theory, and how does spectroscopy come into play? The missing part
of the puzzle is what electrons do when they transition from a near
zero binding energy (for example $n=20$) to a large negative binding
energy (such as $n=1$). For the electron to enter a smaller orbit with
a large negative binding energy it must give up some of its energy. It
does this by radiating a photon with an energy equal to the difference
in energies between the level it was at (initial level) and level it
is going to (final level).
\begin{equation}
  E_\gamma = E_i-E_f =
  \frac{m_ee^4}{8\epsilon_0^2h^2}\left(\frac{1}{n_i^2} - \frac{1}{n_f^2}
  \right)
  \label{equ:hydrogen:energyDifference}
\end{equation}
Here $E_\gamma$ is the energy of the photon, $E_i$ the initial energy
for a principal quantum number $n_i$, and $E_f$ the final energy for a
principal quantum number $n_f$.

You can think of it as the electron having
to pay an entrance fee (the photon) to enter an exclusive club (a more
central radius). The more ``inner circle'' the club (in this case a
smaller radius) the more expensive the cost, and so a more energetic
photon must be given up. Of course in this example, the most exclusive
club is the Bohr radius, $r_1$, and the cost is $13.6$ eV if entering
from the outside.

In the club example, we can tell the difference between a $5$ euro
note and a $10$ euro note by color. We can do the exact same with the
electrons because of Equation \ref{equ:hydrogen:photonEnergy}! A high energy
electron will have a high frequency (for example, blue) while a low
energy electron will have a low frequency (for example, red). What
this means is that when we watch a hydrogen atom being bombarded with
electrons, we should continually see photons being thrown off by the
electrons in their attempts to get to a smaller radius. In the club
analogy, think of a crowd of people jostling to get into the exclusive
section of the club, waving money in their hands to get the attention
of the bouncer.

When we are in a club, we know the admission prices. Here with the
electrons, we also know the admission prices, but now the price is
given by a frequency or wavelength instead of euros.
\begin{equation}
  \begin{aligned}
    E_\gamma &= hf\\
    f &= \frac{m_ee^4}{8\epsilon_0^2h^3}\left(\frac{1}{n_i^2} -
      \frac{1}{n_f^2}
    \right)\\
    f &= \frac{c}{\lambda}\\
    \frac{1}{\lambda} &=
    \frac{m_ee^4}{8c\epsilon_0^2h^3}\left(\frac{1}{n_i^2} -
      \frac{1}{n_f^2} \right)\\
  \end{aligned}
  \label{equ:hydrogen:wavelength}
\end{equation}
In the first step we have just used Equation
\ref{equ:hydrogen:photonEnergy}. In the second step we have used Equation
\ref{equ:hydrogen:energyDifference} to write $E_\gamma$. In the third step we
have just written the normal relation between frequency and
wavelength, $f = v/\lambda$.\footnote{See the waves lab for more
  details if this relation is unfamiliar.} Notice that this equation
is just Equation $1$ of the lab manual. From this we can solve for the
\term{Rydberg constant} which is just the constant of proportionality
in front of this equation.
\begin{equation}
  R = \frac{m_ee^4}{8c\epsilon_0^2h^3} \approx 1.1\times10^7~\mbox{
    m}^{-1} = 0.011\mbox{ nm}^{-1}
  \label{equ:hydrogen:rydbergConstant}
\end{equation}
We have just derived Equation $3$ in the lab manual!

\section{Transition Series}

Now that we know that electrons within the hydrogen atom will emit
photons of a specific energy, frequency, and wavelength, as given in
Equation \ref{equ:hydrogen:wavelength}, we can make an experimental
prediction. We should observe a spectrum of photons with wavelengths
dictated by their initial quantum number $n_i$ and their final number
$n_f$ emitted from hydrogen atoms being bombarded with
electrons. These spectrums, categorized by the final quantum number
$n_f$ are called \term{transition series}; the electron is going
through a series of transitions from energy level to energy
level. Several important transition series are named after their
discoverers: the \term{Lyman series} corresponding to $n_f = 1$, the
\term{Balmer series} corresponding to $n_f = 2$, and the \term{Paschen
  series} corresponding to $n_f = 3$.

\begin{figure}
  \begin{center}
    \small
    \executeiffilenewer{Figures/Hydrogen/balmerSeries.svg}
  {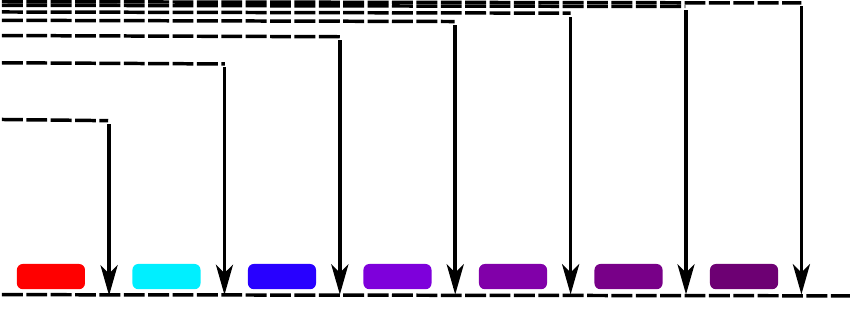}{inkscape-0.48pre1 -z -D --file=Figures/Hydrogen/balmerSeries.svg 
    --export-pdf=Figures/Hydrogen/balmerSeries.pdf --export-latex} \def\svgwidth{\columnwidth}
  \input{Figures/Hydrogen/balmerSeriesLabel.tex}
    \caption{The first nine transitions of the Balmer series. The
      colors approximate the color of the emitted light of each
      transition.\label{fig:hydrogen:balmerSeries}}
  \end{center}
\end{figure}

The most important of these three transition series is the \term{Balmer
  series}, as the wavelengths of the emitted photons are within the
visible spectrum of light for the human eye. As $n_f=2$ we can write
the energy of the emitted photon for an electron transitioning from
principal quantum number $n_i$ using Equation
\ref{equ:hydrogen:energyDifference}.
\begin{equation}
  E_\gamma = 13.6\left(\frac{1}{n_i^2} -
    \frac{1}{4}\right)\mbox{ eV}
  \label{equ:hydrogen:balmerSeries}
\end{equation}
We can make a diagram of the Balmer transitions, as given in Figure
\ref{fig:hydrogen:balmerSeries} with the corresponding approximate color of the
light emitted.\footnote{Color is a very subjective principal, as it
  involves an emitter and an observer. These colors are only meant to
  provide a relative approximation of the color which would be
  observed by the naked eye from the Balmer series. The colors were
  created using the
  \href{http://www.efg2.com/Lab/ScienceAndEngineering/Spectra.htm}{Spectra}
  program.}  The first seven transitions, $n_i = 3$ through $n_i = 9$
have their wavelengths, frequencies, and energies summarized in Figure
\ref{fig:hydrogen:balmerSeries} as well.

From Figure \ref{fig:hydrogen:balmerSeries} we see that only the first seven or
so transitions of the Balmer series should be visible to the naked
eye, and that as $n_i$ increases, the colors of the light are more and
more difficult to distinguish. This means that with minimal diffraction
equipment it is oftentimes only possible to observe the first three
lines, a red line, a bluish green line, and a dark blue or purple
line.

\section{Diffraction}

But what do we mean by ``observe a line''? When electricity is run
through hydrogen gas the gas begins to emit photons from all the
different transitions discussed above. The photons all mix together,
and the experimenter is confronted with a pink glow from the hydrogen
lamp. How is it then possible to separate the photons from each other
so we can observe the various wavelengths? The answer is {\bf
  diffraction}, which while introduced in Chapter
\ref{chp:diffraction}, requires a little more explanation.

When light enters a medium (anything besides a complete vacuum) it
encounters resistance from that medium that slows the light wave
down. The energy of the light wave (made up photons remember) must
stay the same, as energy just can't disappear. Using Equation
\ref{equ:hydrogen:photonEnergy}, we see that if the energy stays the
same, the frequency must also stay the same. This means that if the
velocity of light within the medium decreases then the wavelength of
the light must increase. When the wavelength of light is changed while
the frequency remains constant, the light is diffracted, or bent. More
energetic light is bent less than low energy light, and so the
combined colors in white light are separated. This is the principal by
which the prism works and is described by \term{Snell's law}, derived
in Chapter \ref{chp:optics}, and given by Equation
\ref{equ:optics:snell} which is given again here.
\begin{equation}
\frac{\sin\theta_1}{\sin\theta_2} = \frac{v_1}{v_2}
\end{equation}
Remeber $\theta_1$ and $\theta_2$ are the \term{incidence angle}, the
angle at which the light is incident to the normal of the medium, and
\term{refraction angle}, the angle at which the light is refracted or
bent. The values $v_1$ and $v_2$ are the velocities of light in medium
$1$ and medium $2$.

The same idea is used in \term{diffraction gratings}, but now the {\bf
  diffraction} is accomplished through \term{interference patterns}. A
diffraction grating is made up of many small parallel lines through
which light passes. The various wavelengths interfere with each other
and create bright bands of colors. This is why when looking at a CD,
bright colors are observed. The theory behind diffraction gratings
can take a fair amount of math and waves knowledge to understand, so
an in depth presentation is not given here, but left to the reader to
explore. An excellent book that covers this topic is
\href{http://www.amazon.com/Vibrations-Waves-Mit-Introductory-Physics/dp/0748744479/ref=sr_1_1?ie=UTF8&s=books&qid=1258036503&sr=8-1}{\it
  Vibrations and Waves} by A.P. French. The important relation to
remember for the diffraction grating is,
\begin{equation}
  m\lambda = d\sin\theta_m
  \label{equ:hydrogen:diffractionGrating}
\end{equation}
where $\lambda$ is the wavelength of the light being observed in
meters, $d$ the number of slits per meter, $m$ some integer greater
than zero, and $\theta_m$ the angle at which the light is observed
with respect to the normal of the diffraction grating. What this
equation tells us is that for a specific $\lambda$ we expect to see a
bright bright band of this color at regular intervals in
$\theta$. Subsequently, by observing a color band at some angle
$\theta$ we can determine $\lambda$.

\begin{figure}
  \begin{center}
    \executeiffilenewer{Code/Hydrogen/intensityBalmer.m}{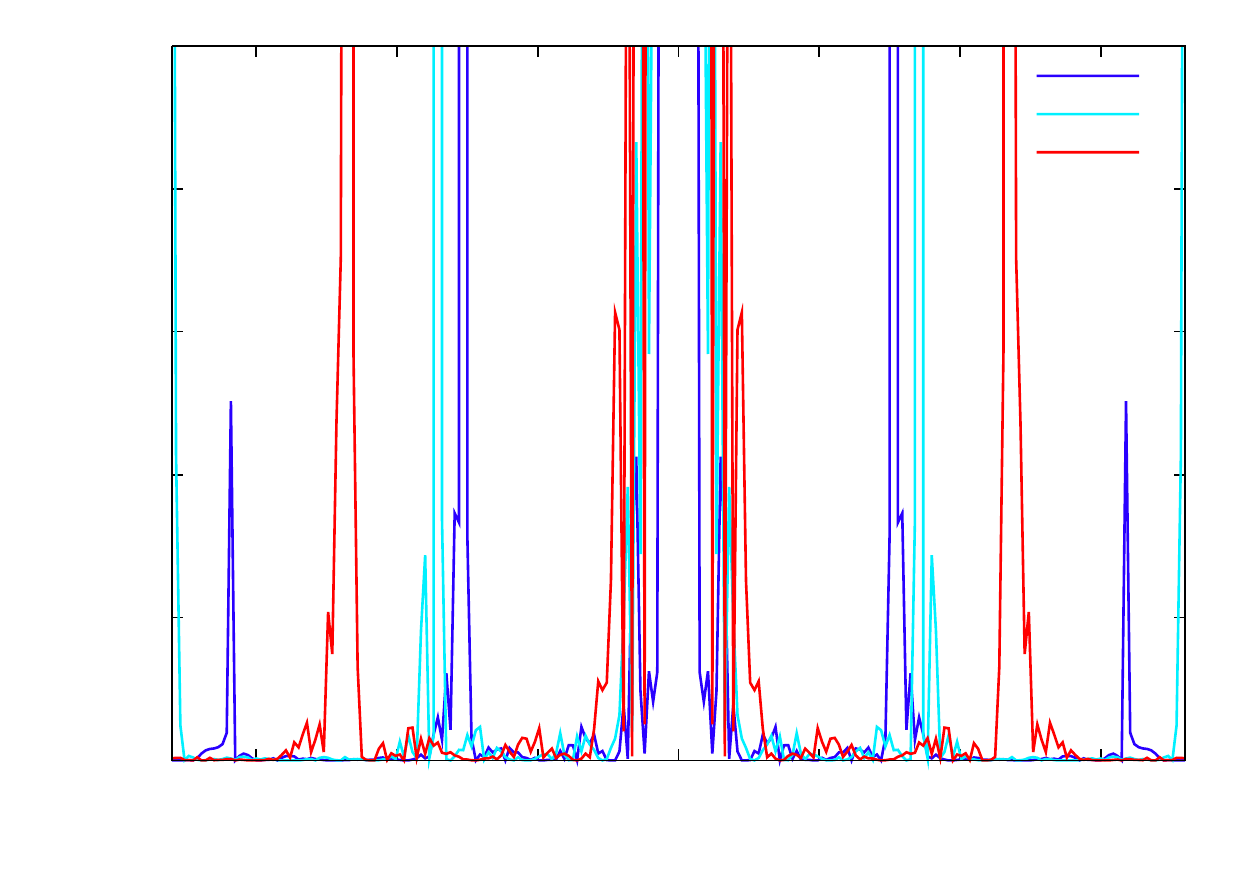}
  {octave --silent --eval "addpath([pwd(),'/Code']); 
    addpath([pwd(),'/Code/Hydrogen']); intensityBalmer(1);"} 
  \setlength{\unitlength}{\columnwidth*\real{0.00013}}
  \input{Figures/Hydrogen/intensityBalmerLabel.tex}
    \caption{The intensity profile for the first three transitions of
      the Balmer series with a diffraction grating of
      $d=0.001/600\mbox{ m/lines}$, $w\approx d$, and $N \approx
      1800$. This plot was calculated using Equation
      \ref{equ:hydrogen:intensity}.\label{fig:hydrogen:intensity}}
  \end{center}
\end{figure}

Figure \ref{fig:hydrogen:intensity} plots the intensity of the first three
Balmer lines, given in Figure \ref{fig:hydrogen:balmerSeries} as they pass
through a diffraction grating. The grating used in this plot has $600$
lines per millimeter and the grating is $3$ cm wide.\footnote{These
  numbers were not chosen at random; they closely model the
  diffraction gratings that are used in this lab.} Every large peak in
the plot corresponds to a specific value for $m$, $\theta$, and
$\lambda$, in Equation \ref{equ:hydrogen:diffractionGrating}. From the plot we
see that if we were using this diffraction grating while observing a
hydrogen lamp, we would expect to see bright bands of color at angles
of $\approx 15^\circ$, $17^\circ$, and $24^\circ$. All of these bands
correspond to $m=1$ in Equation
\ref{equ:hydrogen:diffractionGrating}.\footnote{The equation used to plot
  Figure \ref{fig:hydrogen:intensity} is a little more complicated than
  Equation \ref{equ:hydrogen:diffractionGrating} as now we are looking at the
  intensity of the light. The equation used here is,
\begin{equation}
  I(\theta) = I_0\mathrm{sinc}^2\left(\frac{\pi
      w}{\lambda}\sin(\theta)\right) \left[\frac{\sin\left(\frac{\pi N
          d}{\lambda}\sin\theta\right)}{\sin\left(\frac{\pi
          d}{\lambda}\sin\theta\right)}\right]^2
  \label{equ:hydrogen:intensity}
\end{equation}
where $w$ is the width of the slits, $d$ the separation of the slits,
$N$ the number of slits, and $I_0$ the initial intensity of the light.
}

\section{Experiment}

The goals the experiment associated with this chapter are to observe
the first three electron transitions of the Balmer series, calculate
their wavelengths, verify the Bohr model for the hydrogen atom, and
calculate the Rydberg constant.

Those are quite a few goals to accomplish, and some seem rather
complicated, but the actual experimental procedure is not all that
time consuming. It is the understanding of the theory, explained above
that can be tricky. This lab actually only consists of making three
experimental measurements.

A hydrogen lamp is placed in front of a telescope/collimating
device. All the telescope/collimator does is create a bright band of
light from the hydrogen lamp that we can observe. The light travels
through the telescope to a diffraction grating, and passes through the
diffraction grating. On the other side another telescope, attached to
a scale that reads out angle in degrees, points at the diffraction
grating.

The idea is to look through this telescope at the diffraction grating
and hydrogen lamp, and observe bands of light similar in color to the
first three transitions of Figure \ref{fig:hydrogen:balmerSeries}. The spacing
in $\theta$ and intensity of the light band intervals will be similar
to that of Figure \ref{fig:hydrogen:intensity}. Notice that after a very bright
band there oftentimes will be a few less intense bands of the same, or
similar color. Make sure not to record the angle for these bands as
these are fringe effects.

For each observed band (three for this experiment) an angle $\theta$
is recorded. This $\theta$ is plugged back into Equation
\ref{equ:hydrogen:diffractionGrating} with $m=1$ and $\lambda$ is found. The
values for $1/\lambda^2$ versus $1/n_i^2$ can then be plotted against
each other and a value for the Rydberg constant, given in Equation
\ref{equ:hydrogen:rydbergConstant}, determined from the slope of the
plot. Additionally, the final energy level for the series, $n_f$, can
be determined using the Rydberg constant determined from the slope of
the plot, and the intercept of the plot.

\graphicspath{{Figures/Fluorescence/}}

\chapter{Fluorescence}\label{chp:fluorescence}

The \term{Bohr model}, which was derived from a combination of
classical and quantum theory in Chapter \ref{chp:hydrogen}, seems a
little too good to be true. That's because it is, as was alluded to
with the mention of the \term{atomic orbital model} in the previous
chapter. The Bohr model consists of classical theory with the
introduction of just a single quantization, the \term{principle
  quantum number}, $n$. We know that the model works well for the
hydrogen atom, but it is by an ironic twist of physics that the Bohr
model is right for all the wrong reasons.\footnote{Yes, it really
  does mess with physicists heads when the theory is wrong, but agrees
  with experiment. The real question then is how do you know the
  theory is wrong? The answer is the theory breaks down when the
  experiment is performed in more detail.}

So why exactly is the Bohr model wrong? The first problem is that the
electrons within an atom can move at velocities near the speed of
light, and so Einstein's theory of \term{relativity} must be used
instead. The second problem is that the type of quantum mechanics used
is just a rudimentary form of a much larger and intricate
theory. Remembering back to Chapter \ref{chp:newton}, relativity
describes objects moving very quickly, and \term{quantum mechanics}
describes very small objects. For the hydrogen atom the electrons are
both small and fast, so the must be described using \term{relativistic
  quantum theory}.

In experiment, the Bohr model was found to disagree with a variety of
results. Perhaps one of the more pronounced phenomena is the
\term{Zeeman effect}, which occurs when an atom is placed under a
strong magnetic field. When this occurs, the atom begins to emit many
more wavelengths of light than it should and so we know the atom has
more energy levels than the Bohr model predicts.\footnote{If you would
  like to read more, look at {\it Introduction to Quantum Mechanics}
  by David Griffiths, or check out a lab write-up by Philip Ilten on
  the Zeeman effect at
  \url{http://severian.mit.edu/philten/physics/zeeman.pdf}.} Another
important phenomena is \term{hyperfine splitting} which allows photons
to radiate from hydrogen atoms with a wavelength of $21$ cm. This
wavelength is within the radio range of the electromagnetic spectrum,
and has led to incredible advances in radio astronomy.\footnote{A lab
  report for observing the galactic plane can be found at
  \url{http://severian.mit.edu/philten/physics/radio.pdf}.}

Another phenomena that combines many of the ideas above is
\term{fluorescence}. Not only is fluorescence somewhat challenging to
spell, it is quite challenging to understand. In reality, theory can
describe fluorescence up to a point, but there are so many facets to
the phenomena that there is no complete theory that can accurately
describe the fluorescence of any arbitrary atom. So what exactly is
fluorescence? Fluorescence is when an atom or molecule absorbs a
photon, for example a blue photon, and then after a short time period
releases another photon with less energy, for this example let us say
red.\footnote{This definition of fluorescence is very qualitative, but
  that is only because fluorescence can be somewhat difficult to
  define. There is another type of fluorescence called
  \term{anti-Stokes fluorescence} where the emitted photon has a
  larger energy than the absorbed photon.}

So how exactly does the process of fluorescence occur? The answer is
rather complicated, and first we need some more theory. To begin, we
will look at a simple one-dimensional quantum system in the
\term{simple harmonic oscillator}. This bit of the chapter is not
necessary to understand fluorescence (nor are the next two sections)
so don't panic if it seems a little complicated. Next we will apply
this theory (loosely) to the hydrogen atom, and then see how energy
splittings of the Bohr energy levels can occur. Finally, we will
explore how electrons in molecules can undergo non-radiative
transitions, which can lead to fluorescence.

\section{Simple Harmonic Oscillator}

Very few quantum mechanical systems can be explicitly solved, but one
that can is the case of \term{simple harmonic oscillation} which we
explored classically in Chapter \ref{chp:pendulum}. If you are a
little rusty on the classical derivation, go take a look at Chapter
\ref{chp:pendulum} briefly, at is it will help with the following
theory (although not entirely necessary). The force on a simple
harmonic oscillator is given by \term{Hooke's law},
\begin{equation}
  F = -kx
\end{equation}
where $F$ is force, $k$ is the \term{spring constant}, and $x$ is
position. Both $x$ and $F$ are vector quantities, but because the
problem is one-dimensional, we can indicate direction with just a
positive (pointing towards the right) or negative (pointing towards
the left) sign.

The work done by a force on an object is the integral of the force
over the distance which it was exerted or,
\begin{equation}
  W = \int F\,dx
\end{equation}
where $W$ is work. The change in potential energy for an object after
a force has been applied to it over a certain distance is,
\begin{equation}
\Delta V(x) = -\int F\,dx
\end{equation}
where $V(x)$ is the potential energy of the object at position $x$. If
we move the harmonic oscillator a distance $x$ from equilibrium the
potential energy of the object is,
\begin{equation}
  V(x) = \int_{x_0}^x F\,dx = \frac{1}{2}mkx^2 = \frac{1}{2}mw^2x^2
  \label{equ:fluorescence:potential}
\end{equation}
if we assume the object is at equilibrium for $x_0 = 0$. In the
final step we have used the relation for $k$ in terms of mass of the
object, $m$, and the \term{angular frequency}, $\omega$, given by
Equation \ref{equ:pendulum:springConstant}.

The reason the harmonic oscillator problem is so important is because
of the potential above. This potential is quadratic in $x$, which
means that if an arbitrary potential, $V(x)$, is expanded about a
minimum using a \term{Taylor series} expansion, the potential can be
described very well by simple harmonic motion.\footnote{The Taylor
  expansion of $V(x)$ about $x_0$ is given by $V(x) =
  V(x_0)+V'(x_0)(x-x_0)+V''(x_0)(x-x_0)+\cdots$. But if $x_0$ is at a
  minimum, then $V'(x_0) = 0$ and so the first non-constant term in
  the expansion is given by the quadratic term.} In other words,
simple harmonic motion can be used to approximate much more complex
systems for small oscillations of the system.

So where does quantum mechanics come into all of this? In quantum
mechanics, all particles and objects are represented by waves or
\term{wave packets}. We already discussed this briefly before in
Chapter \ref{chp:diffraction} with the introduction of the \term{de
  Broglie wavelength}. In quantum mechanics, all particles must
satisfy \term{Shr\"odinger's equation}\footnote{This is the
  time-independent Shr\"odinger's equation, the time-dependent
  equation is a bit more complicated, but we don't need to worry about
  that for the example of simple harmonic motion.},
\begin{equation}
  H\psi = E\psi
  \label{equ:fluorescence:shroedinger}
\end{equation}
where $\psi$ is the \term{wavefunction} for a particle, and $E$ is the
energy of the particle. By wavefunction what we mean is the equation
that describes the shape of the particle in space. Oftentimes in
quantum mechanics a Gaussian wavefunction is used, similar the shape
of Figure \ref{fig:uncertainty:normal} in Chapter
\ref{chp:uncertainty}. The square of a wavefunction gives us the
\term{probability density function} of finding the particle (think
back to Chapter \ref{chp:uncertainty}). What this wavefunction tells
us is that the particle is not just in one place, but that it has the
probability of being in a variety of different places. Until we
actually measure where the particle is, it could be anywhere described
by its wavefunction.

The term $H$ in Equation \ref{equ:fluorescence:shroedinger} is called
the \term{Hamiltonian} and is just the potential energy of the
particle less its kinetic energy. Mathematically, the Hamiltonian is
given by,
\begin{equation}
  H = V(x) - \frac{\hbar^2}{2m}\nabla^2 = V(x) -
  \frac{\hbar^2}{2m}\frac{\partial^2}{\partial x^2}
  \label{equ:fluorescence:hamiltonian}
\end{equation}
where the final step is the Hamiltonian for a one-dimensional system
like the simple harmonic oscillator. The term $\hbar$ is the
\term{reduced Planck's constant}, and is just $h/2\pi$. Mathematically
the Hamiltonian is a special object called an \term{operator}. An
operator is a symbol that denotes performing certain steps on whatever
comes after the operator. Operators are nothing new, as a matter of
fact, $\frac{d}{dx}$, which just denotes taking the derivative of a
function, is the \term{differential operator}. What is important to
remember about operators is that they are not necessarily (and usually
are not) \term{commutative}. The order of operators is important and
cannot be switched around.

We can write the Hamiltonian operator in terms of two new operators,
the \term{momentum operator}, $p$, and the position operator, $x$. The
momentum operator is,
\begin{equation}
p = -i\hbar\frac{\partial}{\partial x}
\end{equation}
and the position operator is just $x = x$. Substituting the momentum
operator into the Hamiltonian of Equation
\ref{equ:fluorescence:hamiltonian} yields,
\begin{equation}
  H = V(x)+\frac{p^2}{2m}
\end{equation}
where we have used that $p^2 = -\hbar^2\partial^2/\partial x^2$.

Now we can insert the potential energy for simple harmonic
oscillation, Equation \ref{equ:fluorescence:potential}, into the
Hamiltonian above.
\begin{equation}
  H = \frac{1}{2}m\omega^2 x^2+\frac{p^2}{2m} = \frac{(m\omega x)^2+p^2}{2m}
  \label{equ:fluorescence:hamiltonian1}
\end{equation}
The numerator of this equation looks very similar to a quadratic
equation, $ax^2+bx+c$, without the final two terms. This means that we
should be able to factor it into two new operators. At this point you
might ask why on earth would we want to do that? The answer is read
on, hopefully this step will make more sense after a few more
paragraphs.

We introduce two new operators the \term{creation operator}, $a$, and
\term{annihilation operator}, $a^\dagger$ which are given by,
\begin{equation}
  a = \frac{-ip+m\omega x}{\sqrt{2\hbar m\omega}} ~~~ a^\dagger =
  \frac{ip+m\omega x}{\sqrt{2\hbar m\omega}}
  \label{equ:fluorescence:ladder}
\end{equation}
in terms of $p$ and $x$. It turns out that we can factor the
Hamiltonian of Equation \ref{equ:fluorescence:hamiltonian1} into,
\begin{equation}
  H = \hbar\omega\left(a^\dagger a -\frac{1}{2}\right) 
  \label{equ:fluorescence:hamiltonian2}
\end{equation}
where a few algebraic steps have been left out.\footnote{Check the
  result though if in doubt! Just use Equation
  \ref{equ:fluorescence:ladder}, and see if you can recover
  \ref{equ:fluorescence:hamiltonian2}.} If we plug this back into
Shr\"odinger's equation we get,
\begin{equation}
  \hbar\omega\left(a^\dagger a -\frac{1}{2}\right) \psi = E\psi
  \label{equ:fluorescence:sho}
\end{equation}
for the simple harmonic oscillator.

Now, hopefully this last step will explain why we went to all the
trouble of factoring the Hamiltonian. If we operate on the
wavefunction for our particle, $\psi$ with the creation operator, $a$, we get,
\begin{equation}
  H a \psi = (E+\hbar\omega) a \psi
  \label{equ:fluorescence:creation}
\end{equation}
which means that the wavefunction $a\psi$ also fulfills Shr\"odinger's
equation and has an energy of $E+\hbar\omega$!\footnote{Again, quite a
  few mathematical steps have been left out, but feel free to verify
  the result. Remember that the order of operators is important!} This
means that if we find the lowest energy wavefunction for Equation
\ref{equ:fluorescence:sho}, $\psi_0$, with energy $E_0$, we can find
every possible wavefunction that will fulfill Equation
\ref{equ:fluorescence:sho} by just operating on the ground wavefunction
with the creation operator, $a\psi^0$, however many times is
necessary. This is why $a$ is called the creation operator, it creates
the next wavefunction with a larger energy than the previous
wavefunction.

The annihilation operator does the exact opposite; it finds the next
wavefunction with a lower energy, and if applied to the grounds state,
yields zero, $a^\dagger \psi_0 = 0$. It turns out that the energy of
the lowest ground state is $E_0 = \hbar\omega/2$, and so from this we
can find all the possible energies of a simple harmonic oscillator
using Equation \ref{equ:fluorescence:creation}.
\begin{equation}
  E_1 = (E_0+\hbar\omega) = \frac{3\hbar\omega}{2}, ~~~   E_2 =
  (E_1+\hbar\omega) = \frac{5\hbar\omega}{2}, ~~~ \cdots 
\end{equation}
We can generalize this into,
\begin{equation}
  E_n = \left(n+\frac{1}{2}\right)\hbar\omega
  \label{equ:fluorescence:energy}
\end{equation}
where $E_n$ is the energy of the $n^\mathrm{th}$ wavefunction.

It may have been somewhat difficult to keep track of exactly what was
going on with all the math above but the general idea is relatively
simple. We began with a system, simple harmonic oscillation, and found
the potential energy for the system. We applied Shr\"odinger's
equation to the system, and after a fair bit of mathematical
rigamarole, determined that the energy of the simple harmonic
oscillator must be quantized, and is given by Equation
\ref{equ:fluorescence:energy}. This is the general idea of quantum
mechanics. Find the potential and use Shr\"odinger's equation to
determine the energy levels and wavefunctions for the system. The
method used above is a nice algebraic one, but oftentimes a much more
mathematically intensive \term{power series method} must be used.

\section{Hydrogen Orbitals}

So the previous section was rather complicated, but hopefully it gives
a general idea of how quantum mechanics is approached. As Richard
Feynman, a famous physicist, once said, ``I think I can safely say
that nobody understands quantum mechanics.'' But how does all of this
apply to fluorescence? Let us again turn to the hydrogen atom, like we
did with the Bohr model, but now use a fully quantum mechanical
theory. Don't worry, we aren't going to solve Shr\"odinger's equation
for a three dimensional potential, it's available in many quantum
mechanics text books. Instead, we are going to look at the end result.

The potential for a hydrogen atom is given by the electric potential,
just as it was for the Bohr model in Equation
\ref{equ:hydrogen:electricPotential}.
\begin{equation}
  V(r) = -\frac{q_1q_2}{4\pi\epsilon_0r} =
  \frac{-e^2}{4\pi\epsilon_0r}
  \label{equ:fluorescence:hydrogenPotential}
\end{equation}
Using the potential above, it is possible to
work out (although certainly not in the scope of this chapter!) that
the wavefunctions for the hydrogen atom are given by,
\begin{equation}
  \begin{aligned}
    \psi_{n,l,m_l}(\theta,\phi,r) =&
    \sqrt{\left(\frac{2}{nr_1}\right)^3
      \frac{(n-l-1)!}{2n\left(\left(n-l\right)!\right)^3}}
    e^{\frac{-r}{nr_1}} \left(\frac{2r}{nr_1}\right)^l\\
    &\times \left(L_{n-l-1}^{2l+1}\left(\frac{2r}{nr_1}\right)\right)
    Y_l^{m_l}(\theta,\phi) \\
  \label{equ:fluorescence:hydrogen}
\end{aligned}
\end{equation}
where there are quite a few letters that need
explaining.\footnote{This form is taken from {\it Introduction to
    Quantum Mechanics} by David Griffiths.} Before that however, it is
important to realize that it is not necessary to understand Equation
\ref{equ:fluorescence:hydrogen}, but rather to understand its
implications. The pattern of electrons around a hydrogen atom are not
just simply in spherical orbits like Bohr's model predicted, but
rather, are described by complicated wavefunctions!

So, back to understanding the symbols of Equation
\ref{equ:fluorescence:hydrogen}. First, $\theta$, $\phi$, and $r$, are
just the standard variables for a spherical coordinate system. Next,
there is no longer just one quantum number, $n$, but three! These
quantum numbers are $n$, $l$, and $m_l$ and will be explained more in
the next section. For now, just remember that $0 < l < n$, $-l \leq
m_l \leq l$, and that all three quantum numbers must be
integers. There is only one constant in the equation above and that is
$r_1$ or the Bohr radius, as introduced in Chapter
\ref{chp:hydrogen}. Finally, there are the letters $L$ and $Y$ which
stand for special functions called the \term{Laguerre polynomials} and
\term{Legendre functions}. In two dimensions sine and cosine waves can
be used to describe any function through what is known as a
\term{Fourier decomposition}. These two types of functions serve a
very similar purpose, but now for three dimensions.

\begin{figure}
    \vspace{-1cm}
    \hspace{-0.1\columnwidth}\subfigure[$|\psi_{1,0,0}|^2$]
    {
      \footnotesize
      \executeiffilenewer{Code/Fluorescence/orbitals100.m}{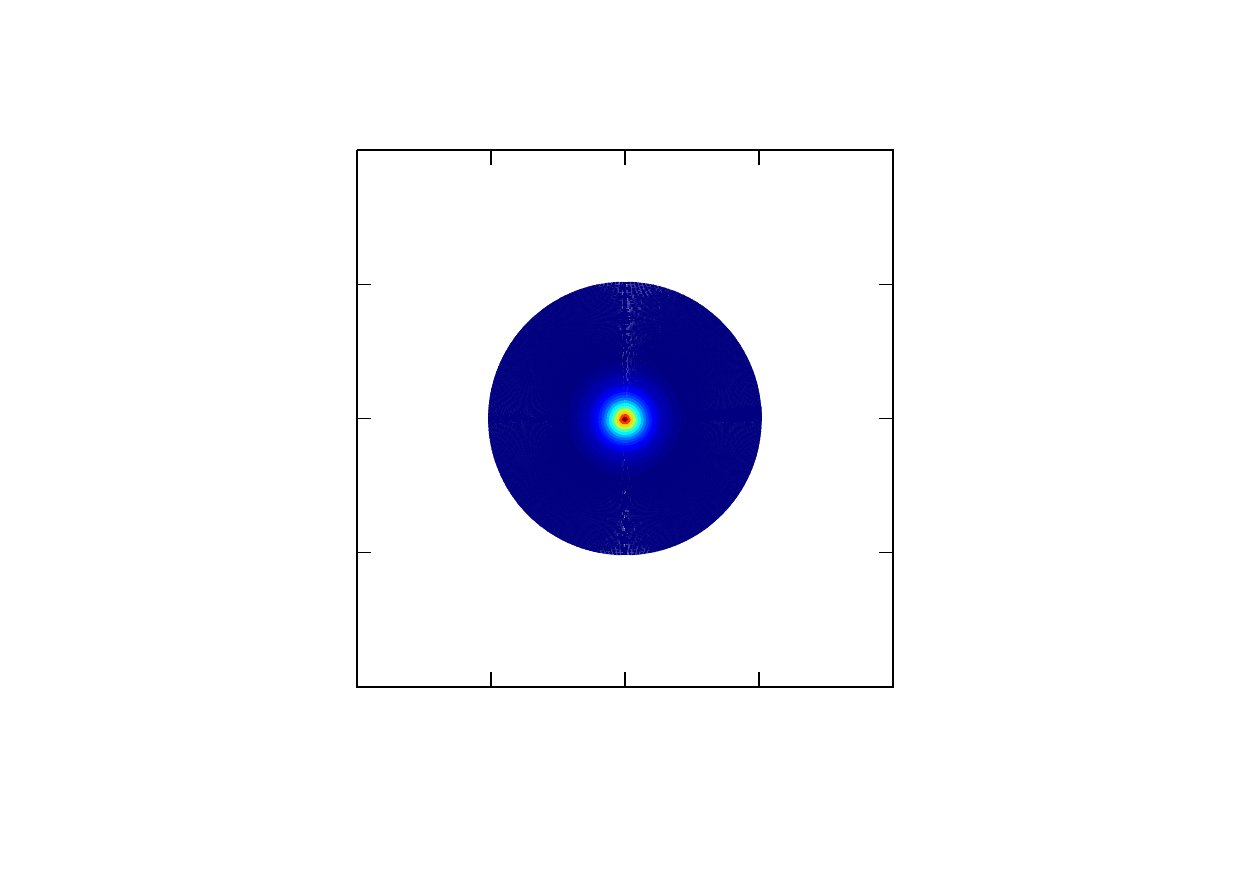}
  {octave --silent --eval "addpath([pwd(),'/Code']); 
    addpath([pwd(),'/Code/Fluorescence']); orbitals100(1);"} 
  \setlength{\unitlength}{0.7\columnwidth*\real{0.00013}}
  \input{Figures/Fluorescence/orbitals100Label.tex}
      \label{fig:fluorescence:orbitals100}
    }
    \nolinebreak
    \hspace{-0.2\columnwidth}\subfigure[$|\psi_{2,0,0}|^2$]
    {
      \footnotesize
      \executeiffilenewer{Code/Fluorescence/orbitals200.m}{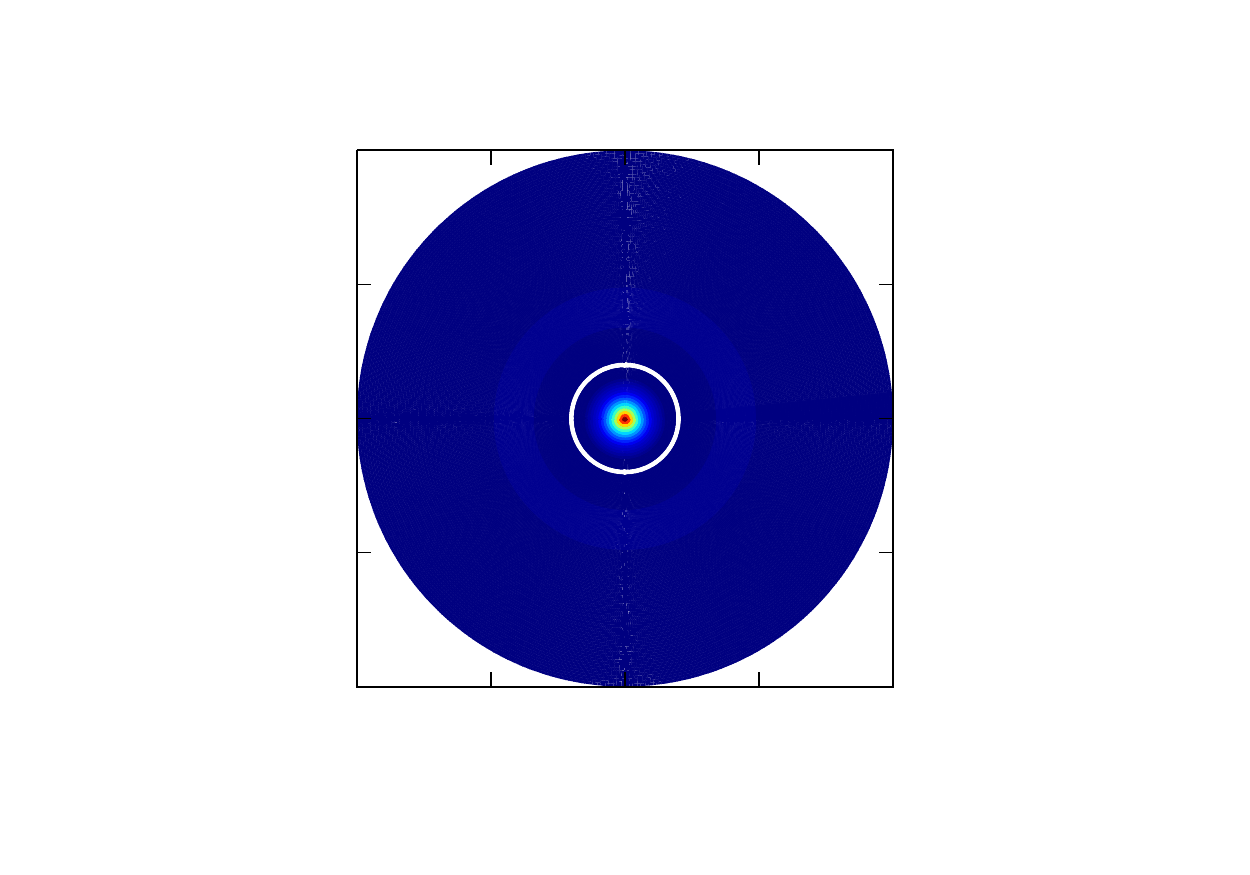}
  {octave --silent --eval "addpath([pwd(),'/Code']); 
    addpath([pwd(),'/Code/Fluorescence']); orbitals200(1);"} 
  \setlength{\unitlength}{0.7\columnwidth*\real{0.00013}}
  \input{Figures/Fluorescence/orbitals200Label.tex}
      \label{fig:fluorescence:orbitals200}
    }
    
    \vspace{-1cm}
    \hspace{-0.1\columnwidth}\subfigure[$|\psi_{2,1,0}|^2$]
    {
      \footnotesize
      \executeiffilenewer{Code/Fluorescence/orbitals210.m}{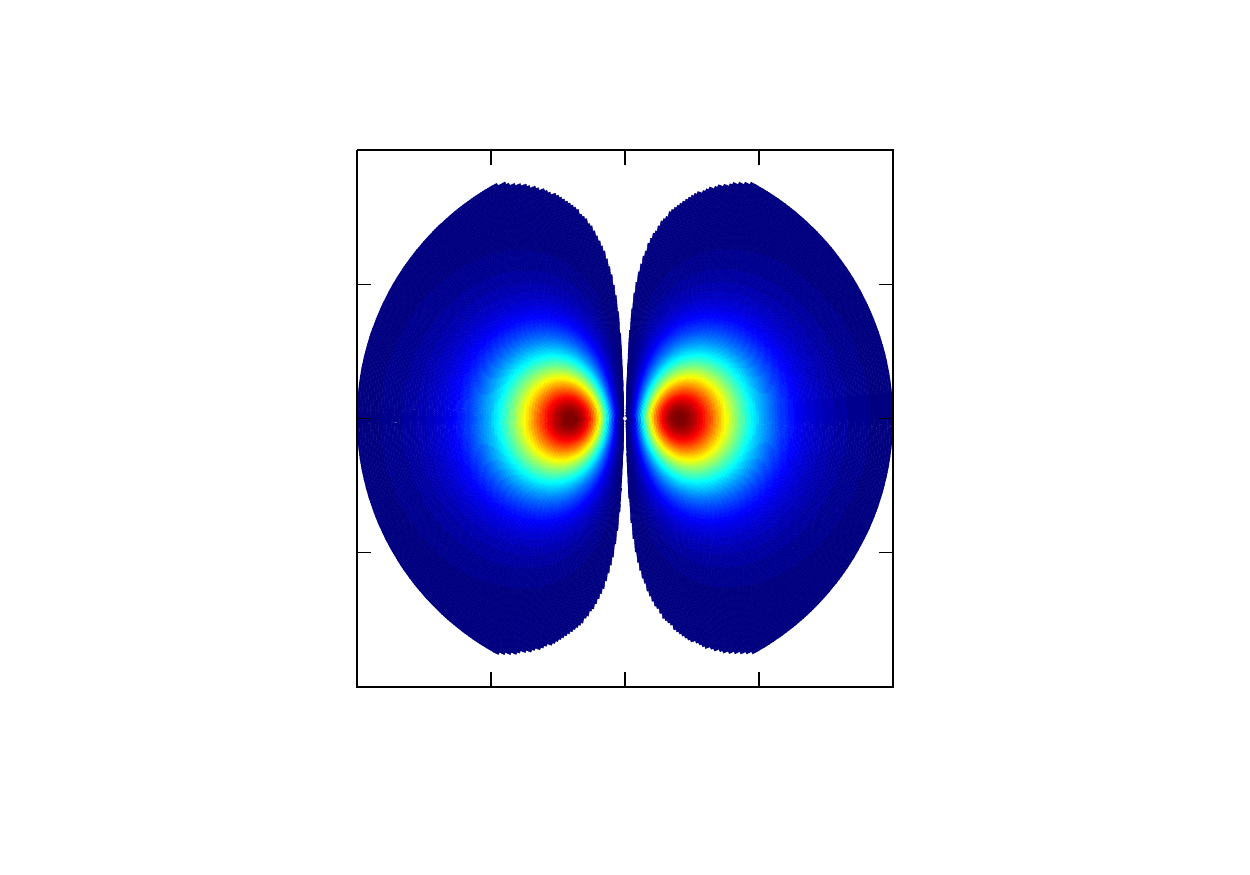}
  {octave --silent --eval "addpath([pwd(),'/Code']); 
    addpath([pwd(),'/Code/Fluorescence']); orbitals210(1);"} 
  \setlength{\unitlength}{0.7\columnwidth*\real{0.00013}}
  \input{Figures/Fluorescence/orbitals210Label.tex}
      \label{fig:fluorescence:orbitals210}
    }
    \nolinebreak
    \hspace{-0.2\columnwidth}\subfigure[$|\psi_{3,0,0}|^2$]
    {
      \footnotesize
      \executeiffilenewer{Code/Fluorescence/orbitals300.m}{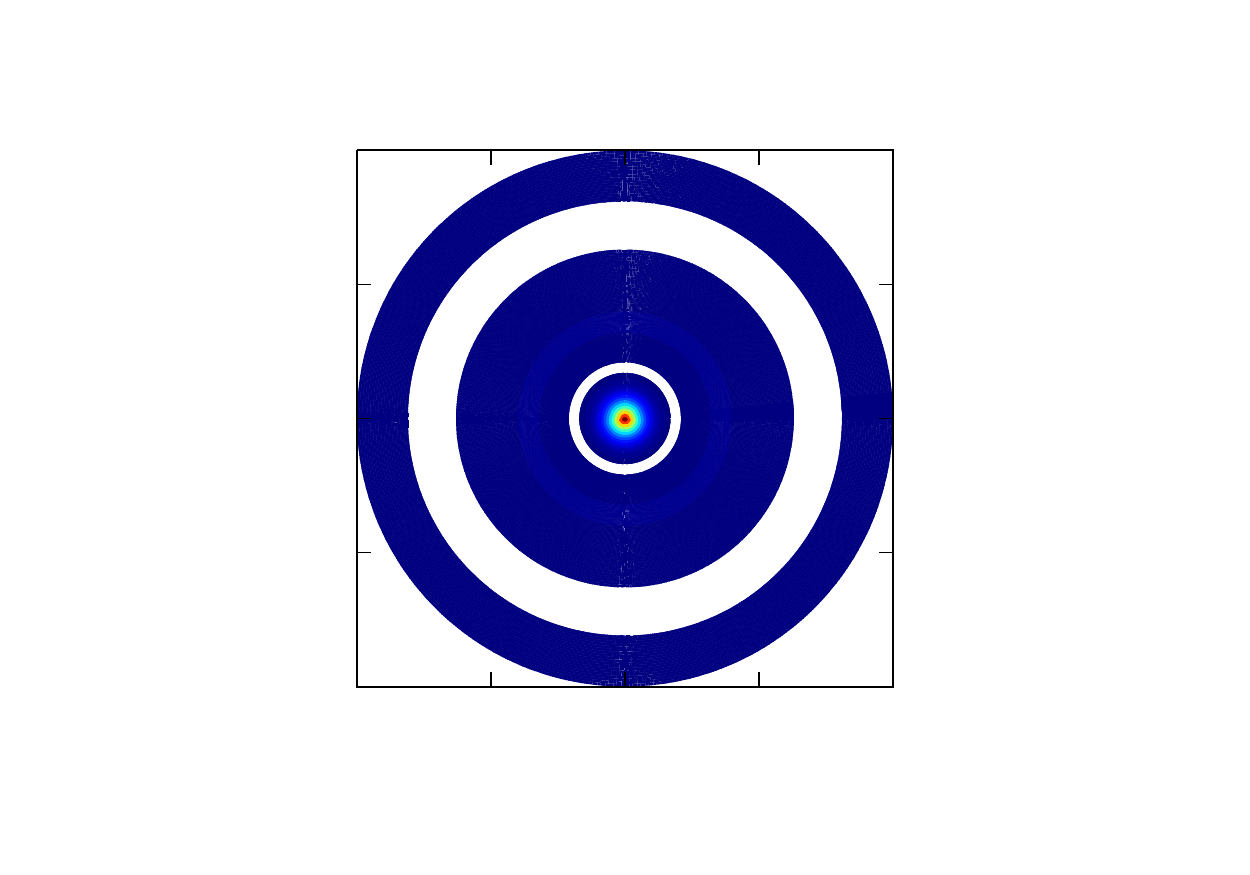}
  {octave --silent --eval "addpath([pwd(),'/Code']); 
    addpath([pwd(),'/Code/Fluorescence']); orbitals300(1);"} 
  \setlength{\unitlength}{0.7\columnwidth*\real{0.00013}}
  \input{Figures/Fluorescence/orbitals300Label.tex}
      \label{fig:fluorescence:orbitals300}
    }
    
    \vspace{-1cm}
    \hspace{-0.1\columnwidth}\subfigure[$|\psi_{3,1,0}|^2$]
    {
      \footnotesize
      \executeiffilenewer{Code/Fluorescence/orbitals310.m}{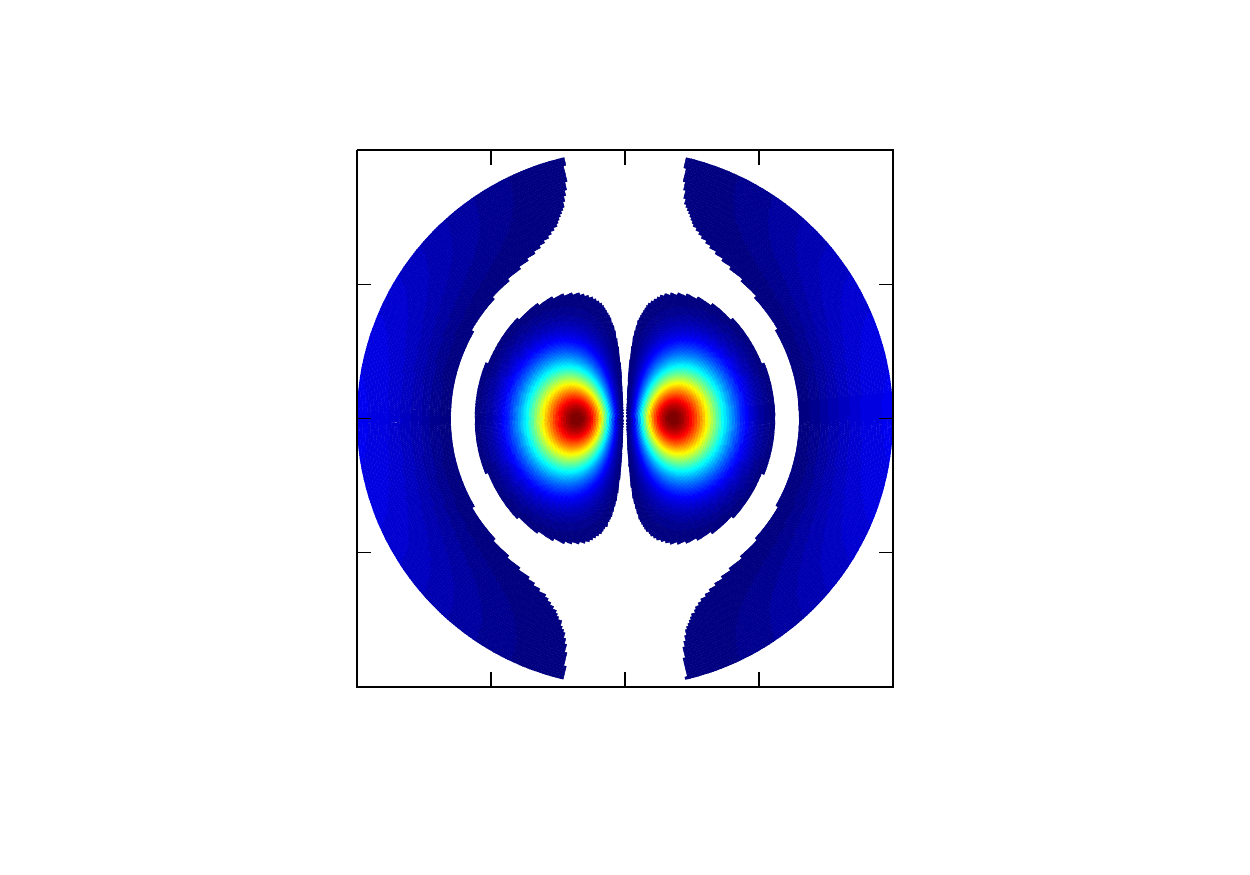}
  {octave --silent --eval "addpath([pwd(),'/Code']); 
    addpath([pwd(),'/Code/Fluorescence']); orbitals310(1);"} 
  \setlength{\unitlength}{0.7\columnwidth*\real{0.00013}}
  \input{Figures/Fluorescence/orbitals310Label.tex}
      \label{fig:fluorescence:orbitals310}
    }
    \nolinebreak
    \hspace{-0.2\columnwidth}\subfigure[$|\psi_{3,2,0}|^2$]
    {
      \footnotesize
      \executeiffilenewer{Code/Fluorescence/orbitals320.m}{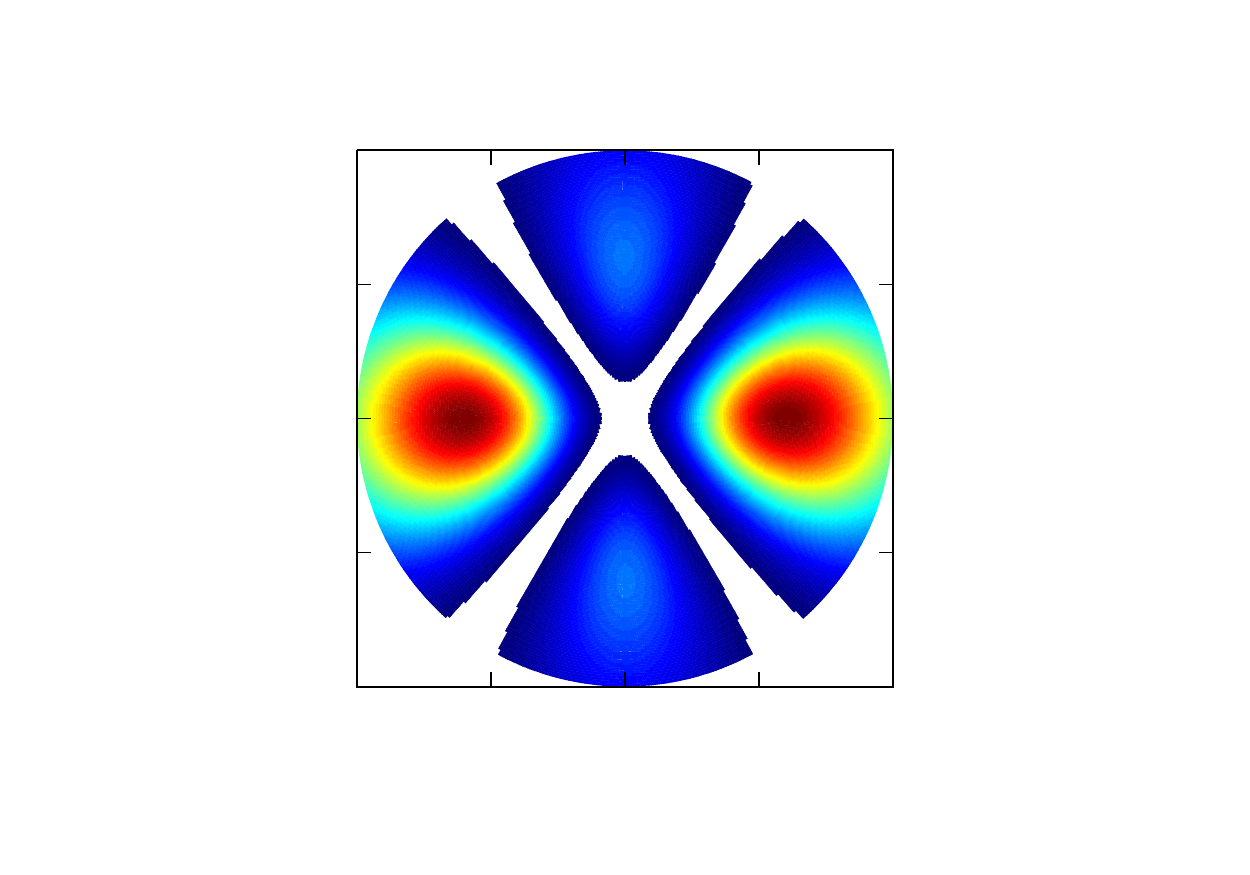}
  {octave --silent --eval "addpath([pwd(),'/Code']); 
    addpath([pwd(),'/Code/Fluorescence']); orbitals320(1);"} 
  \setlength{\unitlength}{0.7\columnwidth*\real{0.00013}}
  \input{Figures/Fluorescence/orbitals320Label.tex}
      \label{fig:fluorescence:orbitals320}
    }
    \caption{Profiles of the probability density clouds describing the
      position of electrons in a hydrogen atom, given the quantum
      numbers $n$, $l$, and $m_l$.\label{fig:fluorescence:orbitals}}
\end{figure}

The square of Equation \ref{equ:fluorescence:hydrogen}, as mentioned
earlier for wavefunctions in general, gives the probability density
functions for the wavefunctions of the hydrogen atom. What this means
is that if we square Equation \ref{equ:fluorescence:hydrogen} for some
$n$, $l$, and $m_l$ and integrate over a volume in the spherical
coordinate system, we will know the probability of finding the
electron within that volume. Visualizing the probability density
functions for the hydrogen atom can be quite challenging because we
are trying to represent the probability for every point in three
dimensional space about the hydrogen atom. The problem of visualizing
the probability density functions for the hydrogen atom is very much
like trying to determine the inside of a fruit cake, without cutting
it. So perhaps the simplest (and most common) solution to visualizing
the probability density functions of the hydrogen atom is to just cut
the cake and look at a single slice.

Figure \ref{fig:fluorescence:orbitals} does just that; it takes a
slice of the probability density functions, $|\psi_{n,l,m_l}|^2$, of
the hydrogen atom for $n=1,2,3$, all possible $l$, and $m_l = 0$. The
results certainly are interesting! In Figure
\ref{fig:fluorescence:orbitals}, the center of each plot corresponds
to the nucleus of the hydrogen atom. Both the $x$ and $y$-axis give
the distance from the nucleus in units of Bohr radii, $r_1$. The color
of the plot corresponds to the value of the probability density
function, red corresponds to a high probability, while blue
corresponds to a low probability.

For the lowest energy level of the hydrogen atom, $\psi_{1,0,0}$, we
see that the electron will be within a sphere of $5$ Bohr radii from
the center of the atom with a nearly $100\%$ probability. This
actually is not very different from the Bohr model, but now instead of
a sharply defined orbit, the electron can reside anywhere within the
blue cloud of Figure \ref{fig:fluorescence:orbitals100}. The next
highest energy level of the hydrogen atom, $\psi_{2,0,0}$, also
exhibits a similar behavior, but now it is more probable for the
electron to be found farther from the nucleus of the atom. Thinking
again in terms of the Bohr model, this makes sense, as we would expect
higher energy levels to have larger radii for their orbits. If we now
look at the probability density function for $\psi_{2,1,0}$, Figure
\ref{fig:fluorescence:orbitals210}, we see a shape that is not even
close to looking spherical, a rather radical departure from the Bohr
model. The same applies to the probability density functions for
$\psi_{3,1,0}$ and $\psi_{3,2,0}$, neither of these are remotely
spherically shaped.

Up until this point we have not discussed what the energy for a given
wavefunction $\psi_{n,l,m_l}$ is. Looking at Figures
\ref{fig:fluorescence:orbitals300},
\ref{fig:fluorescence:orbitals310}, we would certainly expect that the
energy levels of these three wavefunctions should be completely
different, as their probability density functions certainly are. This
however, is not the case, and the energy level for a wavefunction
given by Equation \ref{equ:fluorescence:hydrogen} is only dependent
upon the principle quantum number $n$,
\begin{equation}
  E_n = \frac{-13.6\mbox{ eV}}{n^2}
  \label{equ:fluorescence:hydrogenEnergy}
\end{equation}
and by a strange twist of fate is exactly the same as what we derived
earlier using the Bohr model in Equation
\ref{equ:hydrogen:bohrEnergy}. This yields a rather interesting
result, the electrons described by Figure
\ref{fig:fluorescence:orbitals100} have an energy of $-13.6$ eV,
Figures \ref{fig:fluorescence:orbitals200} and
\ref{fig:fluorescence:orbitals210} both have an energy of $-3.4$ eV,
and Figures \ref{fig:fluorescence:orbitals300},
\ref{fig:fluorescence:orbitals310}, and
\ref{fig:fluorescence:orbitals320} all have the same energy of $-1.5$
eV despite looking completely different.

There is one final note to make about the hydrogen atom, and that is
that oftentimes in chemistry and certain types of physics, the
probability density functions of the hydrogen wavefunctions are called
\term{electron orbitals}. This term is somewhat deceptive; the
electrons do not actually orbit, but rather the probability of finding
an electron is described by the orbital. A type of classification
called \term{spectroscopic notation} is oftentimes used to describe
the different types of orbitals, and is useful to know. The notation
is given by $nl$ where $n$ is just the principle quantum number and
$l$ is also the quantum number from previously, but now is assigned a
letter instead of a number: $l = 0\rightarrow\mathrm{s}$, $l =
1\rightarrow\mathrm{p}$, $l = 2\rightarrow\mathrm{d}$, $l =
3\rightarrow\mathrm{f}$.\footnote{The quantum number $m_l$ can also be
  denote as a subscript to the letter representing $l$, but the naming
  convention for this is rather complicated and is generally not
  used.} For $l$ greater than $3$, the letters are assigned
alphabetically. The first four letters are assigned because they stand
for {\bf s}harp, {\bf p}rincipal, {\bf d}iffuse, and {\bf
  f}undemental, which apparently describe the type of spectroscopic
line given by each orbital. Visually, s orbitals are described by a
spherical shape, p orbitals by a barbell shape, d by a barbell shape
with a ring around it, and f by a barbell with a double ring (not
shown in Figure \ref{fig:fluorescence:orbitals}). As an example of
spectroscopic notation, Figure \ref{fig:fluorescence:orbitals100} is a
$1$s orbital.

\section{Energy Splitting and Quantum Numbers}

In the previous section we needed $3$ quantum numbers to describe the
electron wavefunctions for the hydrogen atom, $n$, $l$, and $m_l$. But
what exactly are these numbers? The quantum number $n$ was first
introduced in Chapter \ref{chp:hydrogen} and is the \term{principal
  quantum number} which describes the energy level of the
wavefunction. The explanation for the quantum numbers $l$ and $m_l$
require a bit more explanation and the introduction of two new quantum
numbers $s$ and $m_s$.

The quantum numbers $l$, $m_l$, $s$, and $m_s$ all help describe the
momentum of a particle, in the case of the hydrogen atom, the electron
orbiting the nucleus. Going back to the planetary model of the Bohr
model we can think of the electron as the earth and the nucleus as the
sun. When the earth orbits the sun it has two types of angular
momentum: angular momentum from orbiting the sun, $L$, and angular
momentum from revolving on its axis, $S$. Similarly, electrons have
orbital angular momentum, $L$, and spin angular momentum, denoted by
the letter $S$. We define the spin and orbital angular momentum in terms of,
\begin{subequations}
  \begin{equation}
    S = \hbar\sqrt{s(s+1)}
  \end{equation}
  \begin{equation}
    L = \hbar\sqrt{l(l+1)}
\end{equation}
\label{equ:fluorescence:spin}
\end{subequations}
where $s$ is called the \term{spin quantum number} and $l$ is called
the \term{orbital quantum number}.\footnote{Oftentimes $l$ is called
  the \term{azimuthal quantum number}} Whenever spin is used in reference
to an electron or particle, we are not referring to the spin angular
momentum of the particle, but rather the spin quantum number.

However, there is a bit of a problem; electrons are considered to be
\term{point-like} objects, and subsequently the idea of spin momenta
and angular momenta doesn't quite work the same way as it does for the
earth. Specifically, the spin quantum number of an electron is
$\frac{1}{2}$. Any type of particle which carries a fractional spin,
like the electron, is called a \term{fermion}, while any particle
which carries an integer spin, such as the photon with spin $0$, is
called a \term{boson}. A particle with spin $s$ can have a \term{spin
  projection quantum number}, $m_s$, from $-s$ to $s$ in integer
steps. For example, the electron can have a spin projection quantum
number of $-\frac{1}{2}$ (spin up) or $+\frac{1}{2}$ (spin down). The
photon can only have a spin projection quantum number of $0$. The same
principle applies for $l$ and $m_l$; $m_l$, the \term{orbital
  projection quantum number} can range from $-l$ to $l$ in integer
steps.\footnote{This quantum number is also called the \term{magnetic
    quantum number} because it describes the magnetic interaction
  between the electron and proton of a hydrogen atom.} The reason both
$m_s$ and $m_l$ are called {\it projection} numbers is because they
represent the projection of the orbital or spin momentum of Equation
\ref{equ:fluorescence:spin} onto an arbitrary axis of the particle.

So why on earth do we need all these quantum numbers?! We just need
$n$ to describe the energy levels of the hydrogen atom, $l$ helps
describe the shape of the electron wavefunctions, but $m_l$, $s$, and
$m_s$ all seem to be overkill. This is because unfortunately the
potential energy for an electron in a hydrogen atom is not given by
just Equation \ref{equ:fluorescence:hydrogenPotential}. As a matter of
fact quite a few other factors that are much less significant than the
Coulomb force must be taken into account, which require the use of all
the additional quantum numbers introduced above.

The first problem with Equation
\ref{equ:fluorescence:hydrogenPotential} is that it does not take into
account the relativistic motion of the electron or the effect of the
proton's magnetic field on the electron (this is called
\term{spin-orbital coupling}). Taking these two effects into
consideration splits the Bohr energy levels of Equation
\ref{equ:fluorescence:hydrogenEnergy} into smaller energy levels using
$m_l$ and $m_s$. This splitting of the energy levels is called the
\term{fine structure} of the hydrogen atom. Additionally, the electric
field that the electron experiences from the proton must be quantized,
and this yields yet another energy splitting called the \term{Lamb
  shift}. Finally, the nucleus of the hydrogen atom interacts with the
electric and magnetic fields of the electrons orbiting it, and so a
final correction called \term{hyperfine splitting} must be made.

This certainly seems like quite a few corrections to Equation
\ref{equ:fluorescence:hydrogenPotential}, and that is because it
is. Sometimes it is easier to think of all the corrections to the
energies in terms of energy level splitting. The difference between
Bohr energies is $\approx 10$ eV, while the difference in fine
splitting is $\approx 10^{-4}$ eV, Lamb shift $\approx 10^{-6}$ eV,
and hyperfine splitting $\approx 10^{-6}$ eV. Really it is not
important to remember this at all. What is important to remember is
that the already rather complicated pattern of electron orbitals shown
in Figure \ref{fig:fluorescence:orbitals} are even more
complicated. The hydrogen atom, the simplest atom we can look at, is
not simple at all!

\section{Non-Radiative Transitions}

As we learned in Chapter \ref{chp:hydrogen}, whenever an electron
makes a transition from one Bohr energy level to another, a photon
is emitted with frequency,
\begin{equation}
  E_\gamma = E_i-E_f \Rightarrow f = \frac{E_i-E_f}{h}
\end{equation}
where Equations \ref{equ:hydrogen:energyDifference} and
\ref{equ:hydrogen:photonEnergy} were combined from Chapter
\ref{chp:hydrogen}. Sometimes, however, electrons can transition from
energy state to energy state without radiating a photon. This is
called a \term{non-radiative transition} and can occur through a
variety of mechanisms.

The first mechanism by which a non-radiative transition can occur in
an atom or molecule is through what is known as \term{internal
  conversion}. In all of the theory above, it has always been assumed
that the nucleus of the hydrogen atom is stationary and not moving. In
reality this is not the case (unless the temperature was absolute
zero) and so another layer of complexity must be added to the already
complex energy structure of the hydrogen atom. When an atom vibrates
it can do so at quantized energy levels known as \term{vibrational
  modes}. These energy levels are layered on top of the already
existing energy levels of the hydrogen atom. When internal conversion
occurs an electron releases its energy to the atom through the form of
a vibration rather than a photon. Usually internal conversion occurs
within the same Bohr energy level (same $n$) because of what is known
as the \term{Franck-Condon principle}. This principle states that a
switch between vibrational modes is more likely when the initial
wavefunction of the electron closely matches the final wavefunction of
the electron, which usually occurs within the same Bohr energy level.

Another possibility for non-radiative transitions is through
\term{vibrational relaxation} which can only occur in a group of atoms
or molecules. This is because the entire sample of molecules has
\term{vibrational energy levels} (essentially the temperature of the
sample). When an atom or molecule undergoes vibrational relaxation it
releases some energy through the decay of an electron to a lower
energy orbital which is absorbed by the sample and so the sample
transitions to a new vibrational energy level.

Finally, another non-radiative transition process can occur through
\term{intersystem crossing}. In the hydrogen atom, the total spin for
a group of electrons can be determined by adding spin quantum numbers,
$s$, if their spins are pointing in the same direction, or subtracting
spin quantum numbers if their spins are pointing in opposite
directions. Electrons try to pair off into groups of two electrons,
with spins pointing in opposite directions, by what is known as the
\term{Pauli exclusion principle}, and so if there is an even number of
electrons, the total spin quantum number is usually zero. This means
that $m_s = 0$, and since there is only one possible value for $m_s$,
this state is called a \term{singlet}. If there is an odd number of
electrons, then generally $s = 1$ and so now $m_s = -1,0,1$ which is
called a \term{triplet}.

Oftentimes an atom will transition from a singlet state to a triplet
state (or vice versa), where a small amount of energy is expended on
flipping the spin of an electron. This type of non-radiative
transition is called an intersystem crossing because the atom
transitions from a singlet system to a triplet system and only occurs
in \term{phosphorescence} or \term{delayed fluorescence}. In the case
of phosphorescence, the electron transitions from a ground singlet
state to an excited singlet state through absorption of a photon. Then
it decays through non-radiative transitions to the triplet
state. Finally, it decays back to the ground singlet state through a
radiative decay. For delayed fluorescence, the same process occurs,
but the electron transitions back to the excited singlet state from
the triplet state before emitting a photon. Both of the processes
occur over a longer time period because transitioning from a triplet
to a singlet does not occur as rapidly as internal conversion or
vibrational relaxation.

\begin{figure}
  \begin{center}
    \executeiffilenewer{Figures/Fluorescence/jablonski.svg}
  {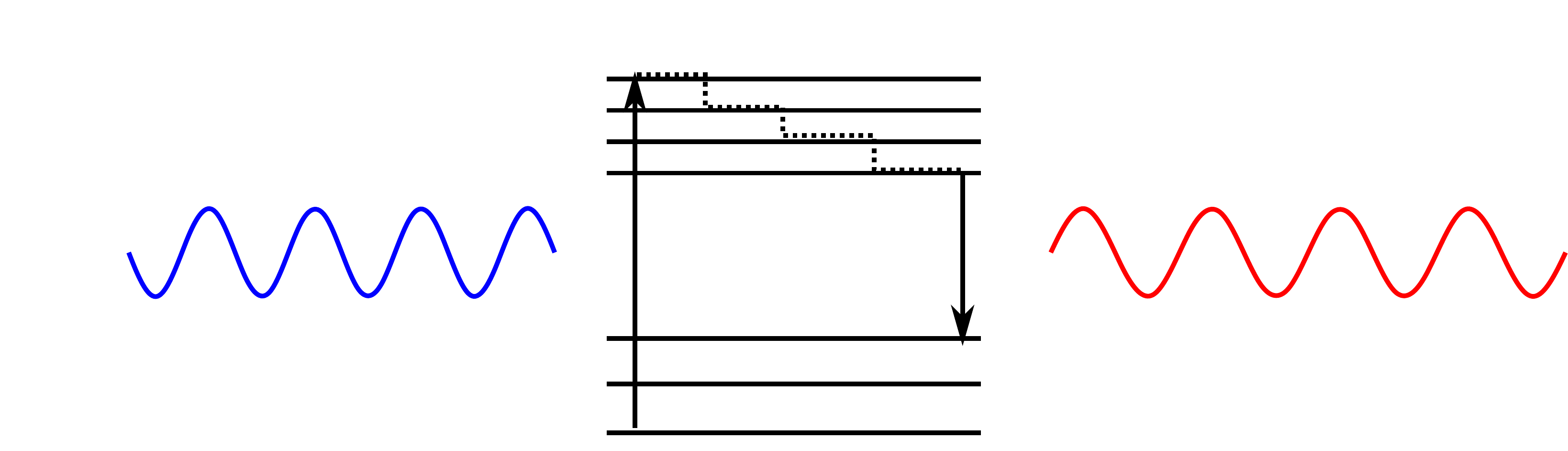}{inkscape-0.48pre1 -z -D --file=Figures/Fluorescence/jablonski.svg 
    --export-pdf=Figures/Fluorescence/jablonski.pdf --export-latex} \def\svgwidth{\columnwidth}
  \input{Figures/Fluorescence/jablonskiLabel.tex}
  \end{center}
  \caption{A Jablonski diagram for an example fluorescence
    process.\label{fig:fluorescence:jablonski}}
\end{figure}

Combining all of these processes with the complicated energy levels of
the hydrogen atom (and even more complicated energy levels for
molecules), makes it very difficult to predict the exact process
through which fluorescence occurs. Oftentimes it is advantageous to
visualize all the processes occurring in fluorescence with what is
known as a \term{Jablonksi diagram}. In the example diagram of Figure
\ref{fig:fluorescence:jablonski} the electron begins in the lowest
Bohr energy level, $n=1$.\footnote{The notation for fluorescence
  states is completely different from either the hydrogen
  wavefunctions using the notation $\psi_{n,l,m_l}$ and spectroscopic
  notation. This is because it is important to differentiate between
  singlet and triplet states of the atom or molecule when dealing with
  fluorescence. For this example however, we will stick with our
  standard notation from before.} Within this energy level are three
vibrational modes of the atom, $v_1$, $v_2$, and $v_3$, and the
electron is at the lowest, $v_1$. Of course, the actual number of
vibrational modes is entirely dependent upon the fluorescing
substance. The electron absorbs a photon (the blue squiggly incoming
line) with energy $E_a$ which boosts the electron into the next Bohr
energy level, $n=2$, and into the fourth vibrational mode, $v_4$, of
this energy level. This excitation of the electron is denoted by the
solid arrow and occurs over a time scale of $10^{-15}$ s. Next, the
electron non-radiatively transitions from $v_4$ to $v_1$ of the energy
level $n=2$. This is given by the dotted line and occurs over a time
period of $10^{-12}$ s. Finally the electron transitions to $v_3$ of
the $n=1$ energy level, the solid black arrow, over a time period of
$10^{-9}$ s and emits a photon with energy $E_e$, the squiggly red
line, where $E_e < E_a$.

\section{Experiment}

In the Figure \ref{fig:fluorescence:jablonski} the absorption of the
photon and non-radiative decay of the electron both occur rapidly in
comparison to the final decay of the electron. This is because after
the electron decays non-radiatively, it is in a \term{metastable}
state, or a state that has a longer lifetime than the previous two
processes. The electron while in the metastable state cannot decay
non-radiatively, as it is in the lowest vibrational mode of the energy
level, and so it must decay radiatively to the next Bohr energy
level. The number of electrons within the metastable state is
dependent upon the rate at which the electrons are able to radiatively
decay. The change in number of electrons is,
\begin{equation}
dN = -\lambda N\,dt
\end{equation}
where $N$ is the number of electrons at a given time $t$, and
$\lambda$ is the decay rate, or fraction of electrons that will
decay. More details on how to solve this type of differential equation
are given in Chapter \ref{chp:beta} for the absorption of
$\beta$-rays.

The general idea however is to integrate both sides and set initial
conditions, which results in,
\begin{equation}
  N = N_0e^{-\lambda t}
  \label{equ:fluorescence:decayRate}
\end{equation}
where again $\lambda$ is the \term{decay rate} and $N_0$ is the
initial number of electrons in the metastable state. The \term{mean
  lifetime} of a particle within the metastable state is then given
by,
\begin{equation}
  \tau = \frac{1}{\lambda}
\end{equation}
and is the expected time an electron would stay within the metastable
state. In Figure \ref{fig:fluorescence:jablonski} the mean lifetime of
the electron in the metastable state is $10^{-9}$ seconds, and so
$\lambda \approx 10^9\mbox{ s}^{-1}$.

Because the number of electrons decaying is directly proportional to
the number of photons being emitted, the intensity is directly
proportional to Equation \ref{equ:fluorescence:decayRate}. In the
experiment for this chapter, a strobe light flashes on two different
types of fluorescent crystals and excites electrons within the
crystals rapidly. The electrons then decay over a short time period
into a metastable state through non-radiative decays. Then, over a
longer time period (although short to the human eye) these electrons
decay radiatively.

A photo-diode is placed in front of the fluorescing crystals and so
after the strobe goes off, it is able to determine the intensity of
the light being emitted from the crystals. This photo-diode is hooked
into an oscilloscope where a plot of intensity on the $y$-axis is made
against time on the $x$-axis. By taking data points from this curve,
it is possible to determine a value for $\lambda$ and subsequently for
$\tau$, the mean lifetime of the electrons in the crystal. Remember
that Equation \ref{equ:fluorescence:decayRate} is exponential, so the
trick of making the equation linear, outlined in Chapter
\ref{chp:waves}, can be applied for better results.

\graphicspath{{Figures/Beta/}}

\chapter{Beta Radiation}\label{chp:beta}

Take a scrawny nerd, add a dash of a scientific experiment, and
bombard the mixture with a large amount of radiation; this is the
recipe for a superhero.\footnote{The two most famous being Spiderman
  and the Incredible Hulk, and possibly Captain America, although his
  origins are somewhat up to debate due to comic book censorship.}
Thanks to popular culture, radiation has a haze of misinformation
surrounding it, so let us strip away any pre-conceived ideas and try
to start back at the beginning with the definition of
radiation. Radiation is literally the emission of rays, but this is
not a very scientifically precise definition, and so let us instead
define radiation as the transfer of energy through a medium or vacuum
by means of waves or subatomic particles.\footnote{It is important to
  make the distinction of {\it subatomic} particles, otherwise we
  could consider rain to be a form of water radiation.} By this
definition sound is radiation, as is well electromagnetic waves, and
so we are constantly being bombarded by radiation of many different
types, yet not very many of us are turning into superheros. That is
because the word radiation is oftentimes used to indicate a more
specific type of radiation, \term{ionizing radiation}, where the wave or
subatomic particle has enough energy to ionize, or strip the electrons
away from atoms. It is this type of radiation that can be harmful to
humans, and has captured the imagination of the public.

There are three types of ionizing radiation, \term{$\alpha$-radiation},
\term{$\beta$-radiation}, and \term{$\gamma$-radiation}. An
$\alpha$-particle is a helium nucleus consisting of two neutrons and
protons, and is liberated from excited heavy nuclei. The $\beta$-ray
is an electron emitted from an excited neutron transitioning to a
proton, while the $\gamma$-ray is just a high energy photon, or light
wave. Because $\alpha$-rays are so much larger than the other two
types of radiation, they cannot penetrate objects as easily and can be
stopped by a piece of paper. On the other hand, $\beta$-rays can be
blocked by a sheet of metal, and $\gamma$-rays such as x-rays can
require a few centimeters of lead.\footnote{Think of this as the
  particles trying to pass through some barrier with holes in
  it. Helium nuclei are huge in comparison to electrons, and so it is
  difficult for them to pass. Electrons are much smaller, and photons
  are even smaller still (although technically electrons are
  considered to be point-like, and subsequently are as small as you
  can get).} This does not, however, indicate the danger of each type
of radiation to the human body. For example, $\alpha$-particles can
cause severe skin damage, while $\beta$-rays from a mild source cause
no significant tissue damage.

\section{Classical Beta Radiation}

The process of $\beta$-radiation has been known for over $100$ years,
yet the implications of the process are still being debated at the
cutting edge of particle physics to this day. Let us first consider
the process of $\beta$ radiation where a neutron turns into a proton
and an electron.
\begin{equation}
  n \rightarrow p^++e^-
  \label{equ:beta:betaDecay}
\end{equation}
In particle physics, mass is not conserved (although energy and
momentum are), but a lone particle cannot decay into particles with
more mass. Here, the neutron has a mass of $939.6$
MeV$/c^2$\footnote{In particle physics we give mass in mega-electron
  volts over the speed of light squared, MeV$/c^2$, but oftentimes
  drop the $c^2$ for convenience. One MeV is equal to
  $1.8\times10^{-30}$ kilograms.}, while the proton has mass of
$938.3$ MeV$/c^2$ and the electron has a mass of $0.5$ MeV$/c^2$, so
the decay of the neutron is allowed.

If we take Equation \ref{equ:beta:betaDecay} and impose conservation of
energy and momentum, and use Einstein's famous relation between
energy, momentum, and mass, we can find the energy of the electron
from the decay. Einstein's relation states,
\begin{equation}
m^2c^4 = E^2-p^2c^2
\label{equ:beta:einstein}
\end{equation}
or that the mass squared of an object is equal to the energy squared
of the object less the momentum squared. If we let the momentum of the
object equal zero, we obtain the famous equation $E=mc^2$. Returning
to the problem at hand, let us consider the decay of a neutron at rest
so $E_n = m_nc^2$ and $p_n = 0$. Now, when the neutron decays, we know
that the electron will have a momentum of $p_e$ and the proton will
have a momentum of $p_p$, but by conservation of momentum $p_e+p_p =
p_n = 0$ and so $p_e = -p_p$. Additionally, by conservation of energy,
we can write,
\begin{equation}
  E_n = E_p+E_e
  \label{equ:beta:totalEnergy}
\end{equation}
where $E_p$ is the energy of the proton and $E_e$ the energy of the
electron.

Using Einstein's relation of Equation \ref{equ:beta:einstein} again, we can
write out the energies of the proton and electron in terms of momentum
and mass.
\begin{equation}
  E_p^2 = m_p^2c^4-p_p^2c^2,~~~  E_e^2 = m_e^2c^4-p_e^2c^2
\end{equation}
Substituting the proton energy into Equation \ref{equ:beta:totalEnergy}
along with the neutron energy, $m_nc^2$, yields a relation between the
momenta, particle masses, and energy of the electron.
\begin{equation}
  m_nc^2 = \sqrt{m_p^2c^4-p_p^2c^2}+E_e
  \label{equ:beta:energySquared}
\end{equation}
Next, we solve for the proton momenta in terms of the electron
energy and mass by conservation of momentum,
\begin{equation}
  p_p^2c^2 = p_e^2c^2 = m_e^2c^4-E_e^2
\end{equation}
and plug this back into Equation \ref{equ:beta:energySquared} to find a
relation with only the masses of the neutron, proton, and electron,
and the energy of the electron.
\begin{equation}
  m_nc^2 = \sqrt{m_p^2c^4-m_e^2c^4+E_e^2}+E_e
\end{equation}
Finally, we solve for the energy of the electron.
\begin{equation}
  \begin{aligned}
    m_n^2c^4+E_e^2-2E_em_nc^2 &= m_p^2c^4-m_e^2c^4+E_e^2 \\
    E_e &= \frac{m_n^2c^4+m_e^2c^4-m_p^2c^4}{2m_nc^2} \\
    E_e &= c^2\left(\frac{m_n^2+m_e^2-m_p^2}{2m_n}\right) \\
    E_e &\approx 1.30~\mathrm{MeV} \\
    \label{equ:beta:betaEnergy}
  \end{aligned}
\end{equation}

Plugging the values for the mass of the neutron, proton, and electron,
yield the last line of Equation \ref{equ:beta:betaEnergy} and provide a
very experimentally verifiable result. The electrons produced from the
decay of a neutron at rest should have an energy of $1.30$ MeV or
$2.1\times10^{-13}$ J. The only problem with this theoretical result
is that when the experiment is performed the results do not match our
theoretical prediction at all! As a matter of fact, the electrons
observed from $\beta$-decay have a wide range of energies ranging from
near $0.5$ MeV up to $1.30$ MeV. So what is the problem with our
theory?

When this problem with $\beta$-decay was first discovered in the
$1930$'s, some physicists wanted to throw out conservation of energy,
just as conservation of mass had been thrown out by Einstein. However,
Paul Dirac and Enrico Fermi proposed a simple and brilliant
alternative. The decay of Equation \ref{equ:beta:betaDecay} is wrong, and
there should be an additional neutral particle on the right hand
side. Additionally, since electrons had been observed up to energies
of Equation \ref{equ:beta:betaEnergy}, the particle must be extremely
light. Considering a \term{three-body decay} instead of a two-body
decay completely changes the theory, and provides a prediction where
the energy of the electron is not fixed, but dependent upon how much
momentum is imparted to the other two particles of the decay.

\section{The Standard Model}

So what is this missing particle? There is a long and interesting
history behind its discovery, but we will skip the history lesson and
move onto what is currently known. The missing particle is an
anti-electron \term{neutrino}, is very light, has no charge, and
interacts only through the  weak force. The statement above
requires a bit of explanation, so let us begin with what is known as
the \term{Standard Model}. The Standard Model is a theory developed
over the past century which describes the most fundamental
interactions (that we know of) between particles. The theory itself
can be rather complicated and consists of local time dependent quantum
field theory. Luckily, understanding the results of the Standard Model
does not require any idea as to what the last sentence meant.

There are three forces described by the Standard Model, the {\bf
  electromagnetic force}, the \term{weak force}, and the \term{strong
  force}. These forces are \term{mediated} or carried out by {\bf
  bosons}. The electromagnetic force is mediated through the photon,
while the weak force is mediated through the \term{weak bosons} and
the strong force is mediated through the \term{gluon}. More generally,
bosons are particles with an integer spin\footnote{See Chapters
  \ref{chp:hydrogen} and \ref{chp:superconductivity} for more
  details.}, but these bosons are special because they carry the
fundamental forces and are themselves fundamental
particles.\footnote{For example, the Cooper pairs of the BCS theory in
  Chapter \ref{chp:superconductivity} are also bosons, but are neither
  fundamental bosons nor carry a fundamental force.} The list of
forces above however is missing a critical component, gravity! This is
because gravity is much weaker than the three forces above at the
particle physics scale, and so we don't yet really know how it
works. Incorporating gravity into the Standard Model is still an open
question.

\begin{figure}
  \begin{center}
    \tiny
    \executeiffilenewer{Figures/Beta/standardModel.svg}
  {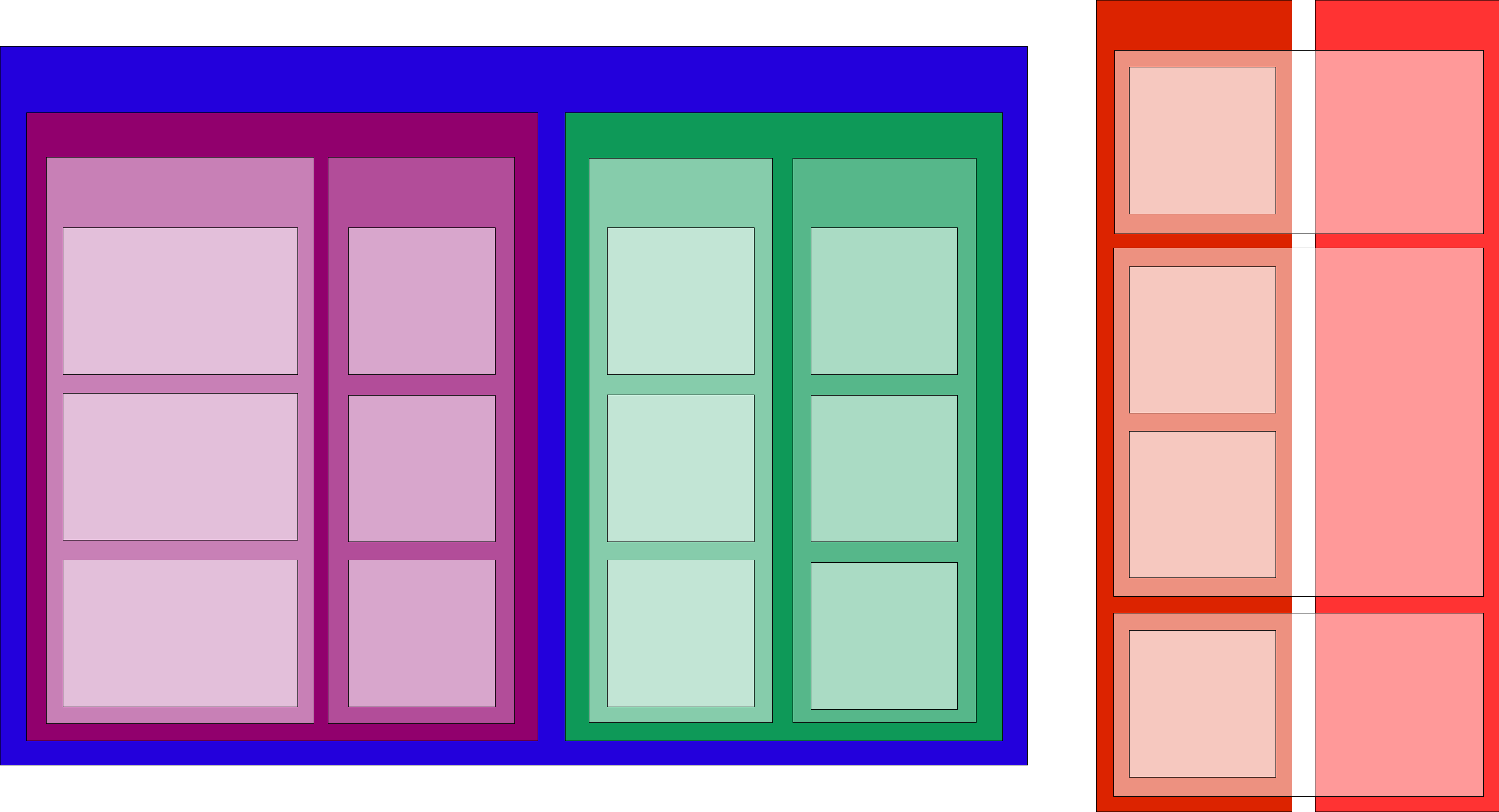}{inkscape-0.48pre1 -z -D --file=Figures/Beta/standardModel.svg 
    --export-pdf=Figures/Beta/standardModel.pdf --export-latex} \def\svgwidth{\columnwidth}
  \input{Figures/Beta/standardModelLabel.tex}
    \caption{A summary of the fundamental particles and forces
      comprising the Standard Model. Each box contains the particle
      name, the particle symbol, and the particle mass, excluding the
      neutrinos for which the masses are unknown. Mass data was taken
      from the Particle Data Group.\label{fig:beta:standardModel}}
  \end{center}
\end{figure}

The right side of Figure \ref{fig:beta:standardModel} summarizes the three
forces and their boson mediators as described above. On the left of
Figure \ref{fig:beta:standardModel} are listed the $12$ fundamental
fermions, or particles with half integer spin, of the Standard
Model. The fermions are further split into two groups, quarks and
leptons. The quarks interact through the strong, weak, and
electromagnetic forces, while the leptons only interact through the
weak and electromagnetic forces. The quarks are divided into three
generations by mass (and by discovery) and have either charge $+2/3$
or $-1/3$. Quarks are bound together into groups called \term{hadrons}
by the strong force. There are two types of hadrons, \term{baryons}
consisting of three quarks, and \term{mesons} consisting of two
quarks. The proton and neutron are both hadrons and baryons, where the
proton is made up of two up quarks and one down quark and the neutron
is made up of one up quark and two down quarks. The LHC is a Large
Hadron Collider because it collides protons with protons.

The leptons do not interact through the strong force like the quarks,
and subsequently do not bind together like the quarks. Individual
quarks are never found in nature, while individual leptons, such as
the electron can be easily found alone. The leptons can also be broken
down into three generations by mass with each generation containing a
charged lepton and a neutral neutrino. The properties of the charged
leptons, electrons, muons, and tau leptons, are well measured while
the properties of the neutrinos are not. This is because the charged
leptons interact through the electromagnetic force, and so they are
easy to detect. The neutrinos, however, only interact through the weak
force, and so they are very difficult to detect. Nearly $50$ trillion
neutrinos pass through the human body every second, but because they
interact so weakly with matter, they have no effect. Experiments to
detect neutrinos require huge detectors of photo-detectors buried deep
underground to filter away extraneous particle noise. This is why
$\beta$-decay is so important to particle physics, even today, as
it helps provide insights into the fundamental nature of neutrinos.

That covers all the basics of the Standard Model, and while it is a
lot to remember, the important things are the three forces, the
difference between fermions and bosons, and the two types of
fundamental fermions. Remembering all the masses and particle names
can be useful, but is not necessary to understand the basic concepts
behind the Standard Model. There is one more important detail to
mention, and that is that for every particle there is an anti-particle
with the opposite charge. These anti-particles are normally designated by a bar
over the symbol of the particle, except for the charged leptons. For
example an anti-up quark has a charge of $+1/3$ and is denoted by the
symbol $\bar{u}$. The anti-electron, more commonly referred to as a
\term{positron} has a charge of $+1$ and is given by the symbol $e^+$.

\section{Feynman Diagrams}

Now that we have the basics of the Standard Model, we can take another
look at the $\beta$-decay process of Equation \ref{equ:beta:betaDecay} and
include an anti-electron neutrino.
\begin{equation}
  n \rightarrow p+e^-+\bar{\nu}_e
  \label{equ:beta:realDecay}
\end{equation}
If we try to perform the same momentum and energy analysis as we did
above we would now have three unknowns, the momenta of the three
particles, and only two equations, one relating the three momenta, and
the other equating the energies. From this we can see that it is
impossible to determine a unique solution. Clearly a new method is
needed to approach this problem.

One of the main features of the Standard Model is that the probability
of a particle being produced or decaying can be calculated, given the
necessary physical constants. The method for performing these
calculations can be very tedious, consisting of performing multiple
integrals over various phase spaces and employing a variety of
mathematical tricks. However, the physicist Richard Feynman developed
a very beautiful method to represent these calculations with diagrams
which allow the reader to understand the underlying physics without
needing to know the math. These graphical representations of
fundamental particle interactions are known as \term{Feynman diagrams}.

\begin{figure}
  \begin{center}
    \subfigure[]
    {
      \begin{fmffile}{Figures/Beta/electronScattering}
    \setlength{\unitlength}{1mm}
    \input{Figures/Beta/electronScattering.fmp}
  \end{fmffile}\executeiffilenewer{Figures/Beta/electronScattering.fmp}{Figures/Beta/electronScattering.1}{cd
    Figures/Beta; mpost electronScattering.mp}
      \label{fig:beta:electronScattering}
    }
    \subfigure[]
    {
      \begin{fmffile}{Figures/Beta/betaDecay}
    \setlength{\unitlength}{1mm}
    \input{Figures/Beta/betaDecay.fmp}
  \end{fmffile}\executeiffilenewer{Figures/Beta/betaDecay.fmp}{Figures/Beta/betaDecay.1}{cd
    Figures/Beta; mpost betaDecay.mp}
      \label{fig:beta:betaDecay}
    }
    \caption{A possible Feynman diagram for electron scattering is
      given in Figure \ref{fig:beta:electronScattering}, while a diagram
      for $\beta$-decay is given in Figure
      \ref{fig:beta:betaDecay}.\label{fig:beta:diagrams}}
  \end{center}
\end{figure}

Take Figure \ref{fig:beta:electronScattering} as an example of a Feynman
diagram where two electrons pass near each other and are repelled. A
Feynman diagram consists of three parts, incoming particles, some
internal structure, and outgoing particles. In Feynman diagrams,
fermions are drawn as solid lines with arrows pointing in the
direction of the particle, weak bosons are indicated by dashed lines,
photons are indicated by wavy lines, and gluons are indicated by loopy
lines.\footnote{Notation in Feynman diagrams does still differ from
  textbook to textbook, but these are the general conventions
  followed. Sometimes diagrams are drawn with time flowing from bottom
  to top, rather from left to right. An arrow pointing outwards on an
  incoming fermion line indicates that the fermion is an
  anti-particle.} Time flows from left to right in Figure
\ref{fig:beta:electronScattering} and so we see there are two incoming
electrons, and two outgoing electrons, with a photon exchanged between
the two in the middle of the diagram, driving the two electrons
apart. It is important to understand that Feynman diagrams do not
indicate the actual trajectory of the particles, and have no
correspondence to actual physical position; they only indicate the
state of the particles at a point in time.

\begin{figure}
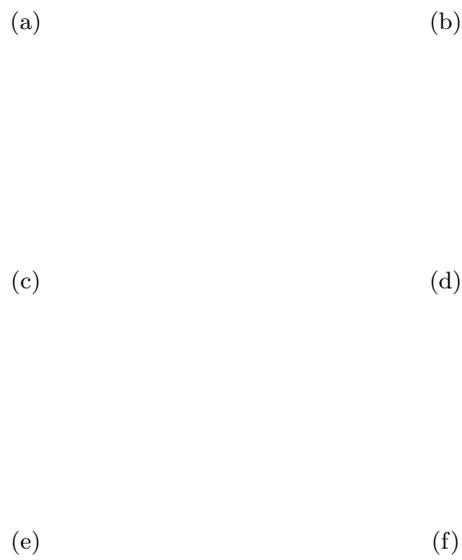

  \begin{center}
    \subfigure[]
    {
      \begin{fmffile}{Figures/Beta/ffpVertex}
    \setlength{\unitlength}{1mm}
    \input{Figures/Beta/ffpVertex.fmp}
  \end{fmffile}\executeiffilenewer{Figures/Beta/ffpVertex.fmp}{Figures/Beta/ffpVertex.1}{cd
    Figures/Beta; mpost ffpVertex.mp}
      \label{fig:beta:ffpVertex}
    }\hspace{2cm}
    \subfigure[]
    {
      \begin{fmffile}{Figures/Beta/ffzVertex}
    \setlength{\unitlength}{1mm}
    \input{Figures/Beta/ffzVertex.fmp}
  \end{fmffile}\executeiffilenewer{Figures/Beta/ffzVertex.fmp}{Figures/Beta/ffzVertex.1}{cd
    Figures/Beta; mpost ffzVertex.mp}
      \label{fig:beta:ffzVertex}
    }\vspace{1cm}

    \subfigure[]
    {
      \begin{fmffile}{Figures/Beta/ffwVertex}
    \setlength{\unitlength}{1mm}
    \input{Figures/Beta/ffwVertex.fmp}
  \end{fmffile}\executeiffilenewer{Figures/Beta/ffwVertex.fmp}{Figures/Beta/ffwVertex.1}{cd
    Figures/Beta; mpost ffwVertex.mp}
      \label{fig:beta:ffwVertex}
    }\hspace{2cm}
    \subfigure[]
    {
      \begin{fmffile}{Figures/Beta/ffgVertex}
    \setlength{\unitlength}{1mm}
    \input{Figures/Beta/ffgVertex.fmp}
  \end{fmffile}\executeiffilenewer{Figures/Beta/ffgVertex.fmp}{Figures/Beta/ffgVertex.1}{cd
    Figures/Beta; mpost ffgVertex.mp}
      \label{fig:beta:ffgVertex}
    }\vspace{1cm}

    \subfigure[]
    {
      \begin{fmffile}{Figures/Beta/gggVertex}
    \setlength{\unitlength}{1mm}
    \input{Figures/Beta/gggVertex.fmp}
  \end{fmffile}\executeiffilenewer{Figures/Beta/gggVertex.fmp}{Figures/Beta/gggVertex.1}{cd
    Figures/Beta; mpost gggVertex.mp}
      \label{fig:beta:gggVertex}
    }\hspace{2cm}
    \subfigure[]
    {
      \begin{fmffile}{Figures/Beta/ggggVertex}
    \setlength{\unitlength}{1mm}
    \input{Figures/Beta/ggggVertex.fmp}
  \end{fmffile}\executeiffilenewer{Figures/Beta/ggggVertex.fmp}{Figures/Beta/ggggVertex.1}{cd
    Figures/Beta; mpost ggggVertex.mp}
      \label{fig:beta:ggggVertex}
    }
    \caption{The basic Feynman diagram vertices. The letter $f$ stands
      for any fermion, $\ell$ for a charged lepton, $\nu$ for a
      neutrino, and $q$ for quarks. The electroweak vertices are given
      in Figures \ref{fig:beta:ffpVertex} through \ref{fig:beta:ffwVertex} while
      the strong force vertices are given in Figures
      \ref{fig:beta:ffgVertex} through
      \ref{fig:beta:ggggVertex}.\label{fig:beta:vertices}}
  \end{center}
\end{figure}

A \term{vertex} is wherever three or more lines connect in a Feynman
diagram. There are six fundamental vertices that can be drawn in
Feynman diagrams using the fermions and bosons of Figure
\ref{fig:beta:standardModel}. Six of these vertices consist of only photons
and the weak bosons and are not of much interest because they are
relatively unlikely to occur. The six remaining vertices are drawn in
the Feynman diagrams of Figure \ref{fig:beta:vertices}. If a Feynman
diagram is drawn with a vertex that is not given in Figure
\ref{fig:beta:vertices}, it cannot happen! Of course, there are many other
rules which dictate what can and cannot be drawn with Feynman
diagrams, but this is the most basic rule.

From the figures above it is clear how Feynman diagrams help visualize
particle interactions, but how do they help calculate decays and
production of particles? Each line of a diagram is assigned a momentum
and polarization vector, and each vertex is assigned a coupling
factor. The diagram is then traced over, and all the mathematical
terms, each corresponding to a line or vertex, are put together into
an integral. The integral is used to calculate what is known as the
\term{matrix element} of the process depicted in the diagram. This
matrix element can then be used to calculate decays and productions
using a relation called \term{Fermi's golden rule}. Neither the details
of this process nor Fermi's golden rule will be given here, but
suffice it to say that this is the deeper mathematical meaning of
Feynman diagrams.

Using the allowed vertices of Figure \ref{fig:beta:vertices} and the decay
process of Equation \ref{equ:beta:realDecay} we can now draw the Feynman
diagram for $\beta$-decay, given in Figure \ref{fig:beta:betaDecay}. To
begin we have the one up and two down quarks of a neutron. One of the
down quarks then radiates a $W^-$ boson and turns into an up
quark. The $W^-$ boson then decays into an electron and an
anti-electron neutrino. We can use the method outlined above for
calculating matrix elements from Feynman diagrams to determine the
decay probability for the neutron. However, without a few months of a
particle physics course the steps might be a bit incomprehensible and
the final result is presented here without
derivation.\footnote{However, if you do want to see the full
  derivation, take a look at the chapter on neutron decays in {\it
    Introduction to Elementary Particles} by David Griffiths.}

\begin{equation}
  PDF(E_e) \approx C_0E_e\sqrt{E_e^2-m_e^2c^4}\left(
    c^2\left(m_n-m_p\right) -E_e\right)^2
  \label{equ:beta:spectrum}
\end{equation}

\begin{figure}
  \begin{center}
    \executeiffilenewer{Code/Beta/spectrum.m}{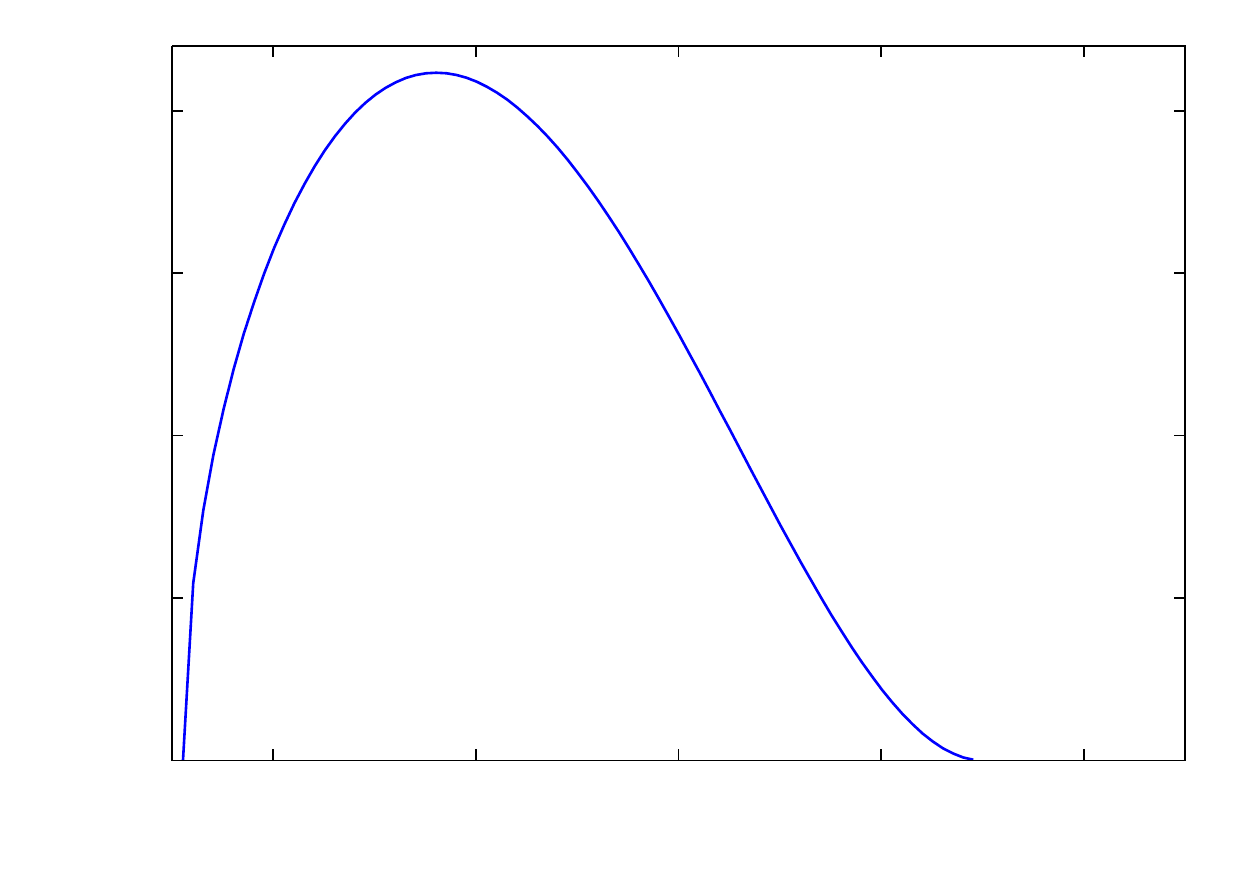}
  {octave --silent --eval "addpath([pwd(),'/Code']); 
    addpath([pwd(),'/Code/Beta']); spectrum(1);"} 
  \setlength{\unitlength}{\columnwidth*\real{0.00013}}
  \input{Figures/Beta/spectrumLabel.tex}
    \label{fig:beta:spectrum}
  \end{center}
\end{figure}

Equation \ref{equ:beta:spectrum} gives the probability density
function for finding an electron with energy $E_e$ from a neutron
decay.\footnote{For more details on probability density functions
  consult Chapter \ref{chp:uncertainty}.} Here, the coefficient $C_0$
is some normalization factor so that the integral of the function is
one over the valid range of the equation. From the square root we see
that the minimum value $E_e$ can have is the mass of the electron,
$m_e$. Additionally, we notice that the maximum energy the electron
can have is when no momentum is imparted to the neutrino, and so the
decay becomes effectively the two body decay of Equation
\ref{equ:beta:betaDecay} with a maximum energy of $1.30$ MeV. Keeping
this in mind we can draw the distribution of the electron energy in
Figure \ref{fig:beta:spectrum}. Notice that the expectation value of
this distribution is $1$ MeV, considerably lower than the $1.3$ MeV
predicted earlier. When compared to experimental data, Equation
\ref{equ:beta:spectrum} and Figure \ref{fig:beta:spectrum} agree very
well.\footnote{See \href{http://prd.aps.org/pdf/PRD/v5/i7/p1628_1}{\it
    Free-Neutron Beta-Decay Half-Life} by C.J. Christensen,
  et. al. published in Physical Review D, Volume 5, Number 7, April 1,
  1972.}

\section{Beta Ray Absorption}

So now we have a theory that accurately predicts the energy spectrum
of electrons from $\beta$-decay, but what about a theoretical model
that predicts the absorption of the electrons? First, let us consider
electrons traveling along the $x$-axis, where they hit some absorber
that is perpendicular to the $x$-axis and parallel to the
$y$-axis. Let us assume the absorber is infinitesimally thin, with
thickness $dx$, so that the absorber is just a single sheet of atoms.

Now if we consider $N$ electrons entering the absorber, we can write
that the number of electrons absorbed is,
\begin{equation}
  dN = -\lambda N dx
  \label{equ:beta:diffPenetration}
\end{equation}
where $dN$ is the just the change in number of electrons, $\lambda$ is
a constant called the \term{absorption coefficient}, $N$ is the number
of electrons, and $dx$ is the distance traveled by the
electrons. Physically, this equation states that the electrons enter
the absorber, and then are reduced by a certain percentage $\lambda$,
per unit distance. If the electrons then travel over a distance $dx$
the total percentage by which $N$ is reduced is just $\lambda dx$.

Equation \ref{equ:beta:diffPenetration} is a separable differential
equation, so we can separate the variables and integrate both sides.
\begin{equation}
  \begin{aligned}
    \frac{1}{N}dN &= -\lambda dx \\
    \int \frac{1}{N}dN &= - \int \lambda dx \\
    \ln N +C_0 &= -\lambda x + C_1 \\
    N &= C_2e^{-\lambda x} \\
  \end{aligned}
\end{equation}
In the first step, the variables have just been separated, while in
the second and third step the indefinite integrals are performed
yielding the integration constants of $C_0$ and $C_1$. In the final
step both sides are exponentiated and all the constants are absorbed
into $C_2$. By letting $x = 0$ we see that $N(0) = C_2$, and so $C_2$
must be the initial number of electrons before any absorption occurs,
or $N_0$.

\begin{equation}
  N = N_0e^{-\lambda x}
  \label{equ:beta:penetration}
\end{equation}

But what is the meaning behind $\lambda$ in Equation
\ref{equ:beta:penetration}? As stated above, $\lambda$ essentially
tells us the percentage of electrons absorbed, per unit distance. From
the previous section on the Standard Model, we know that electrons can
interact through the weak and electromagnetic forces, but primarily
through the electromagnetic force at low energies like these. As
electrons pass through the absorber, they are slowed down by their
electromagnetic interactions with the atoms. Of course, the atoms are
made up of very dense nuclei surrounded by large electron
clouds\footnote{If this is unfamiliar territory, see Chapter
  \ref{chp:hydrogen}.}, so the electrons from the $\beta$-radiation
will mainly interact with the electron clouds of the atoms in a
fashion similar to that of Figure \ref{fig:beta:electronScattering}.

The denser the electron clouds of the absorber, the more likely an
electron from $\beta$-radiation will be absorbed, and so it is clear
that the absorption coefficient should somehow depend on the electron
cloud density, or $\rho_e$. So what is the electron cloud density? Let
us first assume that the absorber is one single element and none of
the atoms are ions, so the number of electrons is equal to the number
of protons, or the \term{atomic number} $Z$ of the element. We now want
to find the number of electrons per unit volume. First we can convert
the number of electrons per atom into the electrons per mole by
multiplying $Z$ with Avogadro's number, $N_A$. Next, we can can
convert this quantity into electrons per gram by dividing by the {\bf
  atomic weight} $M$ of the element which is in grams per
mole. Finally, this can be converted to electrons per volume by
multiplying the previous product with the density of the absorber,
$\rho_a$.

\begin{equation}
  \begin{aligned}
    \underbrace{\left(\frac{\mathrm{electrons}}{\mathrm{cm^3}}\right)}
    _{\begin{array}{c} \rho_e \end{array}} &=
    \underbrace{\left(\frac{\mathrm{electrons}}{\mathrm{atom}}\right)}
    _{\begin{array}{c} Z \end{array}}
    \underbrace{\left(\frac{\mathrm{atoms}}{\mathrm{mole}}\right)}
    _{\begin{array}{c} N_A \end{array}}
    \underbrace{\left(\frac{\mathrm{moles}}{\mathrm{gram}}\right)}
    _{\begin{array}{c} \frac{1}{M} \end{array}}
    \underbrace{\left(\frac{\mathrm{grams}}{\mathrm{cm^3}}\right)}
    _{\begin{array}{c} \rho_a \end{array}} \\
    \rho_e &= \rho_a\left(\frac{Z N_A}{M}\right) \\
    \label{equ:beta:electronDensity}
  \end{aligned}
\end{equation}

The atomic number for carbon is $6$, while the atomic weight is
$12.011$, and so $Z/M \approx 1/2$. Actually, all of the light
elements up to chlorine and argon have a value for $Z/M$ of nearly
$1/2$. Consequently, the electron density of Equation
\ref{equ:beta:electronDensity} is only dependent upon the density of the
absorber for the lighter elements! This means that the absorption
coefficient is linearly proportional to the density of the absorption
material! However, it would be nice to have a slightly better
understanding of what other parameters determine $\lambda$.

Theoretically, determining $\lambda$ is quite challenging as we must
now consider how the electrons are physically absorbed by the electron
clouds. We can instead, just take a qualitative look at the
process. We have already determined how the absorber effects
$\lambda$, but how do the incoming electrons from the $\beta$
radiation effect $\lambda$? If an electron flies by an atom at a very
high velocity (very large $E_e$), the electron is hardly effected, and
passes nearly straight by. However, if the electron has a very small
velocity (very small $E_e$), the electrons of the atom will cause a
much more drastic change in the path of the electron. From this we see
that $\lambda$ should be inversely proportional to the electron energy.

Experimentally, the value for $\lambda$ has been determined to
be\footnote{Taken from
  \href{http://ia331303.us.archive.org/3/items/atomicnucleus032805mbp/atomicnucleus032805mbp.pdf}{\it
    The Atomic Nucleus} by Robley D. Evans, pages 627-629.},
\begin{equation}
  \lambda \approx \rho_a\left(\frac{17}{E_\mathrm{max}^{1.14}}\right)
  \label{equ:beta:coefficient}
\end{equation}
where $E_\mathrm{max}$ is the maximum electron energy, which for the
case of $\beta$ radiation from free neutrons is just given by Equation
\ref{equ:beta:betaEnergy}. Most sources of $\beta$ radiation do not consist
of free neutrons, and so as the electrons leave the nucleus, they must
fight the attractive electromagnetic pull of the protons. This means
that the maximum energy of most electrons from $\beta$ decay are well
below that of Equation \ref{equ:beta:betaEnergy}. The shape of the electron
energies however is very similar to the free neutron of Figure
\ref{fig:beta:spectrum}.

\section{Experiment}

There are two clear predictions about $\beta$-decay derived in the
sections above. The first is that the emitted electrons should not be
at a single energy, but rather over a spectrum of energies. The second
prediction states that the electrons from $\beta$-decay should be
absorbed over distances as described by Equation
\ref{equ:beta:penetration}, where $\lambda$ is directly proportional to the
density of the absorbing material, and is given experimentally by
Equation \ref{equ:beta:coefficient}. Testing the first theoretical
prediction is possible, but outside the scope of this book. The
second prediction is much easier to test, and is done so in three
parts.

In the first part of the experiment for this chapter a Geiger-Muller
tube is calibrated to yield a good count reading from a small source
of $\beta$-radiation. This is done by adjusting the distance of the
tube from the source and changing the bias voltage across the
tube. The source is removed and a background count is made. Next the
source is reinserted, and a radiation count with no absorber is
made. This gives the coefficient $N_0$ in Equation
\ref{equ:beta:penetration}. Next, thin pieces of cardboard and mylar
are placed as absorbers between the source and the tube. The number of
layers is recorded along with rate from the tube. The mass per unit
area of the mylar and cardboard absorbers are found by weighing the
absorber, and dividing by the area of the absorber calculated by
simple geometry. Plots are then made of the number of counts versus
the mass per unit area of the absorbers. Because $\lambda$ is
dependent only on $E_\mathrm{max}$ and $\rho_a$, the plots should
provide the same result assuming theory is correct.

The second part of the experiment then uses the plots just created to
determine the mass per unit area of an irregular cardboard shape
(i.e. the mass per unit area cannot be calculated easily using
geometry) after having determined the count rate of electrons. The
final part of the experiment recasts the data from the first part of
the experiment in terms of count rate and distance absorbed, as
described by Equation \ref{equ:beta:penetration}. A plot of this
relationship is made and a value for $\lambda$ is determined for the
mylar absorber. Using Equation \ref{equ:beta:coefficient} the maximum
electron energy can be determined from the $\beta$-radiation source.
\graphicspath{{Figures/Superconductivity/}}

\chapter{Superconductivity}\label{chp:superconductivity}

While superconductors are rarely encountered in day-to-day life they
are well known by the public, and not just within the physics
community. Superconductors oftentimes play important roles in science
fiction and capture readers' imaginations with their almost mysterious
capabilities, but they are also very real and are used in current
technologies such as Magnetic Resonance Imaging (MRI) and the Large
Hadron Collider (LHC). The discovery of superconductors was made in
$1911$ by Kamerlingh Onnes, yet the theory behind superconductors
remains incomplete to this day. This chapter provides a brief overview
of the physical properties of superconductors along with the theory
behind them and a more in depth look at the Meissner Effect. However,
the information provided here just scrapes the surface of
superconductivity; the experimental and theoretical research done in
this field is extensive.

\section{Superconductors}

So what exactly is a superconductor? As the name implies,
superconductors are very good at conducting; as a matter of fact, they
are \term{perfect conductors}. This means that if a superconductor is
made into a loop, and electricity is run around the loop, the current
will continue to flow forever without being pushed by a battery or
generator. Of course, forever is a rather strong word, and
experimental physicists don't have quite that much patience, but it
has been experimentally shown during experiments over periods of
years, that the current flowing within a superconducting loop has not
degraded enough to be registered by the precision of the instruments
used!\footnote{The resistance of superconductors has been shown to be
  less than $10^{-26}~\Omega$!}

However, superconductors are more than just perfect conductors. They
also exhibit a property called the \term{Meissner effect}, which states
that a magnetic field cannot exist within the interior of the
superconductor. This is an important difference between
superconductors and perfect conductors and will be discussed more in
the Meissner Effect section of this chapter. Just like a square is a
rectangle but a rectangle is not necessarily a square, a
superconductor is a perfect conductor, but a perfect conductor is not
necessarily a superconductor.

There are two fundamental properties that describe superconductors, a
\term{critical temperature} $T_c$, and a \term{critical magnetic field}
(magnitude) $B_c$. If the temperature of a superconductor exceeds the
critical temperature of the superconductor, it will no longer
superconduct, and will transition to a normal state. Similarly, if the
superconductor is subjected to a magnetic field higher than the
critical magnetic field, the superconductor will transition to a
normal state.

\begin{figure}
  \begin{center}
    \executeiffilenewer{Figures/Superconductivity/periodic.svg}
  {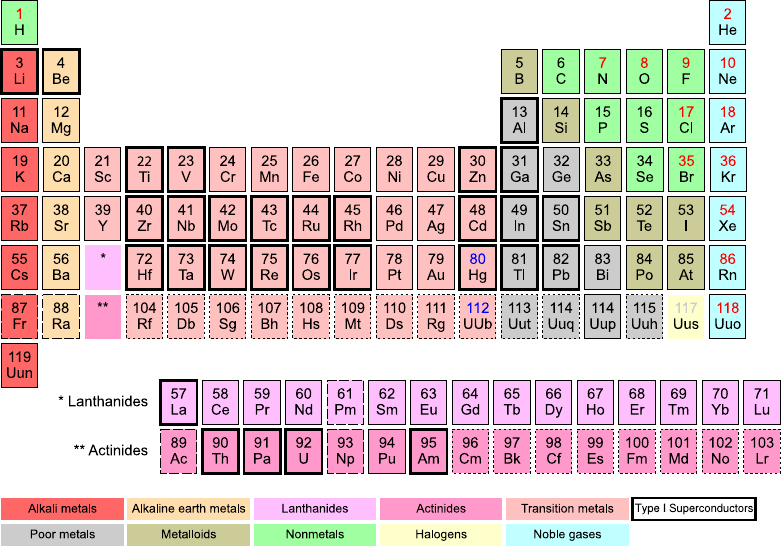}{inkscape-0.48pre1 -z -D --file=Figures/Superconductivity/periodic.svg 
    --export-pdf=Figures/Superconductivity/periodic.pdf --export-latex} \def\svgwidth{\columnwidth}
  \input{Figures/Superconductivity/periodicLabel.tex}
    \caption{All currently known Type I superconductors are outlined
      in bold. This figure was modified from the Wikimedia file
      \href{http://commons.wikimedia.org/wiki/File:Periodic_Table_Armtuk3.svg}{Periodic
        Table Armtuk3.svg}.\label{fig:superconductivity:periodic}}
  \end{center}
\end{figure}

Depending on the type of superconductor, the values for $T_c$ and
$B_c$ can vary greatly. Originally, there were two known types of
superconductors creatively named \term{Type I} and \term{Type II}
superconductors. There are thirty Type I superconductors, all
consisting of pure metals such as gallium, which has a critical
temperature of $T_c \approx 1.1$ K. In general, Type I superconductors
have very low critical temperatures, almost all below $10$
K. Additionally, Type I superconductors have low critical magnetic
fields. For example gallium has a $B_c$ of $\approx 51$ Gauss. All
currently known Type I superconductors are outlined in bold in the
periodic table of Figure
\ref{fig:superconductivity:periodic}.\footnote{Ashcroft and
  Mermin. {\it Solid State Physics}. Brooks/Cole. 1976.}

Type II superconductors consist of alloys, such as niobium and tin,
and have higher critical temperatures, in this example $T_c \approx
17.9$ K.\footnote{Rohlf, James William. {\it Modern Physics from a to
    Z0}. Wiley. 1994.} The highest critical temperature of Type II
superconductors is $23$ K.\footnote{Kittel, Charles. {\it Introduction
    to Solid State Physics}. Wiley. 1996} More recently, \term{high
  temperature superconductors}, neither Type I or Type II, have been
discovered, which, as the name implies, have much higher temperatures
at which they can superconduct.\footnote{More recently being a rather
  relative term. Specifically, in 1988 ceramic mixed oxide
  superconductors were discovered and are now used in commercial
  applications.}

\begin{figure}
  \begin{center}
    \executeiffilenewer{Figures/Superconductivity/ybco.svg}
  {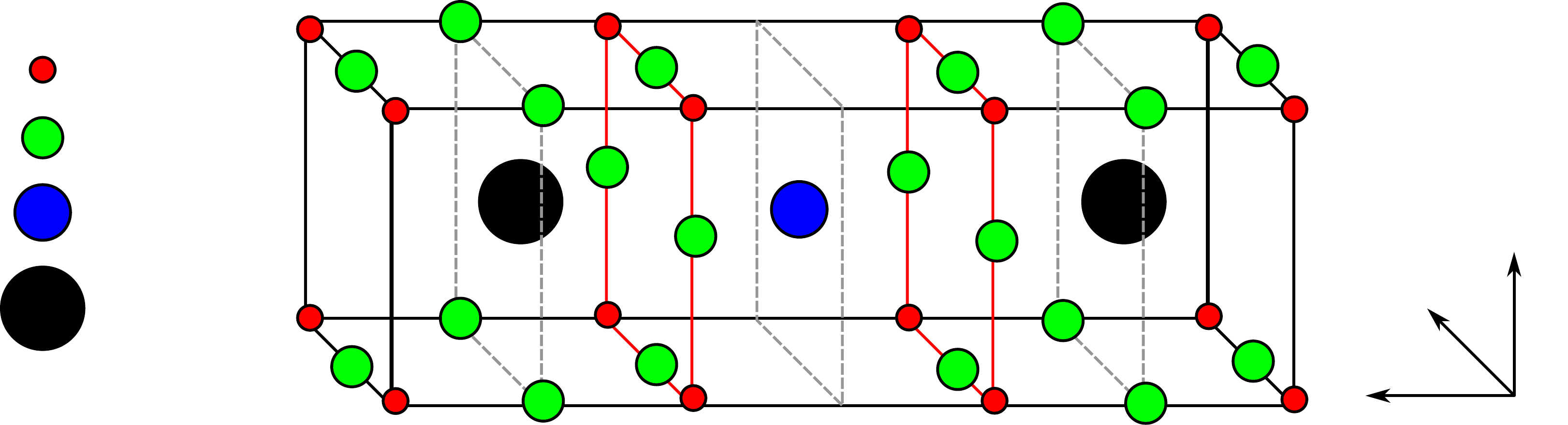}{inkscape-0.48pre1 -z -D --file=Figures/Superconductivity/ybco.svg 
    --export-pdf=Figures/Superconductivity/ybco.pdf --export-latex} \def\svgwidth{\columnwidth}
  \input{Figures/Superconductivity/ybcoLabel.tex}
    \caption{Primitive cell of the $\mathrm{YBa_2Cu_3O_7}$
      superconductor used in this experiment. The superconducting
      currents flow through the planes outlined in
      red.\label{fig:superconductivity:ybco}}
  \end{center}
\end{figure}

These high temperature superconductors are ceramic crystalline
substances, some with critical temperatures as high as $T_c \approx
125$ K. The ceramic $\mathrm{YBa_2Cu_3O_7}$, used in the experiment
associated with this chapter, has a critical temperature of $T_c
\approx 90$, higher than the temperature of liquid nitrogen. The
primitive cell crystal structure of $\mathrm{YBa_2Cu_3O_7}$ is shown
in Figure \ref{fig:superconductivity:ybco}. The superconducting currents flow through
the planes outlined in red.

\section{BCS Theory}

But how exactly do superconductors work? As mentioned previously, the
theory behind superconductors is still developing, although there is a
theory that describes superconductivity for Type I superconductors on
the level of the atom. This theory, or \term{BCS theory}, was developed
by John Bardeen, Leon Cooper, and Robert Schrieffer during the
$1950$'s for which they earned the Nobel prize in $1972$. While the
mathematics behind the theory are very involved, the physical idea
behind the theory is quite elegant.

To fully understand the theory, a few things about conductivity need
to be discussed. The conductivity of a metal arises from the nuclei of
the atoms making up the metal arranging themselves into an \term{ion
  lattice}. The outer electrons of the atoms are not bound tightly to
the nuclei and are able to freely move about in what is known as the
\term{electron gas}. When the electrons in the electron gas scatter off
the nuclei, the electrons lose energy, and this is why resistance
occurs in metals.

\begin{figure}
  \begin{center}
    \subfigure[]{
      \executeiffilenewer{Figures/Superconductivity/phonon1.svg}
  {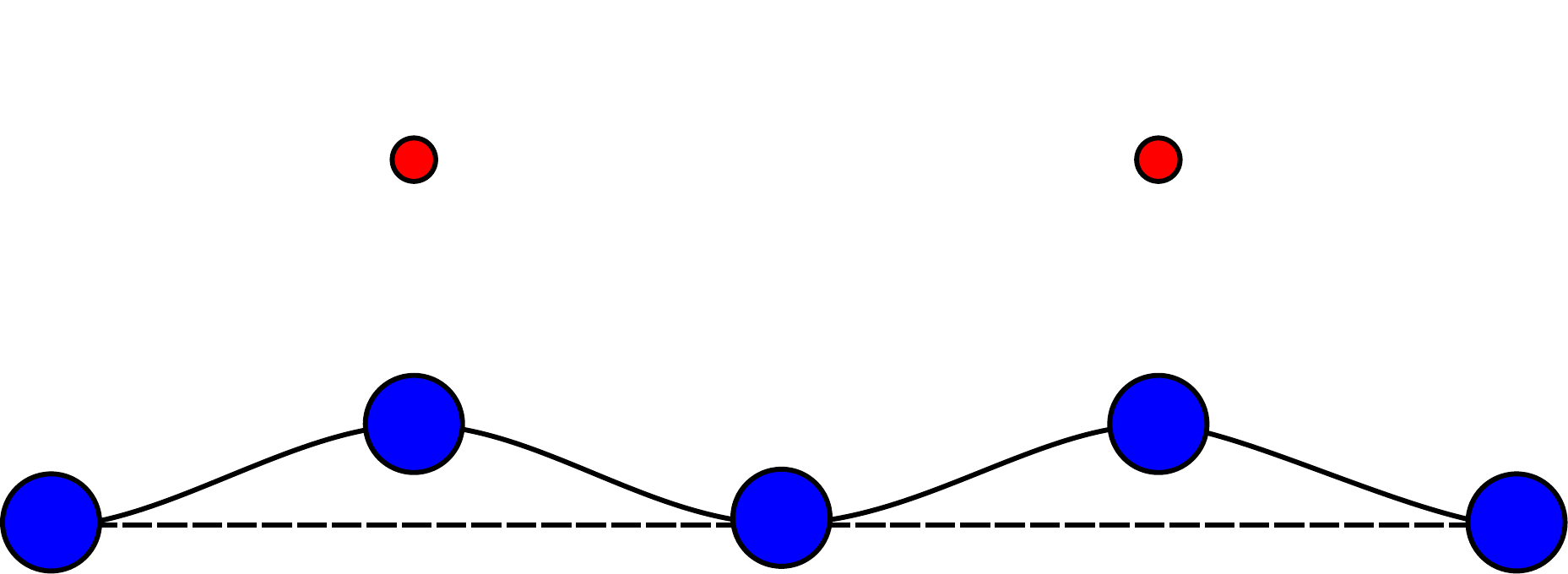}{inkscape-0.48pre1 -z -D --file=Figures/Superconductivity/phonon1.svg 
    --export-pdf=Figures/Superconductivity/phonon1.pdf --export-latex} \def\svgwidth{7cm}
  \input{Figures/Superconductivity/phonon1Label.tex}
      \label{fig:superconductivity:phonon1}
    }
    \subfigure[]{
      \executeiffilenewer{Figures/Superconductivity/phonon2.svg}
  {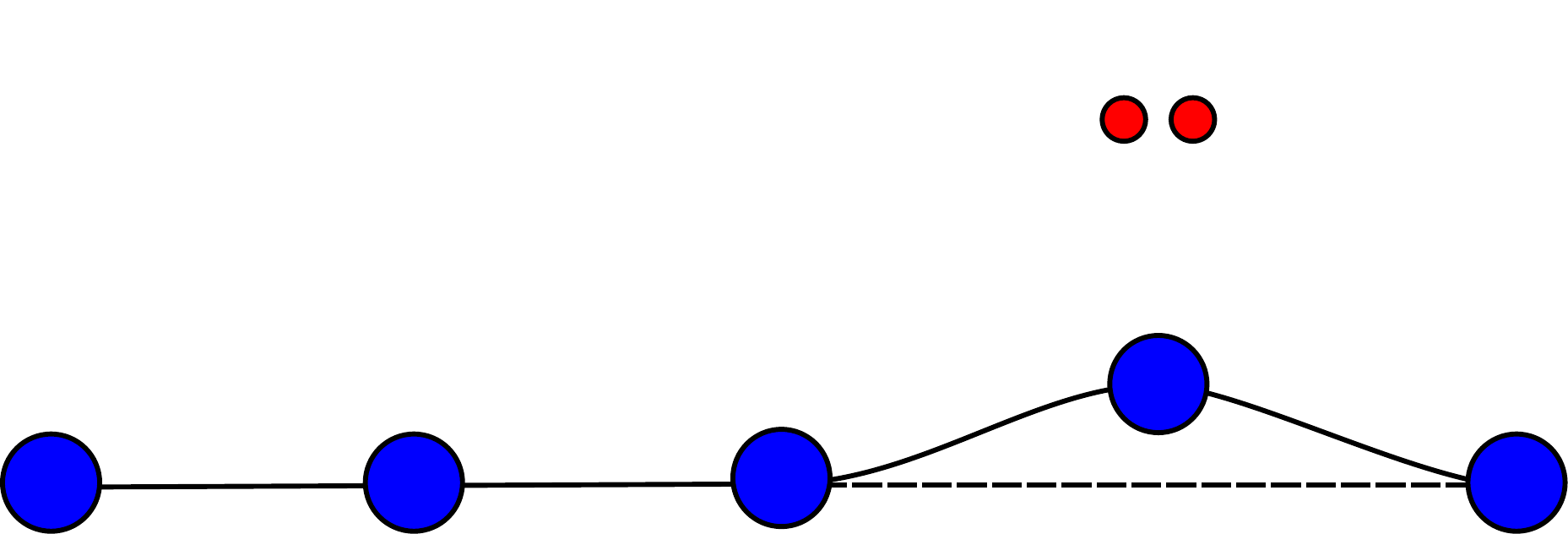}{inkscape-0.48pre1 -z -D --file=Figures/Superconductivity/phonon2.svg 
    --export-pdf=Figures/Superconductivity/phonon2.pdf --export-latex} \def\svgwidth{7cm}
  \input{Figures/Superconductivity/phonon2Label.tex}
      \label{fig:superconductivity:phonon2}
    }
  \end{center}
  \caption{A diagram of the forming of Cooper pairs, necessary for the
    BCS theory. In Figure \ref{fig:superconductivity:phonon1} an electron with spin
    $+\frac{1}{2}$ moves to the right while an electron with spin
    $-\frac{1}{2}$ moves to the left. Each electron is coupled to a
    phonon. In Figure \ref{fig:superconductivity:phonon2}, the electron on the left has
    coupled with the phonon of the electron on the right, and formed a
    Cooper pair with the second electron. }
\end{figure}

When a metal is cooled to a very low temperature, the ion lattice of
the metal becomes more and more like a crystal. As the electrons in
the electron gas move across the lattice, the negative charge of the
electron pulls on the positive charges of the nuclei in the
lattice. This pulls the nuclei towards the electron as the electron
moves by as shown in Figure \ref{fig:superconductivity:phonon1}. As
the electron continues to move, a vibrational wave is formed in the
lattice as more nuclei move towards the electron and the other nuclei
settle back into place. This wave within the lattice is called a
\term{phonon}.\footnote{The reason the name phonon is used is because
  this is a quantized sound wave. The prefix {\it phon} indicates
  sound, such as phonetics, and the prefix {\it phot} indicates light,
  such as Photos. A quantized light wave is called a photon, and so a
  quantized sound wave is called a phonon.}

As a second electron moves across the lattice it becomes attracted to
the phonon trailing the first electron. The second electron is pulled
into the phonon and the two electrons, one with spin $m_s =
+\frac{1}{2}$ and the other with spin $m_s = -\frac{1}{2}$
\term{couple} or join through the shared phonon. This pair of
electrons is called a \term{Cooper pair} and can only form when the
temperature of the metal is less than the binding energy of the Cooper
pair. The spins of particles add together, and so the total spin of
the Cooper pair is zero. This means that the Cooper pair is a boson.

Once a sufficient number of Cooper pairs are formed from the electron
gas, a Bose-Einstein condensate forms from the electrons. This means
that all the Cooper pairs have the lowest possible energy, and so they
no longer are likely to scatter off the ion lattice. As soon as this
occurs, the metal becomes a superconductor, as the Cooper pairs can
now move freely through the ion lattice without scattering.

This theory explains both what the critical temperature of a Type I
superconductor is, and its critical magnetic field. The critical
temperature is reached when the Cooper pairs are no longer broken
apart by the kinetic energy of the nuclei, or when the temperature is
less than the binding energy of the pair. The critical magnetic field
also is reached when the magnetic field no longer breaks apart the
Cooper pairs and allows a Bose-Einstein condensate to form.

\section{Meissner Effect}

While the explanation for the BCS theory given above is fully
qualitative, it is also possible to derive the theory in a more
mathematical fashion. While this will not be done here, as it would
not be helpful to the discussion, there is an important result that
needs to be discussed called the \term{London equations} which are
given in Equation \ref{equ:superconductivity:london}.
\begin{subequations}
  \begin{equation}
    \frac{\partial \vec{I}}{\partial t} =
    \frac{n_se^2}{m_e}\vec{E}
    \label{equ:superconductivity:london1}
  \end{equation}
  \begin{equation}
    \nabla\times\vec{I} = -\frac{n_se^2}{m_e c}\vec{B}
    \label{equ:superconductivity:london2}
  \end{equation}
  \label{equ:superconductivity:london}
\end{subequations}

Here, $\vec{I}$, is the current flowing through the superconductor
(notice that it is a vector quantity and has a direction!), $\vec{E}$
is the electric field in the superconductor, and $\vec{B}$ is the
magnetic field in the superconductor. The letter $e$ is the charge of
the electron, $m_e$ the mass of the electron, and $n_s$ a physical
constant dependent upon the material of the superconductor, and $c$ is
the speed of light.

For those not familiar with the symbol $\nabla$, this is called a
nabla, and when the notation $\nabla\times$ is used in front of a
vector, this means take the \term{curl} of the vector. To find the
direction of the curl of a vector, wrap your right hand around in the
direction the vector is pointing; your thumb then points in the
direction of the curl. This is called the \term{right hand rule}. As an
example, when current is flowing through a wire, there is an
associated magnetic field. The curl of this magnetic field points in
the direction of the current.\footnote{The curl of a vector is also a
  vector and is explicitly calculated by taking the determinant of the
  matrix $ \left[
      \begin{array}{ccc}
        \hat{i} & \hat{j} & \hat{k} \\
        \frac{\partial}{\partial x} & \frac{\partial}{\partial y} &
        \frac{\partial}{\partial z} \\
        v_i & v_j & v_k \\
      \end{array}
    \right]
  $
  where $v_i$ are the components of the vector $\vec{v}$.}

One of Maxwell's equations, specifically Ampere's law, states that the
curl of the magnetic field is equal to the current times the magnetic
constant, $\mu_0$.
\begin{equation}
  \nabla\times\vec{B} = \mu_0\vec{I}
\end{equation}
Substituting in $\vec{I}$ from this equation into Equation
\ref{equ:superconductivity:london2} yields the following differential equation.
\begin{equation}
  \nabla^2\vec{B} = \frac{n_s e^2 \mu_0}{m_e c}\vec{B}
\end{equation}
While many differential equations do not have known solutions, this one
luckily does, as it is just an ordinary differential equation. The
solution is,
\begin{equation}
  B(x) = B_0e^{\frac{-x}{\lambda}}
  \label{equ:superconductivity:penetration}
\end{equation}
where,
\begin{equation}
  \lambda = \sqrt{\frac{m_e c}{n_s e^2 \mu_0}}
\end{equation}
and is called the \term{London penetration depth} and $B_0$ is the
magnitude of the magnetic field at the surface of the
superconductor. The variable $x$ is the distance from the surface of
the superconductor.

The math above can be a little daunting, but it is not necessary to
understand the specifics behind it. The important result to look at is
Equation \ref{equ:superconductivity:penetration}. Interpreting this equation physically,
we see that the magnetic field inside a superconductor decays
exponentially and that past the London penetration depth, the magnetic
field within a superconductor is nearly zero!

This is just the Meissner effect, mentioned at the very beginning of the
chapter. From BCS theory (and a little bit of hand waving around the
math) we have managed to theoretically explain why the Meissner effect
occurs, on a microscopic scale. On a historical note, the London
equations of Equation \ref{equ:superconductivity:london} were developed \term{phenomenologically} before BCS theory. This means that physicists
developed the London equations to model the Meissner effect, which they
did well, but just did not know the physical reason why the equations
were correct.

Now that we have a mathematical understanding of the Meissner effect,
it is time to understand the physical consequences. When a
superconductor is placed in a magnetic field, small \term{eddy
  currents} begin to circulate at the surface of the
superconductor. These currents create magnetic fields that directly
oppose the external magnetic field, canceling it, and keeping the
magnetic field at the center of the superconductor near zero. Because
the superconductor is a perfect conductor, the eddy currents continue
traveling without resistance and can indefinitely oppose the external
magnetic field.

This is why unassisted levitation is possible using superconductors. A
ferromagnet is placed over the superconductor, and through the Meissner
effect the superconductor creates a magnetic field that directly
cancels the magnetic field of the ferromagnet. This in turn causes a
magnetic repulsion that holds the ferromagnet in place against the
force of gravity.

It is important to remember that both the London equations, and BCS
theory are only valid for Type I superconductors. Type II
superconductors allow magnetic fields to pass through \term{filaments}
within the material. Supercurrents surround the filaments in a
\term{vortex state} to produce the \term{mixed-state Meissner effect}
where the external magnetic field is not completely excluded.

\begin{figure}
  \begin{center}
    \subfigure[Perfect Conductor]{
      \executeiffilenewer{Figures/Superconductivity/conductor.svg}
  {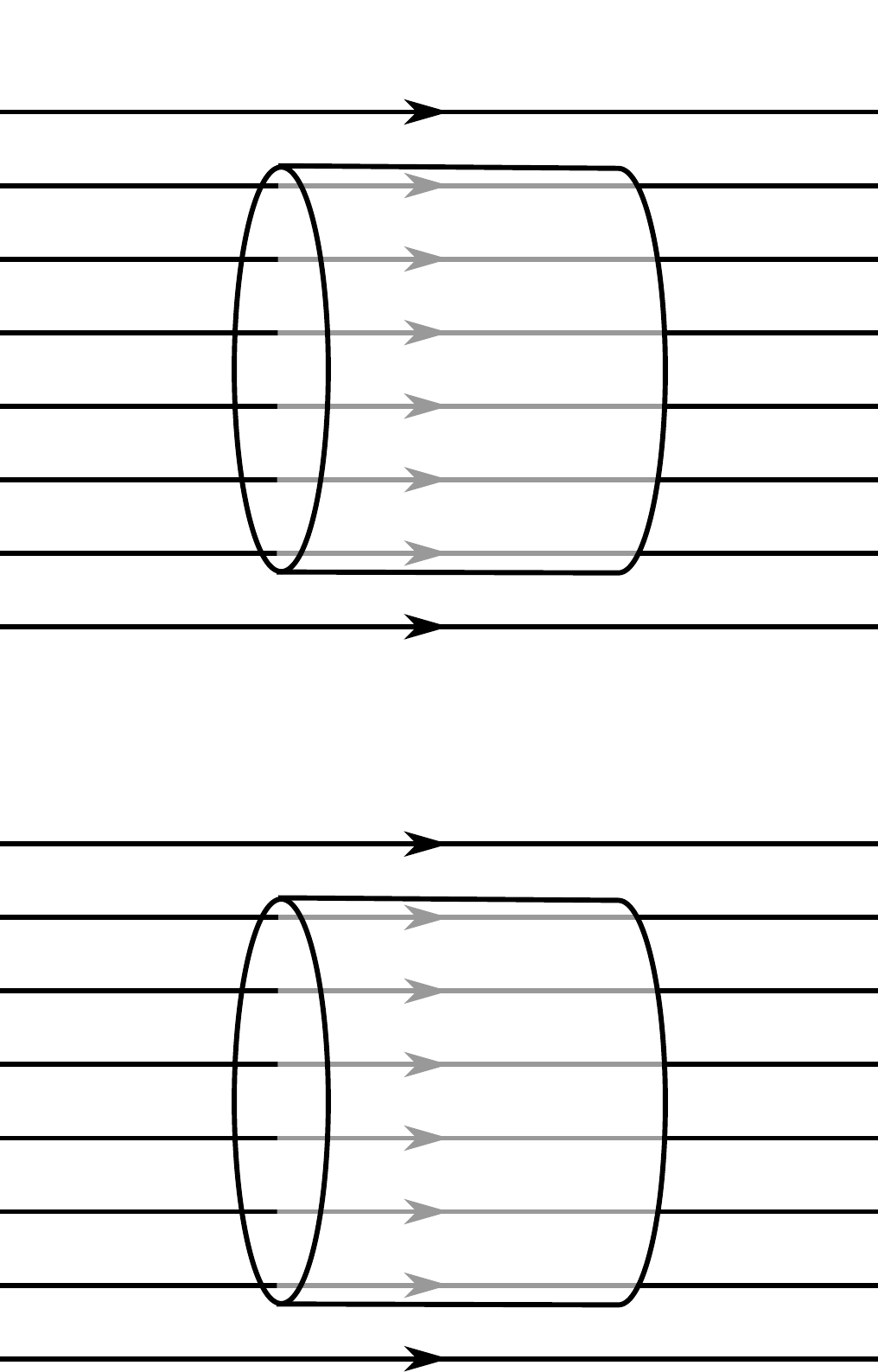}{inkscape-0.48pre1 -z -D --file=Figures/Superconductivity/conductor.svg 
    --export-pdf=Figures/Superconductivity/conductor.pdf --export-latex} \def\svgwidth{4cm}
  \input{Figures/Superconductivity/conductorLabel.tex}
      \label{fig:superconductivity:conductor}
    }
    \subfigure[Type I]{
      \executeiffilenewer{Figures/Superconductivity/type1.svg}
  {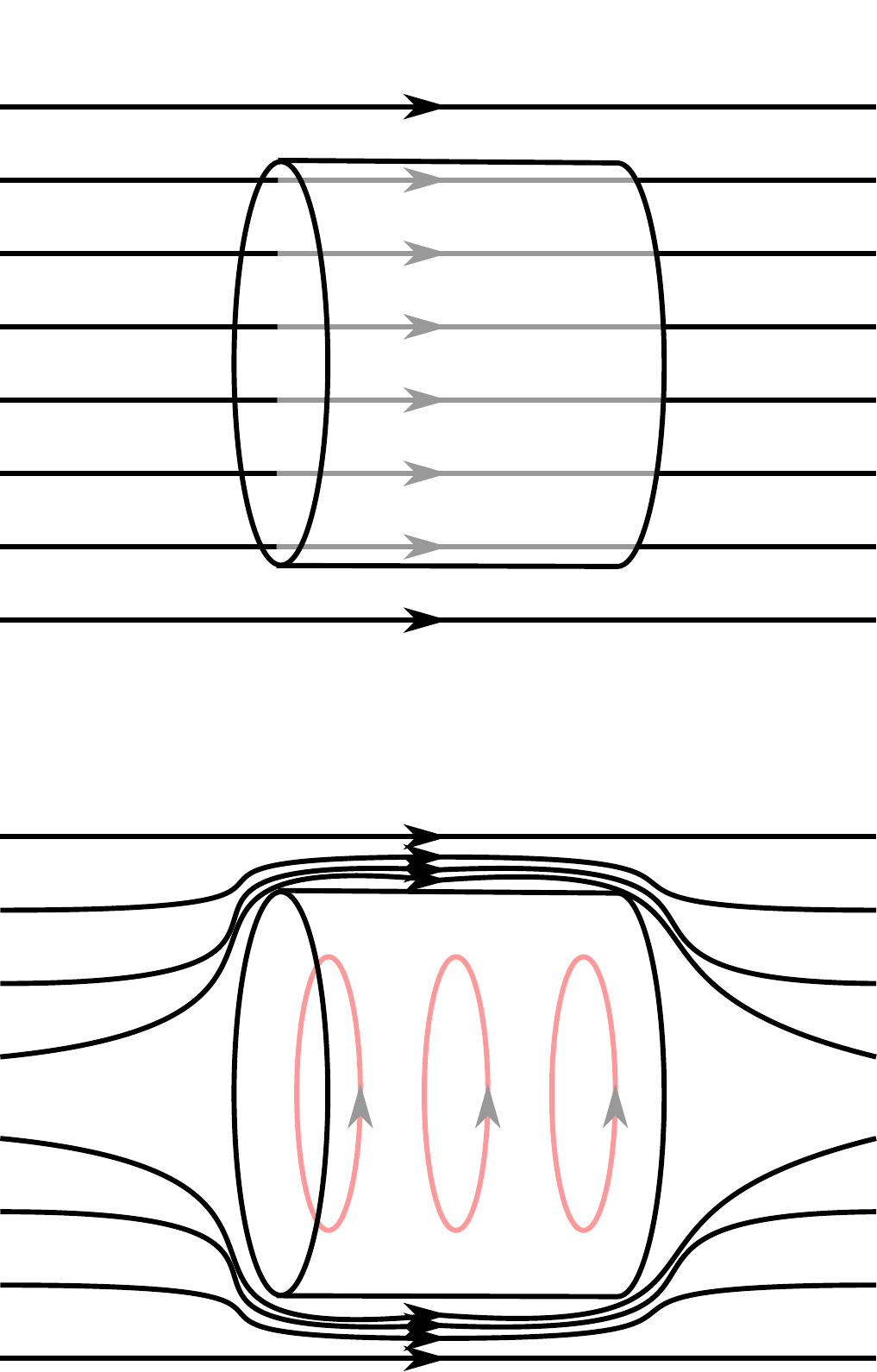}{inkscape-0.48pre1 -z -D --file=Figures/Superconductivity/type1.svg 
    --export-pdf=Figures/Superconductivity/type1.pdf --export-latex} \def\svgwidth{4cm}
  \input{Figures/Superconductivity/type1Label.tex}
      \label{fig:superconductivity:type1}
    }
    \subfigure[Type II]{
      \executeiffilenewer{Figures/Superconductivity/type2.svg}
  {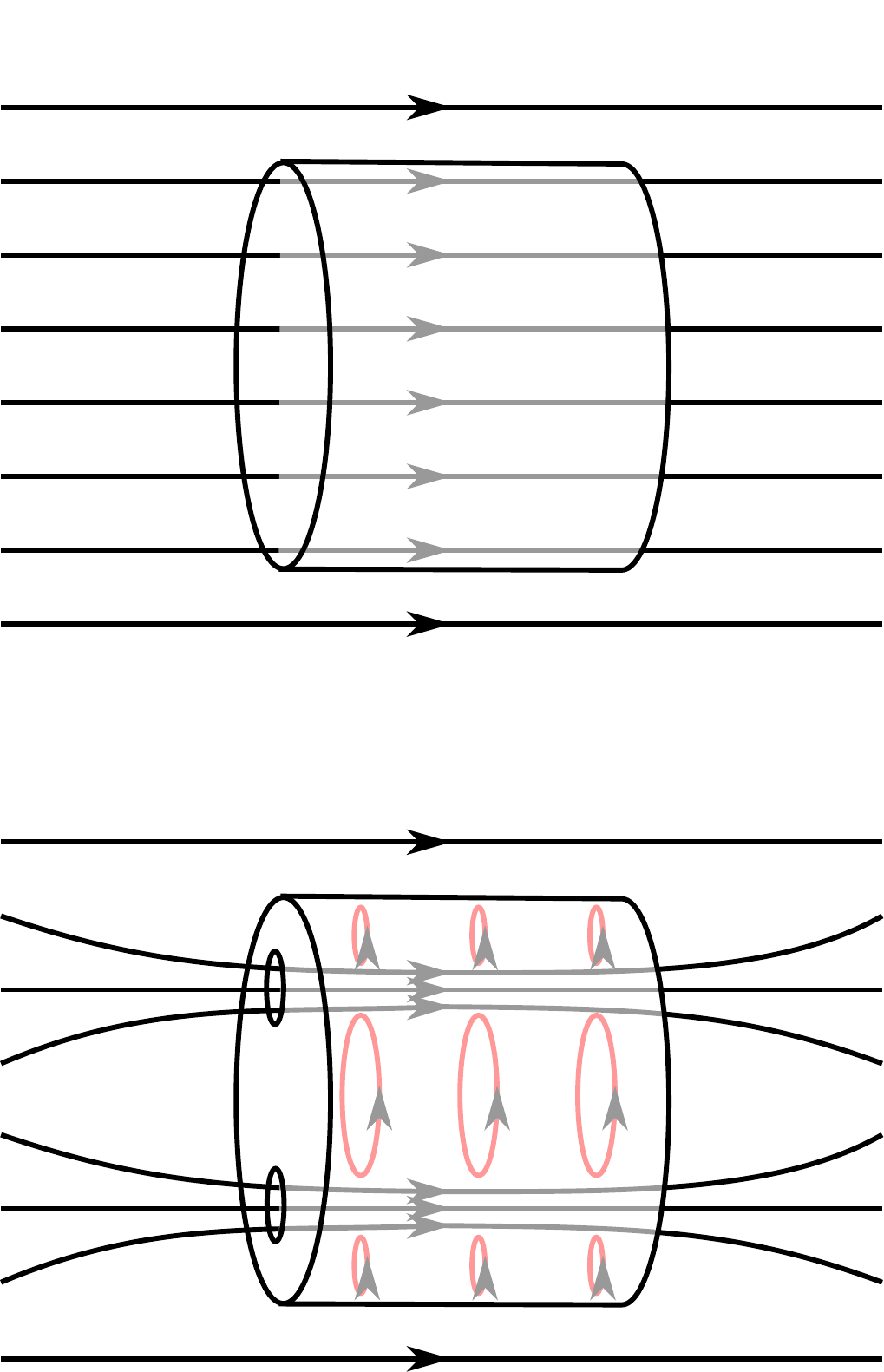}{inkscape-0.48pre1 -z -D --file=Figures/Superconductivity/type2.svg 
    --export-pdf=Figures/Superconductivity/type2.pdf --export-latex} \def\svgwidth{4cm}
  \input{Figures/Superconductivity/type2Label.tex}
      \label{fig:superconductivity:type2}
    }
    \caption{Comparison of the magnetic field lines passing through a
      perfect conductor, Type I superconductor, and Type II
      superconductor above and below the critical temperature. The red
      lines indicate eddy currents which oppose the external magnetic
      field.\label{fig:superconductivity:compare}}
  \end{center}
\end{figure}

Figure \ref{fig:superconductivity:compare} compares what happens if a normally
conducting object is subjected to an external magnetic field and then
transitions to a perfect conductor, Type I superconductor, or Type II
superconductor. In the first scenario, Figure \ref{fig:superconductivity:conductor}, the
magnetic field remains within the perfect conductor. No matter what
external magnetic field is now applied to the perfect conductor, the
internal magnetic field will remain exactly the same. If, for example,
the external magnetic field were shut off, eddy currents would
continue to produce the exact same magnetic field within the center of
the perfect conductor. This phenomena is known as \term{perfect
  diamagnetism}.

In the second scenario, the external magnetic field is completely
excluded from the interior of the Type I superconductor. Any change
within the external magnetic field will trigger eddy currents that
directly oppose the external magnetic field. In the final scenario,
most of the magnetic field is excluded, yet some magnetic field is
still able to pass through the filaments of the Type II
superconductor.

\section{Experiment}

The experiment for this chapter uses the Meissner effect to measure
the critical temperature of a high temperature $\mathrm{YBa_2Cu_3O_7}$
superconductor. The apparatus used consists of a superconductor,
around which is wrapped a solenoid.\footnote{A solenoid is just a
  circular coil of wire. The beauty of solenoids is that they produce
  relatively uniform magnetic fields in their centers.} The experiment
is broken into two steps.

In the first step, the solenoid is cooled to a temperature of $\approx
70$ K using liquid nitrogen. The resistance of the solenoid is
measured using an ohmmeter for different temperatures of the solenoid
(determined by a thermocouple near the solenoid). Because resistance
is caused by electrons scattering off energetic nuclei, we expect the
resistance of the coil to decrease as the temperature decreases. This
portion of the experiment has nothing to do with superconductors.

For the second step we need just a little more theory. When an
external magnetic field is applied to a conductor, it tries to keep
the magnetic field within itself the same, just like with the perfect
conductor. The conductor does this by creating eddy currents which
create magnetic fields that oppose the external magnetic field. Unlike
the case of the perfect conductor, these eddy currents decay over
time due to resistance and eventually the conductor succumbs to the
external magnetic field. This whole process is called \term{inductance}
and is measured by a unit called the \term{henry}. The longer a
conductor fights the external magnetic field, the larger the
inductance.

If we drive an alternating current through the solenoid, we can
measure the inductance using the equation,
\begin{equation}
  \omega L = \sqrt{\left(\frac{V}{I}\right)^2-R^2}
  \label{equ:superconductivity:inductance}
\end{equation}
where $\omega$ is the frequency of the alternating current, $L$ the
inductance of the solenoid, $V$ the root mean square (RMS) voltage in
the solenoid, $I$ the RMS current in the solenoid, and $R$ the
resistance of the solenoid at that temperature. For those
uncomfortable with this equation just being handed down from on high,
try to derive it. The process is not the simplest, but with a little
effort and thinking it can be done. Again, neglecting the derivation,
the inductance for a solenoid is,
\begin{equation}
  L = \frac{\mu N^2 A}{\ell}
  \label{equ:superconductivity:solenoid}
\end{equation}
where $\mu$ is the magnetic permeability of the core of the solenoid,
$N$ the number of turns of wire in the solenoid, $A$ the
cross-sectional area of the solenoid, and $\ell$ the length of the
solenoid.

Looking at Equation \ref{equ:superconductivity:solenoid}, we see that
for a large magnetic permeability, the inductance of the solenoid is
very large, while for a small permeability, the inductance is
small. The core of the solenoid in our experimental setup is just the
superconductor, which when at room temperature, has a relatively
normal value for $\mu$, and so the inductance of the solenoid will be
relatively large. However, below the critical temperature of the
superconductor, the Meissner effect takes hold, and the magnetic field
can no longer pass through the core of the solenoid. Essentially,
$\mu$ has become zero. This means that the inductance of the solenoid
will drop to nearly zero.

Using Equation \ref{equ:superconductivity:inductance}, it is possible to calculate the
inductance for the solenoid using the resistance of the coil,
determined in step one, along with measuring the RMS current and RMS
voltage passing through the coil for various temperatures. From the
explanation above, we expect to see a dramatic jump in inductance at
some point in the graph where the core of the solenoid transitions
from a superconducting state to a standard state. By determining this
jump, we have found the critical temperature of the superconductor!
\addcontentsline{toc}{chapter}{Indices}
\printindex{terms}{Terms}

\end{document}